\documentclass[aps,prx,twocolumn,superscriptaddress,longbibliography]{revtex4-1}

\usepackage[T1]{fontenc}
\usepackage[utf8]{inputenc}
\usepackage{lmodern}
\usepackage[sumlimits]{amsmath} 
\usepackage{amssymb}
\usepackage{yhmath}
\usepackage{graphicx}
\usepackage{epstopdf}
\usepackage{latexsym}
\usepackage{color}
\usepackage{nicefrac}
\usepackage{dsfont} 
\usepackage{wasysym} 
\usepackage{stmaryrd}
\usepackage{hyperref} 
\usepackage{xcolor}
\usepackage{multirow}
\usepackage{soul}
\usepackage{cancel}

\usepackage{esint}
\usepackage{tikz}

\definecolor{airforceblue}{rgb}{0.36, 0.54, 0.66}

\newcommand{\be}{\begin{equation}}
\newcommand{\ee}{\end{equation}}
\newcommand{\bea}{\begin{eqnarray}}
\newcommand{\eea}{\end{eqnarray}}

\usepackage{verbatim}
\usepackage{stmaryrd}
\usepackage[sumlimits]{amsmath} 
\usepackage{dsfont} 
\usepackage{caption}
\usepackage{cancel}
\usepackage{url}

\usepackage{subcaption}

\newcommand{\mb}{\mathbf}
\newcommand{\bs}{\boldsymbol}
\newcommand{\mf}{\mathfrak}
\newcommand{\mc}{\mathcal}
\newcommand{\ms}{\mathsf}
\newcommand{\mt}{\mathtt}
\renewcommand{\ul}{\underline}

\usepackage{scalerel}

\usepackage{accsupp}

\makeatletter
\def\l@subsubsection#1#2{}
\makeatother

\begin{document}

\title{Thermal Conductivity and Theory of Inelastic Scattering of
  Phonons by Collective Fluctuations}

\author{L\'eo Mangeolle}
\affiliation{Universit\'e de Lyon, \'{E}cole Normale Sup\'{e}rieure de
  Lyon, Universit\'e Claude Bernard Lyon I, CNRS, Laboratoire de physique, 46, all\'{e}e
d'Italie, 69007 Lyon}
\author{Leon Balents}
\affiliation{Kavli Institute for Theoretical Physics, University of
  California, Santa Barbara, CA 93106-4030}
\affiliation{Canadian Institute for Advanced Research, Toronto, Ontario, Canada}
\author{Lucile Savary}
\affiliation{Universit\'e de Lyon, \'{E}cole Normale Sup\'{e}rieure de
  Lyon, Universit\'e Claude Bernard Lyon I, CNRS, Laboratoire de physique, 46, all\'{e}e
d'Italie, 69007 Lyon}
\affiliation{Kavli Institute for Theoretical Physics, University of
California, Santa Barbara, CA 93106-4030}

\date{\today}
\begin{abstract}
  We study the intrinsic scattering of phonons by a general quantum degree of freedom,
  i.e.\ a fluctuating ``field'' $Q$, which may have completely general correlations,
  restricted only by unitarity and translational invariance.
  From the induced scattering rates, generalizing the model studied in
  a companion paper, Ref.~\cite{mangeolle2022b}, 
  we obtain the consequences on the thermal conductivity tensor of the phonons.
  We
confirm that, even within our generalized model, the off-diagonal scattering rates
  involve a minimum of three- to four-point correlation functions of
  the $Q$ fields, and discuss the ``semiclassical'' vs ``quantum''
  nature of all contributions.
  We obtain general and explicit forms for these correlations which isolate the contributions to the Hall conductivity,
  and provide a general discussion of the implications of symmetry and
  equilibrium; this elaborates on, and extends, the results of
  Ref.~\cite{mangeolle2022b}. We also extend the discussion and
  evaluation of these two- (diagonal
  scattering) and four-point correlation functions, and hence the
  thermal transport, for the illustrative example of an ordered two
  dimensional antiferromagnet, where the $Q$ field is a composite of
  magnon operators arising from spin-lattice coupling, and confirm numerically that the results, while satisfying all the necessary
  symmetry restrictions, lead to non-vanishing scattering and Hall effects. In particular,
  we investigate, both analytically and numerically, the dependence of such intrinsic scattering
  on a crucial parameter -- the magnon to phonon velocity ratio $\upsilon$. We
  in particular confirm that within some range of $\upsilon$ of order 1
  the skew-scattering mechanism leads to comparable thermal Hall conductivity for thermal currents
  within and normal to the plane of the antiferromagnetism, and discover that the
  temperature scaling
  of the longitudinal conductivity displays a threshold effect and a
non-universal, \emph{continuous} variation of the scaling exponent
with $\upsilon$.
\end{abstract}

\maketitle

\tableofcontents

\section{Introduction}
\label{sec:introduction}

Two-point correlation functions are ubiquitous
in the study of condensed matter systems. They are often the building blocks of
response functions in scattering and other experiments and appear in
Feynman diagrams, as well as Monte Carlo simulations. They are the central elements of linear response
theory, as is evident from Kubo's formula \cite{luttinger1964,chen2020scaling}. They are often independent
of the arbitrary phase choice of the wave function.

Higher-order correlation functions have witnessed renewed interest
recently.  They arise theoretically in the measurement of
chaos.   A particular type of four-point correlation function, the
``out-of-time-ordered''  correlator, has been
shown to be related to the Lyapunov exponent, which measures the
rate at which the result of a measurement diverges after a weak initial
perturbation \cite{swingle}.   Multi-point correlations also naturally describe non-linear response, e.g.\ in non-linear optics such as
second harmonic generation, and in ``multi-dimensional
spectroscopy'' \cite{PhysRevLett.122.257401}.  They may
also arise in scattering
measurements at resonance, such as RIXS
\cite{ament2011,savary2015probing}.   From a statistical point of view, higher
order correlation functions measure the non-Gaussianity of the
distribution of an observable.  The more strongly correlated a state
is, i.e.\ the more it deviates from a free-particle description, the
more significant the non-Gaussianity.  Hence multi-point functions are
essential harbingers of strong correlations.  

In a companion paper \cite{mangeolle2022b}, we present the study of
the thermal conductivity due to phonons {\em linearly} coupled to
another degree of freedom, for example an electronic or a magnetic
one.  We summarize the results of that
paper in this paragraph.  First, it is demonstrated that this coupling
induces two types of scattering of phonons: those which are symmetric
in the sense of respecting detailed balance, and those which are
antisymmetric and obey an ``anti-detailed balance'' relation.  Only
the latter ``skew scattering'' events contribute to a thermal Hall
effect of phonons, as proven by formulating and solving the associated
Boltzmann transport equations.  Finally, the results are applied to an
example calculation of the diagonal and Hall components of the thermal
conductivity for the case of a two-dimensional antiferromagnet.

The purpose of the present paper is to extend the problem of
Ref.~\cite{mangeolle2022b} to the most general case, and to give full
detail of the corresponding scattering contributions and their
derivation. We allow the phonons to be both linearly and quadratically
coupled to the fluctuating degree of freedom, 
i.e.\ with an interaction Hamiltonian density (c.f.\ Eq.~\eqref{eq:main6})
\begin{align}
  \label{eq:10intro}
  H' &= \sum_{n\mathbf
         k}\sum_{q=\pm}a_{n\mathbf{k}}^q Q^q_{n\mathbf{k}} , \nonumber
  \\
  & +  \frac 1 {\sqrt{N_{\rm uc}}} \sum_{n\mathbf k\neq n'\mathbf
      k'}\sum_{q,q'=\pm} a_{n\mathbf{k}}^q a_{n'\mathbf{k}'}^{q'}
      Q^{qq'}_{n\mathbf{k}n'\mathbf k'} ,
\end{align}
where $a_{n\mathbf{k}}^- \equiv a_{n\mathbf{k}}^{\vphantom\dagger},
a_{n\mathbf{k}}^+ \equiv a_{n\mathbf{k}}^{\dagger}$ are the phonon
annihilation and creation operators for the $n^{\rm th}$ phonon mode
with momentum $\mathbf{k}$, and $Q^q_{n\mathbf{k}}$,
$Q^{qq'}_{n\mathbf{k}n'\mathbf k'}$ are the collective fluctuating
``fields'' coupled to the phonons (we discuss even more
general forms in the Appendices).  The quantity $N_{\rm uc}$ is the
number of unit cells in the sample.  We provide a full discussion of
all the different scattering contributions generated by these terms
(up to quartic order in the phonon coupling, see Sec.~\ref{sec:coll-integr-as}), and give a
full exposition of the expressions of the corresponding rates in terms
of correlation functions.  We also give a thorough discussion of the
consequences of symmetries and detailed-balance relations on the Hall
conductivity, and show in particular that the conclusion that two-point
correlations functions do not contribute to a Hall effect, arrived at in
Ref.~\cite{mangeolle2022b} for the linear coupling model, continues to
hold in full generality.  

Up to fourth order in $\lambda$, where $\lambda$ captures the size of
the coupling between one phonon and one $Q$ operator, and terms
involving $p$ phonon operators are assumed to be of order $\lambda^p$,
the longitudinal scattering rate is
\begin{equation}
  \label{eq:111}
  D_{n\mathbf{k}}=D_{n\mathbf{k}}^{(1)}+D_{n\mathbf{k}}^{(2)}+\breve{D}_{n\mathbf{k}}
\end{equation}
where $D^{(1)}$ and $D^{(2)}$ are obtained in our perturbative
expansion at orders $\lambda^2$ and $\lambda^4$, respectively, and
$\breve{D}_{n\mathbf{k}}$ encompasses contributions due to other
scattering processes as well as higher-order terms of the expansion.

The skew scattering rate can be similarly expanded in terms at
different orders in $\lambda$. At fourth order, we find the full set
of scattering rates generalizing the results of
Ref.~\cite{mangeolle2022b} to include the two-phonon couplings and
quantum interference terms:
\begin{eqnarray}
  \label{eq:112}
\mathfrak W^{\ominus,[1,1];[1,1], qq'}_{n \mathbf
  k n'\mathbf k'}  & \sim & 
                            \left\langle [ Q^{-q}_{n\mathbf k} ,Q^{-q'}_{n'\mathbf k'}
                            ]  \{ Q_{n'\mathbf k'}^{q'},Q_{n\mathbf k}^{q}
                            \}\right\rangle \, ,                           
                            \nonumber\\
  \mathfrak W^{\oplus,[1,1];[1,1], qq'}_{n \mathbf
  k n'\mathbf k'}   &\sim& \left\langle \{ Q_{n\mathbf k}^{-q} ,Q^{-q'}_{n'\mathbf k'}
                            \}  \{ Q_{n'\mathbf k'}^{q'},Q_{n\mathbf k}^{q}
         \}\right\rangle 
                            \nonumber\\
  &&\quad- \left\langle [ Q^{-q}_{n\mathbf k} ,Q^{-q'}_{n'\mathbf k'}
                            ]  [ Q_{n'\mathbf k'}^{q'},Q_{n\mathbf k}^{q} ]\right\rangle \, ,
                            \nonumber\\
\mathfrak W^{\oplus,[2];[2], qq'}_{n \mathbf
  k n'\mathbf k'}   &\sim&\left\langle Q_{n\mathbf k n'\mathbf
                           k'}^{-q-q' }
                           Q_{n\mathbf k n'\mathbf
                           k'}^{q'q}\right\rangle \, ,
                           \nonumber \\
   \mathfrak W^{\oplus,[1,1];[2], qq'}_{n \mathbf
  k n'\mathbf k'}   &\sim&\left\langle Q_{n\mathbf k n'\mathbf k'}^{-q,-q'}[
                           Q_{n'\mathbf k'}^{q'}, Q_{n\mathbf k}^{q}]\right\rangle \, ,
                           \nonumber\\
 \mathfrak W^{\ominus,[1,1];[2], qq'}_{n \mathbf
  k n'\mathbf k'}   &\sim&\left\langle Q_{n\mathbf k n'\mathbf k'}^{-q,-q'}\{
                           Q_{n'\mathbf k'}^{q'}, Q_{n\mathbf k}^{q}\}\right\rangle \, ,
                           \nonumber\\
 \mathfrak W^{\ominus,[2,1];[1], qq'}_{n \mathbf
  k n'\mathbf k'}   &\sim&\left\langle Q_{n\mathbf k }^{-q}\{
                           Q_{n'\mathbf k'}^{-q'}, Q_{n\mathbf k n'\mathbf k'}^{qq'}\}\right\rangle \, ,
                           \nonumber\\
\mathfrak W^{\ominus,[1,1,1];[1], qq'}_{n \mathbf
  k n'\mathbf k'}   &\sim&\left\langle Q_{n\mathbf k }^{-q}\left \lgroup
                           Q_{n'\mathbf k'}^{-q'}, Q_{n\mathbf k }^{q} ,Q_{n'\mathbf k'}^{q'}\right\rgroup\right\rangle.
\end{eqnarray}
The full scattering rate is the total of all the contributions, summed
over $\sigma=\pm 1 = \oplus,\ominus$, $q,q'=\pm 1$, and the
$[\cdot],[\cdot']$ indices which denote the ``internal'' states of the
scattering process and will be explained in
Sec.~\ref{sec:t-matrix-elements}.  The $\left \lgroup
  \cdot,\cdot',\cdot''\right \rgroup$
notation in the last term is explained in
Appendix~\ref{sec:computation-at-third}. The first term, $\mathfrak W^{\ominus,[1,1];[1,1], qq'}_{n \mathbf
  k n'\mathbf k'}$ is that discussed in Ref.~\cite{mangeolle2022b}.

Most importantly, we have separated the processes into those which
satisfy detailed ($\sigma=1$) and ``anti-detailed'' ($\sigma=-1$)
balance relations, 
\begin{equation}
  \label{eq:114}
\mathfrak W_{n\mathbf k n'\mathbf k'}^{\sigma,qq'} = \sigma~
e^{-\beta(q\omega_{n\mathbf k}+q'\omega_{n'\mathbf k'})}~\mathfrak
W_{n\mathbf k n'\mathbf k'}^{\sigma,-q-q'},\quad \sigma=\pm\,{\rm or}\,\oplus,\ominus.
\end{equation}
When these rates are used as input to the Boltzmann equation, we
observe that {\em only the anti-detailed balance terms can generate a
  thermal Hall effect.}  A discussion of these (anti-)detailed balance
relations can be found in Secs.~\ref{sec:coll-matr-elem},\ref{sec:relat-deta-balance}.

\begin{figure}[htbp]
  \centering
\includegraphics[width=\columnwidth]{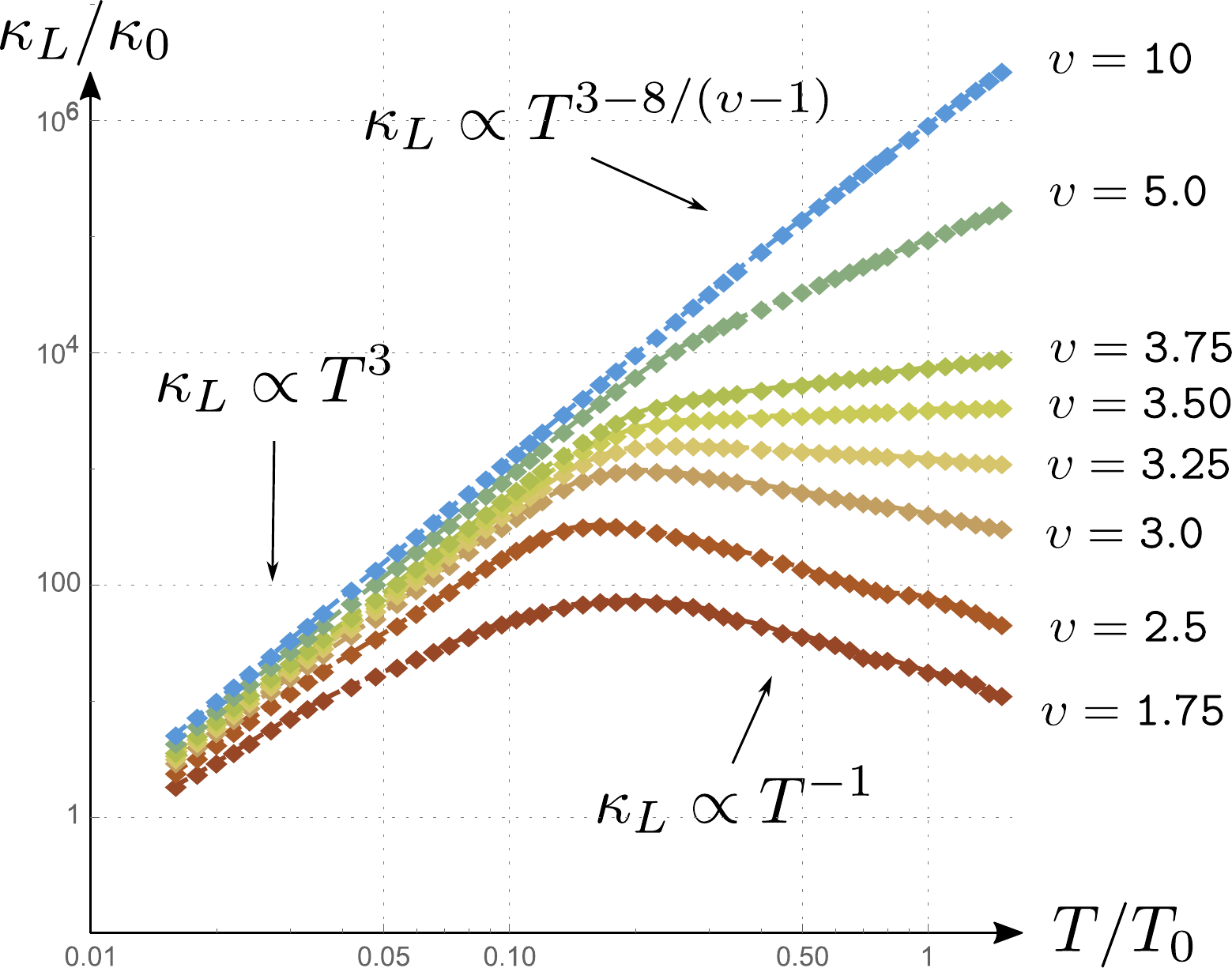}
  \caption{Dependence upon $\upsilon=v_{\rm m}/v_{\rm ph}$  of the longitudinal
    conductivity as a function of
    temperature, in log-log scale, for $\breve D_{n\mathbf k}=\gamma_{\rm ext}=10^{-6}(v_{\rm ph}/\mathfrak{a})$. We highlight the change of scaling behaviors for
    $T>T_\lambda^\star$ (defined in Eq.~\eqref{eq:68}) at $v_{\rm m}/v_{\rm ph}=3$, above which the
    temperature scaling exponent of $\kappa_{\rm L}$ is a continuous
    function of $v_{\rm m}/v_{\rm ph}$, see Eq.~\eqref{eq:104} or Eq.~\eqref{eq:86}. }
      \label{fig:kappaLvmvph}
\end{figure}

  After the derivation of these relations and their relation to the
  thermal conductivity tensor, and a discussion of the consequences of
  symmetry for the latter, we turn to the specific problem of the
  antiferromagnet introduced in Ref.~\cite{mangeolle2022b}.  Notably, we
  significantly extend the treatment there to unveil the dependence of
  the thermal conductivity upon the ratio
  $\upsilon=v_{\rm m}/v_{\rm ph}$ of the magnon and phonon velocities,
  which is striking and non-trivial. In particular $\kappa_{\rm L}$
  has the scaling behavior
  \begin{equation}
    \label{eq:104}
\kappa_{\rm L} \sim
  \begin{cases}
    T^{-1}&\mbox{for }\upsilon<3\\
    T^{3-8(\upsilon-1)^{-1}}&\mbox{for }\upsilon>3
  \end{cases},
\end{equation} exhibiting a threshold effect and a
non-universal, \emph{continuous} variation of the scaling exponent
with $\upsilon$, as shown in Fig.~\ref{fig:kappaLvmvph}.
Furthermore, we find that the Hall resistivity dramatically decreases
as $\upsilon$ increases, due to the
reduction of the allowed phase space for scattering.

The remainder of the paper is organized as follows.
We first (Secs.~\ref{sec:setup},\ref{sec:form-expr-therm})
  provide an expanded derivation of the skew and longitudinal
  scattering rates, 
  including those terms which result from
``higher-order'' $Q$-phonon interactions and are not present in
Ref.~\cite{mangeolle2022b}. We then (Sec.~\ref{sec:disc-symm}) provide a detailed
discussion of the consequences of symmetries and ``detailed
balance''-like relations on the Hall conductivity. The final section (Sec.~\ref{sec:application}) is an
application of the results to an ordered antiferromagnet, as in
Ref.~\cite{mangeolle2022b}, which we expand on in considerably more
detail, both regarding the longitudinal and skew scattering rates. We obtain
analytical results for the longitudinal conductivity $\kappa_L$ and
Hall resistivity $\varrho_H$ in terms of multidimensional integrals,
whose scaling we analyze and verify through numerical
evaluation. Seven appendices provide all details and further
generalizations not given in the main text. 



\section{Setup}
\label{sec:setup}

\subsection{Derivation}
\label{sec:derivation}

The quasiparticle nature of phonons justifies
treating their dynamics within the Boltzmann equation,
\begin{equation}
\label{eq:main1} 
\partial_t\overline{N}_{n\mathbf{k}}+\bs{v}_{n\mathbf{k}}\cdot{\boldsymbol{\nabla}}_\mathbf{r}\overline{N}_{n\mathbf{k}}
= \mathcal C_{n\mathbf{k}}[\{\overline{N}_{n'\mathbf{k}'}\}],
\end{equation}
where $N_{n\mathbf k}({i_p})=\langle i_p|a^\dagger_{n\mathbf
  k}a^{\vphantom{\dagger}}_{n\mathbf k}| i_{ p}\rangle$ is the number of $(n,\mathbf k)$
phonons ($\mathbf{k}$ is the phonon momentum and $n$ an extra phonon
label, containing the band index and polarization) in the $| i_p\rangle$ state, $\overline{N}_{n\mathbf k} =
\sum_{ i_p} N_{n\mathbf k}(i_p)p_{i_p}$ is the average population, and
$\bs{v}_{n\mathbf{k}}={\boldsymbol{\nabla}}_\mathbf{k}\omega_{n\mathbf{k}}$,
with $\omega_{n\mathbf{k}}$ the dispersion of phonons,
is the group velocity of phonons. $\mathcal
C$ is the ``collision integral,'' which captures in particular the
scattering of phonons with other degrees of freedom ($Q$ fields whose coupling to
the phonons is given by $H'$ in Eq.~\eqref{eq:10intro}).  In turn, using Born's approximation, 
we have the following perturbative expansion of the scattering matrix:
\begin{equation}
\label{eq:main7}
  T_{\mathtt {i\rightarrow f}}=T_{\mathtt{fi}}=\langle \mathtt{f}|H'|\mathtt{i}\rangle 
  +\sum_\mathtt{n}\frac{\langle\mathtt{f}|H'|\mathtt{n}\rangle\langle\mathtt{n}|H'|\mathtt{i}\rangle}{E_\mathtt{i}-E_\mathtt{n}+i\eta}+\cdots,
\end{equation}
where the $|\mathtt{i}, \mathtt{f}, \mathtt{n}\rangle$ states are
product states in the $Q$ (index $s$) and phonon (index $p$) Hilbert space,
$|\mathtt{g}\rangle=|g_s\rangle|g_p\rangle$ for
$g=i,f,n$, and $E_{\mathtt{g}}$ is the energy of the unperturbed
Hamiltonians of the $Q$ and phonons in state ${\mathtt{g}}$. $\eta\rightarrow0^+$
is a small regularization parameter. The expression
Eq.~\eqref{eq:main7} can be derived from time-dependent perturbation (scattering)
theory, in which $\eta$ captures causality and the regularizability of
$1/(E_{\mathtt{i}}-E_{\mathtt{n}})$ in the case of a continuous energy
spectrum, appropriate for scattering (unbounded) states which we are
interested in \cite{landau2013quantum}. 

The rate of transitions
from state $\mathtt{i}$ to state $\mathtt{f}$ is obtained using Fermi's golden rule,
\begin{equation}
  \label{eq:main8}
  \Gamma_{\mathtt {i\rightarrow f}} =\frac{2\pi }\hbar
\,|T_{\mathtt{i\rightarrow f}}|^2\delta(E_{\mathtt{ i}}- E_{\mathtt  {f}}).
\end{equation}
Note that $\Gamma_{\mathtt {i\rightarrow f}}$ is a transition rate in
the full combined phonon-$Q$ system.  This in turn determines the
collision integral through the master equation
\begin{equation}
  \label{eq:1}
  \mathcal{C}_{n\mathbf{k}}=\sum_{i_p,f_p}\tilde\Gamma_{i_p\rightarrow f_p}\left(N_{n\mathbf{k}}(f_p)-N_{n\mathbf{k}}(i_p)\right)p_{i_p},
\end{equation}
where $p_{i_p}=\sum_{i_s}p_{\mathtt{i}}$, where $p_{\mathtt{i}}=\frac 1 Z e^{-\beta E_\mathtt{i}}$ is
the probability to find the system in state $\mathtt{i}$, and
$Z$ is the partition function of the two subsystems.  Here
\begin{equation}
  \label{eq:81}
  \tilde\Gamma_{i_p\rightarrow f_p}=\sum_{i_sf_s}\;\Gamma_{\mathtt{i}\rightarrow\mathtt{f}}\,p_{i_s}
\end{equation}
is the transition rate between just phonon states, with
$p_{i_s}=\frac 1 {Z_s} e^{-\beta E_{i_s}}$.

\subsection{Discussion}
\label{sec:comments}

The above approach is ``semiclassical'' in two respects.  First, it
ultimately treats phonons as quasiparticles within a Boltzmann
equation.  This is justified whenever the scattering rate is small
compared to the energy of the particles. 
Second, we use the Fermi's
golden rule relation, Eq.~\eqref{eq:main8}, to determine the scattering
rates.  This approximation leads to slight differences from an exact
calculation of the quantum rates, but preserves all symmetries and
physical processes, and we expect it to capture all the key features
of a fully quantum approach.  We proceed with the T-matrix approach here which has the
advantage of (relative) physical transparency, as every effect can be
directly identified with a scattering process.

One can understand the need for effects beyond the first
Born approximation entirely through the symmetries of the T-matrix.
Specifically, since the time reversal (TR) operator is anti-unitary,
and requires complex conjugation, one can see from
Eq.~\eqref{eq:main7} that under time reversal, ${\rm TR}:T\mapsto T^\dagger$
($\eta \rightarrow -\eta$ under complex conjugation).  Since TR
invariance is sufficient to enforce a vanishing Hall effect, the hermiticity
of $T$ is enough to guarantee a vanishing Hall effect.   From
Eq.~\eqref{eq:main7}, $T$ is indeed always hermitian within the first
Born approximation, because $H'$ itself must be hermitian.  

Finally, we note that we are focusing on collisional effects, i.e.\ on
{\em real} transitions induced by interactions, rather than Berry phase
contributions, which arise from entirely virtual transitions and
manifest as modifications to the semiclassical equations of motion for
phonons, e.g.\ an anomalous velocity.  Formally, real transitions are
captured within the collision integral on the right hand side of the
Boltzmann equation \cite{mori}, while Berry phase contributions enter
the left hand side and in the definition of the currents.  For
phonons, our focus on collisions is justified by strong phase space
constraints on the Berry curvature effects which are typical to
acoustic bosonic modes.  Specifically, as shown in
Ref.~\cite{qin2012berry}, the Berry phase contributions are described
by an emergent vector potential which at small momenta must by
symmetry be at least second order in gradients, making it a formally
``irrelevant'' perturbation to the phonon Lagrangian, and strongly
suppressing its effects at low temperature \cite{ye2021phonon}.

\section{Formal expressions for the thermal conductivity}
\label{sec:form-expr-therm}

\subsection{Formal expressions}
\label{sec:formal-expressions}

To solve Eq.~\eqref{eq:main1}, we expand $\overline{N}_{n\mathbf{k}} = N^{\rm
  eq}_{n\mathbf{k}} + \delta\overline{N}_{n\mathbf{k}}$ around the
equilibrium distribution $N^{\rm
  eq}_{n\mathbf{k}}$, which solves Boltzmann's equation at $\boldsymbol{\nabla}T=\mathbf{0}$, 
keep terms up to linear order in $\delta\overline{N}_{n\mathbf{k}}$ in
the collision integral and for convenience separate the diagonal $D_{n\mathbf{k}}$ and
off-diagonal $M_{n\mathbf{k},n'\mathbf{k}'}$
parts, i.e.\ we write the collision integral
\begin{equation}
\label{eq:main2}
\mathcal C_{n\mathbf{k}} = \sum_{n'\mathbf k'}\big(
                             -\delta_{nn'}\delta_{\mathbf k\mathbf k'} D_{n\mathbf k} +
  M_{n\mathbf k, n'\mathbf k'} \big)\delta\overline{N}_{n'\mathbf{k}'}+O(\delta\overline{N}^2), 
\end{equation}
where by definition $M_{n\mathbf k,n\mathbf k}=0$. The equation $\mathcal C_{n\mathbf{k}}[\{N^{\rm
  eq}_{n'\mathbf{k}'}\}]=0$ ---i.e.\ the collision integral is zero in
equilibrium---should be considered the {\em definition} of the
equilibrium densities $\{N^{\rm
  eq}_{n'\mathbf{k}'}\}$ of the interacting phonons (see Appendix~\ref{sec:energy-shift-phonons}).

Using Fourier's law,
\begin{equation}
  \label{eq:70}
  \mathbf{j}=-\boldsymbol{\kappa}\cdot\boldsymbol{\nabla}T=V^{-1}\sum_{n\mathbf{k}}\overline{N}_{n\mathbf{k}}\boldsymbol{v}_{n\mathbf{k}}\omega_{n\mathbf{k}},
\end{equation}
and formally inverting the collision integral leads to the following
expressions for the longitudinal $\kappa_L^{\mu\mu}$, and Hall
(antisymmetric) $\kappa_H^{\mu\nu}=(\kappa^{\mu\nu}-\kappa^{\nu\mu})/2$ conductivities (along the $\mu$ direction and in the $\mu\nu$ plane, respectively):
\begin{equation}
  \label{eq:2}
  \kappa_{L/H}^{\mu\nu}=\frac{\hbar^2}{k_B T^2} \frac {1}{V} \sum_{n\mathbf{k}n'\mathbf{k}'}J^\mu_{n\mathbf{k}} K^{L/H}_{n\mathbf{k}n'\mathbf{k}'} J^\nu_{n'\mathbf{k}'},
\end{equation}
where $\nu=\mu$ for $\kappa_L$. Assuming $\sum_{n'\mb k'} M_{n\mb k
  n'\mb k'}\ll D_{n\mb k}$, one can effectively invert the
  collision integral to obtain the kernels
\begin{widetext}
  \begin{eqnarray}
    \label{eq:main17}
    K^L_{n\mathbf{k}n'\mathbf{k}'}
    &=& \frac{e^{\beta\hbar\omega_{n\mathbf{k}}}}{D_{n\mathbf{k}}}
       \delta_{nn'}\delta_{\mathbf  k,\mathbf k'} + 
        \frac{e^{\beta\hbar(\omega_{n\mathbf{k}}+\omega_{n'\mathbf{k}'})/2}}
        {2 D_{n\mathbf{k}}D_{n'\mathbf{k}'}}
\left(
\frac{\sinh (\beta\hbar\omega_{n\mathbf{k}}/2)}{\sinh(\beta\hbar\omega_{n'\mathbf{k}'}/2)} M_{n\mathbf{k},n'\mathbf{k}'} + (n\mathbf k \leftrightarrow n'\mathbf k')
        \right), \\
    \label{eq:18}
  K^H_{n\mathbf{k}n'\mathbf{k}'}& =&\frac{e^{\beta\hbar(\omega_{n\mathbf{k}}+\omega_{n'\mathbf{k}'})/2}}{2 D_{n\mathbf{k}}D_{n'\mathbf{k}'}}
\left(
\frac{\sinh (\beta\hbar\omega_{n\mathbf{k}}/2)}{\sinh(\beta\hbar\omega_{n'\mathbf{k}'}/2)}   M_{n\mathbf{k},n'\mathbf{k}'} - (n\mathbf k \leftrightarrow n'\mathbf k')
         \right).
\end{eqnarray}
\end{widetext}
Here we identified the equilibrium phonon current
$J^\mu_{n\mathbf{k}}=N_{n\mathbf{k}}^{\rm eq}\omega_{n\mathbf{k}}v^\mu_{n\mathbf{k}}$, 
and made the ``standard'' approximation
$\boldsymbol{\nabla}_\mathbf{r}\overline{N}_{n\mathbf{k}}\approx\boldsymbol{\nabla}_\mathbf{r}N^{\rm
  eq}_{n\mathbf{k}}$, and looked for a
stationary solution ($\partial_t\overline{N}=0$) to Boltzmann's
equation. While the sign of $\kappa_H$ depends on the details of the
system (see later), the second law of thermodynamics imposes
$\kappa_L>0$. 
Considering
Eq.~\eqref{eq:main17}, we therefore expect $D_{n\mathbf k}>0$. 

Clearly, only contributions to $K^{L/H}_{n\mathbf{k},n'\mathbf{k}'}$
which are symmetric (resp.\ antisymmetric) in exchanging
$(n\mathbf{k}\leftrightarrow n'\mathbf{k}')$ contribute to $\kappa_L$
(resp.\ $\kappa_H$).  The special case of the term diagonal in
$n\mathbf{k},n'\mathbf{k}'$, being symmetric, does not contribute to
the Hall conductivity.  Below we will isolate the correlation functions
of the $Q$ operators which give anti-symmetric (in
$n\mathbf{k}\leftrightarrow n'\mathbf{k}'$) contributions to
$\frac{\sinh
  (\beta\hbar\omega_{n\mathbf{k}}/2)}{\sinh(\beta\hbar\omega_{n'\mathbf{k}'}/2)}
M_{n\mathbf{k},n'\mathbf{k}'}$, and hence contribute to $\kappa_H$.
These correspond to scattering processes which violate detailed
balance.

\subsection{Model}
\label{sec:model}

To describe
the interaction between the phonons and another degree of freedom, we
introduce general coupling terms between phonon annihilation
(creation) 
operators $a^{(\dagger)}_{n\mathbf k}$ and general, for now
unspecified, fields $Q^{\{q_j\}}_{\{n_j,\mathbf k_j\}}$ which
are operators acting in their own Hilbert
space.    In what follows we only
consider the first two terms of the expansion with respect to phonon
operators (see also Eq.~\eqref{eq:10intro}), i.e.\ we write the interaction hamiltonian as
$H'=H'_{[1]}+H'_{[2]}$, where
\begin{eqnarray}
\label{eq:main6}
  H'_{[1]} &=& \sum_{n\mathbf
         k}\sum_{q=\pm}a_{n\mathbf{k}}^q Q^q_{n\mathbf{k}} , \\
  H'_{[2]} &=&  \frac 1 {\sqrt{N_{\rm uc}}} \sum_{n\mathbf k\neq n'\mathbf
      k'}\sum_{q,q'=\pm} a_{n\mathbf{k}}^q a_{n'\mathbf{k}'}^{q'}
      Q^{qq'}_{n\mathbf{k}n'\mathbf k'}\nonumber ,
\end{eqnarray}
and in the following, we consider Eq.~\eqref{eq:main6} as a perturbative
expansion with respect to a small parameter ${\lambda}$, such that
formally $Q_{n\mb k}\sim {\lambda}, Q^{qq'}_{n\mathbf{k}n'\mathbf
  k'}\sim {\lambda}^2$, etc. Note we consider generalizations of this
model in Appendix~\ref{sec:generalizations}.

In the above expression we used $a_{n\mathbf k}^+ \equiv a_{n\mathbf  k}^\dagger$ and $a_{n\mathbf k}^-\equiv a_{n\mathbf k}$. The hermiticity
of $H'$ imposes $Q_{\{n_i\mathbf{k}_i\}}^+\equiv
Q_{\{n_i\mathbf{k}_i\}}^{\dagger}$ and
$Q_{\{n_i\mathbf{k}_i\}}^-\equiv
Q_{\{n_i\mathbf{k}_i\}}^{\vphantom{\dagger}}$, and for many-phonon
terms, we have
$Q^{-q_1,...,-q_M}_{\{n_j\mathbf{k}_j\}}=(Q^{q_1...q_M}_{\{n_j\mathbf{k}_j\}})^\dagger$. The
single-phonon interaction terms, which may physically be seen as single-phonon scattering off the
$Q$ degrees of freedom, corresponds in particular to a coupling
of the $Q$ operators to the strain tensor $\mathcal{E}^{\alpha\beta}(\mathbf{r})$,
\begin{equation}
  \label{eq:9}
\mathcal{E}^{\alpha\beta}(\mathbf{r})=  \frac {i\hbar^{1/2}}
{\sqrt{N_{\rm uc}}}\sum_{\mathbf{k}n}e^{i\mathbf{k}\cdot\mathbf{r}}
\frac
  {\left(k^\alpha\varepsilon^\beta_{\mathbf{k}n}+k^\beta\varepsilon^\alpha_{\mathbf{k}n}\right)} {\sqrt{2M_{\rm uc}\omega_{\mathbf{k}n}}}
  \left(a_{\mathbf{k}n}^{\vphantom{\dagger}}+a_{-\mathbf{k}n}^{\dagger}\right),
\end{equation}
  where $M_{\rm uc}$ is the unit cell mass and $\bs \varepsilon_{n\mb k}$ is the
    polarization vector of the $|n\mb k\rangle$ phonon.
    The two-phonon terms capture quadratic coupling of the lattice
    displacements to the electrons/spins, as is often considered for
    example in treatments of Raman scattering
    \cite{PhysRevLett.96.155901,PhysRevLett.100.145902}. A
      priori, the quadratic terms are much smaller than the linear
      ones, but the former may be important if they give rise to distinct
      effects or contribute at a lower order in perturbation theory
      than the linear ones.

\subsection{Scattering rates}
\label{sec:scattering-rates}

\subsubsection{T-matrix elements}
\label{sec:t-matrix-elements}

The transition matrix elements are
$T_{\mathtt{fi}}=\sum_{l}T_{\mathtt{fi}}^{[l_1,...]}$ (the $l_i$
represent which $H_{[l_i]}$ appear successively in $T$, so that the
number of $l_i$ appearing in $T_{\mathtt{fi}}^{[l_1,...]}$ is the order of the Born approximation used
for that term), where
\begin{widetext}
\begin{eqnarray}
  \label{eq:38}
  T^{[1]}_{\mathtt{i}\rightarrow\mathtt{f}}    &=&\sum_{n\mathbf{k}q}\sqrt{N^i_{n\mathbf{k}}+\tfrac{1+q}2}~\langle
    f_s|Q^{q}_{n\mathbf k}|i_s\rangle~ \mathds{I}({i_p}\overset{q\cdot
                                                       n\mathbf
                                                       k}{\longrightarrow}{f_p}),\\
  \label{eq:38b}
 T^{[2]}_{\mathtt i\rightarrow \mathtt f} &=&\frac 1 {\sqrt{N_{\rm uc}}}\sum_{n\mathbf{k}q, n' \mathbf{k'}q'}\sqrt{N_{n\mathbf k}^i +\tfrac{1+q} 2}\sqrt{N_{n'\mathbf k'}^i +\tfrac{1+q'}2}\langle f_s| Q_{n\mathbf k n'\mathbf k'}^{qq'}| i_s\rangle ~\mathds{I}({ i_p}\overset{q\cdot n\mathbf k}{\underset{q'\cdot n'\mathbf k'}\longrightarrow}{ f_p}),\\
  \label{eq:38c}
  T^{[1,1]}_{\mathtt{i}\rightarrow\mathtt{f}} &=&
\sum_{n\mathbf{k}q,n'\mathbf{k}'q'}\sqrt{N^i_{n\mathbf{k}}+\tfrac{1+q}{2}}\sqrt{N^f_{n'\mathbf{k}'}+\tfrac{1-q'}{2}}\sum_{ m_s}\frac{\langle
    f_s|Q^{q'}_{n'\mathbf k'}|m_s\rangle\langle
    m_s|Q^{q}_{n\mathbf
                                                    k}|i_s\rangle}{E_{i_s}-E_{m_s}-q\omega_{n\mathbf{k}}+i\eta}~\mathds{I}({ i_p}\overset{q\cdot n\mathbf k}{\underset{q'\cdot n'\mathbf k'}\longrightarrow}{ f_p}),
\end{eqnarray}
\end{widetext}
and $T^{[1,2]}_{\mathtt{i}\rightarrow\mathtt{f}}$ and
$T^{[1,1,1]}_{\mathtt{i}\rightarrow\mathtt{f}}$ are given in
Appendices~\ref{sec:comp-at-second} (Eq.~\eqref{eq:42}) and
\ref{sec:computation-at-third} (Eq.~\eqref{eq:362}), respectively. Here, $\mathds{I}({ i_p}\overset{q\cdot n\mathbf k}{\longrightarrow}{
  f_p})$ (resp.\ $\mathds{I}({ i_p}\overset{q\cdot n\mathbf k}{\underset{q'\cdot n'\mathbf k'}\longrightarrow}{ f_p})$) is a large product of delta functions which enforce
$N^f_{n''\mathbf{k}''}=N^i_{n''\mathbf{k}''}$ $\forall
n''\mathbf{k}''\neq n\mathbf{k}$ (resp. $\forall
n''\mathbf{k}''\neq (n\mathbf{k},n'\mb k')$), and
$N^f_{n\mathbf{k}}=N^i_{n\mathbf{k}}+ q$ (resp.\ $N^f_{n\mathbf{k}}=N^i_{n\mathbf{k}}+ q, N^f_{n'\mathbf{k}'}=N^i_{n'\mathbf{k}'}+ q'$). Note that the cases where
$n\mathbf k = n'\mathbf k'$ require a
formal correction. However, at any given order in the ${\lambda}$ expansion, such terms are smaller than all others
 by a factor $1/N_{\rm uc}$, where $N_{\rm uc}$ is the
 number of unit cells, and therefore vanish in the
thermodynamic limit. In what follows we thus use $\sum_{n\mathbf
  k,n'\mathbf k'}$ and $\sum_{n\mathbf k\neq n'\mathbf k'}$
exchangeably, unless we specify
otherwise. 

The scattering rate as given by Eq.~\eqref{eq:main8}, involves the squares of
the elements of the total transition matrix (see Appendices~\ref{sec:comp-at-second},
\ref{sec:computation-at-third} for computational details).
Its full expression to perturbative order $\lambda^4$ is
\begin{equation}
  \label{eq:59}
  \Gamma_{\mathtt{i}\rightarrow \mathtt{f}} = \Gamma^{\texttt{SC}}_{\mathtt{i}\rightarrow \mathtt{f}}
  + \Gamma^{\texttt{Q}1}_{\mathtt{i}\rightarrow \mathtt{f}}+\Gamma^{\texttt{Q}2}_{\mathtt{i}\rightarrow \mathtt{f}},
\end{equation}
where
\begin{eqnarray}
  \label{eq:83}
  \begin{bmatrix}
\Gamma^{\texttt{SC}}_{\mathtt{i}\rightarrow \mathtt{f}}
    &;&  \Gamma_{\mathtt{i}\rightarrow \mathtt{f}}^{\texttt {Q}2}
    &;&  \Gamma_{\mathtt{i}\rightarrow \mathtt{f}}^{\texttt {Q}1}
  \end{bmatrix}= \frac{2\pi}{\hbar}\delta(E_{\mathtt{i}}-E_{\mathtt{f}})\qquad\qquad\nonumber\\
  \times
  \begin{bmatrix}
    |T^{[1]}_{\mathtt{i}\rightarrow\mathtt{f}}|^2+|T^{[1,1]}_{\mathtt{i}\rightarrow\mathtt{f}}|^2+|T^{[2]}_{\mathtt{i}\rightarrow\mathtt{f}}|^2\quad;\\
    2\mathfrak{Re}\left \{(T^{[1,1]}_{\mathtt{i}\rightarrow\mathtt{f}})^*T^{[2]}_{\mathtt{i}\rightarrow\mathtt{f}}
    \right\}\quad;\\
   2\mathfrak{Re}\left\{(T^{[1,2]}_{\mathtt{i}\rightarrow\mathtt{f}})^*T^{[1]}_{\mathtt{i}\rightarrow\mathtt{f}}
      +
      (T^{[1,1,1]}_{\mathtt{i}\rightarrow\mathtt{f}})^*T^{[1]}_{\mathtt{i}\rightarrow\mathtt{f}}\right\}
  \end{bmatrix}.\quad
 \end{eqnarray}
This decomposition into three terms is
discussed in Sec.~\ref{sec:scattering-channels}.

\subsubsection{Collision matrix elements}
\label{sec:coll-matr-elem}

Following Eq.~\eqref{eq:1}, the scattering rates $\Gamma_{\mathtt{i}\rightarrow \mathtt{f}}$
  give access to the collision integral, i.e.\ to $M_{n\mb k,n'\mb k'}$ and $D_{n\mb k}$.
  We decompose the latter as $D_{n\mb k}= D_{n\mb k}^{(1)}+D_{n\mb k}^{(2)}+\breve {D}_{n\mb k}$,
  where $D^{(1)}$ and $D^{(2)}$ are obtained in our perturbative expansion at orders
  $\lambda^{2}$ and $\lambda^{4}$, respectively, and $\breve {D}_{n\mb k}$ encompasses contributions
  due to other scattering processes as well as higher-order terms of the expansion.
  In the following, we also use the ``$[l_i];[l'_j]$'' superscripts to denote a term
  obtained from the product of $T_{\mathtt i\rightarrow \mathtt f}^{[l_i]}$ and
  $T_{\mathtt i\rightarrow \mathtt f}^{[l'_j]}$ within $|T_{\mathtt i\rightarrow \mathtt f}|^2$.
  For instance, at order $\lambda^2$, we have
  \begin{equation}
    \label{eq:113}
    D_{n\mb k}^{(1)}=D_{n\mb k}^{[1];[1]}.
  \end{equation}
Details of the derivation are given in Sec.~\ref{sec:find-title-2} and
Appendix~\ref{sec:1storder}.
  At order $\lambda^4$, the diagonal and off-diagonal contributions to the collision integral
  take the forms
\begin{equation}
  \label{eq:5}
  D^{(2)}_{n\mathbf k} =-\frac 1 {N_{\rm uc}}\sum_{n'\mathbf{k}'}\sum_{qq'}q\left (N^{\rm eq}_{n'\mathbf k'}+\tfrac {q'+1}2\right ) \left [\mathfrak W_{n\mathbf k n'\mathbf k'}^{qq'}\right ],
\end{equation}
and
\begin{equation}
  \label{eq:4}
  M_{n\mathbf{k}n'\mathbf{k}'}= \frac 1 {N_{\rm uc}}\sum_{q,q'=\pm} q \left ( N^{\rm eq}_{n\mathbf k}+\tfrac{q+1} 2 \right ) \left [\mathfrak W_{n\mathbf k n'\mathbf k'}^{qq'}\right ],
\end{equation}
respectively, where $\mathfrak {W}^{qq'}_{n\mb k, n'\mb k'}$ is an
off-diagonal scattering rate which involves two different
phonon states $|n\mb k\rangle$ and $|n'\mb k'\rangle$. More precisely,
$\mf
W^{+,+}$ (resp.\ $\mf W^{-,-}$) corresponds to scattering processes where two
phonons are emitted (resp.\ absorbed), and $\mf W^{+,-},\mf
W^{-,+}$ to processes where one phonon is emitted and one is
absorbed. $D_{n\mb k}$ is
the diagonal scattering rate, i.e.\ it is associated with
variations in $\delta\overline{N}_{n\mathbf{k}}$ only.

We will now decompose the $\mathfrak {W}^{qq'}_{n\mb k, n'\mb k'}$ scattering
  rates into 
  \begin{equation}
    \label{eq:72}
    \mathfrak {W}^{qq'}_{n\mb k, n'\mb k'}=\mathfrak W_{n\mathbf k n'\mathbf k'}^{\oplus,qq'}+\mathfrak
W_{n\mathbf k n'\mathbf k'}^{\ominus,qq'},
\end{equation}
where $\mathfrak
{W}^{\oplus/\ominus,qq'}_{n\mb k, n'\mb k'}$ satisfy detailed ($\sigma=1$) or
``anti-detailed'' ($\sigma=-1$) balance equations
\begin{equation}
  \label{eq:detbal}
\mathfrak W_{n\mathbf k n'\mathbf k'}^{\sigma,qq'} = \sigma~
e^{-\beta(q\omega_{n\mathbf k}+q'\omega_{n'\mathbf k'})}~\mathfrak
W_{n\mathbf k n'\mathbf k'}^{\sigma,-q-q'},\quad \sigma=\pm\,{\rm or}\,\oplus,\ominus.
\end{equation}
Physically, Eq.~\eqref{eq:detbal} expresses ``microscopic'' thermodynamic
equilibrium between the process which takes
$\{N_{n\mathbf{k}}\rightarrow N_{n\mathbf{k}}+q,
N_{n'\mathbf{k}'}\rightarrow N_{n'\mathbf{k}'}+q'\}$ to the ``conjugate''
process taking $\{N_{n\mathbf{k}}\rightarrow
N_{n\mathbf{k}}-q,N_{n'\mathbf{k}'}\rightarrow
N_{n'\mathbf{k}'}-q'\}$, with $q,q'=\pm1$, leaving
$N_{n''\mathbf{k}''}$ unchanged for
$n''\mathbf{k}''\notin\{n\mathbf{k}, n'\mathbf{k}'\}$.
Note
  that this is different from time-reversal symmetry which provides a
  relation between the processes acting on $\{|n_l,\mb k_l\rangle\}$
  phonons to the {\em same} processes acting on the
$\{|n_l,-\mb k_l\rangle\}$ phonons. 

Moreover, since, by construction, the two-phonon scattering rates
satisfy
\begin{equation}
  \label{eq:24}
  \mathfrak{W}^{\sigma,qq'}_{n\mathbf{k}n'\mathbf{k}'}=\mathfrak{W}^{\sigma,q'q}_{n'\mathbf{k}'n\mathbf{k}},
\end{equation}
the following relations also hold:
  \begin{equation}
    \label{eq:55}
    \mathfrak{W}^{\sigma,+-}_{n'\mathbf{k}'n\mathbf{k}}
    = \sigma\, e^{\beta(\omega_{n\mathbf k}-\omega_{n'\mathbf  k}')}
        \mathfrak{W}^{\sigma,+-}_{n\mathbf{k}n'\mathbf{k}'}.
  \end{equation}
  Together, these imply that there are only four independent such scattering
rates between the $|n,\mb k\rangle$ and $|n',\mb k'\rangle$ phonons, namely $\mathfrak{W}^{\sigma,++}_{n\mathbf{k},n'\mathbf{k}'}$ and
$\mathfrak{W}^{\sigma,+-}_{n\mathbf{k},n'\mathbf{k}'}$
with $\sigma=\oplus,\ominus$.

As discussed at length, the first Born approximation alone does not
lead to a nonzero thermal Hall effect, neither do those scattering
rates which satisfy detailed balance as the latter imposes thermal equilibrium
between ``left'' and ``right'' scattering. We find the kernels $K^{L/H}$ defined in Eqs.~(\ref{eq:main17},\ref{eq:18})
in terms of the $\mathfrak W$ scattering rates:
\begin{widetext}
  \begin{eqnarray}
    \label{eq:main15}
     K_{n\mathbf k n'\mathbf k'}^{L}
  &=& 
      \frac{e^{\beta\hbar\omega_{n\mathbf k}}}{D_{n\mathbf k}}
    \Bigg (  \delta_{n,n'}\delta_{\mathbf k,\mathbf k'}
 +
  \frac{ e^{\beta\hbar\omega_{n'\mathbf{k}'}}}{2 N_{\rm uc}
      D_{n'\mathbf{k}'}}\sum_{q=\pm}
      e^{\frac{q-1} 2 \beta\hbar\omega_{n'\mathbf{k}'}}\left\{ \mathfrak W^{\ominus,+,q}_{n\mathbf k,n'\mathbf
   k'}\left(q\coth(\tfrac{\beta\hbar\omega_{n\mathbf k}}2)+\coth(\tfrac{\beta\hbar\omega_{n'\mathbf
      k'}}2)\right) - 2\,\mathfrak W^{\oplus,+,q}_{n\mathbf k,n'\mathbf k'} \right \} \Bigg ),\nonumber\\
    &&\\
\label{eq:main16}
  K^{H}_{n\mathbf{k}n'\mathbf{k}'} &=&
      \frac{e^{\beta\hbar\omega_{n\mathbf{k}}}e^{\beta\hbar\omega_{n'\mathbf{k}'}}}{2 N_{\rm uc} D_{n\mb{k}}D_{n'\mb{k}'}}\sum_{q=\pm}
\mathfrak W^{\ominus,+,q}_{n\mathbf k,n'\mathbf
                                    k'}\,e^{\frac{q-1}2\beta\hbar\omega_{n'\mathbf{k}'}}\left(\coth(\tfrac{\beta\hbar\omega_{n'\mathbf
                                    k'}}{2})-q\coth(\tfrac{\beta\hbar\omega_{n\mathbf k}}{2})\right). 
  \end{eqnarray}
  \end{widetext}
  Incorporating the expression for $D$ in the
denominators of $K^{L,H}$ provides an
expansion up to $O({\lambda}^4)$ of the latter.
We recover, as mentioned before, that the terms in $\mf W^\oplus$ do
not contribute to $K^H$ (they satisfy detailed-balance). The
``anti-detailed-balance'' relations satisfied by the $\mf W^\ominus$
terms do not however prohibit their contribution to $K^L$.
See Sec.~\ref{sec:point-group-symm-1} for a discussion. Inserting
Eq.~\eqref{eq:main15} and Eq.~\eqref{eq:main16} into
Eq.~\eqref{eq:2}, and after some algebra, one obtains the results
for $\kappa_{\rm L,H}$, respectively:
\begin{widetext}
  \begin{align}
    \label{eq:115}
    \kappa_{\rm L}^{\mu\nu}&=\frac{\hbar^2}{k_BT^2}\frac{1}{V} \Bigg \{ \sum_{n\mathbf{k}}\frac{\omega_{n\mathbf{k}}^2
    v^\mu_{n\mathbf{k}}v^\nu_{n\mathbf{k}}}{4D_{n\mathbf{k}}\sinh^2(\beta\hbar\omega_{n\mathbf{k}}/2)}\\
    &+  \sum_{n\mathbf{k}n'\mathbf{k}'}
   J^\mu_{n\mathbf{k}} \frac{ e^{\beta\omega_{n\mb k}/2}}{D_{n\mathbf k}}
\Bigg ( \frac {1} {N_{\rm uc}}
    \sum_{q=\pm}e^{\beta q \omega_{n'\mathbf{k}'}/2}e^{\beta\omega_{n\mb k}/2}
    \Big \{ \mathfrak W_{n\mathbf k n'\mathbf k'}^{\ominus,+q}
    {\rm coth}(\beta\omega_{n'\mb k'}/2) - q\mathfrak W_{n\mathbf k n'\mathbf k'}^{\oplus,+q}
  \Big \} \Bigg ) \frac{e^{\beta\omega_{n'\mb k'}/2}}{D_{n'\mathbf k'}}  J^\nu_{n'\mathbf{k}'}\Bigg \},\nonumber
  \end{align}
  where the first term is the leading-order contribution
  and with $J^\mu_{n\mathbf{k}}=N_{n\mathbf{k}}^{\rm eq}\omega_{n\mathbf{k}}v^\mu_{n\mathbf{k}}$, and
  \begin{equation}
    \label{eq:79}
    \kappa_{\rm H}^{\mu\nu} =  \frac{\hbar^2}{k_B T^2} \frac{1}{V} 
    \sum_{n\mathbf{k}n'\mathbf{k}'}
    J^\mu_{n\mathbf{k}}\frac{e^{\beta\hbar\omega_{n\mathbf{k}}/2} }{D_{n\mathbf{k}}}
    \Bigg (\frac 1 {N_{\rm uc}}\sum_{q=\pm}\frac{\left(e^{\beta\hbar\omega_{n\mathbf{k}}}-e^{q\beta\hbar\omega_{n'\mathbf{k}'} }\right)\,\mathfrak W^{\ominus,+,q}_{n\mathbf k,n'\mathbf
   k'}}{4\sinh(\beta\hbar\omega_{n\mathbf  k}/2)\sinh(\beta\hbar\omega_{n'\mathbf k'}/2)}
\Bigg ) \frac{e^{\beta\hbar\omega_{n'\mathbf{k}'}/2} }{ D_{n'\mathbf{k}'}}J^\nu_{n'\mathbf{k}'}.
    \end{equation}
   \end{widetext}

\subsection{The collision integral as correlation functions}
\label{sec:coll-integr-as}

\subsubsection{Terms at $O(\lambda^2)$}
\label{sec:find-title-2}

The diagonal scattering rate $D^{(1)}_{n\mb k}$, obtained by
  inserting $T^{[1]}_{\mathtt{i}\rightarrow\mathtt{f}}$ from Eq.~\eqref{eq:38} into Eqs.~(\ref{eq:main8}-\ref{eq:81}),
  may now be cast into the form of a correlation function of $Q$ operators.
To do so, we first enforce the energy conservation
$\delta(E_{\mathtt{f}}-E_{\mathtt{i}})$ by writing the latter
as a time integral, i.e.\ use $\int_{-\infty}^{+\infty}{\rm d}t e^{i\omega
  t}=2\pi\delta(\omega)$; we then identify $A(t)=e^{+iHt}Ae^{-iHt}$ and use the
identity $1=\sum_{ f_s}| f_s\rangle\langle f_s|$.
Taking the $Q$s in the initial state to be in thermal equilibrium
$p_{ i_s}=Z_s^{-1}e^{-\beta E_{ i_s}}$, summing over $| i_s\rangle$, identifying $\langle A \rangle_\beta = Z^{-1}\text{Tr}(e^{-\beta
  H}A)$, summing over final phononic states $f_p$ and
taking the average over initial phononic states $i_p$, we obtain the
$|T_{\mathtt{i}\rightarrow\mathtt{f}}^{[1]}|^2$ contribution of $\Gamma^{\mathtt{SC}}$,
\begin{equation}
  \label{eq:84}
  D^{(1)}_{n\mathbf k} = -\frac {1}{\hbar^2} \int dt e^{-i\omega_{n\mathbf{k}}t}\left \langle [Q^{\vphantom{\dagger}}_{n\mathbf k}(t), Q^\dagger_{n\mathbf{k}}(0)]\right\rangle_\beta .
\end{equation}
We now apply the same method to higher orders of the perturbative expansion.

\subsubsection{Terms at $O(\lambda^4)$}
\label{sec:find-titl}

We use the following
time integral representation for the denominators appearing at second
and higher Born orders (using a regularized
definition of the sign function, i.e.\
$\underset{\eta\rightarrow 0}{\lim}{\rm
  sign}(t)e^{-\eta|t|}\rightarrow{\rm sign}(t)$),
\begin{eqnarray}
  \label{eq:85}
  \frac 1 {x \pm i\eta} &=& {\rm PP} \frac 1 x \mp i\pi \delta(x)\\
    &=& \frac 1 {2i} \int_{-\infty}^{+\infty}\text d t_1 e^{it_1 x}\left(\text{sign}(t_1)\pm 1\right).\nonumber
\end{eqnarray}
Using Eqs.~(\ref{eq:38b},\ref{eq:38c}) and Eq.~\eqref{eq:83}, we find the explicit
expressions for the other semiclassical ($\Gamma^{\mathtt{SC}}$) scattering rates as correlation functions of the
$Q$ operators, 
\begin{widetext}
\begin{eqnarray}
  \label{eq:63}
  \mathfrak{W}^{\oplus,[2];[2],qq'}_{n\mathbf{k}n'\mathbf{k}'}
  &=&\frac{2}{\hbar^4}~\fint_{t}\qquad
                                                                  \left\langle
                                                                  Q^{-q,-q'}_{n\mathbf{k}n'\mathbf{k}'}(-t)
                                                                  Q^{q,q'}_{n\mathbf{k}n'\mathbf{k}'}(0)\right\rangle,
                                                                  \\
      \label{eq:101}
  \mathfrak{W}^{\ominus,[1,1];[1,1],qq'}_{n\mathbf{k}n'\mathbf{k}'}
  &=&\frac{2}{\hbar^4}~N_{\rm uc}~\mathfrak{Re}~\fint_{t,t_1,t_2}\left\langle\llbracket
                            Q^{-q}_{n\mathbf{k}}(-t-t_2),Q^{-q'}_{n'\mathbf{k}'}(-t+t_2)\rrbracket\{
                                                                   Q^{q'}_{n'\mathbf{k}'}(-t_1),Q^{q}_{n\mathbf{k}}(t_1)\}\right\rangle,\\
  \label{eq:102}
  \mathfrak{W}^{\oplus,[1,1];[1,1],qq'}_{n\mathbf{k}n'\mathbf{k}'}
  &=&\frac{1}{\hbar^4}~N_{\rm uc}~\fint_{t,t_1,t_2}\left\langle\{
                            \cdot,\cdot\}\{
                                                                  \cdot,\cdot\}-\llbracket
                            \cdot,\cdot\rrbracket\llbracket
                                                                  \cdot,\cdot\rrbracket\right\rangle,
\end{eqnarray}
\end{widetext}
where we use the shorthand notation
\begin{equation}
  \label{eq:137}
  \llbracket A(t_a),B(t_b)\rrbracket={\rm
                                      sign}(t_b-t_a)[A(t_a),B(t_b)],
\end{equation}
and $\fint_{t,\{t_j\}}$, $j=1,..,l$, denotes the set of $1+l$
Fourier transforms evaluated once at
$\Sigma_{n\mb k q}^{n'\mb k'
  q'}=q\omega_{n\mathbf{k}}+q'\omega_{n'\mathbf{k}'}$ and $l$ times at
$\Delta_{n\mb k q}^{n'\mb k'
  q'}=q\omega_{n\mathbf{k}}-q'\omega_{n'\mathbf{k}'}$, i.e.\ $\fint_{t,\{t_j\}}=\int \text dt
\text dt_1.. \text dt_l e^{i\Sigma_{n\mb k q}^{n'\mb k' q'}
  t}e^{i\Delta_{n\mb k q}^{n'\mb k' q'}(t_1+..+t_l)}$. The symbols $\cdot$ must
be replaced by the same set of operators as the expression from the
above. The commutators and
anticommutators ultimately capture antisymmetrization and
symmetrization over the $n\mathbf{k}q\leftrightarrow n'\mathbf{k}'q'$
indices. We provide expressions for
$\mathfrak{W}^{\ominus,[1,1];[2],qq'}_{n\mathbf{k}n'\mathbf{k}'}$,
$\mathfrak{W}^{\oplus,[1,1];[2],qq'}_{n\mathbf{k}n'\mathbf{k}'}$ (from
$\Gamma^{\mathtt{Q2}}$),
$\mathfrak{W}^{[1,2];[1],qq'}_{n\mathbf{k}n'\mathbf{k}'}$ and
$\mathfrak{W}^{[1,1,1];[1],qq'}_{n\mathbf{k}n'\mathbf{k}'}$ (from
$\Gamma^{\mathtt{Q1}}$) in
Appendices~\ref{sec:all-contributions},\ref{sec:computation-at-third},
Eqs.~(\ref{eq:105},\ref{eq:105b},\ref{eq:34},\ref{eq:398}), respectively.

\subsubsection{Scattering channels and conserving approximation}
\label{sec:scattering-channels}

The above terms capture all contributions to the collision
integral arising from the Born expansion of the transition amplitude,
up to perturbative order $\lambda^4$.
  This gives, correspondingly, physical processes in the collision
  integral which contribute up to $O(\lambda^4)$.

In Eq.~\eqref{eq:59}, while $\Gamma_{\mt i \rightarrow
      \mt f}^{\mt {SC}}$ and $\Gamma_{\mt i \rightarrow
      \mt f}^{\mt Q2}$ are ``two-phonon'' terms, the contribution from $\Gamma_{\mt i \rightarrow
      \mt f}^{\mt Q1}$ is
  a ``one-phonon'' term, i.e.\ one where the initial $\mt i$ and final $\mt f$
states differ by only one phonon $|n\mb k\rangle$.
  Physically, this contributes to processes
  which create or annihilate a single phonon, in contrast with the
  $O(\lambda^4)$ processes described so far, which create/annihilate
  two phonons with different quantum numbers.  Because the single
  phonon process is physically distinct from the two-phonon ones, we
  expect that it is independent from the latter in the sense that the
  set of all the $O(\lambda^4)$ single-phonon processes satisfies {\em
    independently} all
  physical constraints such as symmetries and conservation laws.
  Hence  omitting these contributions is a ``conserving
  approximation'' in the traditional sense \cite{kadanoff-baym},
  and we will proceed with
  this omission for the most part in the following.  We however
  include formal expressions for these terms in the appendices.

The remaining contributions in Eq.~\eqref{eq:59} are ``two-phonon''
terms, i.e.\ terms in which the initial $\mt i$ and final $\mt f$ states differ by two
phonons $|n\mb k\rangle, |n'\mb k'\rangle$. The two-phonon,
$O(\lambda^4)$, contributions to the $\mf W$ scattering rates thus read
  \begin{eqnarray}
   \label{eq:99}
    \mf  W^{\ominus,qq'}_{n\mathbf k,n'\mathbf k'}
    &=& \mf W^{\ominus,[1,1];[2],qq'}_{n\mathbf k,n'\mathbf k'} +
     \mf W^{\ominus,[1,1];[1,1],qq'}_{n\mathbf k,n'\mathbf k'},\\
    \mf  W^{\oplus,qq'}_{n\mathbf k,n'\mathbf k'}
    &=& \mf W^{\oplus,[2];[2],qq'}_{n\mathbf k,n'\mathbf k'} +
  \mf  W^{\oplus,[1,1];[2],qq'}_{n\mathbf k,n'\mathbf k'} 
      + \mf W^{\oplus,[1,1];[1,1],qq'}_{n\mathbf k,n'\mathbf k'}.\nonumber
  \end{eqnarray}
  
Another physical distinction between the contributions in
Eq.~\eqref{eq:59} can be
  made according to the ``quantum'' or semiclassical,
  nature of the terms. The one-phonon $\Gamma_{\mt i \rightarrow
      \mt f}^{\mt Q1}$ and two-phonon $\Gamma_{\mt i \rightarrow
      \mt f}^{\mt Q2}$ terms in Eq.~\eqref{eq:59} are ``quantum'' in
    the sense that the physical process corresponding to each contribution therein
    is an interference term between \emph{distinct} scattering channels.
    In particular, in a ``quantum'' term, the number of scattering events
    in the two channels are different. On the contrary, each contribution in
    $\Gamma_{\mt i \rightarrow \mt f}^{\mt {SC}}$ is the probability amplitude
    of one given scattering channel, corresponding physically to the probability amplitude
    of a given scattering process, and in this respect is truly semiclassical.
    As a semiclassical approximation, we will neglect ``quantum'' contributions in the following;
  formal expressions for these terms are nonetheless included in the appendices.
    The only ``semiclassical'' contributions, up to $O({\lambda}^4)$, to the collision integral are
    from the scattering rates shown in Eq.~\eqref{eq:63}.

  Upon applying our results to the case of a staggered antiferromagnet
 in Sec.~\ref{sec:application}, we focus on the lowest-order contributions
    to $K^L_{n\mb k,n'\mb k'}$ and $K^H_{n\mb k,n'\mb k'}$,
    which come from $D^{(1)}_{n\mb k}$ and $\mf
    W^{\ominus,qq'}_{n\mathbf k,n'\mathbf k'}$, respectively.
    Therefore, in Sec.~\ref{sec:application}, we consider only the lowest-order semiclassical contributions
    $D_{n\mb k}\approx D^{(1)}_{n\mb k}+\breve D_{n\mb k}$ and
    $\mf  W^{\ominus,qq'}_{n\mathbf k,n'\mathbf k'} \approx \mf  W^{\ominus,[1,1];[1,1],qq'}_{n\mathbf k,n'\mathbf k'}$.

\subsubsection{Physical interpretation}
\label{sec:phys-interpr}

To leading order, the longitudinal conductivity is controlled by
the diagonal scattering rate, whose main contribution occurs at
order $\lambda^2$.  The latter is given as the first term in Eq.~\eqref{eq:111},
and is shown explicitly in Eq.~\eqref{eq:84}. It is related to the
Fourier transform of the commutator of two
$Q_{n\mathbf k}$ operators at unequal times.  The commutator structure identifies the phonon scattering rate
$D^{(1)}_{n\mathbf k}$ with the spectral function of the $Q_{n\mathbf
  k}$ field at energy $\omega_{n\mb k}$, i.e.\ it captures
the proportion of the energy density contained in the $Q_{n\mb k}$ field located at
$\omega_{n\mb k}$, as expected from (lowest-order) linear response 
\cite{Bruus-Flensberg,PhysRevB.82.134421}.

As mentioned above, the first Born order transition matrices are
hermitian. At second Born's order, the advanced/retarded Green's
function, $1/(E_{\mt
  i/\mt f}-E_{\mt m}\pm i\eta)$, appearing in $T_{\mt i \mapsto \mt
  f}$, splits into
on-shell and off-shell contributions, so that the scattering rate
$\propto |T_{\mt i \mapsto \mt f}|^2$ then
involves the product of two on-shell or two-offshell contributions, as
well as the products of one on-shell and one off-shell one. Because of complex conjugation of one
term upon taking the square modulus of the $T$ matrix, the scattering rates which involve either two
on-shell or two off-shell contributions are blind to the sign of
$\pm i \eta$, i.e.\ to the advanced or retarded nature of
the process, and enforce a detailed-balance relation,
Eq.~\eqref{eq:detbal} with $\sigma=\oplus$. Therefore,
the only scattering rates which can contribute to the Hall
conductivity are those involving one on-shell
(imaginary part) and one off-shell (real part) scattering event, which
translates here into the product of a commutator and an
anticommutator, Eq.~\eqref{eq:101}.

\section{Relations and symmetries}
\label{sec:disc-symm}

In this section, we explore in more detail some physical relations
verified by the scattering rates defined above, and their possible
consequences on the longitudinal and Hall conductivities.

\subsection{Time-reversal symmetry: reversal of the momenta}
\label{sec:time-revers-symm}

\begin{table}[htbp]
  \centering
  \begin{tabular}{c|c}
    \hline\hline
    $Q$ operators &   $\wideparen{Q^q_{n\mb k}}=Q^q_{n,-\mb k}$\\
  &  $\wideparen{Q^{qq'}_{n\mathbf{k},n'\mathbf{k}'}} = Q^{qq'}_{n-\mathbf{k},n'-\mathbf{k}'}$  \\       
scattering rates & $D^{(1)}_{n,\mathbf k}= D^{(1)}_{n,-\mathbf k}$ \\
  & $\mf W^{\sigma,qq'}_{n\mb k,   n'\mb k'}
           =  \sigma\, \mf  W^{\sigma,qq'}_{n-\mb  k, n'-\mb k'}$\\
conjugate process &  $\frac{\mf W^\sigma ({\rm mr}[S])}{\mf W^\sigma (S)}
    = e^{-\delta E_{s \rightarrow p}^{(S)}} \frac {\mf W^\sigma ({\rm pc}[S])}{\mf W^\sigma (S)}
                                = \sigma$ \\
 kernels   &  $K^H_{n\mb k, n'\mb k'}  = - K^H_{n-\mb k, n'-\mb k'}$ \\
    conductivities &  $\bs \kappa_H=\bs 0$       \\
    \hline\hline
  \end{tabular}
  \caption{Relations which hold true in the presence of time-reversal {\em symmetry}.
    The phonon operator relation $\wideparen{a^q_{n\mb
        k}}=a^q_{n,-\mb k}$ holds true even when no time-reversal
    symmetry is present.
    See text for definitions and justifications.}
  \label{tab:TR2}
\end{table}

We investigate the implications of time-reversal (TR) invariance on
our results. In particular, we check explicitly that the Hall
conductivity vanishes in a TR-symmetric system.  It is important to note that, in a time-reversal
invariant system, the scattering rates are a priori not time-reversal
invariant themselves.

We denote with $\wideparen Q$ and $|\wideparen {\mathtt n}\rangle$
the time-reversed of operator
  $Q$ and of state $|{\mathtt n}\rangle$, respectively. Then, because
  of the antiunitarity of the time-reversal operator, for any states
  $\mathtt{n,m}$ and any operator $Q$, we have
  $\langle \wideparen {\mathtt n} | Q^\dagger | \wideparen {\mathtt
    m}\rangle = \langle \mathtt m | \wideparen{Q} | \mathtt
  n\rangle$. Moreover, it is possible to choose a polarization index
  $n$ invariant under TR, whence $\wideparen{a^q_{n\mb  k}}=a^q_{n,-\mb k}$.

Let us now consider what happens in a time-reversal-invariant
system. In that case, the hamiltonian $H'_{[1]}=\sum_{n\mb
  kq}Q^q_{n\mb k}a^q_{n\mb k}$ must be TR-invariant, so that
  $\wideparen {Q^q_{n\mb k}}=Q^q_{n,-\mb k}$.  Similarly,
  TR-invariance of $H'_{[2]}$ (defined in Eq.~\eqref{eq:main6}) entails
  $\wideparen{Q^{qq'}_{n\mathbf{k},n'\mathbf{k}'}}=Q^{qq'}_{n-\mathbf{k},n'-\mathbf{k}'}$.
 
\subsubsection{Consequences for the scattering rates.}

 Following the same steps as those sketched in Sec.~\ref{sec:find-title-2}, and using the fact that
 $E_{\wideparen{\mathtt m}}=E_{\mathtt m}$ for any state $\mt m$ of a TR-symmetric system,
we can show explicitly that, {\em in a time-reversal-invariant
  system}, the following relations for the scattering rates exist:
\begin{eqnarray}
  \label{eq:56}
  D^{(1)}_{n,\mathbf k}&=& D^{(1)}_{n,-\mathbf k}\\
  \mf W^{\sigma,qq'}_{n\mb k,   n'\mb k'}
           &=&  \sigma\, \mf  W^{\sigma,qq'}_{n-\mb  k, n'-\mb k'}.
\end{eqnarray}
The $\sigma$ sign in the second relation 
can be understood as arising from two facts: {\it (1)} schematically, 
$\mf W^\sigma \sim \frac {1}{E_{\mt f}- E_{\mt m}+i\eta}\frac {1}{E_{\mt i}- E_{\mt m}-i\eta}
+ \sigma \, \rm h.c.$ ---which is reflected in the fact that $\mf W^\oplus$ (resp.\ $\mf W^\ominus$)
expressed as an integral, Eqs.~(\ref{eq:63}--\ref{eq:102}), contains an even (resp.\ odd) number of
sign functions---
and {\it (2)} an effect of time-reversal on the $T$-matrix is to exchange denominators
$\frac {1}{E_{\mt f}- E_{\mt m}+i\eta}
\rightarrow \frac {1}{E_{\wideparen{\mt f}}- E_{\wideparen{\mt m}}-i\eta}$ 
(see Sec.~\ref{sec:phys-interpr} for an interpretation of the $+i\eta$ regularization).

\subsubsection{Relation to detailed balance.}
\label{sec:relat-deta-balance}
The decomposition of the scattering rate $\mf W^{qq'}_{n\mb k,   n'\mb
  k'}=\sum_\sigma \mf W^{\sigma,qq'}_{n\mb k,   n'\mb k'}$ into odd
and even terms under the ``conjugation'' (in the
sense of detailed balance, i.e.\ thermodynamic equilibrium) of the
associated scattering processes, is also that of its decomposition
into terms, odd and even under the inversion of momentum, in the
presence of time-reversal symmetry. Indeed, if a scattering process
$S= \big ( \overset{q\cdot n\mathbf k}{\underset{q'\cdot n'\mathbf
      k'}\longrightarrow} \big ) $ transfers an energy
  $\delta E_{s \rightarrow p}^{(S)}=q\omega_{n\mb k}+q'\omega_{n'\mb k'}$ to the phonon system,
 (anti-)detailed balance with the ``conjugate process''  ${\rm pc}[S]
 = \big (\overset{-q\cdot n\mathbf k}{\underset{-q'\cdot n'\mathbf k'}\longrightarrow} \big )$ reads
 $\mf W^\sigma({\rm pc}[S]) = \sigma e^{\delta E_{s \rightarrow p}^{(S)}} \mf W^\sigma (S)$.
  Meanwhile, in the presence of time-reversal symmetry, the momentum-reversal symmetry reads
  $\mf W^\sigma ({\rm mr}[S]) = \sigma \mf W^\sigma (S)$, for the ``momentum-reversed'' process
 $ {\rm mr}[S] = \big (\overset{q\cdot n-\mathbf k}{\underset{q'\cdot n'-\mathbf k'}\longrightarrow} \big )$.
  In other words, in a time-reversal
invariant system, the scattering rate associated with the process
``conjugate'' of a given process $S$ coincides (up to a Boltzmann weight) with that of its
momentum-reversed one:
  \begin{equation}
    \label{eq:65}
    \frac{\mf W^\sigma ({\rm mr}[S])}{\mf W^\sigma (S)}
    = e^{-\delta E_{s \rightarrow p}^{(S)}} \frac {\mf W^\sigma ({\rm pc}[S])}{\mf W^\sigma (S)}
    = \sigma.
\end{equation}
Hence, while $\sigma$ was defined as signature of the behavior
of the scattering rates under ``process conjugation,'' it
is {\em also} that of momentum
reversal in a time-reversal invariant system. \footnote{Notice, however, that this coincidence does not survive the breaking of time reversal symmetry.}

\subsubsection{Consequences for the kernels.}
How is this reflected in the kernels $K^L, K^H$?
Because the relation
$\bs v_{n\mb k}=\bs \nabla_{\mb k}\omega_{n\mb k}= -\bs v_{n,-\mb k}$ holds,
only that component of $K^{L/H}$ which is even upon reversal
of the momenta, $\mb k^({}'{}^)\leftrightarrow -\mb k^({} ' {}^)$, has a non-vanishing contribution
to the sum Eq.~\eqref{eq:2}. A first consequence of this is that, in a TR-invariant system, the identity
\begin{equation}
  \label{eq:58}
  K^H_{n\mb k, n'\mb k'}  = - K^H_{n-\mb k, n'-\mb k'}
\end{equation}
entails $\bs \kappa_H=\bs 0$ -- as per Onsager's reciprocity relations
stating that $\bs \kappa_H$ is TR-odd. Note that $K^L$ in Eq.~\eqref{eq:main15} involves both $\mf
W^{\ominus}$ and $\mf W^{\oplus}$. Therefore, there is no analog to Eq.~\eqref{eq:58} for $K^L$. However, in a TR-invariant system, the $\mf W^{\ominus}$ term
in $K^L$ does not contribute to $\kappa_L$ -- this is consistent with the Onsager-Casimir relations
which state that $\kappa_L$ is TR-even.

This indeed reflects the previous discussion as follows: when time reversal is preserved,
TR-even $\kappa_L$ gets contributions solely from
``detailed-balance-even'' and TR-even $\mf W^\oplus$. On the other hand
TR-odd $\bs \kappa_H$ gets contributions solely from
``detailed-balance-odd'' and TR-odd $\mf W^\ominus$. Since the system
is actually TR-even, $\bs \kappa_H$ vanishes.

\subsection{Point-group symmetries}
\label{sec:point-group-symm-1}

Here we provide some sufficient (but non-necessary) conditions on $K^H_{n\mathbf{k}n'\mathbf{k}'}$ under which the Hall conductivity vanishes.

\subsubsection{Curie relations}
\label{sec:symm-kapp-from}

From Fourier's law $j^\mu=-\kappa^{\mu\nu}\nabla_\nu T$, the Curie
  and Onsager relations provide general constraints on the $\kappa^{\mu\nu}$ coefficients, and in turn on its Hall component $\kappa_H^{\mu\nu}$. In Table~\ref{tab:D4h}, we look at the $D_{4h}=D_4\times\mathbb{Z}_2$ point group---the largest tetragonal point group--- with the associated axes aligned with the orthogonal basis $(\mu,\nu,\rho)$ ($\mu\nu$ is the basal plane and $\rho$ the transverse direction). We can see that if the system is invariant under any one of the transformations $g\in D_{4h}$ which are odd under the $A_{2g}$ representation (i.e.\ $C'_2$, $C''_2$, $\sigma_v$, $\sigma_d$), the Hall conductivity must vanish. 

\begin{table}[htbp]
  \begin{tabular}{c|cccccccccc}
    \hline\hline
    $D_{4h}$ & Id & $C_{4}$ & $C_2$ & $C_2'$ & $C_2''$ & inv & $S_4$
    & $\sigma_h$ & $\sigma_v$ & $\sigma_d$ \\
    \hline
    $A_{2g}$ & $1$ & $1$ & $1$ & $-1$ & $-1$ & $1$ & $1$ & $1$ & $-1$
                              & $-1$\\
    \hline
    $\mu$ & $\mu$ & $\nu$ & $-\mu$ & $\pm\mu$ & $\pm\nu$ & $-\mu$ & $\nu$&
                                                                      $\mu$
                 & $\pm\mu$ & $\pm\nu$ \\
    $\nu$ & $\nu$ & $-\mu$ & $-\nu$ & $\mp\nu$ & $\pm\mu$ & $-\nu$ & $-\mu$
    & $\nu$ & $\mp\nu$ & $\pm\mu$ \\
    $\rho$ & $\rho$ & $\rho$ & $\rho$ &$-\rho$ & $-\rho$ & $-\rho$ & $-\rho$ &
    $-\rho$ & $\rho$ & $\rho$ \\
$\kappa^{\mu\nu}$ & $\kappa^{\mu\nu}$ & $-\kappa^{\nu\mu}$ & $\kappa^{\mu\nu}$
& $-\kappa^{\mu\nu}$ & $\kappa^{\nu\mu}$ & $\kappa^{\mu\nu}$ & $-\kappa^{\nu\mu}$& $\kappa^{\mu\nu}$
& $-\kappa^{\mu\nu}$ & $\kappa^{\nu\mu}$\\
$\kappa_H^{\mu\nu}$ & $\kappa_H^{\mu\nu}$ & $\kappa_H^{\mu\nu}$ &  $\kappa_H^{\mu\nu}$
& $-\kappa_H^{\mu\nu}$ &  $-\kappa_H^{\mu\nu}$ & $\kappa_H^{\mu\nu}$ &$\kappa_H^{\mu\nu}$
    & $\kappa_H^{\mu\nu}$ & $-\kappa_H^{\mu\nu}$ & $-\kappa_H^{\mu\nu}$\\
\hline
cat & {\it (a)} & {\it (d)} & {\it (a)} & {\it (b)} & {\it (c)} & {\it (a)} & {\it (d)} & {\it (a)} & {\it (b)} & {\it (c)} \\
    \hline\hline
  \end{tabular}
  \caption{Elements of the $D_{4h}$ point group aligned along the
    $(\mu,\nu,\rho)$ basis (with $\mu\nu$ the basal plane), their characters in
    the $A_{2g}$ irrep (also labeled $\Gamma_2^+$), and transformations of $\mu,\nu,\rho,\kappa^{\mu\nu},\kappa_H^{\mu\nu}$. The
    lines for $\kappa^{\mu\nu}$ and $\kappa_H^{\mu\nu}$ hold true when the system is invariant under the corresponding
    $D_{4h}$ operation (aligned with the $\mu\nu\rho$ basis). The last line is the ``category'' (cat) to which the operation belongs, as defined in Sec.~\ref{sec:symm-kapp-as}. 
  Here Id is the identity; $C_{4}$ is the $\pi/2$ rotation around the
  $\rho$ axis; $C_{2}$, $C'_{2}$ and $C''_{2}$ are $\pi$ rotations around the
  $\rho$ axis, $\mu$ or $\nu$ axes, and in-plane directions bisecting the $\mu,\nu$ axes, respectively;
  inv is inversion, $S_4$ are $\pi/2$ rotations around the $\rho$ axis followed by a reflection
through the basal $\mu\nu$ plane; $\sigma_h$, $\sigma_v$
and $\sigma_d$ are reflections through the $\mu\nu$ plane ($\sigma_h$), through a
plane containing $\mu$ or $\nu$ and the $\rho$ direction ($\sigma_v$), and
through a plane containing the $\rho$ direction and one bissecting the
$\mu,\nu$ directions ($\sigma_d$), respectively.}
  \label{tab:D4h}
\end{table}

\subsubsection{Symmetry relations on $K^H$}
\label{sec:symm-kapp-as}

We now turn to relations specific to the scattering situation, i.e.\ we analyze under which conditions on $K^H_{n\mathbf{k}n'\mathbf{k}'}$ it befalls that $\kappa^{\mu\nu}_H=0$. We start with the expression of
$\kappa^{\mu\nu}_H$ as a momentum integral, Eq.~\eqref{eq:2}, i.e.\ $\kappa_H^{\mu\nu}\propto\sum_{n\mathbf{k}n'\mathbf{k}'}J_{n\mathbf{k}}^\mu J_{n'\mathbf{k}'}^\nu K^H_{n\mathbf{k}n'\mathbf{k}'}$ and recall $J^\mu_{n\mathbf{k}}=N^{\rm eq}_{n\mathbf{k}}\omega_{n\mathbf{k}}\partial_{k^\mu}\omega_{n\mathbf{k}}$.

If the {\em phonon} system is invariant under a unitary transformation $g$, then $\omega_{n\mathbf{k}}$ is also invariant under this transformation. In turn only $\mu$ in $J^\mu_{n\mathbf{k}}$ transforms nontrivially under $g$. Therefore:
\begin{itemize}
    \item {\em If} the phonon system is invariant under an operation $g\in D_{4h}$ which leaves the $\mu,\nu$ axes invariant, i.e.\ $g=C_2,C_2',{\rm inv},\sigma_h,\sigma_v$, and {\em if} one of the two following conditions, {\it (a)} under $g$ the $\mu\nu$ product is even (i.e.\ $g=C_2,{\rm inv},\sigma_h$) and $K^H_{n\mathbf{k}n'\mathbf{k}'}$ is odd, {\it (b)} under $g$ the $\mu\nu$ product is odd (i.e.\ $g=C_2',\sigma_v$) and  $K^H_{n\mathbf{k}n'\mathbf{k}'}$ is even, is satisfied, {\em then} it follows that $\kappa_{\rm H}^{\mu\nu}=0$. 
    \item Besides, recalling that by construction $K^H_{n\mathbf{k}n'\mathbf{k}'}=-K^H_{n'\mathbf{k}'n\mathbf{k}}$, {\em if} the system is invariant under an operation $g\in D_{4h}$ which exchanges the $\mu,\nu$ axes, i.e.\ $g=C_{4},C_2'',S_4,\sigma_d$, and {\em if} one of the two following conditions, {\it (c)} under $g$ the $\mu\nu$ product is even (i.e.\ $g=C_2'',\sigma_d$) and  $K^H_{n\mathbf{k}n'\mathbf{k}'}$ is even, {\it (d)} under $g$ the $\mu\nu$ product is odd (i.e.\ $g=C_4,S_4$) and $K^H_{n\mathbf{k}n'\mathbf{k}'}$ is odd, is satisfied, {\em then} it follows that $\kappa_{\rm H}^{\mu\nu}=0$.
\end{itemize}
In terms of the behavior of $K^H_{n\mathbf{k}n'\mathbf{k}'}$, this analysis reduces to: {\em if} $g\in D_{4h}$ is a symmetry of the phonon system, and {\em if} $g:K^H_{n\mathbf{k}n'\mathbf{k}'}\mapsto-\chi_{A_{2g}}(g) K^H_{n\mathbf{k}n'\mathbf{k}'}$, where $\chi_{A_{2g}}(g)$ is the character of $g$ in the $A_{2g}$ representation of the $D_{4h}$ point group, {\em then} $\kappa_H^{\mu\nu}=0$. We emphasize that this analysis holds {\em if} the transformation $g$ is a symmetry of the phonon system, and whether or not $g$ is a symmetry of the whole system. For example, we will show explicitly in Sec.~\ref{sec:effect-break-symm} that
  there are cases where, under TR or $\sigma_d$ the system is not invariant, but the kernel $K^H_{n\mb k,n'\mb k'}$ and the phonon system are, and so $\kappa_H=0$. 

Finally, note that the above analysis goes beyond the general predictions from Onsager, which tell us {\em that} $\kappa_H$ vanishes
  in the presence of some symmetries of the {\em whole} system, namely $C'_2$, $C''_2$, $\sigma_v$ or $\sigma_d$ (as well as time-reversal discussed in the previous subsection). Here, not only do we establish relations for the other symmetries in $D_{4h}$ (as symmetries of the {\em phonon subsystem only}), we also show {\em in which way} $\kappa_H$ vanishes, by inspecting the
  behavior of the kernels $K^H_{n\mb k,n'\mb k'}$ under those symmetry
  transformations. In turn, this may for example allow to gather information
  about the {\em system}--- about $K^H_{n\mb k,n'\mb k'}$---from the
  (non-)cancellation of $\kappa_H$.

\section{Application to an ordered magnet} 
\label{sec:application}

We now turn to an application of these general results. There, we keep only the
lowest-order terms in the expressions derived above, as described in Sec.~\ref{sec:phys-interpr},
and we consider an interaction hamiltonian density which contains
single-phonon interactions with a field $Q$ (this is the first term in
Eq.~\eqref{eq:10intro}) of the form
\begin{equation}
  \label{eq:10}
  H' = \sum_{n\mb{k}}  \left( a_{n\mb{k}}^\dagger
    Q_{n\mb{k}}^\dagger+a_{n\mb{k}}^{\vphantom\dagger} Q_{n\mb{k}}^{\vphantom\dagger}\right),
\end{equation}
as obtained from the simplest case of linear coupling to the strain tensor.

We consider an ordered magnetic system, which we take to be a spin-orbit coupled 
N\'eel antiferromagnet with tetragonal symmetry.  For
concreteness, we treat the magnetism as purely two-dimensional,
i.e.\ the full spin+phonon system is described by a stack of two
dimensional antiferromagnets embedded into the three-dimension solid,
so that in particular, we take, when going from the lattice to the
continuum limit
\begin{equation} 
  \label{eq:75}
  \sum_\mathbf{r}\rightarrow\frac{1}{\mathfrak{a}^2}\sum_z\int d^2x,\quad
  \sum_\mathbf{k}\rightarrow\frac{\mathfrak{a}^2}{(2\pi)^2}\sum_{k_z}\int d^2k,
\end{equation}
where $\mathfrak{a}$ is the in-plane lattice spacing.

\subsection{Magnon dynamics}
\label{sec:magnon-dynamics}

\subsubsection{Low-energy field-theoretical description} 
\label{sec:low-energy-field}

We consider a N\'eel antiferromagnet with a two-site magnetic unit
cell, more precisely a bipartite lattice of spins such that
the classical ground state is ordered in an antiferromagnetic
configuration, with a local moment $\mu_0$ oriented in the direction
$\mathbf{n}$, i.e.\ $\mathbf{n}$ is the N\'eel vector which has unit
length in the ordered state at zero field.  Within standard
spin-wave theory, $\mu_0=S$ with $S$ the spin value.   For concreteness, we will
choose the ordering axis at zero field to be aligned along the $\hat{\mb u}_x$ axis (the set $(\hat{\mb u}_x,
\hat{\mb u}_y, \hat{\mb u}_z)$ is an orthonormal cartesian basis)---the results of this
subsection hold regardless of this choice.

A general low energy spin configuration is described by two continuum
fields: the aforementioned N\'eel vector $\mb n(\mb r)$ {\em and} a
uniform magnetization {\em density} $\mb m(\mb r)$, such that
\begin{equation}
  \label{eq:109}
   \mb S_{\mb r}  = 
  (-1)^{\mb r} \mu_0 \mb n(\mb r) + \mathfrak{a}^2 \mb m(\mb r). 
\end{equation}
where $(-1)^{\mb r}$ is a sign which alternates between neighboring
sites (recall we are considering a N\'eel antiferromagnet), and both
continuum fields are assumed to be slowly-varying relative to the
lattice spacing.  Here $\mathfrak{a}$ is the 2d lattice spacing.  We
will assume the non-linear sigma model constraint that the spin length
is fixed to $\mu_0$, which implies that
\begin{equation}
  \label{eq:123}
  |\mb n|^2 + \frac{\mathfrak{a}^4}{\mu_0^2} |\mb m|^2 =
  1, \qquad \mb m \cdot \mb n=0.
\end{equation}
The spin wave expansion consists of expanding these fields around the
zero field ordered state, i.e.\ $\mb n_{\rm ord} = \hat{\mb u}_x,
\mb m_{\rm ord}=\bs 0$.   To linear order around this state, we take
$\mb n = \hat{\mb u}_x+ \bs n$ and $\mb m = \bs m$, where $n_x=m_x=0$,
leaving the remaining degrees of freedom $n_y,n_z,m_y,m_z$.
In terms of the spins, this gives
\begin{align}
  \label{eq:main60}
  \mb S_{\mb r} & = 
  (-1)^{\mb r} \mu_0  \hat{\mb u}_x +\sum_{a=y,z}(
 (-1)^{\mb r} \mu_0 n_a(\mathbf{r}) + \mathfrak{a}^2 m_a(\mathbf{r})) \hat{\mb u}_a.
\end{align}
Because the local moment along the $\hat{\mb u}_x$ axis is non-zero, the low
energy fields satisfy the commutation relations
$[m_y (\mb r),n_z(\mb r')] = -[m_z (\mb r),n_y(\mb r')]=- i\delta(\mb r - \mb r')$. 
The low energy continuum Hamiltonian density for these fields is
\begin{eqnarray}
\mathcal{H}_{\rm NLS} &=& \frac{\rho}{2} \left ( |\ul{\bs \nabla} n_y|^2 + |\ul{\bs \nabla}
    n_z|^2\right )\nonumber\\
&&+ \frac 1{2\chi}(m_y^2+m_z^2)+ \sum_{a,b=y,z}\frac
  {\Gamma_{ab}}2 n_an_b , 
\label{eq:hfreeboson}
\end{eqnarray}
where $\rho$ is the spin stiffness constant, $\chi$ is the spin
susceptibility, $\ul{\bs \nabla}=(\partial_x,\partial_y)$ denotes the in-plane gradient, and the
$\Gamma_{ab}$ are anisotropy coefficients which open a
small spin wave gap (see App.~\ref{sec:derivation-gaps-from}).   For
an approximately Heisenberg system with isotropic exchange constant $J$, we have
within spin wave theory that  $\chi^{-1}\approx 4 J {\mf a}^2$, $\rho
\approx 2 J\mu_0^2$, while $\Gamma_{ab}$ are determined by exchange
anisotropies.    The choice to normalize $\bs m$ as a density while
keeping $\bs n$ dimensionless ensures that $m_{y,z}$
fields are just the canonical momenta conjugate to the
$n_{z,y}$ fields, and hence Eq.~\eqref{eq:hfreeboson} is just a
Hamiltonian density of two
free scalar boson fields.  

The above description is appropriate to describe the ordered phase of
the antiferromagnet, for any value of the spin, provided temperature
is low compared to the N\'eel temperature and any applied magnetic fields
are small compared to the saturation field. These conditions are 
well-satisfied in practice in experiments on many antiferromagnets.
Specifically we will be interested in the case with an applied
magnetic field perpendicular to the axis of the N\'eel vector
(e.g.\ along $z$ or $y$, given the choice in Eq.~\eqref{eq:main60}). In general the field induces
a non-zero uniform magnetization along its direction, e.g.\ for a
$z$-axis field $\langle m_z\rangle \neq 0$. Such a
``spin flop'' configuration is favorable for an antiferromagnet in a
field. 

\subsubsection{Symmetry considerations} 
\label{sec:symm-cons}

Two symmetries clarify the calculations and provide physical insight.
The first is the macroscopic time-reversal symmetry of the zero field
state, which is what makes it an {\em anti}-ferromagnet.
Specifically, the system in zero field is invariant under the
combination of time-reversal symmetry TR and a translation
$T$.  Under this
operation, we see that the continuum fields transform according to
\begin{equation}
  \label{eq:154}
  \mathcal{T} = {\rm TR}\times T: \quad m \rightarrow -m, \quad n \rightarrow n.
\end{equation}
The presence of a staggered magnetization (with any orientation) does
not break this symmetry, but a uniform magnetization does. Note that
the {\em effective} quadratic low energy Hamiltonian, 
Eq.~\eqref{eq:hfreeboson}, is invariant under this symmetry.  This is
true even at non-zero fields, because the low energy Hamiltonian is
quadratic.  Thus effects of time-reversal symmetry breaking will
become evident in terms beyond this form, notably in anharmonic
corrections, and in the spin-lattice coupling itself.   Specifically,
we see that time-reversal symmetry will be effectively broken only by
terms involving an odd number of powers of the $m_a$ fields.  

The second 
important symmetry is one which may be preserved
not only by the underlying exchange Hamiltonian and crystal structure,
but {\em also} by the applied field and the spontaneous ordered moments.
In particular, the latter breaks the original translational symmetry
of the square lattice by a single lattice spacing.  However, a
symmetry may be retained under such a simultaneous translation composed
with a $C_2$ spin rotation around the field axis.  In the presence of
spin-orbit coupling, generically the spin rotation must be accompanied
by a spatial rotation, and the full combined operation is in fact
nothing but a $C_2$ rotation about an axis passing through the
mid-point of a bond of the square lattice. This of course requires the $C_2$ rotation in question to be part of the lattice point group. In our problem, this is true when the field is along $z$ or $y$ (but not for a general orientation in the $yz$ plane).

This odd symmetry is important for simplifying the magnon
interactions.  In particular, if the field axis is along $z$, then we
see that $m_z$ and $n_y$ are both even under this operation, while
$m_y$ and $n_z$ are odd under it (and vice versa if the field is along $y$).  Note that the fields within a
canonically conjugate pair transform the same way under this symmetry.
We take advantage of these facts in the following.
In particular, only $\Gamma_0=\Gamma_{yy}$ and $\Gamma_1 = \Gamma_{zz}$
do not vanish a priori, which ensures that the two valleys ($\ell=0,1$) are exactly decoupled.

\subsubsection{External magnetic field}
\label{sec:extern-magn-field}

At the lowest order, an applied external magnetic field $\mb h$ couples solely to the
$\bs m$ field; this is already taken into account in Eq.\,\eqref{eq:hfreeboson}
where the $\bs m$ fields can acquire a (static) nonzero expectation value
due to the spin alignment with the field. 

Meanwhile, at higher orders the magnetic field also couples to the $\bs n$ field; the main
contribution comes from the square, isotropic coupling $(\bs n \cdot \mb h)^2$.
Due to the $C_2$ symmetry around the field axis ($y$ or $z$), and since first-order terms
of the form $h_a n_b$ are forbidden by translational symmetry, this results in an additional
term
\begin{equation}
  \label{eq:32}
  \mc H_{\rm field} =  \frac \chi 2 \sum_{a=y,z}h_a^2 n_a^2.
\end{equation}
Note this form is valid only when the field is along the $y$ or $z$ axis, not at other angles in the $y-z$ plane (which would violate the $C_2$ symmetry).
The prefactor $\chi/2$ is fixed to match the results obtained from
microscopic calculations in Ref.~\cite{PhysRevB.74.024415}, and we provide an alternative derivation
in App.~\ref{sec:derivation-gaps-from} as well as a more detailed
derivation of the full form of the gap from a microscopic XXZ exchange
model plus a Zeeman coupling to the field in App.~\ref{sec:derivation-gaps-from}.

\subsubsection{Diagonalization}
\label{sec:diagonalization}

We proceed to diagonalize Eq.~\eqref{eq:hfreeboson}, supplemented
  by Eq.~\eqref{eq:32} following the discussion in Sec.~\ref{sec:extern-magn-field}, by
introducing creation and annihilation operators in the standard way
for free fields.  We use the Fourier convention $\phi_{\mb k} =
\frac{1}{\sqrt{V}} \int d{\mb x} \phi({\mb x})e^{-i\mathbf{k}\cdot\mathbf{x}}$ for any continuum
field $\phi$,  where $V$ is the volume of the system.  Then
\begin{align}
  \label{eq:126}
  m_{\mb k}^y & = \sqrt{\frac{\chi\Omega_{{\mb k},0}}{2}} \left(b_{-{\mb
                k},0}^{\vphantom\dagger} + b_{{\mb k},0}^\dagger
                \right), \nonumber \\
  n_{\mb k}^z & = i\frac{1}{\sqrt{2\chi\Omega_{{\mb k},0}}}  \left(b_{-{\mb
                k},0}^{\vphantom\dagger} - b_{{\mb k},0}^\dagger
                \right), \nonumber \\
  m_{\mb k}^z & = \sqrt{\frac{\chi\Omega_{{\mb k},1}}{2}}\left(b_{-{\mb
                k},1}^{\vphantom\dagger} + b_{{\mb k},1}^\dagger 
                \right), \nonumber \\
  n_{\mb k}^y & =-i\frac{1}{\sqrt{2\chi\Omega_{{\mb k},1}}} \left(b_{-{\mb
                k},1}^{\vphantom\dagger} - b_{{\mb k},1}^\dagger
                \right),
\end{align}
where
\begin{align}
  \label{eq:127}
  \Omega_{{\mb k},\ell} = \sqrt{v_{{\rm m}}^2\ul{\mathbf{k}}^2 + \Delta_\ell^2},
\end{align}
with $v_{\rm m}=\sqrt{\rho/\chi}$. The magnon gaps depend on the applied (transverse) magnetic field
in the form
\begin{equation}
  \label{eq:57} 
  \Delta_\ell = \sqrt{\Gamma_\ell /\chi + h_\ell^2},
\end{equation}
with valley index $\ell=0,1$ and where we set $h_0=h_y$ and $h_1 = h_z$.
This reflects the explicit breaking of $O(3)$ rotational symmetry of
the order parameter $\bs n$ by the transverse field. With these definitions, we
obtain
\begin{align}
  \label{eq:140}
  H_{\rm NLS}+H_{\rm field} = \sum_\ell \sum_{\mb k} \Omega_{{\mb k},\ell} b_{{\mb
  k},\ell}^\dagger b_{{\mb k},\ell}^{\vphantom\dagger}.
\end{align}
The $b,b^\dagger$ fields with index $\ell=0$ have
opposite $C_2$ eigenvalue to those with $\ell=1$.  This guarantees
that all terms preserving $C_2$ symmetry must conserve the two boson
flavors modulo 2.

\subsection{Formal couplings}
\label{sec:formal-couplings}

\subsubsection{Definitions}
\label{sec:definitions}

In general we can expand the operator $Q_{n\mb k}$, which couples to
a single phonon, in powers
of the magnon operators,
\begin{align}
  \label{eq:141}
&  Q^q_{n\mathbf k}  =  \sum_{\ell,q_1,z} \mathcal A^{n,\ell | q_1 q}_{\mb
                       k} \, e^{ik_z z} b^{q_1}_{\ell,{\mb k},z}\\
  & + \frac{1}{\sqrt{N_{\rm uc}}}\sum_{\substack{\mathbf p,\ell,\ell'\\q_1,q_2,z} } 
  \!\!\!\mathcal B^{n,\ell_1,\ell_2|q_1 q_2 q}_{\mb k;\mb p} ~ e^{ik_z z} b^{q_1}_{\ell_1,\mathbf
                                                          p+\frac{q}{2}\mb
                                                          k,z}
                                                          b^{q_2}_{\ell_2,-\mb
                                                          p
                                                          +\frac{q}{2}\mb k,z} .\nonumber
\end{align}
Note that while the phonons are three-dimensional excitations, and
hence have a three-dimensional momentum $\mb k$, the spin operators
(and hence magnons) only have two dimensional momenta. We will make
use of the following: $\mb k = \ul{\mb k}+k_z\mathbf{\hat{u}}_z$,
  where $\ul{\mb k}$ is the projection of $\mb k$ onto the $k_z=0$
  plane and $\mathbf{\hat{u}}_z$ is the unit vector along $z$.  A phonon is
coupled to the sum of spin operators in all layers---we have here
introduced the explicit label $z$ for the layer.   Because the
spins in different layers are completely uncorrelated, there are
however no cross-terms involving $b$ operators from different layers,
and in correlation functions the sums over $z$ will collapse to
independent correlators within each layer, which are all identical to
one another.  When possible, we will therefore take $z=0$ and suppress
this index. 

The na\"ive leading term in Eq.~\eqref{eq:141} is the single
magnon one $\mathcal A$, linear in $b_{\ell,\mb k}^{\vphantom\dagger}$ and $b_{\ell,\mb
  k}^\dagger$ operators (notations defined below).  This results in a quadratic mixing term in
the Hamiltonian, hybridizing phonons and magnons.  Being quadratic, it
is trivially diagonalized, and has been considered by several
authors.  Generally, such coupling has little effect except when it
is resonant, i.e.\ near a crossing point of the decoupled magnon and
phonon bands.  Since such a crossing is highly constrained by momentum
and energy matching, it occurs in a narrow region of phase
space, if at all, and is likely to be unimportant for transport.  It
in any case does not give rise to scattering, the focus of this work.
We therefore henceforth neglect the $\mathcal A$ contribution.

Non-trivial scattering processes arise from the second order term in
the magnon field expansion of $Q_{n\mb k}$, parametrized by $\mathcal B$.     
Here as elsewhere we introduce particle-hole indices $q_1,q_2\in
{+,-}$, such that in particular
\begin{align}
  \label{eq:146}
  b_{\ell,\mb p,z}^+ = b^\dagger_{\ell,\mb p,z}, &&  b_{\ell,\mb p,z}^- = b^{\vphantom\dagger}_{\ell,-\mb p,z}.
\end{align}
Notice the minus sign in the momentum in the second relation.  This
means generally that
\begin{equation}
  \label{eq:147}
  \left(b^q_{\ell,\mb p,z}\right)^\dagger = b^{-q}_{\ell,-\mb{p},z}.
\end{equation}
To make the
coefficients unambiguous, we choose the symmetrized form
\begin{equation}
  \label{eq:144}
  \mathcal B^{n,\ell_1,\ell_2|q_1 q_2 q}_{\mb k;\mb p} = \mathcal
  B^{n,\ell_2,\ell_1|q_2 q_1 q}_{\mb k;-\mb p} .
\end{equation}
Demanding that $Q^+_{n\mb k} = (Q_{n\mb k}^-)^\dagger$ implies that
\begin{equation}
  \label{eq:145}
   \mathcal B^{n,\ell_1,\ell_2|q_1 q_2 +}_{\mb k;\mb p} = \left( \mathcal B^{n,\ell_2,\ell_1|-q_2 -q_1 -}_{\mb k;\mb p}\right)^*.
\end{equation}
If the phonon mode $n$ which $Q^q_{n{\mathbf k}}$ is coupled to is
$C_2$ invariant, then only terms with $\ell_1=\ell_2$ are non-zero.
In Sec.~\ref{sec:hamiltonian-density}, we will introduce a concrete and general model of spin-lattice couplings, and see that within this model, almost all interactions obey this selection rule.  In particular, off-diagonal terms with $\ell_1\neq \ell_2$ arise only from the $\Lambda_{6,7}^{(\xi)}$ couplings defined in Eq.\,\eqref{eq:134}, which are furthermore smaller in magnitude than other couplings as they are related to magnetic anisotropy.

\subsubsection{Diagonal scattering rate}
\label{sec:diag-scatt-rate}

Contributions to the first-order longitudinal scattering rate, Eq.~\eqref{eq:84},
can be computed exactly using Wick's theorem.  To do so we use the
free particle two point function, which in the notation of
Eq.~\eqref{eq:146} is
\begin{eqnarray}
  \label{eq:148}
  \left\langle b^{q_1}_{\ell_1,\mb p_1,z_1}(t_1) b^{q_2}_{\ell_2,\mb
    p_2,z_2}(t_2)\right\rangle =
& \delta_{\ell_1,\ell_2}\delta_{z_1,z_2}\delta_{q_1,-q_2}\delta_{\mb p_1 + \mb p_2,\mb
  0}  \nonumber\\
 &\times f_{q_2}(\Omega_{\ell_1,\mb p_1}) e^{-i q_2 \Omega_{\ell_2,\mb p_2}(
  t_1 -t_2)},\nonumber\\
\end{eqnarray}
where $f_q(\Omega) = (1+q)/2 +n_{\rm B}(\Omega)$, where $n_{\rm B}(\Omega)$ is the Bose
distribution. One obtains two contributions,
$D^{(1)}_{n\mb k}=\sum_{s=\pm}D^{(1)|s}_{n\mb k}$, where $D^{(1)|+}$
corresponds to the emission of
two magons and $D^{(1)|-}$ corresponds to the scattering of a magnon
from one state to another: 
 \begin{eqnarray}
   \label{eq:122}
   D^{(1)|+}_{n\mathbf k}&&=\frac{2\pi}{\hbar^2} 
                          \frac{1}{N^{\rm 2d}_{\rm uc}}\sum_{\mathbf {p},\ell,\ell'}  \frac{\sinh(\tfrac
    \beta 2 \hbar\omega_{n\mathbf k})}
           {\sinh(\tfrac
    \beta 2 \hbar\Omega_{\ell,\mb p -\frac{\mb k}{2}})\sinh(\tfrac
    \beta 2 \hbar\Omega_{\ell',-\mb p - \frac{\mb k}{2}})} \nonumber \\
&&
\times\delta(\omega_{n\mathbf k}-\Omega_{\ell,\mb p -\frac{\mb k}{2}}-
           \Omega_{\ell',-\mb p - \frac{\mb k}{2}})
     \left |\mathcal B^{n,\ell,\ell'|++-}_{\mathbf k;\mb p}
    \right |^2, 
 \end{eqnarray}
 and
 \begin{eqnarray}
   \label{eq:143}
 {D}^{(1)|-}_{n\mathbf k}&&=\frac{4\pi}{\hbar^2} 
                          \frac{1}{N^{\rm 2d}_{\rm uc}}\sum_{\mathbf {p},\ell,\ell'}  \frac{\sinh(\tfrac
    \beta 2 \hbar\omega_{n\mathbf k})}
           {\sinh(\tfrac
    \beta 2 \hbar\Omega_{\ell,\mb p -\frac{\mb k}{2}})\sinh(\tfrac
    \beta 2 \hbar\Omega_{\ell',\mb p + \frac{\mb k}{2}})}  \nonumber \\
   &&
\times\delta(\omega_{n\mathbf k}-\Omega_{\ell,\mb p -\frac{\mb k}{2}}+
           \Omega_{\ell',\mb p + \frac{\mb k}{2}})
     \left |\mathcal B^{n,\ell,\ell'|+--}_{\mathbf k;\mb p}
    \right |^2. 
 \end{eqnarray}
 Note that the prefactor involves just the number of two-dimensional
 unit cells in a single layer, $N_{\rm uc}^{\rm 2d} = N_{\rm uc}/N_{\rm
   layers}$, which results because a single sum over $z$ gives a
 factor of the number of layers $N_{\rm layers}$, converting the
 $N_{\rm uc}$ to $N_{\rm uc}^{\rm 2d}$.  One can compare the expressions
 in Eq.~\eqref{eq:122} and Eq.~\eqref{eq:143}, and observe a
 difference of  a factor $2$ in the prefactor, the sign of the second
 $\Omega$ frequency in the delta function, and that of the second to
 last index in $\mathcal{B}$. The squared modulus $|\cdots|^2$ can be
traced back to Fermi's golden rule, and the thermal $\rm
sinh(\cdots)$ factors, which originate from Bose factors,
fall off exponentially at large momenta. Energy 
conservation imposed by the delta functions strongly constrain these
scattering rates.  Specifically, if all magnons have the same velocity
$v_{\rm m}$ and the phonons have an isotropic velocity $v_{\rm ph}$, then we find that
\begin{align}
  \label{eq:53}
& \textrm{supp}\left( D_{n\mb{k}}^{(1)|+}\right)
  \subseteq\left  \{ (\underline{\mb{k}},k_z) \big | (v_{\rm ph}^2-v_{\rm
                             m}^2)|\underline{\mb{k}}|^2+ v_{\rm ph}^2 k_z^2>
                             4\Delta^2\right  \} \nonumber
  \\
                &             \textrm{supp}\left(D_{n\mb{k}}^{(1)|-}\right)
                             \subseteq \left  \{  (\underline{\mb{k}},k_z)\big  | (v_{\rm
                             ph}^2 - v_{\rm
                             m}^2)|\underline{\mb{k}}|^2+ v_{\rm
                  ph}^2 k_z^2<0 \right \}  ,
\end{align}
where $\Delta = {\rm min}(\Delta_0,\Delta_1)$ and ${\rm supp}(D)$ is
the support of $D$. 
It follows that if $v_{\rm m}>v_{\rm ph}$,  $D_{n\mb{k}}^{(1)|+}$ is non-zero in two
regions of large $|k_z|$ bounded by hyperboloid surfaces
  tangent to the $\left \{ \frac{k_z}{|\underline{\mb{k}}|}=
  \frac{\sqrt{v_{\rm m}^2 - v_{\rm ph}^2}}{v_{\rm ph}} \right \}$ cone, while $D_{n\mb{k}}^{(1)|-}$ is
non-zero in the region outside the said cone,
containing large $|\underline{\mb{k}}|$.  The two
regions are mutually exclusive, i.e.\ for any given $\mb{k}$ at most
one of the two rates is non-zero.  For $v_{\rm m}<v_{\rm ph}$, the
constraints are even stronger, and $D_{n\mb{k}}^{(1)|-}=0$ strictly
vanishes, while $D_{n\mb{k}}^{(1)|+}$ is non-zero within an ellipsoid
region containing $\mb{k}=\bs 0$.  The first and
second scenarios are realized in La$_2$CuO$_4$
\cite{bazhenov1996}, and in, e.g., FeCl$_2$ \cite{PhysRevB.8.2130},
respectively.

\subsubsection{Off-diagonal scattering rate}
\label{sec:diag-scatt-rate-1}

Expanding each $Q$ operator in terms of magnon operators in the
four-point correlations, i.e.\ plugging in Eq.~\eqref{eq:141} into
Eq.~\eqref{eq:101}, one can obtain the Hall scattering rate using
Wick's theorem.   We find 
\begin{widetext}
\begin{align}
  \label{eq:110}
  \mathfrak{W}^{\ominus,qq'}_{n\mathbf{k},n'\mathbf{k}'}
   & = \frac{64\pi^2}{\hbar^4}\frac{1}{N_{\rm uc}^{2d}} \sum_{\mathbf{p}}\sum_{\{\ell_i,q_i\}} \mathfrak{D}_{q\mathbf{k}q'\mathbf{k}',\mathbf{p}}^{nn'|q_1q_2q_3,\ell_1\ell_2\ell_3}~\mathfrak{F}_{q\mathbf{k}q'\mathbf{k}',\mathbf{p}}^{q_1q_2q_4,\ell_1\ell_2\ell_3} ~ \mathfrak{Im} \Bigg\{ {\mathcal
    B}^{n\ell_2\ell_3|q_2q_3q}_{\mb k,\mb p +\frac{1}{2}q\mb k + q'\mb
    k'} {\mathcal
    B}^{n'\ell_3\ell_1|-q_3q_1q'}_{\mb k',\mb p+ \frac{1}{2}q'\mb k'}
    \nonumber \\
  & \times \textrm{PP} \Big[\frac{{\mathcal B}^{n\ell_1\ell_4|-q_1q_4 -q}_{\mb k,\mb p +
    \frac{1}{2}q\mb k} {\mathcal B}^{n'\ell_4\ell_2|-q_4-q_2-q'}_{\mb
    k',\mb p + q\mb k +\frac{1}{2}q'\mb k'}}{\Delta_{n\mathbf{k}n'\mathbf{k}'}^{qq'}+ q_1
    \Omega^{\ell_1,-q_1}_{\mb p}-q_2\Omega^{\ell_2,q_2}_{\mb p + q\mb k+q'\mb
    k'}-2q_4 \Omega^{\ell_4,-q_4}_{\mb p +q\mb k}}
    +\frac{{\mathcal B}^{n'\ell_1\ell_4|-q_1-q_4 -q'}_{\mb k',\mb p +
    \frac{1}{2}q'\mb k'} {\mathcal B}^{n\ell_4\ell_2|q_4-q_2-q}_{\mb
    k,\mb p + \frac{1}{2}q\mb k +q'\mb k'}}{\Delta_{n\mathbf{k}n'\mathbf{k}'}^{qq'}- q_1
    \Omega^{\ell_1,-q_1}_{\mb p}+q_2\Omega^{\ell_2,q_2}_{\mb p + q\mb k+q'\mb
    k'}-2q_4 \Omega^{\ell_4,q_4}_{\mb p +q'\mb k'}}\Big]\Bigg\},
\end{align}
\end{widetext}
where we defined $\Omega_{\mb p}^{\ell,q}=\Omega_{\ell,q\mb p}$,
and the product of delta functions $\mathfrak{D}$ and
`thermal factor' $\mathfrak{F}$
\begin{eqnarray}
  \label{eq:17}
  &&\mathfrak{D}_{q\mathbf{k}q'\mathbf{k}',\mathbf{p}}^{nn'|q_1q_2q_3,\ell_1\ell_2\ell_3}
  =\delta\left(
    \Sigma_{n\mathbf{k}n'\mathbf{k}'}^{qq'}+q_1\Omega^{\ell_1,-q_1}_{\mb
     p}+q_2\Omega^{\ell_2,q_2}_{\mb p +q\mb k +  q'\mb  k'}\right)\nonumber\\
 && \delta
       \left(\Delta_{n\mathbf{k}n'\mathbf{k}'}^{qq'}+2q_3\Omega^{\ell_3,-q_3}_{\mb p+q'\mb  k'}
      -q_1\Omega^{\ell_1,-q_1}_{\mb   p}+q_2\Omega^{\ell_2,q_2}_{\mb   p+q\mb k +  q'\mb  k'}\right),\nonumber\\
 && \mathfrak{F}_{q\mathbf{k}q'\mathbf{k}',\mathbf{p}}^{q_1q_2q_4,\ell_1\ell_2\ell_3}
  =q_4 \left(2n_{\rm B}(\Omega^{\ell_3,-q_3}_{\mb p+q'\mb
    k'})+1\right) \nonumber\\
 &&   \left(2n_{\rm B}(\Omega^{\ell_1,-q_1}_{\mb p})+q_1+1\right)\left(2n_{\rm B}(\Omega^{\ell_2,q_2}_{\mb p+q\mb k +q'\mb
    k'})+q_2+1\right),\nonumber\\
\end{eqnarray}
and
$\Sigma^{q,q'}_{n\mathbf{k}n'\mathbf{k}'}=q\omega_{n\mathbf{k}}+q'\omega_{n'\mathbf{k}'}$,
$\Delta^{q,q'}_{n\mathbf{k}n'\mathbf{k}'}=q\omega_{n\mathbf{k}}-q'\omega_{n'\mathbf{k}'}$.
Note that while we described and will use below a continuum formulation of the spin wave theory in Sec.~\ref{sec:magnon-dynamics}, the result in Eq.~\eqref{eq:110} is actually valid at the lattice level, i.e.\ when the full periodic band structure of the magnons is included, as it relies only upon the canonical commutation relations of the magnon operators, and their dispersions and couplings are taken completely arbitrary at this stage.  Therefore this formula could be applied directly in many other circumstances.

We may understand the terms in Eq.~\eqref{eq:110} as follows: the second energy
  conservation delta function comes from Fermi's golden rule; the first
  delta function, and the denominator in the third line, come from $\frac 1 {E_{\mathtt i}-E_{\mathtt n}+i\eta}={\rm PP} \frac 1 {E_{\mathtt i}-E_{\mathtt n}} -i\pi\delta(E_{\mathtt
    i}-E_{\mathtt n})$; while the Bose factors appear
    when evaluating the thermal averages of magnon population numbers, and their product falls off exponentially at large
    momenta. $\mathfrak W_{n\mathbf k n'\mathbf k'}^{\ominus,qq'}$ may
    display divergences when the denominator vanishes. One can
    explicitly check that the detailed balance relation,
    Eq.~\eqref{eq:detbal}, holds, using the properties of the
    $\mathcal{B}$ coefficients, as well as
    $\mathfrak{W}^{\ominus,qq'}_{n\mathbf{k},n'\mathbf{k}'}= \mathfrak{W}^{\ominus,q'q}_{n'\mathbf{k'},n\mathbf{k}}$.

\subsection{Phenomenological coupling Hamiltonian}
\label{sec:phen-coupl-hamilt}

We now propose a symmetry-based phonon-magnon coupling Hamiltonian,
Eq.~\eqref{eq:133}, for the low-temperature ordered phase of a N\'eel
antiferromagnet on lattice made of layers of square lattices, and, as above, we
consider the layers to be {\em magnetically} decoupled. Moreover, for
concreteness, we take the classical ground state to be N\'eel antiferromagnetic
along the $\mb{\hat{u}}_x$ axis, so that all the
point-group symmetries of the crystal are preserved by the magnetic
structure, up to a translation of half a magnetic unit
cell. \footnote{In the absence of a magnetic field, and an ``alternating'' Dzyaloshinskii-Moriya (DM) interaction which we do not consider here.}

\subsubsection{Interaction Hamiltonian density}
\label{sec:hamiltonian-density}

We consider the most general coupling between {\em (1)}
the strain tensor, $\mathcal{E}^{\alpha\beta}=\tfrac 1 2(\partial^\alpha u^\beta
+ \partial^\beta u^\alpha)$, where $\bs u$ is the lattice displacement
field, and {\em (2)} spin bilinears in terms of the $\mb m, \mb n$ fields,
allowed by the symmetries of our tetragonal crystal in its
paramagnetic phase, which has the largest
symmetry group provided by the crystal structure (generated by mirror symmetries
$\mathcal S_x,\mc S_y, \mc S_z$, fourfold rotational symmetry $\mc
C_4^{xy}$, translation and time-reversal).   Since we treat the
magnetism as two dimensional, the coupling Hamiltonian is a sum over
layers and an integral over two dimensional space,
\begin{equation}
  \label{eq:22}
  H'_{\rm tetra} = \sum_z \int \! d^2\mb{x}\, \mathcal{H}'_{\rm tetra}(\mb{r}).
\end{equation}
We use $\mb{r}=(\mb{x},z)$ to denote the three-dimensional coordinate.
The corresponding local 
hamiltonian density reads, with all fields expressed in real space:
\begin{align}
  \label{eq:133}
  & \mathcal{H}'_{\rm tetra}(\mb{r})
    =\\
  & \sum_{\substack{\alpha,\beta \\a,b=x,y,z}}\!\!\!\mathcal{E}^{\alpha\beta}_{\mb{r}}\left.\left(\Lambda_{ab}^{({\rm
      n}),\alpha\beta}{\rm n}_a{\rm  n}_b+\frac{\Lambda_{ab}^{({\rm
        m}),\alpha\beta}}{n_0^2}{\rm
      m}_a{\rm m}_b\right)\right|_{\mb{x},z},\nonumber
\end{align}
where $n_0 = \mu_0/\mathfrak{a}^2$ is the ordered moment density.  Here each $\Lambda^{(\xi)}$ tensor, which we define to be
symmetric in both $ab$ and $\alpha\beta$ variables, has seven independent coefficients, which we
call 
\begin{eqnarray}
  \label{eq:134}
    \Lambda_1^{(\xi)}&=&\Lambda_{xx}^{(\xi),xx}=\Lambda_{yy}^{(\xi),yy}, \nonumber\\
  \Lambda_2^{(\xi)}&=&\Lambda_{yy}^{(\xi),xx}=\Lambda_{xx}^{(\xi),yy},\nonumber\\
\Lambda_3^{(\xi)}&=&\Lambda_{zz}^{(\xi),xx}=\Lambda_{zz}^{(\xi),yy},\nonumber\\
  \Lambda_4^{(\xi)}&=&\Lambda_{xx}^{(\xi),zz}=\Lambda_{yy}^{(\xi),zz},\nonumber\\
  \Lambda_5^{(\xi)}&=&\Lambda_{zz}^{(\xi),zz},\nonumber\\
  \Lambda_6^{(\xi)}&=&\Lambda_{xy}^{(\xi),xy}=\Lambda_{xy}^{(\xi),yx}=\Lambda_{yx}^{(\xi),yx}=\Lambda_{yx}^{(\xi),xy},\nonumber\\
  \Lambda_7^{(\xi)}&=&\Lambda_{xz}^{(\xi),xz}=\Lambda_{xz}^{(\xi),zx}=\Lambda_{zx}^{(\xi),zx}=\Lambda_{zx}^{(\xi),xz},\nonumber\\
&=&\Lambda_{yz}^{(\xi),yz}=\Lambda_{yz}^{(\xi),zy}=\Lambda_{zy}^{(\xi),zy}=\Lambda_{zy}^{(\xi),yz},
\end{eqnarray}
and all other $\Lambda^{(\xi),\alpha\beta}_{ab}$ are zero.

In Appendix~\ref{sec:micr-deriv-coupl}, we provide a microscopic derivation
of these coupling constants starting from a spin hamiltonian
  $H_{\rm spin}=\sum_{\mb r,\mb r'}S^a_{\tilde{\mb r}}J^{ab}(\tilde {\mb r}-\tilde {\mb r}')S^b_{\tilde{\mb r}'}$
  on the \emph{distorted} lattice, with $\tilde {\mb r}=\mb r + \bs
  u(\mb r)$ ($J^{ab}$ is the exchange parameter between the $a$ and
  $b$ spin components, and depends a priori exponentially on the
  distance between the two sites).
  Expanding of the magnetic exchange at linear order in the
  displacement $\bs
  u(\mb r)$ (away from the position $\mathbf{r}$ of the
  atoms in the absence of a phonon) results in a magnetoelastic coupling
of the form Eq.~\eqref{eq:133}, with coefficients $\Lambda_{ab}^{(\xi),\alpha\beta}$ expressed
in terms of spatial derivatives of the magnetic exchange $J^{ab}$.

Within this microscopic approach $\Lambda_{6,7}^{(\xi)}$ are related to the spatial
derivatives of symmetric off-diagonal exchange $J^{xy},J^{xz},J^{yz}$,
while $\Lambda_{1,2}^{(\xi)}-\Lambda_3^{(\xi)}$ and
$\Lambda_4^{(\xi)}-\Lambda_5^{(\xi)}$ are associated with the spatial
derivatives of XXZ exchange anisotropy $J^{xx,yy}-J^{zz}$.
Finally, note that in Eq.~\eqref{eq:133} bilinears of the ${\rm m}_a {\rm n}_b$ kind, arising from e.g.\ alternating DM
interactions $J^D_{ij}$ i.e.\ such that $J^D_{\mb r,\mb r+\hat{\mb u}_a}=-J^D_{\mb r-\hat{\mb u}_a,\mb r}$ with $a=x,y$, could also
contribute to the thermal Hall conductivity
\cite{PhysRevLett.123.167202}, but are not allowed in the single-site
(paramagnetic) Bravais lattice we consider here.

\subsubsection{Expansion}
\label{sec:expansion}

We now carry out an expansion of the $\mb{m,n}$ fields in two
steps. First we expand around the {\em zero-field}, {\em zero-net-magnetization} N\'eel-ordered
configuration ($\mb n_{\rm ord} = \hat{\mb u}_x, \mb m_{\rm
  ord}=\bs 0$), assuming deviations are small and satisfy
Eq.~\eqref{eq:123}.  One thereby expresses ${\rm n}_x$ and ${\rm m}_x$
in terms of the free fields ${\sf m}_{y/z},{\sf n}_{y/z}$ as, in real space:
\begin{align}
  \label{eq:129}
  {\rm n}_x & = 1 - \frac 1 2 \sum_{b=y,z}\left({\sf
              n}_b^2+\frac 1 {n_0^2}{\sf
              m}_b^2\right),\nonumber \\
  {\rm m}_x& = -\sum_{b=y,z} {\sf m}_b{\sf n}_b,
\end{align}
which are correct to second order in the free fields (this constitutes
a non-linear correction to Eq.~\eqref{eq:main60}).
In a second step, we include a net magnetization and expand
$\vec{{\sf m}}$ around it, i.e.\ write
${\sf m}^\alpha = m^\alpha_0 + m^\alpha$ where $m^\alpha_0$ is the sum
of both a possible spontaneous magnetization and response to the external
magnetic field. This two-step expansion physically assumes
$m\ll m_0\ll n_0$.   Using these forms in
Eq.~\eqref{eq:133}, we obtain the spin-lattice coupling to second order
in the free field fluctuations:
\begin{equation}
  \label{eq:142}
  \mathcal{H}'_{\rm
    tetra}(\mathbf{r})\approx\sum_{\alpha\beta}\mathcal{E}^{\alpha\beta}_{\mathbf{r}}
  \sum_{a,b=y,z}\sum_{\xi,\xi'=0,1}\lambda^{\alpha\beta}_{ab;\xi\xi'} n_0^{-\xi-\xi'}\eta_{a\xi{\mathbf{r}}}\eta_{b\xi'\mathbf{r}},
\end{equation}
where $\eta_{a0}=n_a$ and $\eta_{a1}=m_a$ and with 
\begin{eqnarray}
  \label{eq:15}
  \lambda_{ab;\xi\xi}^{\alpha\beta}&=&\Lambda_{ab}^{(\xi),\alpha\beta}-\delta_{ab}\Lambda_{xx}^{(0),\alpha\beta},\\
  \lambda_{ab;01}^{\alpha\beta}=\lambda_{ba;10}^{\alpha\beta} & = &
                                        \frac{-1}{n_0}\left[
                                        m^a_0 \Lambda^{(1), \alpha\beta}_{bx} +
                                        \delta_{ab} m^{\overline a}_0
                                        \Lambda_{\overline a x}^{(1), \alpha\beta} + m^b_0
                                        \Lambda^{(0), \alpha\beta}_{ax}\right], \nonumber
\end{eqnarray}
where $\overline{y}=z$, $\overline{z}=y$ and we have associated
$\xi={\rm n}\Leftrightarrow \xi=0$ and
$\xi={\rm m}\Leftrightarrow \xi=1$ in $\Lambda^{(\xi)}$.

These relations are satisfied for any
$\Lambda_{ab}^{(\xi),\alpha\beta}$ in Eq.~\eqref{eq:133} (i.e.\ not
necessarily satisfying the constraints Eq.~\eqref{eq:134}), but do
assume a N\'eel moment along the $x$ direction, and a net moment in
the $yz$ plane.  Note that, while the bare (not linearized)
interactions in Eq.~\eqref{eq:133} did not couple the ${\rm n}_a$ and
${\rm m}_b$ fields, such a coupling is present in the linearized
$\lambda_{ab;01}$ coefficient (i.e.\ that coupling $n_a$ and $m_b$).  We can see immediately from Eq.~\eqref{eq:15}
that this coupling 
vanishes in the absence of ``anisotropic''
  couplings $\Lambda_{6,7}^{(\xi)}$. Importantly, it also vanishes in the absence of any uniform
magnetization. This is a consequence of macroscopic time-reversal
symmetry, Eq.~\eqref{eq:154}.  Conversely, $\lambda_{ab;01}$ is the
{\em only} term in our low energy description of the coupled
spin-lattice system which is odd under this effective time-reversal
symmetry.  Consequently, time-reversal odd effects like skew
scattering {\em must} involve at least one factor of this coupling.
This will appear explicitly at the end of the next subsection.

\subsubsection{In terms of the eigenbosons, $b,b^\dagger$}
\label{sec:terms-eigenbosons-b}

We now seek to identify the $\mathcal{B}$ coefficients as defined in
Eq.~\eqref{eq:141} (with the convention Eq.~\eqref{eq:146}). To do so,
we use the Eq.~\eqref{eq:126} representation of the $m_a,n_a$ fields
in terms of the $b$ bosons, which diagonalize the pure magnetic
Hamiltonian, and plug in their expressions into
Eq.~\eqref{eq:142}. This involves a unitary transformation which can
be defined as (using $a=1\Leftrightarrow a=y$ and $a=2\Leftrightarrow a=z$)
\begin{equation}
  \label{eq:35}
  \eta_{a\xi\mathbf{r}}=\sum_\mathbf{p}\sum_{\ell=0,1}\sum_{q=\pm}U_{a\xi\ell q}(\mathbf{p})b^q_{\ell\mathbf{p}}e^{i\mathbf{p}\cdot\mathbf{r}},
\end{equation}
with
\begin{align}
  \label{eq:43a}
      U_{a\xi\ell q}(\mathbf{p})&=-\delta_{a-1,\ell-\overline{\xi}\;{\rm mod}2}F_{\xi q\ell}(\mathbf{p}),\\
        \label{eq:43b}
F_{\xi q\ell}(\mathbf{p})&=(iq)^{\overline{\xi}}(-1)^{\overline{\xi}\ell}(\chi\Omega_{\ell\mathbf{p}})^{\xi-\frac{1}{2}}.
\end{align}
We defined $\overline{\xi}=1-\xi$, i.e.\ $\overline{0}=1,\overline{1}=0$, and
$\widetilde{\xi}=2\xi-1$, i.e.\ $\widetilde{0}=-1,\widetilde{1}=1$.
We also used relation for the valley $\ell=\delta_{a-1,\xi}$, and conversely 
$a=1+\widetilde{\xi}\ell +\overline{\xi}$.
Now inserting this expression into Eq.~\eqref{eq:142}, and collapsing
the $a,b$ sums, we obtain
\begin{eqnarray}
  \label{eq:45}
  \mathcal{H}'_{\rm tetra}&=&\sum_{\alpha\beta} \sum_{\mb{p}_1,\mb{p}_2}
  \mathcal{E}^{\alpha\beta}_{\mathbf{r}} \sum_{q_1q_2} \sum_{\ell_1\ell_2=0,1}
                              \sum_{\xi\xi'=0,1} n_0^{-\xi-\xi'} \lambda_{\ell_1-\bar{\xi},\ell_2-\bar{\xi}';\xi\xi'}^{\alpha\beta}\nonumber\\
  &&
  F_{\xi q_1 \ell_1}(\mb{p}_1)
  F_{\xi' q_2\ell_2}(\mb{p}_2) b_{\ell_1\mb{p}_1}^{q_1} b_{\ell_2\mb{p}_2}^{q_2}e^{i(\mb{p}_1+\mb{p}_2)\cdot\mb{r}}.
\end{eqnarray}

We similarly express the local
strain in terms of its constituent Fourier modes, which are
proportional to the phonon creation/annihilation operators, as
discussed in detail in Appendix~\ref{sec:strain-tensor}.  Putting in
these two ingredients, some algebra (shown also in
Appendix~\ref{sec:strain-tensor}) finally yields, if we define
$\hat{\lambda}_{\xi\xi'}^{\ell\ell';\alpha\beta}=\lambda^{\alpha\beta}_{\ell-\bar{\xi}\;{\rm
  mod}2,\ell'-\bar{\xi}'\;{\rm mod}2;\xi\xi'}$,
\begin{widetext}
\begin{equation}
  \label{eq:119}
   \mathcal{B}^{n,\ell_1\ell_2|q_1q_2q}_{\mathbf{k};\mathbf{p}} =
 \frac{iq}{2\sqrt{2M_{\rm uc}}} 
  \sum_{\xi\xi'} n_0^{-\xi-\xi'} \mathcal{L}_{n\mathbf{k};\xi,\xi'}^{q,\ell_1,\ell_2}\; F_{\xi q_1 \ell_1}\left(\mb{p}+\frac{q}{2}\mb{k}\right)
  F_{\xi' q_2\ell_2} \left(-\mb{p}+\frac{q}{2}\mb{k}\right)
\end{equation}
\end{widetext}
where 
\begin{equation}
  \label{eq:155}
  \mathcal{L}_{n\mathbf{k};\xi,\xi'}^{q,\ell_1,\ell_2}=\sum_{\alpha,\beta=x,y,z}
  \hat{\lambda}^{\ell_1\ell_2;\alpha\beta}_{\xi\xi'}\;
  \frac{k^\alpha(\varepsilon_{n\mathbf{k}}^\beta)^q+k^\beta(\varepsilon_{n\mathbf{k}}^\alpha)^q}
  {\sqrt{\omega_{n\mathbf{k}}}}.
\end{equation}
Eq.~\eqref{eq:119} may now be inserted into Eq.~\eqref{eq:110}.
Note that $i^{\overline{\xi}}$ in $F$
plays an important role as discussed in
  Sec.~\ref{sec:effect-break-symm}. 
  
  Finally, note that the only
coefficients $\lambda_{ab,\xi\xi'}$ which contribute to
$\mathcal{B}^{\ell_1\ell_2}$ {\em with $\ell_1\neq\ell_2$} (i.e.\ to
``intervalley hopping'' recalling $\mathcal{B}^{\ell_1\ell_2}$ is the coefficient of
$b^{q_1}_{\ell_1}b^{q_2}_{\ell_2}$ in $Q^q$) are those
which satisfy $\delta_{ab}+\delta_{\xi\xi'}=1$---see App.~\ref{sec:contrib-to-intervalley} for details. Such coefficients involve only the $\Lambda^{(\xi)}_{6,7}$ couplings, which are typically much smaller than
  $\Lambda^{(\xi)}_{1..5}$. Therefore a good approximation is to consider only those contributions to the
  scattering rates Eqs.~(\ref{eq:122},\ref{eq:143},\ref{eq:110})
  with the smallest possible number of intervalley hoppings.
  Now, the forms $D^{(1)}\sim \mc B^{\ell_1 \ell_2}\mc B^{\ell_2 \ell_1}$
  and $\mf W^\ominus \sim \mc B^{\ell_1 \ell_2} \mc B^{\ell_2 \ell_3} \mc B^{\ell_3 \ell_4} \mc B^{\ell_4 \ell_1} $
  impose that intervalley hopping can only happen an even number of times in $D^{(1)}$ and $\mf W^\ominus$.
 Because $D_{n\mb k}^{(1)}$ is a priori nonzero even when $\Lambda_{6,7}=0$,
  we discard the subdominant, of order $\left (
    \tfrac{\Lambda_{6,7}}{\Lambda_{1..5}} \right)^2$, contributions from $\ell_1\neq\ell_2$ upon
  calculating $D_{n\mb k}^{(1)}$. On the other hand, a
  nonzero
  \begin{equation}
\mf W^{\ominus,{\rm eff},qq'}_{n\mb k,n'\mb k'} := 
\frac 1 2 \left ( \mf W^{\ominus,qq'}_{n\mb k,n'\mb k'}+\mf  W^{\ominus,qq'}_{n-\mb k,n'-\mb k'} \right )
\end{equation}
requires either (or both) nonzero $\Lambda_{6,7}$. The first nonzero term with
  $\ell_1=\ell_2=\ell_3=\ell_4$ in turn occurs at order $\left (
    \tfrac{\Lambda_{6,7}}{\Lambda_{1..5}} \right)^1$, and therefore corrections due to $\ell_i\neq\ell_j$
  are another order $\left ( \tfrac{\Lambda_{6,7}}{\Lambda_{1..5}}
  \right)^1$ smaller for $\mf W^{\ominus,{\rm eff},qq'}_{n\mb k,n'\mb
    k'}$.
We use this approximation in what follows, i.e.\ in
Secs.~\ref{sec:solut-delta-funct} and
\ref{sec:numerical-results}.

\subsubsection{Effective breaking of symmetries}
\label{sec:effect-break-symm}

\paragraph{Time reversal.}
We now briefly comment on the relation between 
the ``effective'' time-reversal of the spin system $\mathcal{T}$ and the transport properties of the
phonon system.
Indeed, it is obvious from Eqs.~\eqref{eq:122} and Eq.~\eqref{eq:110} that if all
the $\mc B$ coefficients satisfy
\begin{equation}
  \label{eq:66}
  \mathcal{B}^{n,\ell_1\ell_2|q_1q_2q}_{-\mathbf{k};-\mathbf{p}}\overset ? =\left
  (\mathcal{B}^{n,\ell_1\ell_2|q_1q_2q}_{\mathbf{k};\mathbf{p}}\right
)^*,
\end{equation}
 then $D^{(1)}_{n-\mb k}=D^{(1)}_{n\mb k}$ and $\mf{W}^{\ominus,qq'}_{n-\mathbf{k},n'-\mathbf{k}'}
=-\mf {W}^{\ominus,qq'}_{n\mathbf{k},n'\mathbf{k}'}$, i.e.\ the phonon
collision integral is effectively time-reversal symmetry preserving, 
as discussed in Sec.~\ref{sec:time-revers-symm}.
Therefore, no phonon Hall effect follows if the spin-phonon coupling satisfies Eq.~\eqref{eq:66}.

Which terms in Eq.~\eqref{eq:142} are compatible with an effective
time-reversal symmetry breaking? 
By direct inspection of Eq.~\eqref{eq:155}, one finds that
$i\mathcal{L}_{n-\mathbf{k};\xi,\xi'}^{q,\ell_1,\ell_2}
= \left (i\mathcal{L}_{n\mathbf{k};\xi,\xi'}^{q,\ell_1,\ell_2} \right
)^*$. Thus, only those terms in Eq.~\eqref{eq:119} with
$i^{\overline{\xi}+\overline{\xi}'}=i$ may satisfy
$\mathcal{B}^{n,\ell_1\ell_2|q_1q_2q}_{-\mathbf{k};-\mathbf{p}}\neq\left
  (\mathcal{B}^{n,\ell_1\ell_2|q_1q_2q}_{\mathbf{k};\mathbf{p}}\right
)^*$. All others are such that $\mathcal{B}^{n,\ell_1\ell_2|q_1q_2q}_{-\mathbf{k};-\mathbf{p}}
=\left(\mathcal{B}^{n,\ell_1\ell_2|q_1q_2q}_{\mathbf{k};\mathbf{p}}\right )^*$.

The breaking of effective time-reversal in the phonon system
 thus relies upon the presence of spin-phonon couplings where
$\xi'=\overline{\xi}$, i.e.\ $\lambda_{ab,01}$ and $\lambda_{ab,10}$
(henceforth denoted ``$\lambda_{nm}$'') coefficients; 
this is consistent with the argument in Sec.~\ref{sec:expansion}, based on macroscopic
time-reversal $\mc T$, Eq.~\eqref{eq:154}. Morevoer, going back to Sec.~\ref{sec:symm-kapp-as}, we see that if $\bs m_0\neq \bs 0$ but
  $\Lambda_{6}^{(\xi)}=0=\Lambda_{7}^{(\xi)}$, then the kernel $K^H_{n\mb k, n'\mb k'}$ is invariant under momentum reversal; and so $\kappa_H=0$,
  even though the system breaks $\mc T$.

\paragraph{$\sigma_d$ operation.}

Here we briefly study the $\sigma_d$ operation, i.e.\ a mirror transformation
  through the plane containing the $\hat z$ and $\tfrac{\hat x+\hat y}{\sqrt 2}$ directions.
  The system, having antiferromagnetic ordering along the $x$ axis as well as possibly $m_0^y\neq 0$,
  explicitly breaks this symmetry.
  However, if $\Lambda_1^{(\xi)}=\Lambda_2^{(\xi)}$ and $\Lambda_7^{(\xi)}=0$, then $\sigma_d$ is preserved
  at the level of the kernel $K^H_{n\mb k, n'\mb k'}$, whence
  $\kappa_H=0$.
  This illustrates the importance of knowing the action of $D_{4h}$ operations
  upon the {\em kernels} $K^H_{n\mb k, n'\mb k'}$, because some symmetries which are
  explicitly broken globally might fail to be {\rm effectively} broken in phonon scattering.

\subsection{Solutions of the delta functions}
\label{sec:solut-delta-funct}

Each contribution to the scattering rate involves a momentum integral
over an integrand which contains either a single delta function or a
product of two delta functions.  These express energy conservation
constraints, which must be solved to carry out the integration.  The
argument of each delta function, which must be set to zero, is of the form
\begin{equation}
  \label{eq:60a}
\varpi - \Omega_{\ell,\mathbf{p}} -s
\Omega_{\ell,\mathbf{p}-\mathbf{k}} = 0,
\end{equation}
where $s=\pm1$. Using the continuum form of the magnon dispersion,
$\Omega_{\ell,\mathbf{p}} = \sqrt{v_{\rm m}^2 |\ul{\mathbf{p}}|^2 +
  \Delta_\ell^2} = v_{\rm
  m}\sqrt{|\ul{\mathbf{p}}|^2+\delta_\ell^2}=v_{\rm m}\hat{\Omega}_{\ell,\mathbf{p}}$, where
$\delta_\ell = \Delta_\ell/v_{\rm m}$, we can rewrite
this as
\begin{equation}
  \label{eq:62a}
  \sqrt{|\ul{\mathbf{p}}|^2+\delta_\ell^2} +s
  \sqrt{|\ul{\mathbf{p}}-\ul{\mathbf{k}}|^2+\delta_\ell^2} = a,
\end{equation}
where $a=\varpi/v_{\rm m}$ and $s=\pm1$.

The existence and type of solutions depend
on the value of $a^2-\ul{k}^2$, where $\ul{k}^2=k_x^2+k_y^2$. When they exist, the
solutions are conics, as is summarized in Table~\ref{tab:solutions}.
\begin{table}[htbp]
  \centering
  \begin{tabular}{c|ccc}
\hline\hline
    & $a^2-\ul{k}^2<0 $ & $\quad 0<a^2-\ul{k}^2<4\delta_\ell^2\quad$ &
                                                                   $4\delta_\ell^2<a^2-\ul{k}^2$ \\
\hline
$s=+$&no solutions & no solutions & ellipse\\
$s=-$&half-hyperbola & no solutions & no solutions\\
\hline\hline
  \end{tabular}
  \caption{Solutions to a single delta function of the form
    $\delta(\varpi-\Omega_{\ell,\mathbf{p}}-s\Omega_{\ell,\mathbf{p}-\mathbf{k}})$,
    with $s=\pm1$, as a function of the value of $a^2-k^2$, where
    $a=\varpi/v_{\rm m}$, $\ul{k}^2=k_x^2+k_y^2$, and
    $\Omega_{\ell,\mathbf{p}}=v_{\rm m}\sqrt{\ul{p}^2+\delta_\ell^2}$. The
    necessary existence conditions described in
    this Table are captured by the equation $s(a^2 - \ul{k}^2) >
    4\delta^2_\ell(s+1)/2$.}
  \label{tab:solutions}
\end{table}

It is then best to introduce coordinates $p_\parallel,p_\perp$ which are
along the principal axes of the hyperbola/ellipse:
\begin{equation}
\label{eq:87}
\ul{\mathbf{p}} = p_\parallel \widehat{\mathbf{\ul{k}}} + p_\perp \hat{\mathbf{z}}\times \widehat{\mathbf{\ul{k}}},
\end{equation}
where we define
$\widehat{\mathbf{\ul{k}}}=(k_x\mathbf{\hat{x}}+k_y\mathbf{\hat{y}})/\ul{k}$
(note the denominator $\ul{k}$ which differs from $k$ when $k_z\neq0$), and we can define the major $\bar{a}$ and minor $\bar{b}$
semi-axes, or conversely, of the conics:
\begin{equation}
  \label{eq:150}
  \bar{a}=\frac{|a|}{2}\sqrt{1-\frac{4\delta_\ell^2}{a^2-\ul{k}^2}},\quad
  \bar{b}=\frac{1}{2}\sqrt{|a^2-\ul{k}^2-\delta_\ell^2|}.
\end{equation}
An immediate consequence is that, in the case of the ellipse
$-\bar{a}\leq p_\parallel-\frac{\ul{k}}{2}\leq\bar{a}$ and $-\bar{b}\leq
p_\perp\leq\bar{b}$, while in the case of the half-hyperbola: $p_\parallel\geq\frac{\ul{k}}{2}+\overline{a}$.

Both Eq.~\eqref{eq:62a} and a pair of such equations may be solved analytically, but the solutions are analytically
complicated. We provide their details in Appendix~\ref{sec:solv-delta-funct}, and give
here only the final results.

\subsubsection{Diagonal scattering rate}

We have, in particular, the following compact form for $D_{n\mathbf{k}}^{(1)|s}$,
with $s=\pm$,
\begin{eqnarray}
  \label{eq:24a}
D^{(1)|s}_{n\mathbf{k}}&=&\frac{(3-s)\pi \mathfrak{a}^2}{(2\pi)^2\hbar^2 } 
                          \sum_{\ell,\ell'}  \int \! d^2\mb{p}\, \frac{\sinh(\tfrac
    \beta 2 \hbar\omega_{n\mathbf k})}
           {\sinh(\tfrac
    \beta 2 \hbar\Omega_{\ell,\mb p })\sinh(\tfrac
  \beta 2 \hbar\Omega_{\ell',\mb p - \mb k})}  \nonumber\\
  &&
\quad\delta(\omega_{n\mathbf k}-\Omega_{\ell,\mb p }-s
           \Omega_{\ell',\mb p - \mb k})
     \left |\mathcal B^{n,\ell,\ell'|+s-}_{\mathbf k;-\mb p+\frac{\mathbf{k}}{2}}
     \right |^2,
\end{eqnarray}
where we converted the {\em two-dimensional} momentum sum to an integral using
$\sum_{\mb{p}} \rightarrow N_{\rm uc}^{\rm 2d} \mathfrak{a}^2 \int \!
\frac{d^2\mb{p}}{(2\pi)^2}$, where $N_{\rm uc}^{\rm 2d}
\mathfrak{a}^2$ is the area of the sample in the $xy$ plane.

From now on, in this paragraph and the following, we make use of the approximation
$\mc B^{\ell \neq\ell'}\rightarrow 0$, as explained previously. Then, collapsing the delta function (to avoid clutter, we identify $y=p_\perp$):
\begin{widetext}
\begin{equation}
  \label{eq:31}
  D^{(1)|s}_{n\mathbf k}= \frac{(3-s) \mathfrak{a}^2\sinh(\tfrac
    \beta 2 \hbar\omega_{n\mathbf k})}{4\pi v_{\rm m}\hbar^2 } 
                          \int_{-\infty}^{+\infty} \text dy\sum_\eta{\rm   f}_\eta^s(y)J_\eta^s(y)
     \sum_{\ell}  \frac{\left |\mathcal B^{n,\ell,\ell|+s-}_{\mathbf k;-\mb p^{(\eta)}_{\ell,\mathbf{k}}(y)+\frac{\mathbf{k}}{2}}
     \right |^2}
           {\sinh(\tfrac
    \beta 2 \hbar\Omega_{\ell,\mb p^{(\eta)}_{\ell,\mathbf{k}}(y) })\sinh(\tfrac
    \beta 2 \hbar\Omega_{\ell,\mb p^{(\eta)}_{\ell,\mathbf{k}}(y) - \mb k})},
\end{equation}
\end{widetext}
where
\begin{equation}
  \label{eq:36}
\begin{cases}
  {\rm f}_\eta^{s=1}(y) = \Theta(\bar{b}-|y|)\,\Theta(a^2-\ul{k}^2-4\delta_\ell^2)\\
  {\rm f}_\eta^{s=-1}(y) = \delta_{\eta,1}\,\Theta(\ul{k}^2-a^2)
\end{cases},
\end{equation}
where $a=\omega_{n\mathbf{k}}/v_{\rm m}$, $\eta=\pm1$ and
\begin{equation}
  \label{eq:37}
 J_D^s(y)=\left|\sum_{r=\pm}\frac{s^{(r-1)/2}rc_r(y)}{\sqrt{c_r(y)^2+y^2+\delta_\ell^2}}\right|^{-1},
\end{equation}
with
\begin{equation}
  \label{eq:39}
  c_\eta(y)=\frac{1}{2}\left(\ul{k}+\eta\,a\sqrt{1-4\frac{\delta_\ell^2+y^2}{a^2-\ul{k}^2}}\right),
\end{equation}
and
\begin{equation}
  \label{eq:40}
\mb
  p^{(\eta)}_{\ell,\mathbf{k}}(y)=c_\eta(y)\widehat{\mathbf{\ul{k}}}+y\mathbf{\hat{z}}\times \widehat{\mathbf{\ul{k}}},
\end{equation}
i.e.\ we identified $p_\parallel$ and $p_\perp$ in Eq.~\eqref{eq:87}
with $c_\eta(y)$ and $y$, respectively. At this point it may be comforting to check dimensions.  Noting that
$y$ has dimensions of momentum, i.e.\ inverse length, and $\mathcal{B}$
has dimensions of energy, i.e.\ inverse time, one can indeed see that
$D$ in Eq.~\eqref{eq:31} has proper dimensions of a rate.

\subsubsection{Off-diagonal scattering rate}

In this case, we must solve a {\em pair} of conic equations simultaneously,
which takes the form:
\begin{equation}
  \label{eq:183}
  \begin{cases}
    \varpi_1 - \Omega_{\ell,\mathbf{p}} -s_1
    \Omega_{\ell,\mathbf{p}-\mathbf{k}_1} = 0\\
    \varpi_2 - \Omega_{\ell,\mathbf{p}} -s_2
    \Omega_{\ell,\mathbf{p}-\mathbf{k}_2} = 0,
    \end{cases}
  \end{equation}
  i.e.
  \begin{equation}
    \label{eq:184}
    \begin{cases}
    \sqrt{|\ul{\mathbf{p}}|^2+\delta_\ell^2} +s_1
  \sqrt{|\ul{\mathbf{p}}-\ul{\mathbf{k}}_1|^2+\delta_\ell^2}= a_1\\
    \sqrt{|\ul{\mathbf{p}}|^2+\delta_\ell^2} +s_2
  \sqrt{|\ul{\mathbf{p}}-\ul{\mathbf{k}}_2|^2+\delta_\ell^2} = a_2,
    \end{cases}
  \end{equation}
where $a_i=\varpi_i/v_{\rm m}$. Indeed, the integrals which occur in the second order scattering rates involve
pairs of delta functions, whose arguments are of the form
considered above, with in Eq.~\eqref{eq:183}, 
$\varpi_1=-q_1\Sigma_{n\mathbf{k}n'\mathbf{k}'}^{qq'}$,
$\varpi_2=-q_1q'\omega_{n'\mathbf{k}'}$, $s_1=q_1q_2$, $s_2=-q_1q_3$,
$\mathbf{k}_1=-q\mathbf{k}-q'\mathbf{k}'$,
$\mathbf{k}_2=-q'\mathbf{k}'$, $\delta_\ell=\Delta_\ell/v_{\rm m}$.  In this case, each of the two delta function
constraints defines a half-hyperbola or an ellipse in the $\mathbf{p}$
plane, and the integrand is confined to the intersections of these two
curves.  Consequently, the integral will be collapsed to a discrete
set of points.  It is straightforward to see geometrically that the
intersection of two curves of these types is, except for the
degenerate cases in which the two curves are identical, a set of at most four
points. The two simultaneous equations can be
  solved analytically, but the solutions are algebraically
  complicated and we give here only the results and leave details to the Appendices.
\begin{widetext}
Collapsing the delta functions as explained in Appendix~\ref{sec:solv-delta-funct}, we
can write:
\begin{eqnarray}
  \label{eq:47}
  \mathfrak{W}^{\ominus,qq'}_{n\mathbf{k},n'\mathbf{k}'}
  &=&\frac{4\mathfrak{a}^2}{v_{\rm m}^3\hbar^4 }\sum_{j}\sum_{\ell,\{q_i\}}
      J_{\mathfrak{W}}(\mathbf{p}_j)\hat{\mathcal{F}}^{\ell,\ell,\ell|q_4,q_1,q_2}_{\mathbf{p}_j,q\mathbf{k},q'\mathbf{k}'}\nonumber\\
  &&\mathfrak{Im}\Bigg\{{\mathcal
    B}^{n\ell\ell|q_2q_3q}_{\mb k,\mb p_j +\frac{1}{2}q\mb k + q'\mb  k'} {\mathcal  B}^{n'\ell\ell|-q_3q_1q'}_{\mb k',\mb p_j+ \frac{1}{2}q'\mb   k'}
     ~ {\rm PP}\Bigg[\frac{{\mathcal B}^{n\ell\ell|-q_1q_4 -q}_{\mb k,\mb p_j +
    \frac{1}{2}q\mb k} {\mathcal B}^{n'\ell\ell|-q_4-q_2-q'}_{\mb
    k',\mb p_j + q\mb k +\frac{1}{2}q'\mb
     k'}}{\frac{q_1q\omega_{n\mathbf{k}}}{v_{\rm m}}+
    \hat{\Omega}_{\ell,\mb p_j}-q_1q_4 \hat{\Omega}_{\ell,\mb p_j +q\mb k}}+\frac{{\mathcal B}^{n'\ell\ell|-q_1-q_4 -q'}_{\mb k',\mb p_j +
    \frac{1}{2}q'\mb k'} {\mathcal B}^{n\ell\ell|q_4-q_2-q}_{\mb
    k,\mb p_j + q'\mb k' +\frac{1}{2}q\mb
     k}}{\frac{q_1q'\omega_{n'\mathbf{k}'}}{v_{\rm m}}+
    \hat{\Omega}_{\ell,\mb p_j}+q_1q_4 \hat{\Omega}_{\ell,\mb p_j +q'\mb k'}}\Bigg]\Bigg\},\nonumber
\end{eqnarray}
where $\hat{\Omega}=\Omega/v_{\rm m}$, and
\begin{equation}
  \label{eq:44}
  \hat{\mathcal{F}}^{\ell_3,\ell_1,\ell_2|q_4,q_1,q_2}_{\mathbf{p},q\mathbf{k},q'\mathbf{k}'}
 = q_1 q_4    \left(2n_{\rm  B}(\Omega_{\ell_3,\mb  p+q'\mb  k'})+1\right) \left(2n_{\rm B}(\Omega_{\ell_1,\mb
  p})+q_1+1\right) \left(2n_{\rm B}(\Omega_{\ell_2,\mb p+q\mb k +q'\mb
    k'})+q_2+1\right)
\end{equation}
is a product of thermal factors and where, when they exist, the solutions, $j=0,..,3$ take the form
\begin{equation}
  \label{eq:48}
  \mathbf{p}_j=t_{\lfloor j/2\rfloor}\mathbf{v}_{\lfloor j/2\rfloor}+u_{\lfloor j/2\rfloor}^{(\widetilde{j\;[2]})}\mathbf{w}_{\lfloor j/2\rfloor},
\end{equation}
where, for $i=0,1$
$\mathbf{v}_i=a_2\ul{\mathbf{k}}_1+(-1)^ia_1\ul{\mathbf{k}}_2$,
$\mathbf{w}_i=\mathbf{\hat{z}}\times \mathbf{v}_i$ (note that
  $\mathbf{v}_i=\ul{\mathbf{v}}_i$, $\mathbf{w}_i=\ul{\mathbf{w}}_i$
  and $\mathbf{p}_j=\ul{\mathbf{p}}_j$ are all in-plane vectors), $t_i$ and
$u_i^{(\pm)}$ are given in Appendix~\ref{sec:solv-delta-funct}
(also recall we defined $\widetilde{0}=-1$, $\widetilde{1}=1$, $x\,[2]$ is
$x$ mod $2$, and $\lfloor x\rfloor$ denotes the floor of $x$), and 
\begin{equation}
  \label{eq:49}
J_\mathfrak{W}(\mathbf{p}_j) = \left|s_1\frac{\ul{\mathbf{k}}_1\wedge
      \ul{\mathbf{p}}_j}{\hat{\Omega}_{\ell,\mathbf{p}_j}
     \hat{\Omega}_{\ell,\mathbf{p}_j-\mathbf{k}_1}}+s_2\frac{\ul{\mathbf{p}}_j\wedge
     \ul{\mathbf{k}}_2}{\hat{\Omega}_{\ell,\mathbf{p}_j}
     \hat{\Omega}_{\ell,\mathbf{p}_j-\mathbf{k}_2}}
   +s_1s_2\frac{-\ul{\mathbf{k}}_1\wedge
      \ul{\mathbf{k}}_2+\ul{\mathbf{p}}_j\wedge \ul{\mathbf{k}}_2-\ul{\mathbf{p}}_j\wedge \ul{\mathbf{k}}_1}{\hat{\Omega}_{\ell,\mathbf{p}_j-\mathbf{k}_1} \hat{\Omega}_{\ell,\mathbf{p}_j-\mathbf{k}_2}}\right|^{-1},
\end{equation}
\end{widetext}
where $\ul{\mathbf{V}}_1\wedge
      \ul{\mathbf{V}}_2=V_1^xV_2^y-V_2^yV_1^x$ for any in-plane
      vectors $\ul{\mathbf{V}}_{1,2}$. Coefficients $t_{0,1}$ are always well defined, but {\em for each} $i$, $u_i^{(\pm)}$ are the solutions
to a quadratic equation which has zero, one or two solutions, whether
the discriminant $\mathsf{d}_{u,i}$ thereof is negative, zero, or positive.

Necessary (but not sufficient) conditions of existence of solutions are: {\em (i)} the
existence of {\em both} conics, cf.~Table~\ref{tab:solutions}, {\em
  (ii)} $\mathsf{d}_{u,0}\geq0$ and/or $\mathsf{d}_{u,1}\geq0$, {\em (iii)} when $s_1$ and/or $s_2$ is
negative, the $\mathbf{p}_j$ must lie on the $\eta_{1,2}=1$ branch of the 1
and/or 2 hyperbola. Even with these constraints, spurious solutions exist, so that one must 
check that the solutions Eq.~\eqref{eq:48} also satisfy the equations for the given
values of $a_1,a_2,\ul{\mathbf{k}}_1,\ul{\mathbf{k}}_2,q,q',q_i$.

\subsection{Scaling and orders of magnitude}
\label{sec:orders-magnitude}

In this subsection, we discuss the temperature dependence and
magnitude of the magnonic contributions to the different phonon
scattering rates, which determine the phonon thermal conductivity and
thermal diffusivity tensors. Since we consider a low-energy
  continuum theory (without a momentum cutoff) in which the dispersion of the phonons is linear, these hold
  only in the low-temperature limit, i.e.\ for $T\ll \hbar v_{\rm ph}/(\mathfrak{a}k_{\rm B})$.
Similarly, we consider the low-energy dispersion of magnons, so our results are valid for $T\ll \hbar v_{\rm m}/
    (\mathfrak{a}k_{\rm B})\sim J/k_{\rm B}$. In Table~\ref{tab:scaling}, we summarize
some of the relations derived in this section.
\begin{table}[htbp]
  \centering
  \begin{tabular}{c|ccccc}
    \hline\hline
   quantity & $\tau^{-1}$ & $\kappa_{\rm L}$ & $\mathfrak{W}^{\ominus}$ &
                                                                    $\tau_{\rm
                                                                    skew}^{-1}$
    & $\varrho_{\rm H}$ \\
  $T$-scaling  &
    $T^{d+2x}$ & $T^{3-d-2x}$ & $T^{d-1+3x}$ & $T^{d+2+3x}$ & $T^{d-1+3x}$ \\
 Eq.~ref &  \eqref{eq:152} & \eqref{eq:74}& \eqref{eq:16} & \eqref{eq:19} & \eqref{eq:69} \\
     \hline\hline
  \end{tabular}
  \caption{Scaling relations derived in
    Sec.~\ref{sec:orders-magnitude} and the corresponding equation
    number where they appear. Note that these were obtained within a low-energy approach which omit
      in particular larger-$k$ deviations away from the acoustic
      phonon linear dispersion limit
    and other higher-$T$ effects such as Umklapp \cite{tritt}.}
  \label{tab:scaling}
\end{table}

\subsubsection{Longitudinal scattering rate: Role of anisotropies and scaling exponent}
\label{sec:long-scatt-rate-1}

First we consider the leading magnonic contributions to the
longitudinal scattering rate, $D^{(1)}_{n\mb{k}}$.  The typical
magnitude of this quantity for $|\mb{k}|\sim k_BT/v_{\rm ph}$ sets the
basic rate $1/\tau$.   This rate has been studied previously in
classic work on the phonon-magnon coupling in antiferromagnets.  
Reference~\cite{cottam_spin-phonon_1974} finds that $1/\tau
\sim T^5$ (for the moment we give only the $T$ dependence under the
above condition, and do not give the prefactor), for a model of
exchange-striction in a Heisenberg antiferromagnet in three
dimensions. This should be recovered from our formalism.

A general estimate can be obtained from
Eqs.~(\ref{eq:122},\ref{eq:143}).  To evaluate it requires, in addition
to the dispersion relations, the phonon-magnon couplings
$\mathcal{B}$, which are given in Eq.~\eqref{eq:119}.  At the level of
temperature scaling for typical thermal momenta, for temperatures
well above the magnon gap, $v_{\rm m} k \gg \Delta$, we may replace
$k \sim k_B T/v_{\rm ph}$, $\omega \sim v_{\rm ph} k \sim k_B T$ and
$\Omega\sim k_B T$ (the latter is true if the ratio between $v_{\rm m}$ and
$v_{\rm ph}$ is order one).   Noting that $\tilde{\xi}$
and $\tilde{\xi'}$ in Eq.~\eqref{eq:119} equal $\pm 1$, we see that a
general phonon-magnon coupling is a sum of three contributions,
\begin{equation}
  \label{eq:153}
  \mathcal{B} \sim \left(\frac{k_B T}{Mv_{\rm ph}^2}\right)^{\frac{1}{2}}n_0^{-1}
\left(
    \lambda_{mm} \frac{\chi k_BT}{n_0} + \lambda_{mn} + \lambda_{nn} \frac{n_0}{\chi
      k_B T}\right).
\end{equation}
Here, as above, we label generic N\'eel-N\'eel vector couplings
  $\lambda_{nn}\equiv\lambda_{ab,00}$, net
  magnetization-magnetization couplings
  $\lambda_{mm}\equiv\lambda_{ab,11}$ and ``cross''
  N\'eel-magnetization couplings $\lambda_{mn}\equiv\lambda_{ab,10}$.

Depending upon which of these terms is dominant, the temperature
dependence of $\mathcal{B} \sim T^{1/2+x}$, with
$x=-1,0,1$ corresponding to the $\lambda_{nn}$, $\lambda_{mn}$ and
$\lambda_{mm}$ terms, respectively.    We can then estimate the scattering rate
by converting the momentum sum over $\mb{p}$ to a $d$-dimensional
integral ($d$ is the spin-exchange dimensionality) and
recalling $|p|\sim T$.  We see therefore that
\begin{equation}
  \label{eq:152}
  \frac{1}{\tau} \sim T^{d-1} |\mathcal{B}|^2 \sim T^{d+2x}.
\end{equation}
A priori, the dominant contributions would arise from terms with
$x=-1$, which have the smallest power of temperature, which would give
$1/\tau \sim_? T^{d-2} \sim T$ in $d=3$ dimensions.  This {\em does
  not} agree with Ref.\,\cite{cottam_spin-phonon_1974}.  Instead, one
notices that what one might expect to be the subdominant contribution
from $x=+1$, which gives $1/\tau \sim T^{d+2}$ in general dimensions,
does agree with the classic theory for $d=3$.

Why is this the case?  The resolution lies in the fact that
Ref.~\cite{cottam_spin-phonon_1974} assumes isotropic Heisenberg
interactions, and is carried out in zero magnetic field.  As a
consequence, the Hamiltonian has SU(2) symmetry, and Goldstone's
theorem protects the gaplessness of the magnon modes {\em even in the
  presence of strain}.  In particular, because even an arbitrarily
strained lattice must preserve the gapless magnons in this case, the
spin-lattice coupling, Eq.~\eqref{eq:133} must be spin-rotationally
invariant, and moreover its quadratic expansion, Eq.~\eqref{eq:142},
must vanish for a magnon configuration which is a small rotation of
the N\'eel order, which corresponds to either $n_y$ or $n_z$ non-zero
and spatially constant.  This means that the non-zero terms in
Eq.~\eqref{eq:142} involve only $\xi,\xi'=m$ and not $n$ (in a
treatment including higher order terms, spatial gradients $\nabla n$
would appear, but these scale in the same manner as $m$).  One can
indeed check in Eq.~\eqref{eq:15} that when the interactions
$\Lambda^{({\rm m/n}),\alpha\beta}_{ab}$ are isotropic ($\propto
\delta_{ab}$), $\lambda_{nn}$ vanishes, and $\lambda_{mn}$
vanishes at zero field when the uniform magnetization $m_0^a=0$.
Taking the $\lambda_{mm}$ contribution in Eq.~\eqref{eq:153} gives $x=+1$ in
Eq.~\eqref{eq:152} as needed for agreement with earlier work.

What is the physics of the different values of $x$?  We see that
stronger effects (smaller powers of temperature) arise from coupling
to $n$ than to $m$.  This is a fundamental
property of antiferromagnets: fluctuations of the order parameter $n$ are
stronger and more long-ranged than those of the uniform magnetization
$m$, which is naturally suppressed when antiferromagnetic interactions
dominate.  Thus larger effects would be expected from coupling of
strain to the staggered magnetization than to the uniform one, as the
formula indeed shows.

How is this reflected in $\kappa_L(T)$?
The last step from the scattering time $\tau$ to the longitudinal conductivity $\kappa_L$
  is a standard one \cite{tritt,carruthers}. The sum over phonon momentum $\mb k$ in
  the first term of Eq.~\eqref{eq:115} is converted
  to a three-dimensional integral (the {\em magnon} momentum integral was $d$-dimensional,
  with $d=2$ in the case of a layered antiferromagnet).
  
  For temperatures $k_BT \gg \Delta$, the scaling for the temperature dependence
  of the longitudinal conductivity is
  \begin{equation}
    \label{eq:74}
    \kappa_L \sim T^{3-d-2x}.
  \end{equation}
  As can be seen from Eq. \eqref{eq:153}, a crossover between the low-temperature
  $x=-1$ and the high-temperature $x=+1$ behaviors occurs at $T^\star_\lambda$, 
  \begin{equation}
    \label{eq:68}
    k_B T^\star_\lambda
  \sim \frac{n_0}{\chi}\sqrt{\frac{\lambda_{nn}}{\lambda_{mm}}}.
\end{equation}
Eq.~\eqref{eq:68} assumes that the intermediate behavior $x=0$, due to the $\lambda_{mn}$ coupling
which is proportional to both anisotropic exchanges and the net magnetization,
is negligible; this is consistent with our numerical results shown in
Sec.~\ref{sec:results-kappa_l}. 
The above results, Eqs.~(\ref{eq:74},\ref{eq:68}), also assume that $D^{(1)}_{n\mb k}$
  is the dominant scattering rate contributing to the longitudinal inverse scattering time $D_{n\mb k}$. (The role of $\breve D_{n\mb k}$ is considered in more detail in Sec.~\ref{sec:results-kappa_l}.)
  However, many more scattering processes, such as boundary or impurity scattering,
  which in Eq.~\eqref{eq:111} are encompassed as $\breve D_{n\mb k}$,
  ontribute (through Matthiessen’s rule) to the phonon relaxation. Thus, $\kappa_L$ should be considered a probe of the \emph{full} $D_{n\mb k}$.

  \subsubsection{Longitudinal scattering rate:
    Role of the gap and magnetic field dependence}
\label{sec:long-scatt-rate-2}

Since we have seen that the assumption of isotropic interactions
suppresses the coupling to the staggered magnetization, this
discussion suggests that breaking of spin-rotation symmetry should
greatly enhance phonon scattering.  While this may indeed be the case,
we should note a subtlety: although spin anisotropy indeed allows such
coupling, it also allows the formation of a magnon gap ---enlarged by the
presence of an external magnetic field,
$\Delta_\ell = \sqrt{\Gamma_\ell /\chi + h_\ell^2}$.
Which behavior should be expected from the combination of these two effects?

Regardless of the form of coupling (scaling exponent $x$),
if $k_BT \ll \Delta$, magnon-phonon scattering will become energetically unavailable.
More precisely, $D^{(1)|+}$, corresponding to the
process whereby a phonon excites two magnons, is exponentially suppressed
due to the required rest energy $2\Delta$, while $D^{(1)|-}$, corresponding to
the process whereby a phonon scatters a magnon, is exponentially suppressed due
to the exponential decrease of all magnon populations at temperatures below the gap.
Therefore $D^{(1)}$ as a whole is exponentially suppressed if $k_BT \ll \Delta$;
We check this behavior numerically in Sec.~\ref{sec:results-d_nmb-k}.

Thus, a crossover in the behavior of $\kappa_L(T)$ occurs at temperature
$T^\star_\Delta \sim \Delta/k_B$. Below $T^\star_\Delta$, the phonon thermal
conductivity is mostly due to {\em other} scattering effects, which are captured by
$\breve D_{n\mb k}$ in this work. For constant $\breve D_{n\mb k}$, this yields
$\kappa_L \sim T^3$. Above $T^\star_\Delta$, phonon-magnon scattering becomes
available, and is enhanced by anisotropic coupling; provided this is the dominant effect,
the resulting thermal conductivity behavior is $\kappa_L\sim T^{3-d-2x}$ with $x=-1$
which, for $d=2$ (two-dimensional magnons), is the {\em same power} of temperature
as that obtained with only constant $\breve D_{n\mb k}$. However, the proportionality constant is
larger with phonon-magnon scattering than without, which, for sufficiently strong anisotropic
couplings (i.e.\ sufficiently large $\lambda_{nn}$), may lead to a ``bump'' in $\kappa_L(T)$,
as we indeed numerically see in Sec.~\ref{sec:results-kappa_l}.

Remarkably, this effect depends on the external magnetic field through
the width of the magnon gap (recall the latter is field dependent),
and may be an important feature of $\kappa_L(\mb h,T)-\kappa_L(\bs 0,T)$.
For the sake of completeness, we note that types of dependences on the magnetic field may arise
at temperatures where the scaling exponent $x=0$ plays a role, because
the $\lambda_{mn}$ coupling depends explicitly on the net magnetization $m_0$ in (see Eq.~\eqref{eq:15}).
It is however not clear how this contribution could
become non-negligible in any range of temperatures, and the gap dependence $\Delta(\mb h)$
is arguably the main culprit as regards the dependence on $\mb h$ of the longitudinal conductivity.

\subsubsection{Transverse scattering: scaling exponent}
\label{sec:skew-scattering-1}

We can now apply similar reasoning to the transverse/Hall scattering
rate $\mathfrak{W}^\ominus$ from Eq.~\eqref{eq:110}.  Obviously if
temperature is sufficiently low, i.e.\ below magnon gaps, the result
will be exponentially suppressed.  Of greater interest is the energy regime
above the magnon gaps, in which we may assume acoustic linearly
dispersing magnons (and phonons).  We proceed by
counting the obvious factors of momentum and energy, and by assuming the
relevant momentum scales are set by dimensional analysis,
i.e.\ $\mb{k},\mb{k}'\sim k_B T/v_{\rm m}$ etc.  Inspection of
Eq.~\eqref{eq:110} shows one  sum over magnon momentum $\mb{p}$, which
converts to an integration in the thermodynamic limit, two energy
delta functions, and one energy denominator, which, using the
aforementioned momentum scaling implies that
\begin{equation}
  \label{eq:3}
  \mathfrak{W}^\ominus \sim T^{d-3} \mathcal{B}^4.
\end{equation}
Here we considered the {\em magnon} momentum integration as $d$-dimensional,
as in the previous discussion of longitudinal scattering rates.

Now to proceed we must estimate the contribution of the four
$\mathcal{B}$ factors.  To do so, we need to consider the effective
time-reversal symmetry $\mathcal{T}$.  This symmetry must be broken
to obtain a non-zero effective skew-cattering rate, $\mf W^{\ominus,{\rm eff},qq'}_{n\mb k,n'\mb k'}$,
which in particular is odd
under $\mathcal{T}$.  As discussed in Secs.~\ref{sec:expansion}
and \ref{sec:terms-eigenbosons-b}, under
$\mathcal{T}$ the $\lambda_{mm}$ and $\lambda_{nn}$ couplings are
even while only the $\lambda_{mn}$ couplings are odd; therefore
$\mathfrak{W}^{\ominus,\rm eff}$ must contain an odd number of factors of
$\lambda_{mn}$.  Furthermore, in the low field regime we consider
here, $\mathcal{T}$ symmetry breaking happens through the development
of a small uniform magnetization, hence $\lambda_{mn} \propto m_0$, which
in turn is linearly proportional to the applied field (see Eq.~\eqref{eq:15}).
Consequently, to obtain the linear-in-field Hall scattering rate, we should keep
just one (and not three, the other available odd number) factors of
$\lambda_{mn}$.  Therefore, we may use Eq.~\eqref{eq:153} to estimate
\begin{equation}
  \label{eq:16}
  \mathfrak{W}^{\ominus,\rm eff} \sim T^{d-1} \lambda_{mn} \left(
    \lambda_{mm} T + \lambda_{nn} T^{-1}\right)^3 \sim T^{d-1+3x}.
\end{equation}
Here, as in Sec.~\ref{sec:long-scatt-rate-1}, $x=+1$ obtains in a large parameter
  region where $\frac{\lambda_{mm}}{\lambda_{nn}}\ll \big ( \frac{n_0}{\chi k_B T}\big )^2$,
while $x=-1$ results if $\lambda_{nn}$ is non-zero and dominant in a low-temperature regime where
the magnon gap remains negligible.

It is by no means clear {\em how} the latter regime would be achieved,
and if we assume that the $x=+1$ case dominates, then it is interesting
to see that $\mathfrak{W}^{\ominus,\rm eff}$ in Eq.~\eqref{eq:16} scales like
$T^{d+2}$, which is the {\em same power} of temperature as the magnon
contribution to the longitudinal scattering rate in Eq.~\eqref{eq:152}.  

This scaling is a bit surprising, as we should expect that the
transverse is smaller than the longitudinal scattering, since it comes
from a higher order term.  To resolve this, we should consider more
carefully the relationship of $\mathfrak{W}^{\ominus,\rm eff}$ to a ``skew scattering rate''.  In particular, one should note that
$\mathfrak{W}_{n\mb{k}n\mb{k}'}^{\ominus,\rm eff}$ enters the collision term via
a {\em sum} over $\mb{k}'$, which converts to an integral over
$\mb{k}'$ in the thermodynamic limit.  Therefore the measure of this
integral, which is expected to be dominated by $|\mb{k}'| \sim
k_BT/v_{\rm m}, k_BT/v_{\rm ph}$, contributes an additional factor of $T^3$
(since phonons are always three-dimensional).
Thus it would be more correct to estimate the skew scattering rate as
\begin{equation}
  \label{eq:19}
  \frac{1}{\tau_{\rm skew}} \sim  T^3 \mathfrak{W}^{\ominus,\rm eff}  \sim T^{d+2+3x}.
\end{equation}
For $x=1$ and $d=2$, this scales  as $T^7$ which is indeed small
compared to the $T^4$ predicted in the same regime for the
longitudinal scattering.

Additionally, we highlight in Sec.~\ref{sec:results-mf-womin}, through numerical evaluations,
the strong momentum-orientation dependence of $\mathfrak{W}^\ominus$.

\subsubsection{Transverse scattering: thermal Hall resistivity}
\label{sec:transv-scatt-therm}

We would like to emphasize that within any scattering mechanism of
phonon thermal Hall effect, the skew scattering rate is a more fundamental
measure of chirality of the phonons than the thermal Hall {\em
  conductivity}.  This is because the Hall conductivity inevitably
involves the combination of the skew and longitudinal scattering rates
(in the form $\tau^2/\tau_{\rm skew}$), and the longitudinal
scattering rate of phonons has many other contributions that do not
probe chirality, and may have complex dependence on temperature and
other parameters that obscure the skew scattering. The
scaling of the temperature dependence of $1/\tau_{\rm skew}$ given
above is a much more reliable prediction than any corresponding one
made for $\kappa_H$ for this reason, and we do not quote the latter
here. Instead, to extract the
skew scattering rate, one should look at the thermal Hall {\em
  resistivity}, $\varrho_H$, which is simply proportional to $1/\tau_{\rm skew}$,
at least in the simplest view where the angle-dependence of the
longitudinal scattering does not spoil its cancellation. 

We define the thermal Hall resistivity tensor as usual by the matrix inverse,
$\bs{\varrho} = \bs{\kappa}^{-1}$.
In particular, considering the simplest case of isotropic
$\kappa^{\mu\mu}\rightarrow \kappa_L$ and $\kappa_{L} \gg
\kappa^{\mu\neq\nu}$, one thus has
\begin{equation}
  \label{eq:rho12}
  \varrho^{\mu\nu}_H = \frac {\varrho_{\mu\nu}-\varrho_{\nu\mu}} 2
  \approx \frac{-\kappa_{\mu\nu}+\kappa_{\nu\mu}}{2\kappa_{L}^2}
  = - \frac{\kappa^{\mu\nu}_{H}}{\kappa_{L}^2}.
\end{equation}
The quantity $\varrho^{\mu\nu}_{H}$ is independent of the scale of the longitudinal
scattering, in the sense that under a rescaling $D_{n\mb{k}}\rightarrow \zeta
D_{n\mb{k}}$, then $\varrho^{\mu\nu}_{H}$ is unchanged 
(see indeed Eqs.~(\ref{eq:115},\ref{eq:79}) for an explicit check at leading perturbative order).

As explained before, let us further assume that $D_{n\mb{k}} = 1/\tau$
is $(n,\mb{k})$---independent, e.g.\ as if the case if dominated by some extrinsic effects. 
 In that case, we can extract the longitudinal dependence 
from the transverse conductivity kernel, and redefine
$ \tilde{K}^H_{n\mb{k}n'\mb{k}'} =\tau^{-2}
K^H_{n\mb{k}n'\mb{k}'} $ which is now independent of
the longitudinal scattering rate $\tau^{-1}$. Besides, to leading order one has simply
$ K_{n\mathbf k n'\mathbf k'}^{L}
  = 
  \tau e^{\beta\hbar\omega_{n\mathbf k}}  \delta_{nn'}\delta_{\mb{k}\mb{k}'}$,
from which, assuming $\omega_{n\mathbf{k}}=v_{\rm ph}|\mathbf{k}|$ and
$N^{\rm eq}=n_B$, we have simply
\begin{equation}
  \label{eq:rho17}
 \kappa_L = \tau \frac{\hbar^2}{k_B T^2}\frac{1}{V} \sum_{n\mb{k}} e^{\beta
        \hbar \omega_{n\mb{k}}}\left(J^\alpha_{n\mb{k}}\right)^2 
= \tau c_v \frac{v_{\rm ph}^2}{3},
\end{equation}
where by construction the result does not depend on the chosen direction $\alpha$ of the current
(for instance $\alpha=x,y,z$).
This is the well-known relation between the thermal conductivity $\kappa_L$
and the thermal capacity
\begin{equation}
  \label{eq:rho1}
  c_v=\frac{\partial}{\partial T} \left[\frac{1}{V} \sum_{n\mb{k}} N^{\rm eq}_{n\mb{k}}
  \hbar\omega_{n\mb{k}}\right]
=k_{B}\frac{2\pi^2}{5}\left(\frac{k_{B} T}{\hbar v_{\rm ph}}\right)^3
\end{equation}
of the phonon gas. 
Consequently, Eq.~\eqref{eq:rho12} evaluates to
\begin{eqnarray}
  \label{eq:rho16}
  \varrho_{H}^{\mu\nu}
  &=& - k_B^{-1}\left(\frac{15v_{\rm ph}}{2\pi^2}\right)^2
      \left(\frac{\hbar}{k_B T}\right)^8\nonumber\\
  &&\quad \times \frac{V}{(2\pi)^6}
    \sum_{n\mathbf{k}n'\mathbf{k}'}J^\mu_{n\mathbf{k}}
    \tilde{K}^{H}_{n\mathbf{k}n'\mathbf{k}'} J^\nu_{n'\mathbf{k}'}  .
\end{eqnarray}
This expression does not depend on $\tau$, which justifies studying
$\varrho_{H}$ instead of $\kappa_{H}$. From it and Eq.~\eqref{eq:main16},
one can readily derive the scaling relation
\begin{equation}
  \label{eq:69}
  \varrho_H \sim \mf W^{\ominus,\rm eff} \sim T^{d-1+3x},
\end{equation}
which we check numerically in Sec.~\ref{sec:results-rho_h}.

\subsubsection{Detailed scaling analysis of the longitudinal conductivity}
\label{sec:new}

The scaling with temperature described above, obtained by
replacing every momentum scale $\hbar v_{\rm m,ph}k$, $\hbar v_{\rm m,ph}\ul{k}$,
$\hbar v_{\rm m,ph}p$ by that of the temperature, $k_{\rm B}T$, are
expected to be valid for $\upsilon = v_{\rm m}/v_{\rm ph}$ of ``order
one''.  Here we investigate more carefully the dependence of these
quantities on $\upsilon$ when the latter becomes large.  Surprisingly,
we show below that the prediction of Eq.~\eqref{eq:74} for the high
temperature scaling of $\kappa_L$ breaks down already for $\upsilon>3$,
giving way to a continuously variable power law exponent.
For the
thermal Hall resistivity, we find that the temperature exponent,
Eq.~\eqref{eq:69}, remains independent of the velocity ratio. 

To obtain these results, we analyze the full integral expressions
directly, distinguishing momentum and temperature dependencies.
Various technical details are provided in Appendix~\ref{sec:gener-forms-scal}.

To analyze $\kappa_L$, we start by writing the expression for the diagonal scattering rate,
Eq.~\eqref{eq:31}, in dimensionless form, in terms of a scaling
parameter $\varkappa$ and dimensionless variable $\tilde{y}$,
\begin{equation}
  \label{eq:43}
  \varkappa=\frac{\hbar v_{\rm
               ph}k}{k_{\rm B}T},
  \quad\tilde{y}=y/k.
\end{equation}
We assume the relevant momentum and energy scales are large compared
to any gaps,  $\delta_\ell \ll k, k_B T/v_{\rm m}$, and therefore in
the following set $\delta_\ell \rightarrow 0$.  Then we can obtain
scaling forms, 
\begin{eqnarray}
  \label{eq:77}
&&c_\eta(y)=k\,\tilde{c}_\eta(\tilde{y}),\\
&&\Omega_{\ell,\mb p^{(\eta)}_{\ell,\mathbf{k}}(y) },\Omega_{\ell,\mb
  p^{(\eta)}_{\ell,\mathbf{k}}(y) - \mb k}=v_{\rm
                                              m}\,k\,\tilde{\Omega}_\ell^{\pm
                                              \eta}(\tilde{y}),\nonumber\\
&&\mathcal{B}^{n,\ell,\ell|+s-}_{\mathbf{k};-\mathbf{p}+\mathbf{k}/2}=k^{3/2}\,\tilde{\mathcal{B}}_{mm}(\tilde{y})+k^{1/2}\,\tilde{\mathcal{B}}_{mn}(\tilde{y})+k^{-1/2}\,\tilde{\mathcal{B}}_{nn}(\tilde{y}),\nonumber
\end{eqnarray}
and the precise functional forms of $\tilde{c}_\eta(\tilde{y})$, $\tilde{\Omega}_\ell^{\pm
                                              \eta}(\tilde{y})$ and $\tilde{\mathcal{B}}_{\xi\xi'}$ are given in
Appendix~\ref{sec:gener-forms-scal}, Eqs.~(\ref{eq:106}-\ref{eq:100}). We find, after making the
change of variables in the integral of Eq.~\eqref{eq:31} from $y$ to
$\tilde{y}$,
\begin{equation}
  \label{eq:80}
  D^{(s)}_{n\mb{k}}(T)= \sum_{x=-1}^1 k^{2+2x}{\rm
  F}^{(s)}_{x}(\varkappa,\upsilon,\theta),
\end{equation}
where the scaling functions ${\rm F}^{(s)}_x(\varkappa,\upsilon,\theta)$ are
\begin{eqnarray}
  \label{eq:78}
  {\rm  F}^{(s)}_{x}(\varkappa,\upsilon,\theta) &&= \frac{(3-s) \mathfrak{a}^2\sinh(\varkappa/2)}{4\pi v_{\rm m}\hbar^2 }\int_{-\infty}^{+\infty} \!\!\!\!\text d\tilde{y}\sum_\eta\tilde{\rm
     f}_\eta^s(\tilde{y})\tilde{J}_D^s(\tilde{y})\nonumber 
  \\
  && \times  \sum_{\ell}  \frac{\tilde{\mathcal C}_{x}(\tilde{y})}
           {\sinh(\frac{\upsilon\varkappa}{2} \tilde{\Omega}^{+\eta}_\ell(\tilde{y})) \sinh(\frac{\upsilon\varkappa}{2} \tilde{\Omega}^{-\eta}_\ell(\tilde{y}))}.
\end{eqnarray}
Here $\tilde{\mathcal{C}}_x(\tilde{y})$ are quadratic combinations of
the original $\tilde{\mathcal{B}}_{mm}, \tilde{\mathcal{B}}_{mn},$ and
$\tilde{\mathcal{B}}_{nn}$ coefficients given in
Appendix~\ref{sec:gener-forms-scal}.  Eq.~\eqref{eq:80} agrees with the scaling behavior given in
Eq.~\eqref{eq:152} (with $d=2$).  In
Appendix~\ref{sec:gener-forms-scal} we derive the behavior of the
scaling functions ${\rm F}_x^{(s)}$ at small and large $\varkappa$,
which will be useful in the following.  

The scaling form of the scattering rate is input to the thermal
conductivity.  To see the implication, we presume for simplicity the total
scattering rate $D_{n\mb{k}} = \sum_s D_{n\mb{k}}^{(s)} + \breve{D}$
is dominated by a single value of $x$ .   We note in passing that when
the gaps are zero, only one value of $s$ contributes here: $s=-1$ for
$\upsilon>1$ (the case of most interest), and $s=1$ for $\upsilon<1$.  Then 
\begin{align}
  \label{eq:90}
  D_{n\mb{k}} =  \left(\frac{k_{\rm B}T}{\hbar v_{\rm ph}}\right)^{2+2x}\left(\varkappa^{2+2x}{\rm
  F}_{x}(\varkappa)+\check{D}(T)\right).
\end{align}
Here we defined $\check{D}(T) = \left(\frac{\hbar v_{\rm ph}}{k_{\rm
      B}T}\right)^{2+2x}\breve{D}$, which is temperature-dependent.
In particular for $x=0,1$, it becomes very small at high temperature.
Inserting this into Eq.~\eqref{eq:115}, turning the sum over $\mathbf{k}$
into an integral, and using spherical coordinates, we obtain
\begin{eqnarray}
  \label{eq:89}
  \kappa_{\rm L}^{\mu\mu}&\sim&\frac{k_{\rm
  B}}{\hbar^2}\left(\frac{\hbar v_{\rm ph}}{k_{\rm
  B}T}\right)^{2x-1}\\
  && \int_0^{+\infty} \text d \varkappa\int_{0}^{\pi}\text d \theta \int_0^{2\pi}\text d\phi\frac{ \hat k^\mu \hat  k^\mu\sin\theta \varkappa^4 \sinh^{-2}(\varkappa/2)}
   {\varkappa^{2x+2}{\rm
  F}_{x}(\varkappa,\upsilon,\theta)+\check{D}(T)}\nonumber.
\end{eqnarray}

Now we are in a position to investigate the temperature dependence of
the conductivity.    To simplify the discussion, we restrict the
remainder of this section to the case
$x=1$, since ${\rm F}_1$ is the largest contribution when spin-orbit
coupling is weak, and is also enhanced at high temperature.   First consider the low temperature limit.  Then $\check{D}(T)$
becomes large at low $T$, and we can simply replace the denominator of
the integrand in Eq.~\eqref{eq:89} by $\check{D}(T)$.  This is just
the extrinsic limit in which the constant $\breve{D}$ scattering
dominates and one recovers the $T^3$ dependence of the thermal
conductivity arising from the phonon heat capacity.

Next we turn to the higher temperature limit.  There, the parameter
$\check{D}(T)$ becomes small, and might na\"ively be neglected.
Dropping this term in Eq.~\eqref{eq:89}, the sole remaining
temperature dependence is in the prefactor, and agrees with what was
found earlier in Eq.~\eqref{eq:74} (for $d=2$).  This procedure is
valid provided the integral in Eq.~\eqref{eq:89} converges for
$\check{D}(T)=0$.  To check this, we must consider the potential
divergences at small and large $\varkappa$.  At small $\varkappa$, the
integrand behaves like $1/(\varkappa^{2} {\rm F}_1(\varkappa))$.  As
shown in Sec.~\ref{sec:gener-forms-scal}, $F_1(\varkappa)$ grows as
small $\varkappa$ (see Eq.~\eqref{eq:96}), ensuring there is no
divergence.  The large $\varkappa$ limit is more
problematic.  This is because although the $\sinh^{-2}(\varkappa/2)$
factor decays exponentially, the factor ${\rm F}_x$ in the denominator
also decays exponentially.  Specifically, we show in
Sec.~\ref{sec:gener-forms-scal} that Eq.~\eqref{eq:78} implies
\begin{equation}
  \label{eq:92}
  {\rm F}_x(\varkappa,\upsilon,\theta) \underset{\varkappa\gg1}{\sim}
  {\rm \bar{F}}(\upsilon,\theta)e^{-\alpha(\upsilon,\theta)\varkappa},
\end{equation}
where the function
$\alpha(\upsilon,\theta)=\frac{1}{2}(\upsilon|\sin\theta|-1)$ (see
Eq.~\eqref{eq:97}), and ${\rm \bar{F}}(\upsilon,\theta)$ a constant.
This implies an
exponential {\em growth} of $1/{\rm F}_x$ with $\varkappa$ when
$\check{D}$ is neglected.  For $\upsilon>3$, the integral becomes
divergent for $\check{D}=0$, and the na\"ive scaling fails.

To see what happens for $\upsilon>3$, we deduce from the above
discussion that the integral in Eq.~\eqref{eq:89} becomes dominated in
this case by large $\kappa \gg 1$.  Then we approximate
$\sinh\frac{\varkappa}{2}\sim e^{\varkappa/2}$, and use
the asymptotic form of ${\rm F}_x$ in Eq.~\eqref{eq:92}.  We must then
distinguish two cases. If $\alpha<1$, the $\varkappa$ integral
converges, even for $\check{D}(T)\rightarrow0$, and we obtain, in the
latter limit, $\kappa_{\rm L}\sim T^{-1}$. When $\alpha>1$, we must be
more careful. Successively
performing the changes of variables $u=e^{-\varkappa}$,
$u=v\check{D}(T)^{\alpha(\upsilon,\theta)^{-1}}$ and
$v=w[-\alpha(\upsilon,\theta)^{-1}\ln\check{D}(T)]^{-\alpha(\upsilon,\theta)^{-1}}$,
and a saddle-point procedure assuming $\alpha>1$, we arrive at (see Appendix~\ref{sec:gener-forms-scal})
\begin{equation}
  \label{eq:93}
  \kappa_{\rm L}\sim\frac{1}{T}\check{D}(T)^{\alpha_0^{-1}-1}\left(-\alpha_0^{-1}\ln\check{D}(T)\right)^{4-\alpha_0^{-1}}I_0(\upsilon)
\end{equation}
where $\alpha_0(\upsilon)=\alpha(\upsilon,0)=(\upsilon-1)/2>1$, $I_0(\upsilon)=\int_0^{+\infty}\text d
w\frac{1}{w^{\alpha_0}F_0(\upsilon,0)+1}$. This entails, up to
logarithmic corrections, the result quoted in Eq.~\eqref{eq:104},
\begin{equation}
  \label{eq:86}
  \kappa_{\rm L} \sim
  \begin{cases}
    T^{-1}&\mbox{for }\upsilon<3\\
    T^{3-8(\upsilon-1)^{-1}}&\mbox{for }\upsilon>3
    \end{cases}.
\end{equation}
We see that the $1/T$ high-temperature behavior of $\kappa_{\rm L}$
changes for $\upsilon>3$ to a power law with an exponent that
continuously depends upon $\upsilon$, and even changes sign: for $\upsilon
> 11/3$, the conductivity {\em increases} with increasing temperature
at high $T$.   We
indeed recover this nontrivial feature numerically, see Fig.~\ref{fig:kappaLvmvph}.

Note that this behavior is all obtained within the linearized
phonon and magnon models, and thus eventually changes when the temperature
exceeds for example the Debye energy.




\subsection{Numerical results}
\label{sec:numerical-results}

\subsubsection{Implementation}
\label{sec:implementation}

Details about the numerical implementation are given in Appendix~\ref{sec:numer-impl}.
In short, we use C together with {\it (i)} the Cubature library to perform the one-dimensional momentum integrals (appearing in the definitions of
$D^{(1)|s}_{n\mb k}$, Eq.~\eqref{eq:31}), {\it (ii)} the Cuba library
\cite{cubapaper} to perform multi-dimensional integrals
(three-dimensional for
$\kappa_L^{\mu\mu}$, first term in Eq.~\eqref{eq:115},
and in six-dimensional for $\varrho_H^{\mu\nu}$, Eq.~\eqref{eq:rho16}).

\subsubsection{Choice of parameters}
\label{sec:choice-parameters}

\paragraph{Polarization vectors} In Eq.~\eqref{eq:155}, $\mathcal{L}$ is the trace over the product of the coupling
matrix $\boldsymbol{\lambda}$, with matrix elements $\lambda^{\alpha\beta}$,
and that, $\boldsymbol{\mathcal{S}}$, which determines the structure
of the strain tensor and has matrix elements
\begin{equation}
  \label{eq:20}
  \mathcal{S}^{q;\alpha\beta}_{n\mathbf{k}}
  =\frac{k^\alpha(\varepsilon_{n\mathbf{k}}^\beta)^q+k^\beta(\varepsilon_{n\mathbf{k}}^\alpha)^q}
  {\sqrt{\omega_{n\mathbf{k}}}}.
\end{equation}
Values of $(n,\mb k)$ such that this factor vanishes correspond to phonons which
are not coupled to the magnons, and whose longitudinal conductivity is solely driven
by $\breve{D}_{n\mb k}$, i.e.\ other scattering effects. While this may indeed happen in practice,
to highlight the effects of phonon-magnon scattering we choose a basis of polarization vectors
$(\bs \varepsilon_{0,\mb k}, \bs \varepsilon_{1,\mb k}, \bs
\varepsilon_{2,\mb k})$ such that this is never the case (at least for $\alpha=\beta$, as with $\Lambda_{1..5}$
which are much larger than $\Lambda_{6,7}$).

These polarization vectors enforce $\bs \varepsilon_{n,-\mb k}= \bs \varepsilon_{n\mb k}^*=-\bs \varepsilon_{n\mb k}$
(so that $\mathcal{S}^{q;\alpha\beta}_{n,-\mathbf{k}}
=\mathcal{S}^{q;\alpha\beta}_{n\mathbf{k}}=-\mathcal{S}^{-q;\alpha\beta}_{n\mathbf{k}}$) as well as
the tetragonal symmetry of the crystal, as required by the general theory of elasticity \cite{maradudin};
explicit expressions are given in  App.~\ref{sec:choice-2}.

\paragraph{Extrinsic phonon scattering rate} For similar reasons, the extrinsic phonon scattering rate is taken to be 
$\breve{D}_{n\mb k}\rightarrow\gamma_{\rm ext}$, a constant
independent of $(n,\mb k)$ and small compared with the typical $D_{n\mb
  k}$ as soon as $T>T^\star_{\Delta}$ (see Sec.~\ref{sec:long-scatt-rate-2}).
In very clean monocrystals and in the absence of any other phonon
scattering events,
$\gamma_{\rm ext}\sim v_{\rm ph}/L$ reduces to the rate at which phonons bounce
off the boundaries of the sample (of size $L$).

\paragraph{Phonon dispersion} The phonon dispersion relation is chosen linear, $n$-independent
and isotropic, $\omega_{n\mb k}=v_{\rm ph}|\mb k|$, so that the different
regimes of scaling exponents $x$ appear clearly.

\subsubsection{Units and numerical values}
\label{sec:units-numer-valu}

We express our numerical results in units where $\hbar^{\rm code}=1$,
$k_B^{\rm code}=1$,
$v_{\rm ph}^{\rm code}=1$ and with unit lattice spacing $\mf{a}^{\rm code}$.
Then, the mass of the unit cell $M_{\rm uc}$ is expressed in
units of $M_0=\frac{\hbar}{v_{\rm ph}\mf a}$ and is typically
large---of the order $M_{\rm uc}\sim 10^4M_0$.
$T$ is expressed in units of
$T_0=\frac{\hbar v_{\rm ph}}{\mf a k_B}$ and should verify $T/T_0 \lesssim 1$
so that the assumption of linearly dispersing phonons is
correct. Correspondingly, we can define an energy $\epsilon_0=k_{\rm
  B}T_0$, and the isotropic part of the exchange $J$ is expressed in
units of $\epsilon_0$.

The magnon velocity is fixed according to linear spin wave theory,
which gives $v_{\rm m} = 2\sqrt{d} J S \mathfrak{a}/\hbar$, with $J$ the isotropic magnetic exchange constant.
We take $d=2$ and $S=1/2$; moreover, it is known that for $S=1/2$ there is a renormalization
factor $Z \approx 1.2$ enhancing the velocity, so that $v_{\rm
  m}/v_{\rm ph} =\sqrt{2}Z J/\epsilon_0 \approx 1.7 J/\epsilon_0$ in our units.
Since, for isotropic exchange, $\chi = \frac{1}{4\mf a^2 J}$, we also
take $\chi_{\rm code}^{-1}=4(J/\epsilon_0)$.

Spin-phonon couplings $\Lambda_{1..7}$ are expressed in units of
$\epsilon_0/\mathfrak{a}=\hbar v_{\rm ph}/\mf a^2$.
We describe a possible microscopic mechanism for spin-strain coupling in App.~\ref{sec:micr-deriv-coupl},
where we show that $\Lambda_{1..5}$ typically arise as derivatives of the isotropic magnetic exchange constants.
Since the latter ultimately arises from the overlap of atomic wavefunctions, which vary over distances of the order
$a_B$ the Bohr radius, we expect $\Lambda_{1..5} \sim J/{a_B}$.
Meanwhile $\Lambda_{6,7}$ come from anisotropic exchanges and are thus
expected to be considerably smaller.

Since the differences $\Lambda^{(\xi)}_{1,2}-\Lambda^{(\xi)}_{3}$ and $\Lambda^{(\xi)}_{4}-\Lambda^{(\xi)}_{5}$
are due to anisotropic exchanges, they are chosen a fraction of a $\Lambda^{(\xi)}_{1..5}$.
Since these magnetoelastic couplings typically arise as derivatives of magnetic exchange, we also
take $\Lambda^{(\rm m)}_{i} \approx -\Lambda^{(\rm n)}_{i}$ for $i=1..7$;
see App.~\ref{sec:micr-deriv-coupl} for a detailed derivation.

Scattering rates $D_{n\mb k}$ and $\gamma_{\rm ext}$ are expressed in units of
$\gamma_0=v_{\rm ph}/\mf a$, and we assume $\gamma_{\rm ext}$ to be small, of the
order of $1/L$ with $L$ the size of the sample---typically
$\gamma_{\rm ext}\sim 10^{-7}v_{\rm ph}/\mathfrak{a}$.
Finally, thermal conductivities are expressed in units of $\kappa_0=k_B v_{\rm
  ph}/\mf a^2$. 

For numerical calculations, we kept most dimensionless materials parameters (e.g.\ the ratio of $v_{\rm m}$ and $v_{\rm ph}$) fixed and constant, with the values given in  Table~\ref{tab:table-params}.   Those parameters for which we explore a given range of values are given in the captions of the figures in the following subsections. The fixed values are loosely inspired by Copper Deuteroformate Tetradeuterate (CFTD), a square lattice S=1/2 antiferromagnet which has been intensively studied via neutron scattering \cite{christensen2007quantum,dalla2015fractional,ronnow2001spin} due to its convenient scale of exchange which suits such measurements. For our purposes, CFTD has the desirable attribute that the magnon and phonon velocities are comparable (based on an estimate of the sound velocity from the corresponding hydrate \cite{kameyama1973elastic}), which creates a significant phase space for magnon-phonon scattering.  By contrast, in La$_2$CuO$_4$, $v_{\rm m}$ is much larger than $v_{\rm ph}$.  

\begin{table}[htbp]
  \centering
  \begin{tabular}{ccccccccc}
    \hline\hline
    $v_{0}$ & $\mathfrak{a}_0$ & $\gamma_0$ & $M_0$ & $T_0$ & $\epsilon_0$ & $\chi^{-1}_0$
  & $\Lambda_0$  & $\kappa_0$ \\
    $v_{\rm ph}$ & $\mathfrak{a}$ & $\frac{v_{\rm ph}}{\mathfrak{a}}$
                                            &  $\frac{\hbar}{v_{\rm
                                              ph}\mathfrak{a}}$ & $\frac{\hbar v_{\rm ph}}{\mathfrak{a}k_{\rm B}}$ & $k_{\rm B}T_0$
     &
 $\epsilon_0\mathfrak{a}^2$ 
& $\frac{\epsilon_0}{\mathfrak{a}}$ & $\frac{k_{\rm B}v_{\rm ph}}{\mathfrak{a}^2}$ 
    \\
    \hline\hline
  \end{tabular}
  \caption{Table of units of velocity $v_0$, distance $\mathfrak{a}_0$,
    rate $\gamma_0$, mass $M_0$, temperature $T_0$,
    energy $\epsilon_0$, inverse susceptibility $\chi^{-1}_0$,
    coupling $\Lambda_0$ and thermal conductivity $\kappa_0$ used in Table~\ref{tab:table-params}.}
  \label{tab:units}
\end{table}

\begin{table}[htbp]
  \begin{tabular}{ccccccccccc}
\hline\hline
  $\,v_{\rm m}\,$ & $\,v_{\rm ph}\,$ & $\chi^{-1}$ & $n_0$ &
                                                             $\mathfrak{a}$ & $M_{\rm uc}$ & $m_0^x$
  &  $m_0^y$ & $m_0^z$ & $\Delta_0$ & $\Delta_1$ \\ \hline
  $\mt{\upsilon}$ & $\mt{1.0}$ & $\mt{2.08\upsilon}$ & $\mt{1/2}$ & $\mt{1.0}$ & $\mt{8\cdot 10^3}$ &  $\mt{0}$
  &  \begin{tabular}{@{}c@{}} $\mt{0.0}$ \\ $\mt{0.05}$\end{tabular} & \begin{tabular}{@{}c@{}} $\mt{0.05}$ \\ $\mt{0.0}$\end{tabular} & $\mt{0.2}$ & $\mt{0.04}$ \\
\hline\hline
  \end{tabular}
\begin{tabular}{c|ccccccc}
\hline\hline
  $\xi$ & $\Lambda^{(\xi)}_1$ &  $\Lambda^{(\xi)}_2$ & $\Lambda^{(\xi)}_3$ &
  $\Lambda^{(\xi)}_4$ & $\Lambda^{(\xi)}_5$ & $\Lambda^{(\xi)}_6$ & $\Lambda^{(\xi)}_7$\\ \hline
${\rm n} = 0$  & $\mt{4.8 \upsilon}$ & $\mt{4.0 \upsilon}$ & $\mt{5.6 \upsilon}$& $\mt{4.0 \upsilon}$ & $\mt{4.8 \upsilon}$ &$\mt{0.24 \upsilon}$ & $\mt{0.32 \upsilon}$ \\ \hline
 ${\rm m} = 1$ & $\mt{-4.0 \upsilon}$ & $\mt{-4.8 \upsilon}$& $\mt{-5.6 \upsilon}$ & $\mt{-4.8 \upsilon}$ & $\mt{-4.0 \upsilon}$ &$\mt{-0.32 \upsilon}$ & $\mt{-0.24 \upsilon}$\\ \hline\hline
\end{tabular}
\caption{Numerical values of the fixed parameters used in all
  numerical evaluations, expressed in the units given in
  Table~\ref{tab:units}. The upper and lower entries for $m_0^y$ and
  $m_0^z$ correspond to the two cases for calculating $\varrho_H^{xy}$
  and $\varrho_H^{xz}$, respectively. $\upsilon=v_{\rm m}/v_{\rm ph}$ is a dimensionless multiplier
  used to reproduce the effect of a varying magnetic exchange scale,
  which mainly impacts $v_{\rm m},\chi,\Lambda_i^{\xi}$.
    In the plots, we use the values $\upsilon=\mt{2.5},\mt{5},\mt{10}$.}
\label{tab:table-params}
\end{table}

Finally, note that the following scaling relations for $\kappa_L$,
\begin{equation}
  \label{eq:76}
  \kappa_L\left (\{\Lambda_{1,..,7}^{(\xi)}\},\gamma_{\rm ext}\right )= \zeta_0^{2}
    ~ \kappa_L\left (\{\zeta_0
    \Lambda_{1,..,7}^{(\xi)}\},\zeta_0^2\gamma_{\rm ext} \right ),
\end{equation}
and for $\varrho_H$,
\begin{align}
  \label{eq:52}
  &\varrho_H \left (\{\Lambda_{1,..,5}^{(\xi)}\},
    \{\Lambda_{6,7}^{(\xi)}\}, m_0^{y,z} \right ) \nonumber\\
  &\quad=
    \zeta_1^{-3}\zeta_{2}^{-1}\zeta_3^{-1} \varrho_H \left (\{\zeta_1 \Lambda_{1,..,5}^{(\xi)}\},
    \{\zeta_2 \Lambda_{6,7}^{(\xi)}\}, \zeta_3 m_0^{y,z} \right ) \nonumber\\
& \qquad+O \left ((m_0^{y,z})^3 (\Lambda_{6,7}^{(\xi)})^3\right ), 
\end{align}
hold for any rescaling factors $\zeta_{0,..,3}$. Eqs.~(\ref{eq:76},\ref{eq:52}) make it possible to extrapolate
results from our calculations for values of the parameters which are not explicitly
explored in Table~\ref{tab:table-params} and Figs.~\ref{fig:Dnk-plots}(a), \ref{fig:Dnk-plots}(b),
\ref{fig:kappaLplots}(a), and \ref{fig:kappaLplots}(b).

\subsubsection{Results for $\kappa_L$}
\label{sec:results-kappa_l}

\begin{figure*}[htbp]
  (a)
\begin{subfigure}[b]{0.46\textwidth}
  \includegraphics[width=\columnwidth]{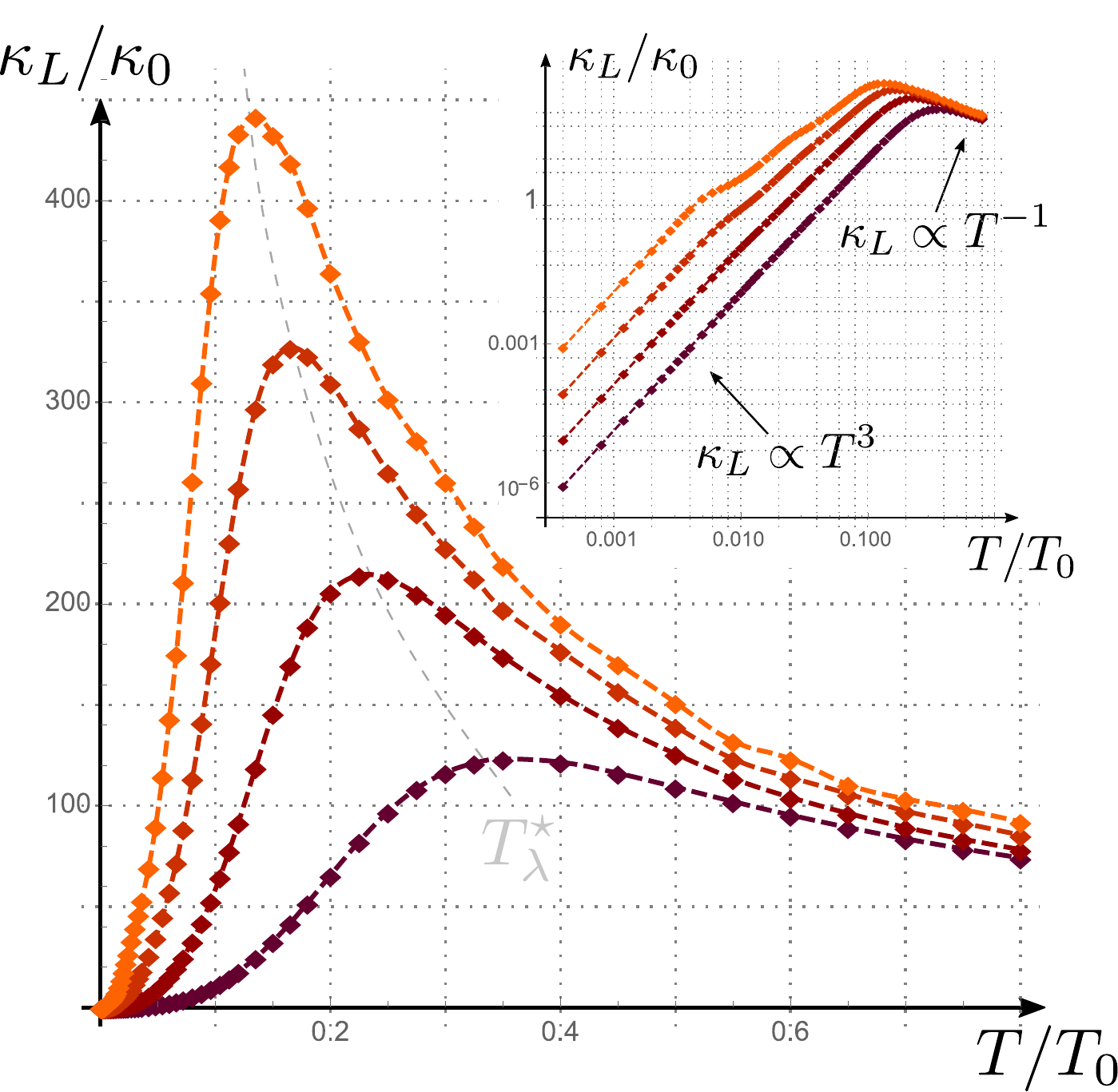}
\end{subfigure}
\hfill
(b)
\begin{subfigure}[b]{0.46\textwidth}
  \includegraphics[width=\columnwidth]{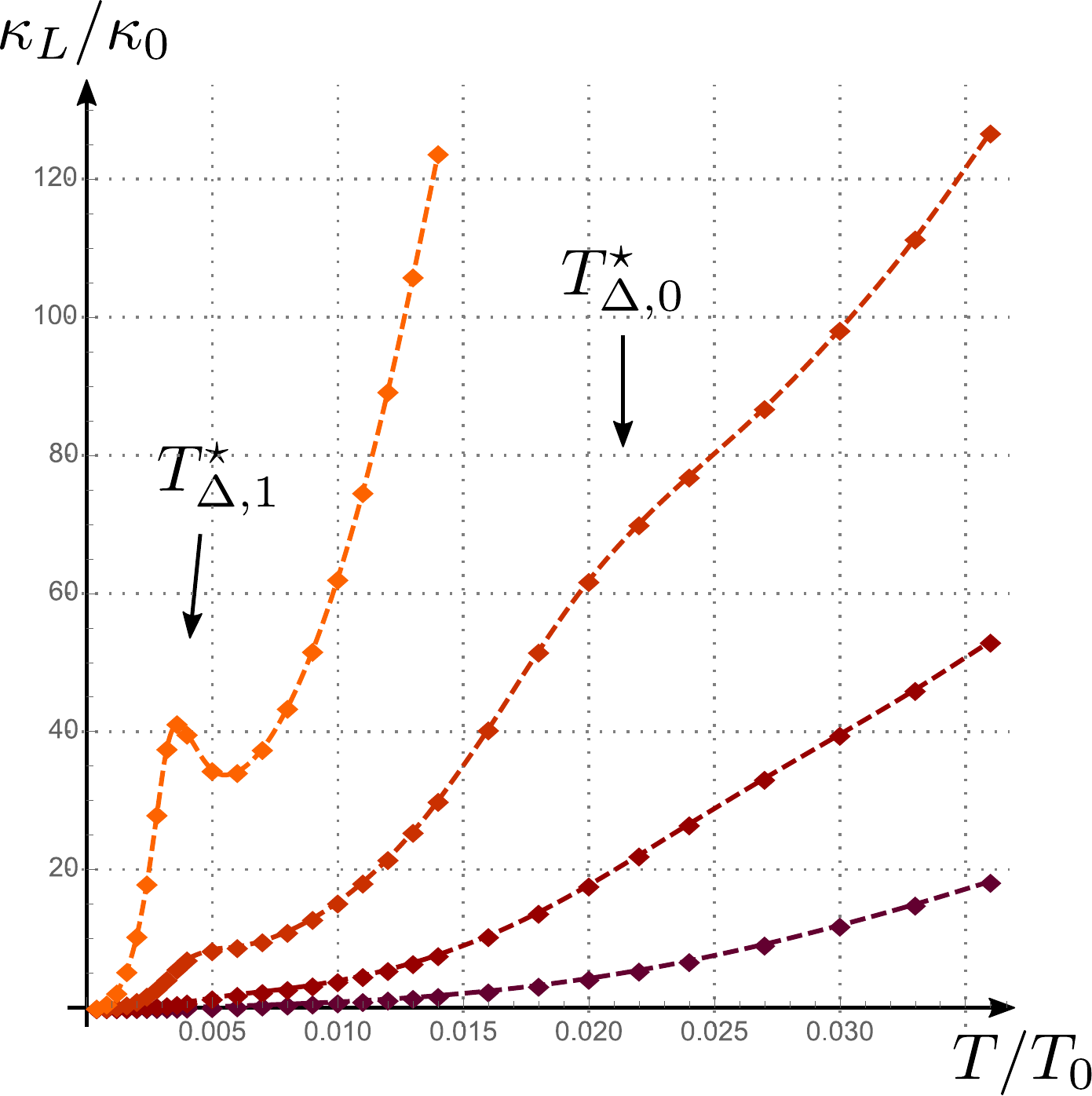}
  \end{subfigure}
  \caption{Longitudinal thermal conductivity $\kappa_L$ (in units of $\kappa_0=k_{\rm B}v_{\rm ph}/\mathfrak{a}^2$) with respect to temperature $T$ (in units of $T_0=\hbar v_{\rm ph}/(\mathfrak{a}k_{\rm B})$),
    for four different values of 
    $\gamma_{\rm ext}$. (a) $\gamma_{\rm ext}= 1\cdot 10^{-z}(v_{\rm ph}/\mathfrak{a}), z \in \llbracket 4, 7\rrbracket$,
    from darker $(z=4)$ to lighter $(z=7)$ shade. The dashed gray line indicates the evolution of the crossover temperature $T^\star_\lambda$ as a function of $\gamma_{\rm ext}$. Inset: log-log plot; the scaling behaviors are consistent with the analysis
  presented in the text. The inset is reproduced in App.~\ref{sec:app-figures-1}. (b) $\gamma_{\rm ext}= 1\cdot 10^{-z}(v_{\rm ph}/\mathfrak{a}), z \in \llbracket 6, 9\rrbracket$,
    from darker $(z=6)$ to lighter $(z=9)$ shade. The two crossover temperatures $T^\star_{\Delta,1}$ and $T^\star_{\Delta,0}$ are defined in the text up to a prefactor; here we identify the corresponding features in $\kappa_L$ but do not indicate specific values of $T$. See App.~\ref{sec:app-figures-1} for a log-log plot.}
     \label{fig:kappaLplots}
\end{figure*}

Numerical results for $\kappa_L(T)$ are displayed in
Figs.~\ref{fig:kappaLplots}(a), \ref{fig:kappaLplots}(b) at fixed $\upsilon=v_{\rm
  m}/v_{\rm ph}=2.5$.

Fig.~\ref{fig:kappaLplots}(a) shows plots of $\kappa_L(T)$ for several values
of the extrinsic scattering $\gamma_{\rm ext}$ and fixed $\upsilon=v_{\rm m}/v_{\rm ph}=2.5$.
This figure exhibits all the behaviors described in
Secs.~\ref{sec:long-scatt-rate-1}, \ref{sec:long-scatt-rate-2},
with the extra feature that here there are two crossovers
temperatures, $T^\star_{\Delta,0}$ and $T^\star_{\Delta,1}$
defined by the two different magnon gaps $\Delta_0,\Delta_1$ whose
values we give in Tab.~\ref{tab:table-params}.
These are more clearly visible in Fig.~\ref{fig:kappaLplots}(b),
where we show $\kappa_L(T)$ in a small window of low temperatures and for smaller values of $\gamma_{\rm ext}$. 

Four scaling regimes can then be identified:
\paragraph{} For $T\lesssim T^\star_{\Delta,1}$, only extrinsic scattering contributes to the full phonon scattering rate, 
  and $\kappa_L \propto T^3/\gamma_{\rm ext}$.
\paragraph{} For $T^\star_{\Delta,1} \lesssim T \lesssim T^\star_{\Delta,0}$, both the extrinsic and the $x=-1$ phonon-magnon
  (only in the $\ell=1$ valley) scattering rates contribute with the same scaling exponent,
  yielding $\kappa_L\propto T^3$ with a smaller proportionality coefficient than in the first regime.
\paragraph{} For $T^\star_{\Delta,0} \lesssim T \lesssim T^\star_\lambda$, both the extrinsic and the $x=-1$ phonon-magnon
  (now in both valleys)  scattering rates contribute with the same scaling exponent,
  yielding $\kappa_L\propto T^3$ with yet a smaller proportionality coefficient.
 \paragraph{} For $T>T^\star_\lambda$, the $x=+1$ phonon-magnon
 scattering rate is dominant and yields $\kappa_L\propto T^{-1}$.
Note that $T_\lambda ^\star$ is defined in Eq.~\eqref{eq:68} in the $\breve D_{n\mb k}=0$ case; here by $T_\lambda ^\star$ we mean the more general crossover temperature in the presence of a finite $\breve D_{n\mb k}=\gamma_{\rm ext}$.
 
The exponents quoted above are found with very good accuracy from a log-log scale plot (see inset of Fig.~\ref{fig:kappaLplots}(a) and Appendices), regardless of the value of $\gamma_{\rm ext}$; in that sense these exponents are universal. The influence of (non-universal) $\gamma_{\rm ext}$ on the results of Fig.~\ref{fig:kappaLplots}(a) is essentially threefold:
\begin{itemize}
\item Since the full phonon scattering rate is
$D_{n\mb k}=\gamma_{\rm ext}+D^{(1)}_{n\mb k}$, unsurprisingly $\kappa_L(T)$ is always a decreasing function of $\gamma_{\rm ext}$.
\item The ``bumps'' at $T\sim T^\star_{\Delta,\ell}$ come from the
  fact that the $x=-1$ phonon-magnon scattering rate is much
larger than $\gamma_{\rm ext}$ as soon as the gap permits this scattering process; therefore, for large enough $\gamma_{\rm ext}$,
this feature disappears. More precisely, one should compare
  $\gamma_{\rm ext}$ with $D_{nn,\ell}:=\eta^2 f^2 \Delta_\ell/M_{\rm uc}$,
  where the dimensionless parameters $\eta,f$ are defined by $\lambda_{nn}\simeq \eta \lambda_{mm}$
  and $\Lambda_{1..5}\simeq f J/\mf a$. The first bump is noticeable iff $\gamma_{\rm ext}\lesssim D_{nn,1}$,
  and the second bump is noticeable iff ${\rm max}(\gamma_{\rm ext}, D_{nn,1})\lesssim D_{nn,0}$.
\item Since $\gamma_{\rm ext}$ and the $\lambda_{nn}$ coupling lead to the same scaling exponent,
the $T\sim T^\star_\lambda$ crossover results from a competition between $\lambda_{mm}$ on the one hand and
$(\gamma_{\rm ext},\lambda_{nn})$ on the other; thus the larger
$\gamma_{\rm ext}$, the greater the dependence of $T^\star_\lambda$ on
$\gamma_{\rm ext}$, and $T^\star_\lambda(\gamma_{\rm ext})$ is an increasing function of $\gamma_{\rm ext}$.
\end{itemize}

Finally, Fig.~\ref{fig:kappaLvmvph} shows plots of $\kappa_L(T)$ for several values
of the velocity ratio $v_{\rm m}/v_{\rm ph}=\upsilon$ at fixed
$\gamma_{\rm ext}=10^{-6}$. In particular, we recover, at
  $T\gg T^{\star}_\lambda$, the particularly non-trivial behavior described in
Sec.~\ref{sec:new}; namely that for all $\upsilon$ values greater than
$\upsilon=3$ the high-temperature behavior of $\kappa_{\rm L}$ goes like
$T^{3-8(\upsilon-1)^{-1}}$ (Eq.~\eqref{eq:86}) and the exponent indeed changes signs at $\upsilon
= 11/3$, i.e.\ the conductivity {\em increases} with increasing
temperature for $\upsilon>11/3$.

\subsubsection{Results for $\varrho_H$}
\label{sec:results-rho_h}

\begin{figure}[htbp]
  \centering
\includegraphics[width=\columnwidth]{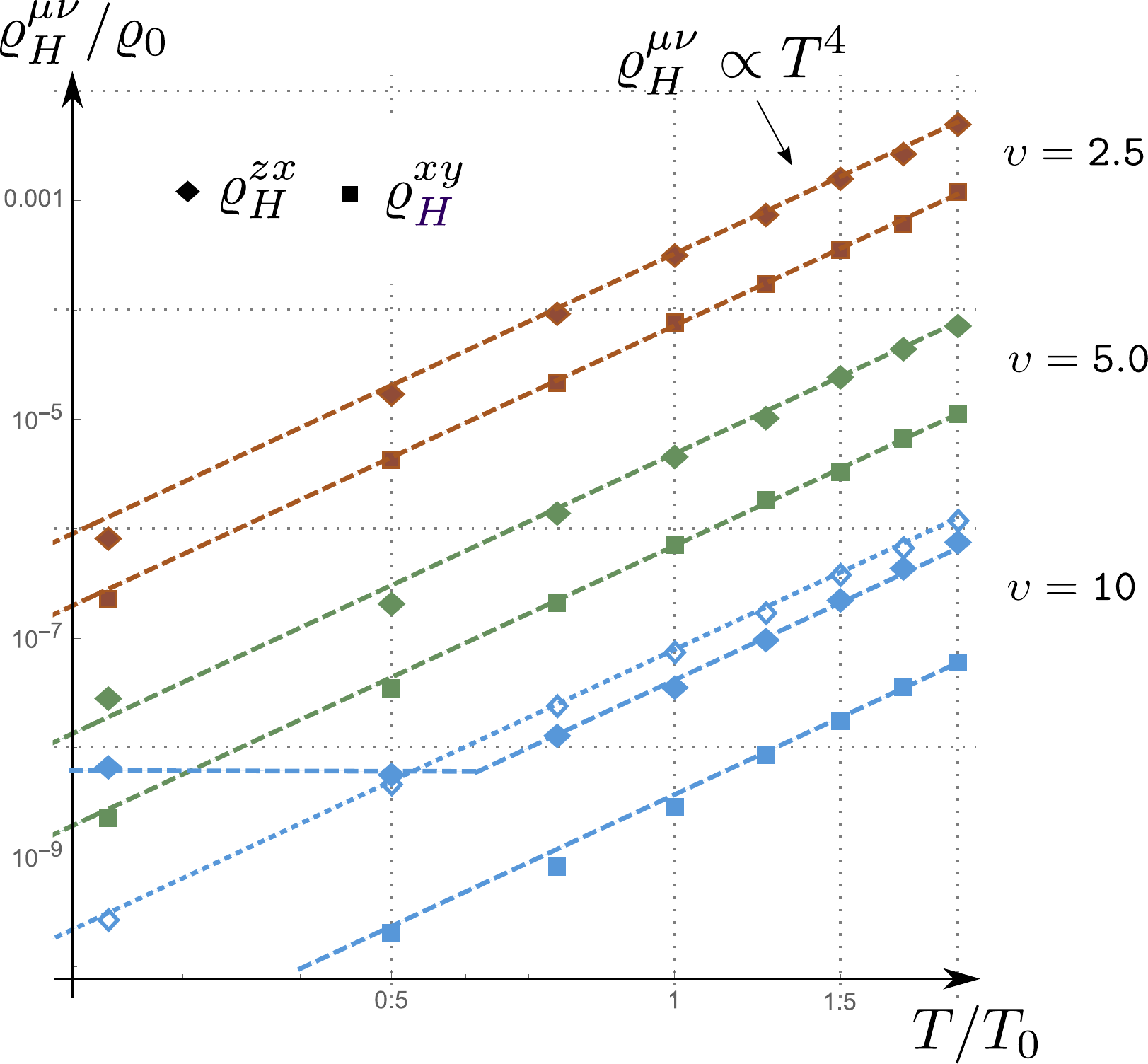}
\caption{Hall resistivities $\varrho_{\rm H}^{xy}(T)/\varrho_0$ and
  $\varrho_{\rm H}^{xz}(T)/\varrho_0$ for three values of
  $\upsilon =v_{\rm m}/v_{\rm ph}$.  The low temperature saturation of
  $\varrho_{\rm H}^{xz}(T)/\varrho_0$ observed for $\upsilon=10$ is
  due to the non-negligible contributions of the $\lambda_{nn}$ term
  in that range.  This is confirmed by the data shown in empty blue
  diamonds, which is the result of calculations at the same value of
  $v_{\rm m}/v_{\rm ph}=10$ and approximately the same coupling
  constants as the full blue diamonds, except that $\lambda_{nn}=0$
  there (the values of $\lambda_{mm}$ are also slightly different but,
  importantly, the values of $\lambda_{mn}$ are unchanged).}
  \label{fig:rhoHloglog}
\end{figure}

We evaluated numerically $\varrho_H^{\mu\nu}$ for both $\mu\nu = xy$
and $xz$, in both cases with a net magnetization $\bs m_0$ oriented
along $\rho$, the axis perpendicular to the Hall plane
$\mu\nu$. Results are presented in Fig.~\ref{fig:rhoHloglog}.  Here
the dashed straight lines on the double logarithmic scale indicate the expected $T^4$ scaling.

This behavior is consistent with the arguments given in Sec.~\ref{sec:transv-scatt-therm}, especially Eq.~\eqref{eq:69}, with $d=2$-dimensional magnons and scaling exponent $x=+1$, corresponding to the temperature regime where isotropic exchange dominates over the phonon-magnon coupling. We expect from Eq.~\eqref{eq:16} that deviations from this scaling behavior would be observed at lower temperatures, not investigated here.

We emphasize that the numerical values of $\varrho_H^{xy}$ and
$\varrho_H^{zx}$ are of the same order of magnitude. This is
remarkable in a layered system which has entirely different magnon
dynamics in the $xy$ and $xz$ planes, in this case where magnons are
explicitly two-dimensional, carrying energy only within $xy$
layers. It can be understood from the fact that here {\em phonons} are
isotropic, carrying energy in all three directions, and that {\em
  including} $\mc T$-odd scattering exists in all directions,
therefore allowing a Hall effect in both the $xy$ and $xz$
directions.


Numerically evaluating the dependence on $\upsilon=v_{\rm m}/v_{\rm ph}$
of $\varrho_{\rm H}$, we find that $|\varrho_H^{xy}/\varrho_H^{xz}|<1$
for all the values of $\upsilon$ studied, and that the prefactors of
$\varrho_H^{xy}, \varrho_H^{xz}$ are rapidly suppressed for large
$\upsilon$, as shown in Fig.~\ref{fig:rhoHloglog}.  From a simple
analysis, we expect $|\varrho_H^{xy}/\varrho_H^{xz}|\propto 1/\upsilon$
for large $\upsilon$.  The reason for the overall suppression of the
Hall resistivity with increasing $\upsilon$ is also clearly due to the
diminishing phase space for scattering, but we have not obtained the
exact dependence analytically.

Finally, we note that for our {\em choice} of antiferromagnetic order along
the $\hat x$-axis in this model and within linearized
spin-wave theory, $\varrho_H^{yz}=0$.

\subsubsection{Results for $D_{n\mb k}$}
\label{sec:results-d_nmb-k}

\begin{figure*}[htbp]
  \centering
        (a)
  \begin{subfigure}[b]{0.46\textwidth}
  \includegraphics[width=.825\columnwidth]{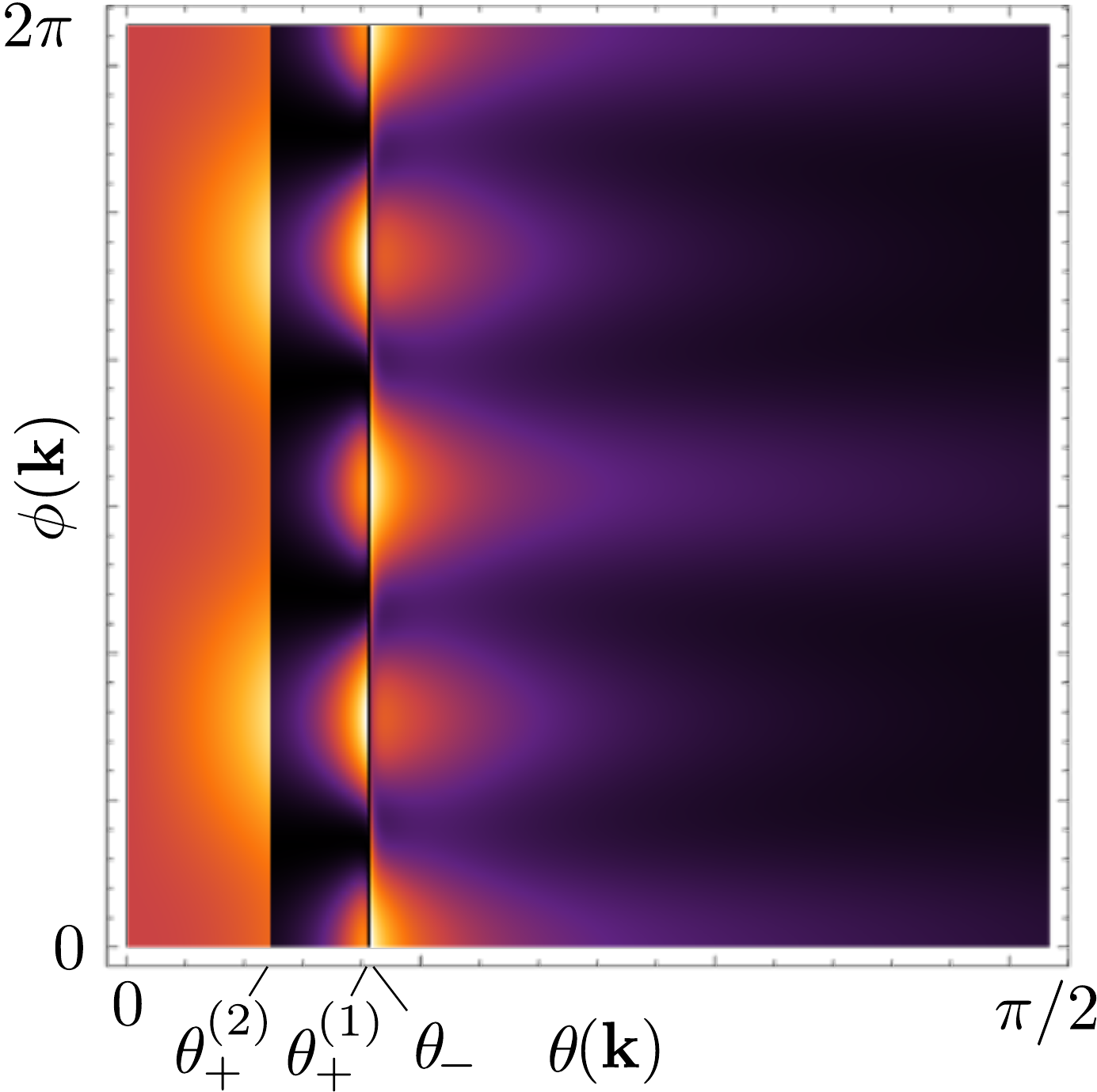}
   \includegraphics[width=.115\columnwidth]{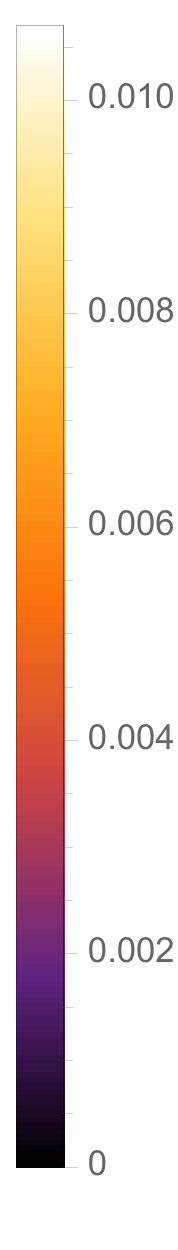}
   \label{fig:Dnk-angle-angle}
 \end{subfigure}
     (b)
 \begin{subfigure}[b]{0.46\textwidth}
  \includegraphics[width=.825\columnwidth]{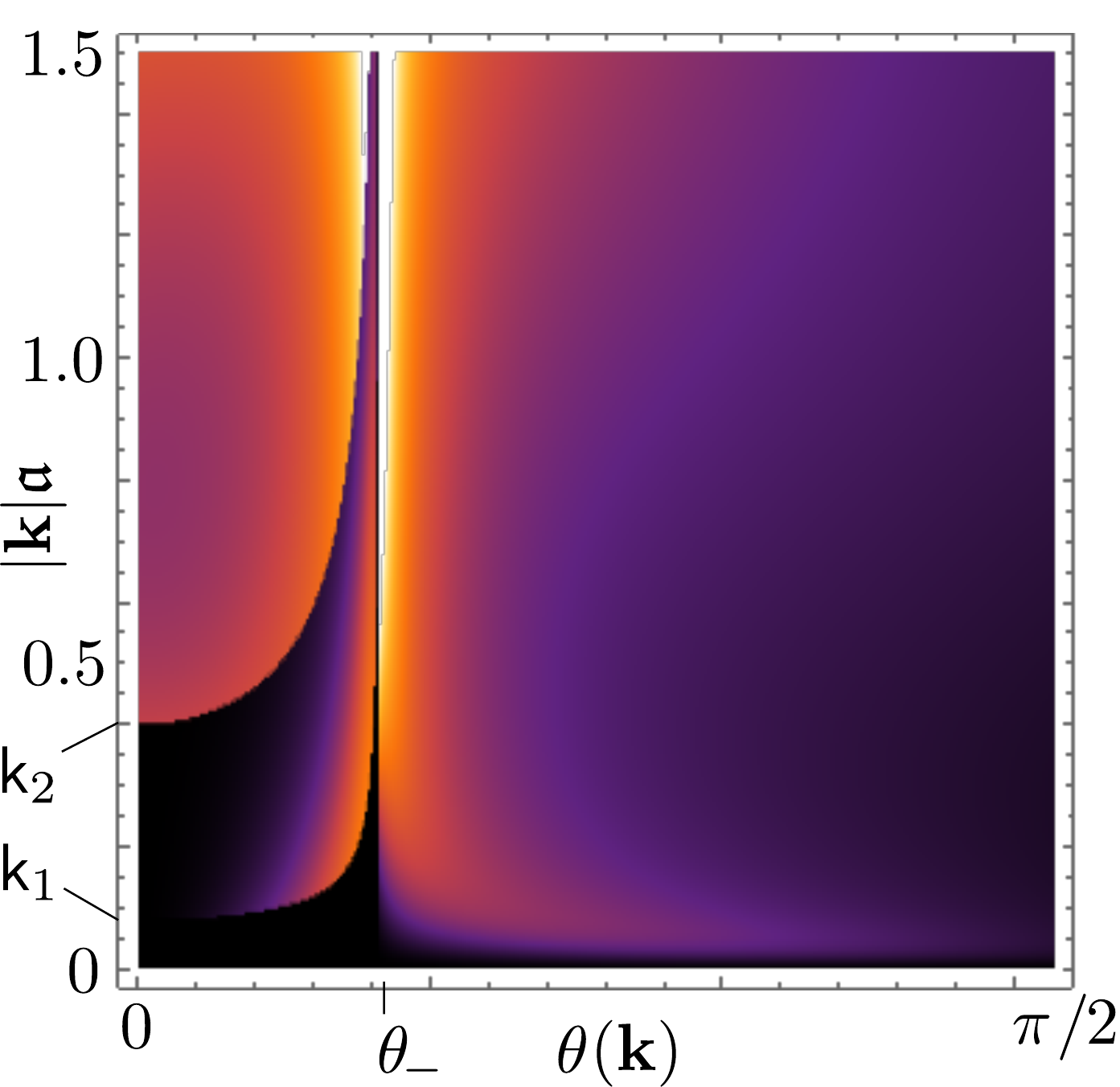}
  \hfill \includegraphics[width=.125\columnwidth]{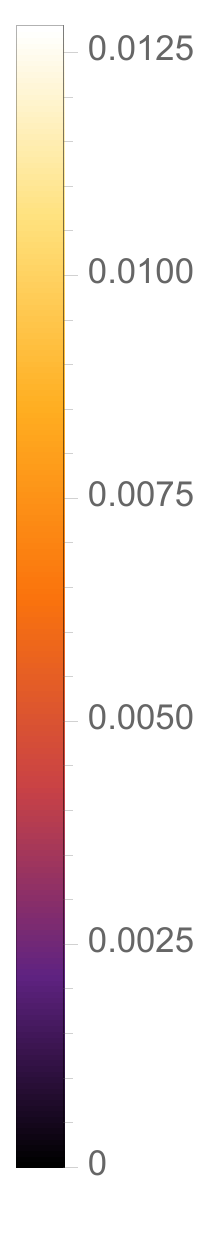}
  \label{fig:Dnk-angle-norm}
  \end{subfigure}
  \caption{Diagonal scattering rate $D_{n\mb k}^{(1)}/\gamma_0$ at fixed temperature $T=0.5T_0$ and in polarization
    $n=0$. (a) as a function of $\theta(\mb k)={\rm arccos}(k_z/|\mb
    k|) \in [0,\pi/2]$ (horizontal axis) and $\phi(\mb k)={\rm
      Arg}(k_x+\text ik_y) \in [0,2\pi]$ (vertical axis) for fixed $|\mb k|=0.5/\mathfrak{a}$. (b) as a function of $\theta(\mb k)={\rm arccos}(k_z/|\mb
    k|) \in [0,\pi/2]$ (horizontal axis) and $|\mb k|\mf a$ (vertical axis) for fixed $\phi(\mb k)=0$. Other parameter values are explored in App.~\ref{sec:app-figures-1}.}
    \label{fig:Dnk-plots}
\end{figure*}

Fig.~\ref{fig:Dnk-plots}(a) shows the angular dependence of
$D_{n\mb k}$. Throughout this section, we use
$\ul{\mathbf{k}}=\ul{k}(\cos\phi\mathbf{\hat{u}}_x+\sin\phi\mathbf{\hat{u}}_y)$,
with $\phi\in[0,2\pi[$, 
and $k_z=k\cos\theta$ with $\theta\in[0,\pi]$. Note that, in turn, $\ul{k}^2=k^2\sin^2\theta$. Also, since all the results are invariant under 
$k_z\rightarrow -k_z$ i.e. $\theta\rightarrow \pi-\theta$, we plot results for $\theta\in[0,\pi/2]$ only.

\paragraph{In-plane $\phi(\mb k)$ angular dependence.} We see from
Fig.~\ref{fig:Dnk-plots}(a) that phonon-magnon scattering is typically larger for
values of $\phi(\mb k)$ associated with high-symmetry axes of the
system, i.e.\ $\phi=0\;{\rm mod}[\pi/2]$.
This is inherited from the structure of $\mc L = {\rm Tr}[\bs \lambda^T \bs{\mc S}]$
in Eq.~\eqref{eq:155}, which enters the magnetoelastic coupling,
Eq.~\eqref{eq:133}. The latter is by definition invariant under all symmetries
of the crystal, so that components of the strain tensor couple to functions
of the magnetization fields $\bs m,\bs n$ with the same symmetries.

Now, while the symmetry group of the crystal structure is tetragonal,
the $C_4$ symmetry is spontaneously broken
by the antiferromagnetic order along the $x$ axis, while the $C_2$ and mirror symmetries
are preserved when the magnetic field is along the $z$ axis.
More precisely, {\em how} does the $C_4$ symmetry (as acting on the $\alpha,\beta$ indices)
break? Since the $\bs{\mc S}$ factor in Eq.~\eqref{eq:155} has the same structure as the
strain tensor itself, it preserves $C_4$; therefore the latter can only be broken in the $\bs \lambda$ factor.
Let us focus on the $\lambda_{mm},\lambda_{nn}$ cases, since these coefficients can be nonzero in the absence
of a net magnetization $m_0$. A broken $C_4$ symmetry then means that
$ 0 \neq \lambda'_{\xi,a} := \lambda^{xx}_{aa;\xi\xi} - \lambda^{yy}_{aa;\xi\xi}$.
By inspection of Eqs.~\eqref{eq:134} and
\eqref{eq:15}, one sees that there are two ways the latter
can be nonzero: {\em (1)}
in the $(\xi={\rm n})$ channel, $\lambda'_{{\rm n},a}$ is proportional to anisotropic exchanges;
and {\em (2)} in the $(\xi={\rm m})$ channel, $\lambda'_{{\rm m},a}$ contains both
isotropic and anisotropic exchange constants, and is consequently much larger than $\lambda'_{{\rm n},a}$.
From this analysis, it follows that the deviation from $C_4$ symmetry
as captured in $D^{(1)}_{n\mathbf{k}}$ by $\lambda'_{\xi,a}$ is largest for values of $|\mb k|$ 
where the $\lambda_{mm}$ contributions dominate over the
$\lambda_{nn}$ ones, i.e.\ at large $|\mb k|$ (recall Eq.~\eqref{eq:153}).
One can check that this is indeed the case, as is shown in
Appendix~\ref{sec:app-figures-1}.

\paragraph{Out-of-plane $\theta(\mb k)$ angular dependence.} The
out-of-plane angular dependence illustrates quite clearly the
dynamical constraints satisfied by $D_{n\mb k}$, as outlined in
Sec.~\ref{sec:diag-scatt-rate}.
By inspection of Eq.~\eqref{eq:53}, we define
  \begin{align}
    \label{eq:73}
    \theta_- &= {\rm arctan}\left(v_{\rm ph}^{-1}\sqrt{v_{\rm ph}^2-v_{\rm m}^2}\right),\\
\theta_+^{(1)}(|\mb k|) &= {\rm arcsin}\left ( v_{\rm m}^{-1}\sqrt{v_{\rm ph}^2- 4\Delta'/|\mb k|^2}\right ),\\
\theta_+^{(2)}(|\mb k|) &=  {\rm arcsin}\left ( v_{\rm m}^{-1}\sqrt{v_{\rm ph}^2- 4\Delta/|\mb k|^2}\right ),
  \end{align}
  where $\Delta' = {\rm max}(\Delta_0,\Delta_1)$. Note that outside
  the domain of definition of $\sqrt{\dots}$, by continuity one fixes $\theta_+^{(1,2)}:=0$.
The figure Fig.~\ref{fig:Dnk-plots}(a) can then be divided in four areas as follows:
\begin{itemize}
\item The vertical black band at angles $\theta(\mb k) \in [\theta_+^{(1)}(|\mb k|),\theta_-]$ corresponds to values of $(k_z,|\ul{\mb k}|)$ such that energy and momentum
conservation cannot be satisfied simultaneously because of the magnon gap
$\Delta$; therefore
$D_{n\mb k}^{+}=0=D_{n\mb k}^{-}$.
\item For angles $\theta(\mb k)>\theta_-$, scattering of the
``ph+m $\rightarrow$ m'' type becomes possible, i.e.\
$D^->0$. Meanwhile, following Eq.~\eqref{eq:53}, $D^+=0$.
\item Conversely, for $\theta(\mb k)<\theta_+^{(2)}(|\mb k|)$, scattering of the ``ph $\rightarrow$ m+m''
type becomes possible, i.e.\ $D^+>0$, while $D^-=0$.
\item For $\theta \in [\theta_+^{(2)},\theta_+^{(1)}]$, scattering of the ``ph $\rightarrow$ m+m''
type is possible only in the valley with the smallest gap, while in the other no scattering can happen;
therefore, in that region $D^+>0$ but its value drops (without vanishing a priori) at the interface
$\theta(\mb k)=\theta_+^{(2)}(|\mb k|)$.
\end{itemize}

\paragraph{Dependence on $|\mb k|$.} In Fig.~\ref{fig:Dnk-plots}(b),
we show the dependence of $D_{n\mathbf{k}}$ as a function of the norm
$|\mathbf{k}|$ and the out-of-plane angle $\theta$. This plot displays
divergences near the singular lines $\theta_+^{(1,2)},\theta_-$, which can be attributed
to the thresholds for magnon scattering just above the gaps. 

The angular width $\delta\theta(\mb k)$ of the two black and darker regions bounded from the right by $\theta_-$,
where scattering is forbidden in at least one of the two valleys, varies with $|\mb k|$.
From Eq.~\eqref{eq:53}, we see that this width scales like $\delta
\theta \sim (\Delta_\ell/v_{\rm ph}|\mb k|)^2$.
These regions extend down to $|\mb k|=0$, reflecting the fact that
phonons with too little energy are unable to excite magnon pairs. The momentum magnitude thresholds for the excitation of magnon pairs are naturally given by ${\sf k}_1=2\Delta /v_{\rm ph}$ and ${\sf k}_2=2\Delta'/v_{\rm ph}$.

\subsubsection{Results for $\mf W^\ominus_{n\mb k n'\mb k'}$}
\label{sec:results-mf-womin}

\begin{figure}[htbp]
  \centering
  \includegraphics[width=.825\columnwidth]{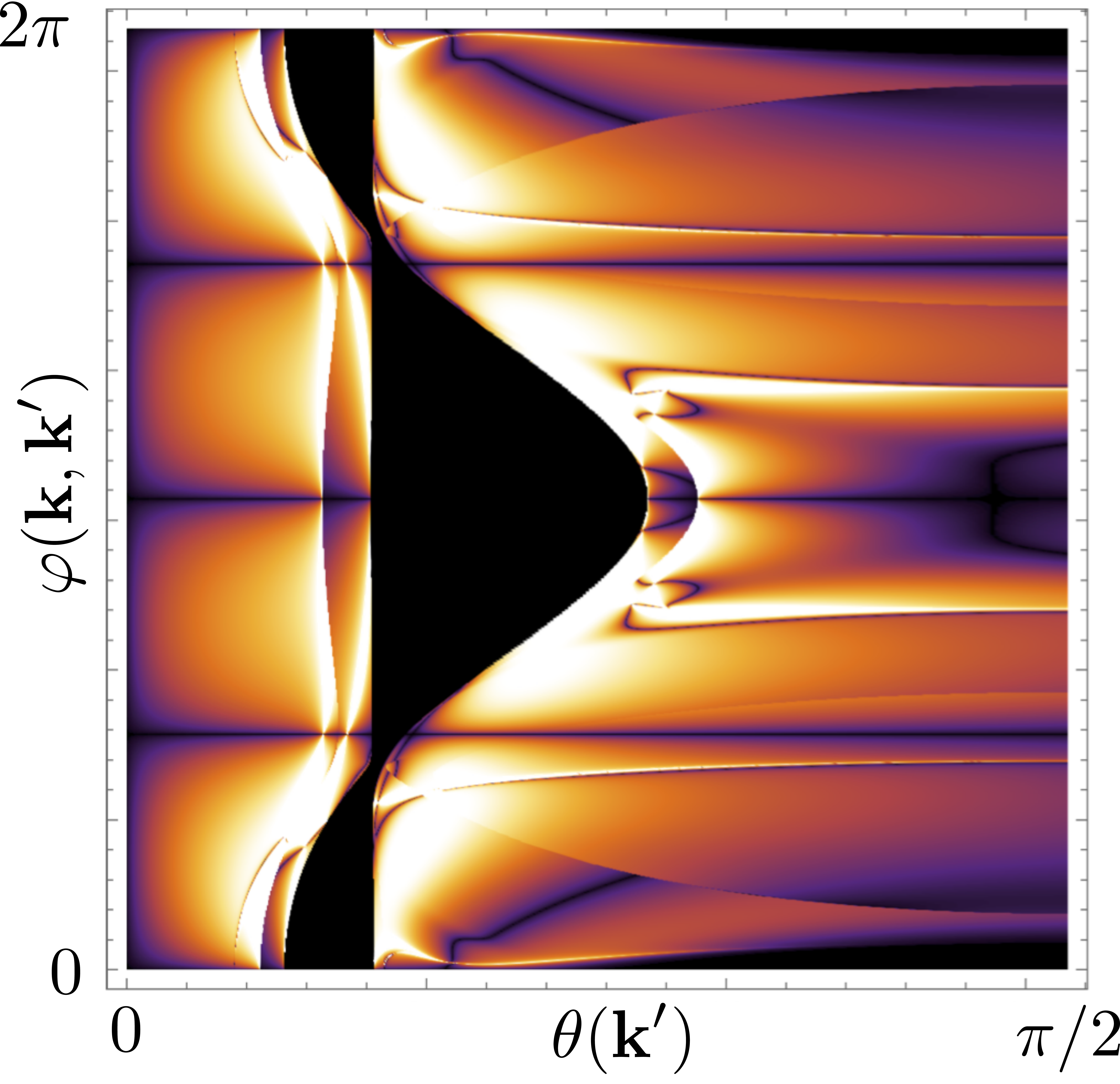}
  \hfill \includegraphics[width=.16 \columnwidth]{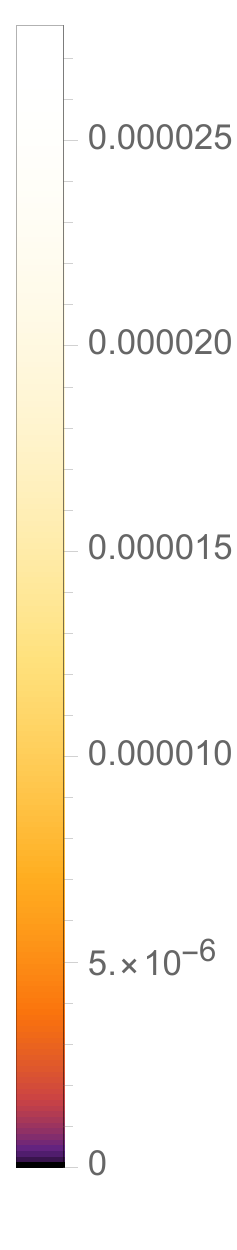}
  \caption{Skew-scattering rate $\mf W^{\ominus,--}_{n\mb k n'\mb k'}/\gamma_0$ 
  as a function of $\theta(\mb k')\in [0,\pi/2]$ (horizontal axis) and 
  $\varphi(\mb k,\mb k')=\phi(\mb k')-\phi(\mb k)$ (vertical axis) for fixed $|\mb k'|=\mt{0.8}/\mf a$, $k_x=0$,
  $k_y = \mt{0.2}/\mf a$, $k_z = \mt{0.1}/\mf a$, $\bs m_0=\mt{0.05}~\hat{\bs z}$
  and temperature $T=\mt{0.5}~T_0$. Other parameter values are
  explored in App.~\ref{sec:app-figures-1}. Note that the colorbar is not scaled linearly.}
  \label{fig:WH}
\end{figure}

Although the angular dependences of the $\mf W^{\ominus,qq'}_{n\mb k n'\mb k'}$ skew-scattering rates
are more intricate than those of $D_{n\mb k}$, a few general remarks can be made. In particular, in Fig.~\ref{fig:WH}, where we plot $\mathfrak{W}^{\ominus,-+}$ as a function of $\theta$ and $\varphi$ at fixed $|\mb k'|=\mt{0.8}/\mf a$, $k_x=\mt{0.2}/\mf a$,
  $k_y = 0$, $k_z = \mt{0.1}/\mf a$ (and temperature $T=\mt{0.5}T_0$), we have: 
\begin{itemize}
    \item Although $k_z\neq 0$, we can still take advantage of the $k'_z \leftrightarrow -k'_z$ symmetry, and it is sufficient to consider $\theta(\mb k')\in [0,\pi/2]$. This comes from the fact that, for purely planar magnons, the phonon momenta $k_z,k'_z$ are not coupled. Meanwhile there is a priori no $\varphi(\mb k,\mb k')\leftrightarrow -\varphi(\mb k,\mb k')$ symmetry except when $\ul{\mb k}$ is along one of the high-symmetry axes of the crystal, as is the case here (cf.\ $k_y=0$).
    \item The vertical black line at $\theta(\mb k')=\theta_-$ can still be identified, and corresponds to magnons being gapped as in $D^{(1)}_{n\mb k}$. However, in $\mf W^\ominus$, the width and position of the gapped (black) zone now depend also on $\varphi(\mb k,\mb k')$, due to the second energy conservation constraint in $\mf W^\ominus$ (a feature absent in $D^{(1)}$ where there is only one energy constraint).
    \item In Appendix~\ref{sec:app-figures-1}, we explore other orientations of in-plane $(k_x,k_y)$, and show that the features of $\mf W^{\ominus,qq'}_{n\mb k n'\mb k'}$ quoted above still hold. This is consistent with the above observations being consequences of the energy conservation constraints, which depend only of the {\em relative} angle $\phi(\mb k)-\phi(\mb k')$ since both phonon and magnon dispersions are isotropic in the $xy$ plane.
    \item In Fig.~\ref{fig:WH}, $\mf W^{\ominus,-+}_{n\mb k n'\mb k'}$ also seems to vanish along certain special lines, especially those located at $\varphi(\mb k,\mb k')=0,\pi/2,\pi,3\pi/2,2\pi$. These features are {\em not} independent of the orientation $\phi(\mb k)$; in fact they are salient features of the in-plane momenta being along the high-symmetry axes of the crystal. Thus, they do not result from energy conservation constraints, but from subtle effects in the structure of Eq.~\eqref{eq:110}.
\end{itemize}
Finally, we point out that the values of $\mf W^{\ominus}$ in Fig.~\ref{fig:WH} are small compared to the values of $D^{(1)}$ obtained for similar values of momenta. This can be understood from the combination of {\em (1)} the anti-detailed-balance structure of $\mf W^{\ominus}$, from which it follows that $\mf W^{\ominus,qq'}_{n\mb k,n'\mb k'} + \mf W^{\ominus,qq'}_{n-\mb k,n'-\mb k'}=O(\bs m_0)$ as shown in Sec.~\ref{sec:time-revers-symm}, and {\em (2)} the $C_2$ symmetry of the system around the $\hat z$ axis, which (since for planar magnons $k_z \leftrightarrow -k_z$ is a symmetry) entails $\mf W^{\ominus,qq'}_{n\mb k,n'\mb k'} = \mf W^{\ominus,qq'}_{n-\mb k,n'-\mb k'}$. Thus $\mf W^{\ominus,qq'}_{n\mb k,n'\mb k'}=O(\bs m_0)$ itself. This, together with the analysis given in Sec.~\ref{sec:effect-break-symm} showing that terms which are odd in $\bs m_0$ are also proportional to anisotropic couplings, implies that $\mf W^{\ominus,qq'}_{n\mb k,n'\mb k'}$ is indeed typically much smaller than $D^{(1)}_{n\mb k}$.

\subsection{Discussion of the results in absolute scales}

Here we discuss the absolute scales of $\kappa_L$, $\varrho_H$ and $\kappa_H$ we obtain using the parameter values from Table~\ref{tab:table-params} and those in the figure captions. First it is instructive to estimate the basic scales for thermal conductivity and temperature derived from phonons, which define the scales for our numerical plots.  Using the phonon velocity for CFTD, $v_{\rm ph}^{\rm CFTD}=4\cdot 10^{3}$~m$\cdot$s$^{-1}$ and its in-plane lattice parameter $\mathfrak{a}^{\rm CFTD}=5.7\cdot10^{-10}$~m, we find (see Table~\ref{tab:units}),
\begin{itemize}
    \item $\kappa_0^{\rm CFTD}=0.17$~W$\cdot$K$^{-1}\cdot$m$^{-1}$,
    \item $T_0^{\rm CFTD}=54$~K,
    \item $\gamma_0=7.0\cdot10^{12}$~Hz.
\end{itemize}
Note that these scales do not vary greatly for many materials.  For example, in La$_2$CuO$_4$, we find $\kappa_0^{\rm LCO} = 0.38$~W$\cdot$K$^{-1}\cdot$m$^{-1}$ and $T_0^{\rm LCO} =80$~K.  Importantly, the scale $\kappa_0$ is order one in SI units, which allows a roughly direct comparison with most data.

Next we can use the actual computed values to see what this mechanism predicts for the ``test'' material CFTD.  We have at $T\approx 0.5T_0\approx27$~K,
\begin{itemize}
    \item $\kappa_L^{\rm CFTD}\approx125\kappa_0\approx 22$~W$\cdot$K$^{-1}\cdot$m$^{-1}$ for any of the $\gamma_{\rm ext}$ values presented in Fig.~\ref{fig:kappaLplots}(a),
    \item for $\gamma_{\rm ext}=10^{-4}\gamma_0=7.0\cdot10^{8}$~Hz, $T_{\lambda}^{\star,{\rm CFTD}}\approx0.3T_0\approx16$~K,
    \item for $\gamma_{\rm ext}=10^{-7}\gamma_0=7.0\cdot10^{5}$~Hz, $T_{\lambda}^{\star,{\rm CFTD}}\approx0.1T_0\approx5.4$~K,
    \item $\varrho_H^{\rm CFTD}\approx 2\cdot10^{-5}\varrho_0\approx 1.2\cdot10^{-4}$~K$\cdot$m$\cdot$W$^{-1}$,
    \item $|\theta_H^{\rm CFTD}|\approx 2.6\cdot10^{-3}$,
    \item $|\kappa^{\rm CFTD}_H|\approx 5.8\cdot10^{-2}$~W$\cdot$K$^{-1}\cdot$m$^{-1}$.
\end{itemize}
Note that $\kappa_L$, $\kappa_H$ and $\theta_H$ all depend on the choice of values for $\gamma_{\rm ext}$.

\section{Conclusions}
\label{sec:conclusions}

\subsection{Summary of results and method}
\label{sec:summ-results-meth}

In this paper, we studied the problem of scattering of phonons due to
a weak {\em intrinsic} (i.e.\ without disorder) coupling to a
fluctuating field $Q$, which is itself a quantum mechanical degree of
freedom.  Using the T-matrix formalism, we derived the scattering
rates of phonons up to fourth order in coupling.  The result is
expressed generally, without any assumptions on the nature of the
fluctuating field (i.e.\ it can be highly non-Gaussian), in terms of
correlation functions of $Q$.  Using these scattering rates in the
Boltzmann equation leads to general expressions for the thermal
conductivity tensor, and, when symmetry allows, a non-vanishing
thermal Hall effect.  A central result is that the skew scattering of
phonons (which we define sharply as a scattering component which obeys
an {\em anti-}detailed balance relation), and hence the thermal Hall
conductivity, is proportional to a four-point correlation function of
$Q$, which we give explicitly.  We highlight throughout the various
constraints due to symmetry (both exact and approximate), unitarity,
and thermal equilibrium.

As an illustration of the method, we applied these results to the case
where the fluctuating field $Q$ arises from spin wave (magnon)
excitations of an ordered two-sublattice antiferromagnet.  We model
the latter via standard spin wave theory, for which phase space
constraints imply that the dominant contribution arises from bilinears
in the creation/annihilation operators of the spin waves.  We obtain a
general formula for the second order and fourth order scattering rates
in terms of the dispersion of phonons and magnons, and the
spin-lattice coupling constants.  To obtain concrete results, we focus
in particular on the limit in which the relevant magnons are acoustic,
and we assume tetragonal symmetry and two-dimensionality of the
magnons (but we retain the three dimensionality of the phonons).  Under these
assumptions we obtain all the (seven) symmetry-allowed spin-lattice
coupling interactions, and calculate the second order and fourth order
scattering rates, and thereby the thermal conductivity, including a
phenomenological parallel scattering rate of phonons due to other
mechanisms, e.g.\ boundary and impurity scattering.  The final formulae
are evaluated via numerical integration for representative model
parameters.  We observe a number of distinct scattering regimes, which
we identify with features in the longitudinal thermal conductivity.
We obtain a non-vanishing thermal Hall effect, in agreement with
general symmetry arguments.  Please see Sec.~\ref{sec:application} for
details. 

\subsection{About (anti-)detailed balance}
\label{sec:about-anti-detailed}

The detailed-balance and anti-detailed-balance relations,
Eq.~\eqref{eq:114}, played an important role in our discussion of the
thermal Hall effect.  A few comments on their nature and implications are appropriate.

{\em Quasi-equilibrium assumption:} These detailed-balance relations arise
(as generalizations of the Kubo-Martin-Schwinger relations \cite{martin1959theory})
from the assumption that the $Q$ fields relax to equilibrium between two scattering events.
This is typical in a linear response regime, when transport is dominated by the contribution
  of well-defined quasiparticles and drag effects are negligible.

{\em Role of the self-energy:} Within our treatment, the relations
we obtain rely on the fact that, \emph{at the order considered}, the
equilibrium phonon distribution is the unperturbed one
(this is shown in particular in Appendix~\ref{sec:energy-shift-phonons}).
At a general order in perturbation theory, this is not guaranteed a priori,
because the phonons are renormalized by an interaction-induced self-energy whose real part
shifts the dispersion relation and hence the equilibrium populations.
However, within the quasi-particle picture, it seems likely that this
assumption of a preserved spectrum of bare phonons is not necessary,
and the (anti-) detailed-balance relations should hold for the \emph{renormalized} phonon quasiparticles.

{\em No two-point contributions to the Hall effect:} The
detailed-balance relations enforce that all terms involved in the
calculation of the thermal Hall conductivity which contain two-point correlation
functions of the $Q$ fields cancel each other and therefore provide no
contributions to $\kappa_H$.
This relies on the linear-response limit: the cancellation occurs when we expand the collision integral to linear order
in the $\delta N$ out-of-equilibrium populations.

\subsection{Relation to other work}
\label{sec:relation-other-work}

While we are not aware of any general results on the intrinsic phonon
Hall conductivity due to scattering, there are a number of
complementary theoretical papers as well as some prior work which
overlap a small part of our results.  The specific problem of phonons
scattering from magnons was studied long ago to the leading second
order in the coupling by Cottam \cite{cottam_spin-phonon_1974}.  That work, which assumed the isotropic
SU(2) invariant limit, agrees with our calculations when these
assumptions are imposed.  The complementary mechanism of intrinsic
phonon Hall effect due to phonon Berry curvature was studied by many
authors \cite{qin2012berry,saito2019berry,zhang2010topological,zhang2021phonon}, including how the phonon Berry curvature is
induced by spin-lattice coupling in Ref.~\cite{ye2021phonon}.  The
majority of recent theoretical work has concentrated on {\em
  extrinsic} effects due to scattering of phonons by defects \cite{sun2021large,guo2021extrinsic,guo2022resonant,flebus2021charged}.  The
pioneering paper of Mori {\em et al.\ }\cite{mori} in particular recognized the
importance of higher order contributions to scattering for the Hall
effect, and is in some ways a predecessor to our work.

\subsection{General observations}
\label{sec:general-observations}

While often times scattering is regarded as a process which destroys
coherence and suppresses interesting dynamical phenomena, our work
reveals that higher order scattering probes highly non-trivial
structure of correlations.  Due to the constraints of detailed
balance, the skew scattering, appropriately defined, contains only contributions of $O(Q^4)$ and no terms of lower order in $Q$, and so can in principle directly reveal subtle
structures in the quantum correlations, without a need for
subtraction.  Measurements of such skew scattering of phonons---which
{\em a priori} include but are not limited to the thermal Hall effect---might therefore be considered a probe of the quantum material
hosting those phonons.  Taking advantage of this potential opportunity
is a challenge to experiment, as well as to theory, which should
interpret the results and predict systems to maximize the effects.

We would like to comment on the analysis of thermal Hall effect
experiments in quantum materials.  As is well-known, thermal Hall
conductivity is generally a small effect.  In particular, the
dimensionless measure of the Hall angle,
$\theta_H = \textrm{tan}^{-1}( \kappa_{\rm H}^{xy}/\kappa^{xx})$ is always
much less than $\pi/2$ by two or more orders of magnitude, even in
systems where thermal Hall effect is lauded as ``huge''. (An
actually large
thermal Hall angle ($\theta_H=O(1)$) is obtained only the quantum thermal Hall regime
when phonons are ballistic and edge states dominate over the bulk
phonon contributions, which is extraordinarily difficult to achieve.) 
For small $\theta_H$, the skew scattering contributions are
perturbative to the thermal conductivity, i.e.\ proportional to the
latter rate $1/\tau_{\rm skew}$.  Dimensional reasoning implies that
therefore $\kappa_H \sim \tau^2/\tau_{\rm skew}$, where $\tau$ is
the standard, non-skew scattering time.  This means that the thermal
Hall {\em conductivity} has a very strong dependence on $\tau$, which
is often sample-dependent and of course grows with sample quality,
implying that the thermal Hall conductivity is larger in cleaner
samples.

This dependence also means that $\kappa_H$ itself, as well as the
dimensionless Hall angle $\theta_H \sim \kappa_H/\kappa_L$ depend not
only on the skew scattering but also the ordinary scattering.  Since
the latter receives contributions from many different mechanisms,
which may themselves have strong temperature and field dependence,
neither $\kappa_H$ itself nor $\kappa_H/\kappa_L$ are ideal quantities
to examine to probe the physics of skew scattering.  Instead, we
suggest that the thermal Hall {\em resistivity},
$\varrho_H \equiv -\kappa_H/\kappa_L^2$, is the quantity which is most
easily interpreted physically.  This quantity is independent of the
non-skew scattering, at least when the latter is largely
momentum-independent, and is always independent of the overall scale of
non-skew scattering.  The temperature and field dependence of $\varrho_H$
is generally expected to be simpler than that of the other quantities,
at least when phonon skew scattering is the dominant mechanism for the
Hall effect.  This expectation is true not only when the skew
scattering is intrinsic, as studied here, but also for extrinsic skew
scattering due to defects.

\subsection{Future directions}
\label{sec:future-directions}

Our general formalism can be applied very broadly.  In particular,
because it does not require any assumptions on the nature of the $Q$
correlations, it may be applied directly to exotic states, to quantum
or classical critical points, or to situations in which the $Q$ field
is a composite operator.  We will present an application to fermionic
systems, including the spinon Fermi surface spin liquid, in an
upcoming paper.  Apart from other specific applications which may be
easily imagined, it would also be interesting to explore further how
general properties of four-point correlations of $Q$ may be detected
via phonon skew scattering.  In particular, the correlations which
enter the scattering rates are not obviously time-ordered, and we
wonder if these might contain some information on many-body chaos
(Ref.~\cite{swingle}).

Despite the generality of our formulation, it is still specialized
in several ways.  We consider only scattering contributions to the
phonon Boltzmann equation.  In general the interactions with fields
$Q$ will  both induce scattering and modify the dynamics of the
phonons in a non-dissipative way, e.g.\ induce phonon Berry phases \cite{ye2021phonon}.
While we believe it is usually the case that scattering is dominant, a
more complete treatment including both effects would be of interest.
Furthermore, in this paper we fully ``integrate out'' the electronic
degrees of freedom, and follow the distribution function of the
phonons only.  More generally, there are coupled modes of phonons and
electronic states, and one can consider the distributions for these
coupled modes.  One expects such effects are important largely when
there are resonances between phonons and electronic excitations.  All
these problems could be addressed via a Keldysh treatment of coupled
quantum kinetic equations, which is an interesting subject for future
work.

\acknowledgements

We thank Mengxing Ye for
valuable discussions, as well as Xiao Chen and Jason Iaconis for a
collaboration on a related topic. We also sincerely acknowledge Roser
Valent\'i for her encouragements and enthusiasm. The premises
of this project were funded by the Agence Nationale de la Recherche through Grant ANR-18-ERC2-0003-01 (QUANTEM). The bulk of this project was funded by the European Research Council (ERC) under the European Union’s Horizon 2020 research and innovation program (Grant agreement No.\ 853116, acronym TRANSPORT). L.B.\ was supported by the DOE, Office of Science, Basic Energy Sciences under Award No.\ DE-FG02-08ER46524.  It befits us to acknowledge the hospitality of
the KITP, where part of this project was carried out, funded under NSF Grant
NSF PHY-1748958.

\bibliography{phononpaper.bib}

\appendix
\label{sec:appendices}

\section{Strain tensor}
\label{sec:strain-tensor}


In Sec.~\ref{sec:phen-coupl-hamilt} we employ a continuum model of the
spin-phonon system.  The phonons themselves correspondingly derive
from the theory of continuum elasticity, which has the Hamiltonian density
\begin{equation}
  \label{eq:157}
  \mathcal{H}_{\textrm{el}} = \frac{1}{2\rho}\Pi_\mu^2 +
  \frac{1}{2}C_{\alpha\beta\gamma\delta} \mathcal{E}^{\alpha\beta} \mathcal{E}^{\gamma\delta}.
\end{equation}
Within this appendix, $\rho$ is the mass density (we will take $\rho
V=M_{\rm uc}N_{\rm uc}$) and $C_{\alpha\beta\gamma\delta}$ is a
rank four tensor of elastic constants, which can be taken to satisfy
$C_{\alpha\beta\gamma\delta} =
C_{\beta\alpha\gamma\delta}=C_{\alpha\beta\delta\gamma}=C_{\gamma\delta\alpha\beta}$.
The canonical variables of this classical field theory are the
displacement field $u_\mu$ and its canonically conjugate momentum
$\Pi_\mu$.  Due to translational and rotational symmetry, the Hookian
potential energy is expressed solely through the strain tensor,
\begin{equation}
  \label{eq:strain129}
\mathcal{E}^{\alpha\beta}(\mathbf{R})=\frac{1}{2}\left(\partial_\alpha
    u_\beta+\partial_\beta u_\alpha\right).
\end{equation}
By construction the strain is a symmetric tensor in its two indices,
i.e.\ $\mathcal{E}^T=\mathcal{E}$.  Define the Fourier transforms
\begin{align}
  \label{eq:158}
  u_\mu(\mb{x}) = \frac{1}{\sqrt{V}} \sum_{\mb{k}}
  e^{i\mb{k}\cdot\mb{x}} u_{\mu,\mb{k}},
 && \Pi_\mu(\mb{x}) = \frac{1}{\sqrt{V}} \sum_{\mb{k}}e^{i\mb{k}\cdot\mb{x}} \Pi_{\mu,\mb{k}}.
\end{align}
Here since $u_\mu(\mb{x})$ and $\Pi_\mu(\mb{x})$ are real fields, we
have $u_{n,-\mb{k}}=u_{n\mb{k}}^*$ and
$\Pi_{n,-\mb{k}}=\Pi_{n\mb{k}}^*$.  The Fourier space fields satisfy the commutation relations
\begin{align}
  \label{eq:162}
  \left[\Pi_{\mu,\mb{k}},u_{\nu,\mb{k}'}\right] =  i \delta_{\mu\nu} \delta_{\mb{k}+\mb{k}',\mb{0}}.
\end{align}
We obtain
\begin{align}
  \label{eq:159}
  H_{\textrm{el}} = \sum_{\mb{k}} \left\{
  \frac{1}{2\rho}\Pi_{\mu,-\mb{k}} \Pi_{\mu,\mb{k}} +
  \frac{1}{2} \mathcal{K}_{\alpha\beta}(\mb{k}) u_{\alpha,-\mb{k}} u_{\beta,\mb{k}}\right\},
\end{align}
with
\begin{align}
  \label{eq:160}
  \mathcal{K}_{\alpha\beta}(\mb{k}) = C_{\alpha\gamma\beta\delta} k_\gamma k_\delta.
\end{align}
The matrix $\mathcal{K}_{\alpha\beta}$ is by construction real and symmetric, and
hence has real eigenvalues $\mathcal{K}_n$, which additionally must be positive for
stability.  We define the eigenvalues and eigenvectors
$\varepsilon_n^\alpha$ via
\begin{align}
  \label{eq:161}
  \mathcal{K}_{\alpha\beta}(\mb{k}) \varepsilon_{n}^\beta(\mb{k}) =
  \mathcal{K}_n(\mb{k}) \varepsilon_{n}^\alpha(\mb{k}),
\end{align}
with $\varepsilon_n^\alpha(-\mb{k}) = (\varepsilon_n^\alpha(\mb{k})
)^*$ and the standard normalization $\sum_\alpha (\varepsilon_n^\alpha(\mb{k}))^*
\varepsilon_{n'}^\alpha(\mb{k}) = \delta_{nn'}$.  Now we make the
change of basis
\begin{align}
  \label{eq:163}
  u_{\mu\mb{k}} = \sum_n \varepsilon_n^\mu(\mb{k}) u_{n\mb{k}}, &&   \Pi_{\mu\mb{k}} = \sum_n \varepsilon_n^\mu(\mb{k}) \Pi_{n\mb{k}}, 
\end{align}
which gives
\begin{align}
  \label{eq:164}
   \left[\Pi_{n\mb{k}},u_{n'\mb{k}'}\right] = i \delta_{nn'} \delta_{\mb{k}+\mb{k}',\mb{0}},
\end{align}
and
\begin{align}
  \label{eq:165}
  H_{\textrm{el}} = \sum_{n,\mb{k}} \left\{
  \frac{1}{2\rho}\Pi_{n,-\mb{k}} \Pi_{n,\mb{k}} +
  \frac{1}{2} \mathcal{K}_{n}(\mb{k}) u_{n,-\mb{k}} u_{n,\mb{k}}\right\}.
\end{align}
Now we can finally define creation/annihilation operators
\begin{align}
  \label{eq:166}
  u_{n\mb{k}} & = \frac{1}{\sqrt{2}}\frac{1}{(\rho\mathcal{K}_n)^{1/4}}
                (a_{n\mb{k}}^{\vphantom\dagger}+a_{n,-\mb{k}}^\dagger), \nonumber \\
         \Pi_{n\mb{k}} & = i\frac{1}{\sqrt{2}}(\rho\mathcal{K}_n)^{1/4}
  (a_{n\mb{k}}^{\vphantom\dagger}-a_{n,-\mb{k}}^\dagger),
\end{align}
with canonical boson operators
\begin{align}
  \label{eq:167}
  \left[ a_{n\mb{k}}^{\vphantom\dagger},a_{n'\mb{k}'}^\dagger\right] = \delta_{nn'}\delta_{\mb{k},\mb{k}'},
\end{align}
and the Hamiltonian
\begin{align}
  \label{eq:168}
  H_{\textrm{el}} = \sum_{n\mb{k}} \omega_{n\mb{k}} a_{n\mb{k}}^\dagger a_{n\mb{k}}^{\vphantom\dagger},
\end{align}
and
\begin{align}
  \label{eq:169}
  \omega_{n\mb{k}} = \sqrt{\frac{\mathcal{K}_n(\mb{k})}{\rho}}.
\end{align}
Having finally arrived at the canonical phonon operators, we recombine
the several steps of the above procedure to obtain the expression for
the displacement field,
\begin{align}
  \label{eq:170}
  u_\mu(\mb{x}) = \frac{1}{\sqrt{V}} \sum_{n\mb{k}}
  \frac{1}{\sqrt{2\rho \omega_{n\mb{k}}}}\left(a_{n\mb{k}}^{\vphantom\dagger} +
  a_{n,-\mb{k}}^\dagger\right) \varepsilon_{n\mb{k}}^\mu e^{i\mb{k}\cdot\mb{x}}.
\end{align}
Now we can use the definition in Eq.~\eqref{eq:129} of the strain to
obtain
\begin{align}
  \label{eq:171}
 & \mathcal{E}^{\mu\nu}(\mb{x})  = \\
  &\frac{1}{\sqrt{V}} \sum_{n\mb{k}}
  \frac{1}{\sqrt{2\rho \omega_{n\mb{k}}}}\left(a_{n\mb{k}}^{\vphantom\dagger} +
  a_{n,-\mb{k}}^\dagger\right) \frac{i}{2}\left( k^\mu\varepsilon_{n\mb{k}}^\nu +k^\nu\varepsilon_{n\mb{k}}^\mu \right)e^{i\mb{k}\cdot\mb{x}}.
\end{align}
Now let us consider the coupling of the strain to the continuum spin
fluctuations, Eq.~\eqref{eq:142} of the main text.  The full
spin-lattice coupling in three dimensions is written as 
\begin{align}
  \label{eq:172}
  H'_{\rm s-l} = \sum_z \int\! dxdy\, \mathcal{E}^{\alpha\beta}(\mb{x})
 \lambda^{\alpha\beta}_{ab;\xi\xi'}n_0^{-\xi-\xi'}\eta_{a\xi{\mathbf{x}}}\eta_{b\xi'\mathbf{x}}.
\end{align}
Note the sum over discrete 2d layers.  We now insert the Fourier
expansion of the strain from Eq.~\eqref{eq:171} and the corresponding
Fourier expansion of the magnetic fluctuations, which we repeat here:
\begin{align}
  \label{eq:173}
  \eta_{a\xi\mb{x}} = \frac{1}{\sqrt{A_{2d}}} \sum_{\underline{\mb{q}}}
  \eta_{a\xi\underline{\mb{q}},z} e^{i\underline{\mb{q}}\cdot\underline{\mb{x}}}.
\end{align}
In this equation, and in the rest of this section, we are careful to
denote two-dimensional vectors with an underline.  Since magnetic
fluctuations in different layers are taken as independent, we do not
introduce a $z$-component of the wavevector for the magnons, and
simply leave $z$ explicitly as a layer index for these fields.  Note
also the prefactor Eq.~\eqref{eq:173} therefore involves the square
root of the two dimensional area of a single plane, $A_{2d}$.

With
this in mind, we obtain from Eq.~\eqref{eq:172} 
\begin{align}
  \label{eq:174}
  &H'_{\rm s-l} = \frac{1}{\sqrt{V}}\sum_z   \sum_{\mb{k},\underline{\mb{p}}} 
  \lambda^{\mu\nu}_{ab;\xi\xi'} n_0^{-\xi-\xi'} \frac{1}{\sqrt{2\rho \omega_{n\mb{k}}}}
                                                                               e^{ik_z z} \\
  & \times \frac{i}{2}\left( k^\mu\varepsilon_{n\mb{k}}^\nu +k^\nu\varepsilon_{n\mb{k}}^\mu \right)\left(a_{n\mb{k}}^{\vphantom\dagger} +
  a_{n,-\mb{k}}^\dagger\right)
  \eta_{a\xi,\underline{\mb{p}}-\frac{1}{2}\underline{\mb{k}},z}
  \eta_{b\xi',-\underline{\mb{p}}-\frac{1}{2}\underline{\mb{k}},z}.\nonumber
\end{align}
From here we can see that 
\begin{align}
  \label{eq:175}
 Q_{n\mb{k}}= \frac{i}{2\sqrt{V}}&\frac{\left(
                k^\mu\varepsilon_{n\mb{k}}^\nu
                +k^\nu\varepsilon_{n\mb{k}}^\mu \right)}{\sqrt{2\rho
                \omega_{n\mb{k}}}}\sum_{\xi\xi',ab}\lambda^{\mu\nu}_{ab;\xi\xi'} 
                n_0^{-\xi-\xi'} \nonumber \\
  & \times \sum_z   \sum_{\underline{\mb{p}}} 
  e^{ik_z z}
  \eta_{a\xi,\underline{\mb{p}}-\frac{1}{2}\underline{\mb{k}},z}
  \eta_{b\xi',-\underline{\mb{p}}-\frac{1}{2}\underline{\mb{k}},z}.
\end{align}

\begin{widetext}
Next we use Eq.~\eqref{eq:35} to express this in terms of canonical
bosons: 
\begin{align}
  \label{eq:176}
  Q_{n\mb{k}}  = \frac{i\left(
                k^\mu\varepsilon_{n\mb{k}}^\nu
                +k^\nu\varepsilon_{n\mb{k}}^\mu \right)}{4\sqrt{2V\rho\, \omega_{n\mb{k}}}}\sum_z
                \sum_{\underline{\mb{p}}} \sum_{qq'}\sum_{\xi\xi'
                ab}&
 \lambda^{\mu\nu}_{ab;\xi\xi'} 
                n_0^{-\xi-\xi'}
                e^{ik_z z}
                (-1)^{\overline{\xi}(\delta_{a-1,\xi}+\frac{1+q}{2})+\overline{\xi}'(\delta_{b-1,\xi'}+\frac{1+q'}{2})}i^{\overline{\xi}+\overline{\xi}'}\nonumber\\
  &\times
                (\chi\Omega_{\delta_{a-1,\xi},\mb{p}-\frac{1}{2}\mathbf{k}})^{\widetilde{\xi}/2}(\chi\Omega_{\delta_{b-1,\xi'},-\mb{p}-\frac{1}{2}\mathbf{k}})^{\widetilde{\xi}'/2}
                b^q_{\mb{p}-\frac{1}{2}\mathbf{k},\delta_{a-1,\xi},z}
                                                    b^{q'}_{-\mb{p}-\frac{1}{2}\mathbf{k},\delta_{b-1,\xi'},z}.
\end{align}
We now define $\ell_1 = \delta_{a-1,\xi}=0,1$ and
$\ell_2=\delta_{b-1,\xi'}$, which is inverted by
$a=1+\tilde{\xi}\ell_1+\bar{\xi}$ and
$b=1+\tilde{\xi}'\ell_2+\bar{\xi}'$.  This gives
\begin{align}
  \label{eq:177}
  Q_{n\mb{k}} = \frac{i\left(
                k^\mu\varepsilon_{n\mb{k}}^\nu
                +k^\nu\varepsilon_{n\mb{k}}^\mu \right)}{4\sqrt{2V\rho\, \omega_{n\mb{k}}}}\sum_z
                \sum_{\underline{\mb{p}}} \sum_{qq'}\sum_{\xi\xi'
                \ell_1\ell_2}&
 \lambda^{\mu\nu}_{\tilde{\xi}\ell_1+\bar{\xi}+1,\tilde{\xi}'\ell_2+\bar{\xi}'+1;\xi,\xi'}
                n_0^{-\xi-\xi'}
                e^{ik_z z}
                (-1)^{\overline{\xi}(\ell_1+\frac{1+q}{2})+\overline{\xi}'(\ell_2+\frac{1+q'}{2})}i^{\overline{\xi}+\overline{\xi}'}\nonumber\\
  &\times
                (\chi\Omega_{\ell_1,\mb{p}-\frac{1}{2}\mathbf{k}})^{\widetilde{\xi}/2}(\chi\Omega_{\ell_2,-\mb{p}-\frac{1}{2}\mathbf{k}})^{\widetilde{\xi}'/2} 
                b^q_{\mb{p}-\frac{1}{2}\mathbf{k},\ell_1,z}
                                                    b^{q'}_{-\mb{p}-\frac{1}{2}\mathbf{k},\ell_2,z}.
\end{align}
\end{widetext}

From here, we recognize that $Q_{n\mb{k}}=Q^-_{n\mb{k}}$ in
Eq.~\eqref{eq:141}, and thereby extract $\mathcal{B}$.  We use $V\rho
= N_{\rm uc} M_{\rm uc}$.

\section{General hydrodynamics of phonons}

Our goal is to derive the thermal current carried by the phonons,
\begin{equation}
\label{eq:app1}
    j^\mu = \frac 1 V \sum_{n\mathbf k} \overline{N}_{n\mathbf k} v_{n\mathbf k}^\mu \omega_{n\mathbf k},
\end{equation}
in order to extract the thermal conductivity tensor. This requires
knowledge of the average phonon populations $\overline{N}_{n\mathbf{k}}$, which, in presence of a gradient of temperature, differ from their equilibrium values. These populations can be obtained by solving Boltzmann's equation 
\begin{equation}
\label{eq:app2}
  \partial_t\overline{N}_{n\mathbf{k}}+\bs {v}_{n\mathbf{k}}\cdot{\boldsymbol{\nabla}}_\mathbf{r}\overline{N}_{n\mathbf{k}}= \mathcal C_{n\mathbf{k}}[\{\overline{N}_{n'\mathbf{k}'}\}],
\end{equation}
where the collision integral $\mathcal
C_{n\mathbf{k}}[\{\overline{N}_{n'\mathbf{k}'}\}]$ depends on the
populations in {\em all} $(n',\mathbf k')$ states. To solve this
equation, we expand the out-of-equilibrium populations around their
equilibrium value as $\overline{N}_{n\mathbf{k}} = N^{\rm
  eq}_{n\mathbf{k}} + \delta\overline{N}_{n\mathbf{k}}$. Within linear
response, the perturbations can be considered small and we may expand the collision integral
\begin{equation}
\label{eq:app3}
\mathcal C_{n\mathbf{k}}[\{\overline{N}_{n'\mathbf{k}'}\}] = O_{n\mathbf k} + \sum_{n'\mathbf k'} C_{n\mathbf k n'\mathbf k'} \delta\overline{N}_{n'\mathbf{k}'} + O(\delta\overline{N}^2),
\end{equation}
around its value $O_{n\mathbf k}$ in equilibrium.  Since the thermal
current must vanish in equilibrium, $O_{n\mathbf k}$ must be zero (we
go back to this statement in Sec.~\ref{sec:detailed-balance-1}). In
Eq.~\eqref{eq:app3}, the ``collision matrix'' $C_{n\mathbf k n'\mathbf
  k'}$ is defined as the first-order Taylor coefficient, and one
neglects the quadratic order in the perturbation. Formally inverting
the collision matrix in the stationary Boltzmann equation (i.e.\
Eq.~\eqref{eq:app1} with $\partial_t\overline{N}=0$) leads to 
\begin{equation}
\label{eq:app4}
 \delta\overline{N}_{n\mathbf{k}}=C^{-1}_{n\mathbf k n'\mathbf k'} \frac{\omega_{n'\mathbf k'}}{k_B T^2}(N^{\rm eq}_{n'\mathbf{k}'})^2
 v^\nu_{n'\mathbf k'}\partial_\nu T.
\end{equation}
From Eq.~\eqref{eq:app4} and Fourier's law, we can identify the
components of the thermal conductivity tensor:
\begin{eqnarray}
\label{eq:app5}
  \frac{\kappa^{\mu\nu}\pm\kappa^{\nu\mu}} 2 &=&-\frac{1}{2k_BT^2}\frac 1 V
      \sum_{n\mathbf{k}n'\mathbf{k}'}\omega_{n\mathbf{k}}\omega_{n'\mathbf{k}'}v^\mu_{n\mathbf{k}}v^\nu_{n'\mathbf{k}'}\\
  &&\cdot\left(
    C_{n\mathbf{k},n'\mathbf{k}'}^{-1}\,
     e^{\beta\omega_{n'\mathbf{k}'}}(N^{\rm eq}_{n'\mathbf{k}'})^2 \pm
     (n\mathbf k \leftrightarrow n'\mathbf k')\right).
    \nonumber
\end{eqnarray}
This expression shows that a nonzero phonon Hall conductivity requires
the factor in the second line to be nonzero, which is equivalent to
\begin{equation}
\label{eq:app6}
 C_{n\mathbf{k},n'\mathbf{k}'}\,
 e^{\beta\omega_{n'\mathbf{k}'}}(N^{\rm eq}_{n'\mathbf{k}'})^2 \neq
 C_{n'\mathbf{k}',n\mathbf{k}}\, e^{\beta\omega_{n\mathbf{k}}}(N^{\rm eq}_{n\mathbf{k}})^2,
\end{equation}
where the constraint is now on $C_{n\mathbf{k},n'\mathbf{k}'} $
instead of its inverse. In other words, only the antisymmetric in $n\mathbf{k}\leftrightarrow
n'\mathbf{k}'$ part of $C_{n\mathbf{k},n'\mathbf{k}'}\,
 e^{\beta\omega_{n'\mathbf{k}'}}(N^{\rm eq}_{n'\mathbf{k}'})^2$
 contributes to the Hall conductivity.

In order to proceed further analytically, and invert the scattering
matrix, we the separate diagonal from the off-diagonal parts in $C_{n\mathbf{k}n'\mathbf k'} = -\delta_{nn'}\delta_{\mathbf k,\mathbf k'} D_{n\mathbf k} + M_{n\mathbf k n'\mathbf k'}$, 
and assume that $D_{n\mb k}\gg \sum_{n'\mb k'}M_{n\mb kn'\mb k'}$. This ought to be the case whenever the
interactions are small, and/or if other damping processes are large. Then, $
C^{-1}_{n\mathbf{k}n'\mathbf k'} \approx  -\delta_{nn'}\delta_{\mathbf
  k,\mathbf k'}{D^{-1}_{n\mathbf k}}-M_{n\mathbf k n'\mathbf
  k'}{D^{-1}_{n\mathbf k}}{D^{-1}_{n'\mathbf k'}}$. The antisymmetry
in $n\mathbf{k}\leftrightarrow n'\mathbf{k}'$ condition for the Hall
conductivity mentioned above leads to the fact that the diagonal term
contributes to the longitudinal conductivity, but not to the Hall
part, and translates into 
\begin{eqnarray}
\label{eq:app7}
  \frac{\kappa^{\mu\nu}-\kappa^{\nu\mu}}2&=:&\frac{1}{2k_B T^2}\frac 1 V 
                                              \sum_{n\mathbf{k}n'\mathbf{k}'}\frac{\omega_{n\mathbf{k}}v^\mu_{n\mathbf{k}}}{D_{n\mathbf k}}\frac{\omega_{n'\mathbf{k}'}v^\nu_{n'\mathbf{k}'}}{D_{n'\mathbf k'}}\\
 &&\cdot\left [M_{n\mathbf{k},n'\mathbf{k}'}\,
    e^{\beta\omega_{n'\mathbf{k}'}}(N^{\rm eq}_{n'\mathbf{k}'})^2 - (n\mathbf k \leftrightarrow n'\mathbf k') \right ].
 \nonumber
\end{eqnarray}
The longitudinal conductivity is
\begin{eqnarray}
\label{eq:app8}
 \kappa^{\mu\mu}
&=&
\frac{1}{k_B T^2}\frac 1 V \sum_{n\mathbf{k}n'\mathbf{k}'}\omega_{n\mathbf{k}}\omega_{n'\mathbf{k}'}v^\mu_{n\mathbf{k}}v^\mu_{n'\mathbf{k}'}\\
&&\cdot \left [\frac {\delta_{nn'}\delta_{\mathbf k,\mathbf k'}}
   {D_{n\mathbf k}} + \frac{M_{n\mathbf k,n'\mathbf k'}}{D_{n\mathbf
   k}D_{n'\mathbf k'}} \right ] e^{\beta\omega_{n'\mathbf{k}'}}(N^{\rm
   eq}_{n'\mathbf{k}'})^2. \nonumber
\end{eqnarray}

Note that we will include all other (diagonal) scattering processes not taken into
account here (e.g.\ boundary scattering, scattering by impurities,
phonon-phonon sccattering etc.) by adding a phenomenological relaxation
rate $\breve D_{n\mb k}$ to the diagonal of the scattering
matrix.

\section{From interaction terms to the collision integral}
\label{sec:from-interaction-terms}

\subsection{General method and definitions}

We now aim at deriving an expression for the collision integral of Boltzmann's equation using kinetic theory methods. The probability for the system to be found in a given quantum state $|\mathtt i\rangle = |i_p\rangle|i_s\rangle$ is governed by the master equation
\begin{equation}
\label{eq:app9}
  \partial_t p_{i_p i_s} = \sum_{f_p f_s}  \left[ \Gamma_{f_p
      f_s \rightarrow i_p
    i_s} \, p_{f_p f_s} - \Gamma_{i_p i_s \rightarrow f_p
    f_s} \, p_{i_p i_s} \right],
\end{equation}
where we will compute the transition rates $\Gamma_{i_p i_s \rightarrow f_p f_s}$ using scattering theory. The probability of a phonon state $|i_p\rangle$ is then obtained by summing over all possible spin configurations of the system,
$p_{i_p} = \sum_{i_s} p_{i_p i_s}$. Assuming the phonon and spin
probabilities are independent, i.e.\ $p_{i_p i_s} = p_{i_p} p_{i_s}$, and defining the transition rates $\Gamma_{f_p \rightarrow i_p} = \sum_{i_s f_s} \Gamma_{
f_p  f_s \rightarrow i_p  i_s}\, p_{f_s}$  between phonon states only,
we obtain the master equation for the probabilities of phonon
states. We may in turn express the collision integral in the RHS of Boltzmann's equation, which is given by the time evolution of the populations in each phonon state $|i_p\rangle$ through the definition $\mathcal C_{n\mathbf{k}}[\{\overline{N}_{n'\mathbf{k}'}\}]=
 \sum_{i_p} N_{n\mathbf k}({i_p})~\partial_t p_{i_p}$, in terms of transition rates between phonon states:
\begin{equation}
\label{eq:app10}
 \mathcal C_{n\mathbf{k}}[\{\overline{N}_{n'\mathbf{k}'}\}]=\sum_{i_p,f_p} \Gamma_{i_p \rightarrow  f_p}
 \left ( N_{n\mathbf k}(f_p)- N_{n\mathbf k}(i_p) \right )p_{i_p},
\end{equation}
where $N_{n\mathbf k}(i_p)=\langle i_{p}|a^\dagger_{n\mathbf k}a_{n\mathbf k}^{\vphantom{\dagger}}|i_{p}\rangle$
is the number of $(n,\mathbf k)$ phonons in the $|i_p\rangle$ state and $\overline{N}_{n\mathbf k}
= \sum_{i_p} N_{n\mathbf k}(i_p) p_{i_p}$ is the average population. The only phonon states involved in the sums are those whose populations of $(n,\mathbf k)$ phonons are different.
Now, in order to obtain the scattering rates between spin-phonon states, we use Fermi's golden rule
\begin{equation}
\label{eq:app11}
    \Gamma_{i_pi_s \rightarrow f_pf_s} ={2\pi}~|T_{i_s i_p\rightarrow f_s f_p}|^2~\delta(E_{ i_p i_s}- E_{f_p f_s}),
\end{equation}
where the factor $N_{\rm uc}$ ensures that $\Gamma_{i_pi_s \rightarrow f_pf_s}$ is a finite quantity in the
thermodynamic limit, consistent with the choice of $H'$ as a hamiltonian \emph{density}.
We use Born's expansion of the scattering matrix
\begin{eqnarray}
\label{eq:app12}
 T_{i_s i_p\rightarrow f_s f_p}&=&\langle f_s f_p|H'|i_s i_p\rangle\\
  &&+\sum_{n_s n_p}\frac{\langle f_s f_p|H'| n_s n_p\rangle\langle n_s n_p|H'|i_s i_p\rangle}{E_{i_s i_p}-E_{n_s n_p}+i\eta}+\cdots, \nonumber
\end{eqnarray}
where $H'$ is the (perturbative) interaction hamiltonian between the phonons
and the $Q$ fields, and the $\eta\rightarrow 0^+$ appearing in the denominator of the second-order
term ensures causality, which will prove crucial in the following.

To describe the interaction between phonon and spin degrees of freedom, we introduce general coupling terms between phonon creation-annihilation operators $a^{(\dagger)}_{n\mathbf k}$ and general, for now unspecified, fields $Q^{\{q_j\}}_{\{n_j\mathbf k_j\}}$ which depend on the spin structure:
denoting $a_{n\mathbf k}^+ \equiv a_{n\mathbf
  k}^\dagger$ and $a_{n\mathbf k}^-\equiv a_{n\mathbf k}$, and
similarly for the $Q$ operators, $Q_{\{n_i\mathbf{k}_i\}}^+\equiv
Q_{\{n_i\mathbf{k}_i\}}^{\dagger}$ and
$Q_{\{n_i\mathbf{k}_i\}}^-\equiv
Q_{\{n_i\mathbf{k}_i\}}^{\vphantom{\dagger}}$, we write the couplings
\begin{eqnarray}
\label{eq:app13}
 H'_{[1]} &=& \sum_{n\mathbf k}\sum_{q=\pm} a_ {n\mathbf k}^q Q^q_{n\mathbf k}\\
 H'_{[2]} &=& \frac 1{\sqrt{N_{\rm uc}}} \sum_{n\mathbf k,n'\mathbf k'}\sum_{q,q'=\pm}a_{n\mathbf k}^q a_{n'\mathbf k'}^{q'} Q^{qq'}_{n\mathbf k n'\mathbf k'},
\end{eqnarray}
where $Q^{qq'}_{n\mathbf k n'\mathbf k'} = Q^{q'q}_{n'\mathbf k'
  n\mathbf k}=(Q^{-q-q'}_{n\mathbf k n'\mathbf k'})^\dagger $ ensures
the hermiticity of $H'_{[2]}$. Here and throughout the manuscript, a
square bracket index, e.g.\ $[p]$ denotes the number of interacting
phonons.

By definition the term $H'_{[p]}$ 
involves $p$ phonon creation-annihilation operators, and as such typically
arises from microscopic models as the $p$th spatial derivative of
orbital overlaps. Consequently, we assume $H'_{[2]}$ to be of the same
order of magnitude as the square of $H'_{[1]}$, that is to say,
$Q^q_{n\mathbf{k}}\sim {\lambda}$, $Q^{qq'}_{n\mathbf{k}n'\mathbf
  k'}\sim {\lambda}^2$ with ${\lambda}$ a small parameter.
In this paper, we keep only the first two
terms of the expansion (i.e. we take $H'=H'_{[1]}+H'_{[2]}$).

\subsection{Computation at first Born order}
\label{sec:1storder}

In this subsection we consider only the first term of Born's
expansion. The transition rates associated with $H'=H'_{[1]}+H'_{[2]}$
at this order derive from the matrix elements:
\begin{equation}
\label{eq:app14}
    T^{[1]}_{\mathtt{i}\rightarrow\mathtt{f}}    = \sum_{n\mathbf{k}q}\sqrt{N^i_{\mathbf{k},n}+\tfrac{q+1}2}~\langle
    f_s|Q^{q}_{n\mathbf k}|i_s\rangle~ \mathds{I}({i_p}\overset{q\cdot n\mathbf k}{\longrightarrow}{f_p}),
    \end{equation}
\begin{eqnarray}
  T^{[2]}_{\mathtt i\rightarrow \mathtt f}
  &=& \frac 1 {\sqrt{N_{\rm uc}}}\sum_{n\mathbf{k}, n' \mathbf{k'}}\sum_{qq'}\sqrt{N_{n\mathbf k}^i +\tfrac{q+1} 2}\sqrt{N_{n'\mathbf k'}^i +\tfrac{q'+1}2}\nonumber\\
     &&\cdot\langle f_s| Q_{n\mathbf k n'\mathbf k'}^{qq'}| i_s\rangle ~\mathds{I}({ i_p}
\overset{q\cdot n\mathbf k}{\underset{q'\cdot n'\mathbf k'}\longrightarrow}{ f_p}),
\end{eqnarray}
where $\mathds{I}({ i_p}\overset{q\cdot n\mathbf k}{\longrightarrow}{
  f_p})$ means that the only difference between $| i_p \rangle$ and $|
f_p\rangle$ is that there is $q=\pm1$ more phonon of species
$(n,\mathbf k)$ in the final state. Note that the cases where
$n\mathbf k = n'\mathbf k'$ require a
formal correction. However, at any given order in the ${\lambda}$ expansion, such terms are smaller than all others
 by a factor $1/N_{\rm uc}$, where $N_{\rm uc}$ is the
 number of unit cells, and therefore vanish in the
thermodynamic limit. In this article we thus take $\sum_{n\mathbf
  k,n'\mathbf k'}$ and $\sum_{n\mathbf k\neq n'\mathbf k'}$
exchangeably, unless we specify otherwise.

We then compute the squared matrix element. ``Cross terms'' such as
$\langle \mathtt i|H'_{[2]}|\mathtt f\rangle\langle\mathtt
f|H'_{[1]}|\mathtt i\rangle$ (which are of order ${\lambda}^3$)
vanish because $\langle  i_p|\widehat A| i_p\rangle=0$ for any
operator $\widehat A$ containing an odd number of phonon
creation-annihilation operators. 
At order ${\lambda}^2$, there thus remains only $\langle \mathtt i|H'_{[1]}|\mathtt
f\rangle\langle\mathtt f|H'_{[1]}|\mathtt i\rangle$, and at order
${\lambda}^4$, only $\langle
\mathtt i|H'_{[2]}|\mathtt f\rangle\langle\mathtt f|H'_{[2]}|\mathtt
i\rangle$.

\subsubsection{Terms at $O(\lambda^2)$}
\label{sec:at-order-lambdabar2}

At order ${\lambda}^2$, we have therefore
\begin{eqnarray}
\label{eq:app15}
  \left|T^{[1]}_{\mathtt{i}\rightarrow\mathtt{f}}\right|^2 &=& 
\sum_{n\mathbf{k}q}\left(N^i_{\mathbf{k},n}+\tfrac{q+1}{2}\right)~\mathds{I}({i_p}\overset{q\cdot n\mathbf k}{\longrightarrow}{f_p})\nonumber\\
&&\qquad\langle
i_s|Q^{-q}_{n\mathbf k}|f_s\rangle\langle
f_s|Q^{q}_{n\mathbf k}|i_s\rangle.
\end{eqnarray}
We then enforce the energy conservation
$\delta(E_{\mathtt{f}}-E_{\mathtt{i}}) =
\delta(q\omega_{n\mathbf{k}}+E_{f_s}-E_{i_s})$ by writing the latter
as a time integral, i.e.\ $\int_{-\infty}^{+\infty}{\rm d}t e^{i\omega
  t}=2\pi\delta(\omega)$, identify $A(t)=e^{+iHt}Ae^{-iHt}$, use the
identity $1=\sum_{ f_s}| f_s\rangle\langle f_s|$, and take the spins
in the initial state to be in thermal equilibrium $p_{
  i_s}=Z^{-1}e^{-\beta E_{ i_s}}$. Finally summing over $| i_s\rangle$
and identifying $\langle A \rangle_\beta = Z^{-1}\text{Tr}(e^{-\beta
  H}A)$, we find
\begin{eqnarray}
\label{eq:app16}
  W^{[1];[1]}_{n\mathbf k q} &=& 2\pi \sum\nolimits_{ f_s,i_s}\langle
i_s|Q^{-q}_{n\mathbf k}|f_s\rangle\langle
                                f_s|Q^{q}_{n\mathbf k}|i_s\rangle p_{i_s}\nonumber\\
  &&\qquad \times \delta(q\omega_{n\mathbf{k}}+E_{ f_s}-E_{ i_s})\nonumber\\
&=& \int_{-\infty}^\infty \! \text dt\, e^{-i q\omega_{n\mathbf{k}} t}  \left\langle
Q^{-q}_{n\mathbf k}(t)~Q^{q}_{n\mathbf k}(0)\right\rangle_\beta.
\end{eqnarray}
Note that this calculation, in a time-reversal symmetric system, leads to the extra symmetry
$W^{[1];[1]}_{n\mathbf k q} = W^{[1];[1]}_{n-\mathbf k q}$.

The scattering rate between phonon states, for the one-phonon interaction term at first Born's order, then reads
\begin{equation}
\label{eq:app17}
  \Gamma^{[1];[1]}_{ i_p \rightarrow  f_p}= 
\sum_{n\mathbf k q}
\left (N^i_{n\mathbf{k}} + \tfrac{q+1}{2}\right )
W^{[1];[1]}_{n\mathbf k q}~\mathds{I}({ i_p}\overset{q\cdot n\mathbf k}{\longrightarrow}{ f_p}).
\end{equation}

To arrive at the collision integral, the final step involves summing over final phononic states $ f_p$ and
taking the average over initial phononic states $i_p$. We find, the
contributions to $\mathcal{C}$ at order ${\lambda}^2$ to be:
\begin{eqnarray}
\label{eq:app18}
O^{[1];[1]}_{n\mathbf k} &=& \sum_{q=\pm} q~
 W^{[1];[1]}_{n,\mathbf{k}, q}
 \left ( N^{\rm eq}_{n,\mathbf{k}}
 + \tfrac{q+1}{2} \right ), \\
 -D^{[1];[1]}_{n\mathbf k} &=& \sum_{q=\pm} q~
 W^{[1];[1]}_{n,\mathbf{k}, q}.
\end{eqnarray}
We will address the constant term $O^{[1];[1]}_{n\mathbf k}$ (expected to be zero)
in more detail in Sec.~\ref{sec:detailed-balance-1}. The collision matrix is
clearly diagonal, i.e.\ $M^{[1];[1]}_{n\mathbf k,n'\mathbf k'}
=0$. Therefore this ${\lambda}^2$ contribution to $\mathcal{C}$ may contribute to the longitudinal conductivity, but not to
the Hall conductivity.

\subsubsection{Terms at $O(\lambda^4)$}
\label{sec:at-order-lambdabar4}

We address the $O(\lambda^4)$ term in a similar
fashion. There, the energy conservation reads $\delta(E_{\mathtt{f}}-E_{\mathtt{i}}) =\delta(q\omega_{n\mathbf{k}}+q'\omega_{n'\mathbf{k'}}+E_{f_s}-E_{i_s})$, and we find
\begin{eqnarray}
\label{eq:app19}
  \Gamma^{[2];[2]}_{ i_p \rightarrow  f_p} &=& \frac 1 {2 N_{\rm uc}} \sum_{n\mathbf k, n'\mathbf k'}\sum_{q,q'=\pm}
 W^{[2];[2]}_{n\mathbf{k}q,n'\mathbf{k}'q'} ~\mathds{I}({ i_p}\overset{q\cdot n\mathbf k}{\underset{q'\cdot n'\mathbf k'}\longrightarrow}{ f_p})\nonumber\\
 &&\cdot \left(N_{n\mathbf{k}}^i+\tfrac{q+1}2\right) \left(N_{n'\mathbf{k}'}^i+\tfrac{q'+1}2\right),
\end{eqnarray}
where
\begin{eqnarray}
\label{eq:app20}
W^{[2];[2]}_{n\mathbf{k}q,n'\mathbf{k}'q'}&=& 2\int_{-\infty}^{+\infty}\text d t ~ e^{-i(q\omega_{n\mathbf{k}}+q'\omega_{n'\mathbf{k}'})t}\\
&&\times\left \langle Q_{n\mathbf k n' \mathbf{k}'}^{-q-q'}(t)~Q_{n\mathbf k n'\mathbf k'}^{qq'}(0) \right \rangle_\beta,\nonumber
\end{eqnarray}
and
$W^{[2];[2]}_{n\mathbf{k}q,n'\mathbf{k}'q'}=W^{[2];[2]}_{n'\mathbf{k}'q',n\mathbf{k}q}$,
by definition. The resulting collision integral, up to linear order in
the perturbed populations $\delta\overline N$ contains the following contributions:
\begin{eqnarray}
\label{eq:app21}
 O^{[2];[2]}_{n\mathbf k}&=& \frac 1 {N_{\rm uc}} \sum_{n'\mathbf k'}\sum_{q,q'=\pm} q W^{[2];[2]}_{n\mathbf{k}q,n'\mathbf{k}'q'}\\
 &&\nonumber \cdot
 \left(N_{n\mathbf{k}}^{\rm eq}+\tfrac{q+1}2\right) \left(N_{n'\mathbf{k}'}^{\rm eq}+\tfrac{q'+1}2\right),
 \end{eqnarray}
 \begin{equation}
   -D^{[2];[2]}_{n\mathbf k}= \frac 1 {N_{\rm uc}}\sum_{n'\mathbf k'}\sum_{q,q'=\pm}q~\left ( N^{\rm eq}_{n'\mathbf k'}+\tfrac{q'+1}2\right )
   W^{[2];[2]}_{n\mathbf kq,n'\mathbf k'q'},
 \end{equation}
 \begin{equation}
 M^{[2];[2]}_{n\mathbf k,n'\mathbf k'} =  \frac 1 {N_{\rm uc}}\sum_{q,q'=\pm}q~\left (N^{\rm eq}_{n\mathbf k}+\tfrac{q+1}2 \right )~W^{[2];[2]}_{n\mathbf kq,n'\mathbf k'q'}.
\end{equation}
As above, we will address the constant term in
Sec.~\ref{sec:detailed-balance-1}. The diagonal contribution is of order
${\lambda}^4$, and we therefore expect it to be subdominant compared
with the ${\lambda}^2$ contribution from the previous
section. Finally, the off-diagonal contribution is nonzero. However,
we will show that
its contribution to $C_{n\mathbf{k},n'\mathbf{k}'}\,
 e^{\beta\omega_{n'\mathbf{k}'}}(N^{\rm eq}_{n'\mathbf{k}'})^2$  is
 purely symmetric under $n\mb k\leftrightarrow n'\mb k'$ and therefore
contributes only to the symmetric off-diagonal conductivity but not to
the Hall one $-$ see Eq.~\eqref{eq:app6}.

\subsubsection{Detailed balance}
\label{sec:detailed-balance-1}

First, we notice that a change of variables ${i_s}\leftrightarrow{f_s}$ in Eq.~\eqref{eq:app16} leads to the detailed-balance relation
\begin{equation}
\label{eq:app28}
W^{[1];[1]}_{n\mathbf k, q}= W^{[1];[1]}_{n\mathbf k, -q}e^{-q\beta\omega_{n\mathbf k}}.
\end{equation}
An immediate consequence is that $O^{[1];[1]}_{n\mathbf k}=0$ if we take the equilibrium phonon
population $N^{\rm eq}_{n\mathbf k}$ to be Bose-Einstein's
distribution, as was physically required. Similarly, for the two-phonon interactions at first order, we find the detailed-balance relation
\begin{equation}
\label{eq:app31}
W^{[2];[2]}_{n\mathbf{k}q,n'\mathbf{k}'q'} = e^{-\beta(q\omega_{n\mathbf k}+q'\omega_{n'\mathbf k'})} W^{[2];[2]}_{n\mathbf{k}-q,n'\mathbf{k}'-q'}.
\end{equation}
Again, taking $N^{\rm eq}_{n\mathbf k}$ to be Bose-Einstein's
distribution implies $O^{[2];[2]}_{n\mathbf k}=0$. Moreover, the detailed-balance relation also implies
\begin{equation}
\label{eq:app34}
 M^{[2];[2]}_{n\mathbf k, n'\mathbf k'}e^{\beta\omega_{n'\mathbf k'}}(N^{\rm eq}_{n'\mathbf k'})^2 = (n\mathbf k \leftrightarrow  n'\mathbf k'),
\end{equation}
i.e.\ there are no antisymmetric
contributions, and hence no thermal Hall effect at first Born's
order. While we proved this explicitly for the one-phonon and
two-phonon cases, this is true in general (along with $O_{n\mathbf
  k}=0$) for any number of phonon creation-annihilation operators at
first order in Born's expansion (see Sec. \ref{sec:no-phonon-hall}).

\subsubsection{Extra structure}
\label{sec:extra-structure}

Independently, by writing
\begin{equation}
\label{eq:app29}
Q^{-q}_{n\mathbf{k}}(t)Q^q_{n\mathbf{k}}(0) =
\frac 1 2 [Q^{-q}_{n\mathbf{k}}(t),Q^q_{n\mathbf{k}}(0)]
+ \frac 1 2 \{Q^{-q}_{n\mathbf{k}}(t),Q^q_{n\mathbf{k}}(0)\}
\end{equation}
it is straightforward to show that only the commutator term contributes to $W^{[1](1)}_{n,\mathbf k, q}-W^{[1](1)}_{n,\mathbf k, -q}$.
The final expression for the diagonal of the collision matrix
Eq.~\eqref{eq:app18} takes the form of the spectral function:
\begin{equation}
\label{eq:app30}
 D^{[1];[1]}_{n\mathbf k} = - \int_{-\infty}^{+\infty}\text dt e^{-i\omega_{n\mathbf k}t}\left \langle [Q^{-}_{n\mathbf k}(t),
Q^+_{n\mathbf k}(0)]\right\rangle_\beta.
\end{equation}
In the two-phonon case, such a commutator structure does not naturally appear, and
\begin{eqnarray}
\label{eq:app33}
 &&D^{[2];[2]}_{n\mathbf k}~=~- \frac{2}{N_{\rm uc}}\sum_{n'\mathbf k'}\sum_{q,q'=\pm}q~\left (N^{\rm eq}_{n'\mathbf k'}+\tfrac{q'+1}2\right )\\
 &&\times\int_{\mathbb R}\text d t ~ e^{-i(q\omega_{n\mathbf{k}}+q'\omega_{n'\mathbf{k}'})t}\left \langle  Q_{n\mathbf kn'\mathbf k'}^{-q-q'}(t)Q_{n\mathbf k n'\mathbf k'}^{qq'}(0) \right \rangle_\beta,\nonumber
\end{eqnarray}
at order ${\lambda}^4$ and first Born's order.

\subsection{Energy shift of the phonons}
\label{sec:energy-shift-phonons}

We now address the constant term $O_{n\mathbf k}$ appearing in the collision integral,
which must vanish because there is no current in equilibrium.
Its cancellation is equivalent to a redefinition of the energies of the phonons,
due to their interaction with the $Q$ degrees of freedom.
This energy shift corresponds to the real part of the associated self-energy.
Consequently, the equilibrium phonon populations $N^{\rm eq}_{n'\mathbf k'}$ are
a priori not equal to $N^{\rm BE}_{n'\mathbf k'}$, the Bose-Einstein populations
for the {\em unperturbed} phonon energies.

In this subsection, we show that the energy shift, although a priori nonzero,
does not alter the results which we obtained for the thermal conductivities, up to the order $\lambda^4$ in our perturbative expansion.
To understand this, we decompose
\begin{equation}
  \label{eq:121}
  O_{n\mathbf k}[N^{\rm eq}_{n'\mathbf k'}]=O^{(1)}_{n\mathbf k}[N^{\rm eq}_{n'\mathbf k'}]
  + O^{(2)}_{n\mathbf k}[N^{\rm eq}_{n'\mathbf k'}]+\mc O(\lambda^6)
\end{equation}
where, as for $D_{n\mb k}$ elsewhere in this paper,
the upper index $O^{(p)}$ indicates a term of order $\lambda^{2p}$.

We have shown in Sec.~\ref{sec:detailed-balance-1} that $O^{(1)}_{n\mathbf k}[N^{\rm BE}_{n'\mathbf k'}]=0$.
However, $O^{(2)}_{n\mathbf k}[N^{\rm BE}_{n'\mathbf  k'}]\neq 0$ a priori, so that an energy shift is actually required
to cancel the equilibrium current. We thus consider the {\em physical requirement},
$O_{n\mathbf k}[N^{\rm eq}_{n'\mathbf k'}]=0$, to be an equation on
the unknown $N^{\rm eq}_{n'\mathbf k'}$.

Now expanding $N^{\rm eq}_{n'\mathbf k'}=N^{\rm BE}_{n'\mathbf k'}+\delta N^{\rm eq}_{n'\mathbf k'}$
(with $\delta N^{\rm eq}_{n'\mathbf k'}$ at least of order $\lambda^2$),
this equation becomes

\begin{eqnarray}
   \label{eq:130}
  0  &=& O^{(1)}_{n\mathbf k}[N^{\rm BE}_{n''\mathbf k''}]
         + \sum_{n'\mb k'}\delta N^{\rm eq}_{n'\mathbf k'}\partial_{N^{\rm eq}_{n'\mathbf k'}}
         O^{(1)}_{n\mathbf k}[N^{\rm BE}_{n''\mathbf k''}]\nonumber\\
 &+& O^{(2)}_{n\mathbf k}[N^{\rm BE}_{n''\mathbf k''}] + \mc O({\lambda}^6)
\end{eqnarray}

At order ${\lambda}^2$, one recovers $O^{(1)}_{n\mathbf k}[N^{\rm
  BE}_{n'\mathbf k'}]=0$, as is required by detailed balance (see Sec. \ref{sec:no-energy-shift}).

At order ${\lambda}^4$, formally inverting this linear equation, one obtains
\begin{eqnarray}
  \label{eq:309}
  \delta N^{\rm eq}_{n\mathbf k}
  &=& - \sum_{n'\mb k'} \left ( \partial_{N^{\rm eq}_{n_j\mb k_j}}
      O^{(1)}_{n_i\mathbf k_i}[N^{\rm BE}_{n''\mathbf k''}] \right )^{-1}\Big|_{n\mb k,n'\mb k'}\nonumber\\
   &&\qquad \times   O^{(2)}_{n'\mathbf k'}[N^{\rm  BE}_{n''\mathbf k''}]\qquad+ \mc O({\lambda}^4).
\end{eqnarray}
This correction Eq.~\eqref{eq:9} to the phonon equilibrium populations is of order
${\lambda}^2$.

This ensures that using the approximate populations $N^{\rm eq}_{n'\mathbf k'}=N^{\rm BE}_{n'\mathbf k'}$
leads to a correct estimation of $D_{n\mb k}^{(1)}$ the lowest-order contribution to $D_{n\mb k}$,
of order $\lambda^2$. However, the next-order contribution $D_{n\mb k}^{(2)}\sim\lambda^4$
can only be estimated correctly if one adds to it the correction that $\delta N^{\rm eq}_{n\mathbf k}$
brings to $D_{n\mb k}^{(1)}$.
Similarly, using the approximate populations $N^{\rm eq}_{n'\mathbf k'}=N^{\rm BE}_{n'\mathbf k'}$
leads to a correct estimation of the lowest-order contribution, of order $\lambda^4$, to
$M_{n\mb kn'\mb k'}$ as expressed in the main text. Corrections of order $\lambda^6$, not considered
in the present work, would require that the population corrections $\delta N^{\rm eq}_{n\mathbf k}$
be taken into account.

\subsection{Computation at second Born order}
\label{sec:comp-at-second}

As discussed at length, the first Born approximation alone does not
lead to a nonzero thermal Hall effect. Here we compute that which
appears when the Born expansion is taken up to the second Born
order. More precisely, we consider all possible terms up to second
Born order that lead to an off-diagonal scattering rate of order at
most ${\lambda}^4$. This includes terms like $\langle \mathtt f |H'_{[1]}|\mathtt n\rangle\langle \mathtt n|H'_{[1]}|\mathtt i\rangle$ 
as well as $\langle \mathtt f |H'_{[1]}|\mathtt n\rangle\langle
\mathtt n|H'_{[2]}|\mathtt i\rangle$, but not $\langle \mathtt i
|H'_{[2]}|\mathtt n\rangle\langle \mathtt n|H'_{[2]}|\mathtt i\rangle$
since this term is already of order ${\lambda}^4$ (thus contributes
to $|T_{\mathtt i \rightarrow \mathtt f}|^2$ at order ${\lambda}^5$
at least).

\subsubsection{Term with one-phonon interactions only}
\label{sec:terms-with-h_1}

The first of these terms reads
\begin{eqnarray}
\label{eq:app23}
T^{[1,1]}_{\mathtt{i}\rightarrow\mathtt{f}} &=& 
\sum_{n\mathbf{k},n'\mathbf{k}'}\sum_{q,q'=\pm}\sqrt{N^i_{n\mathbf{k}}+\tfrac{1+q}{2}}\sqrt{N^f_{n'\mathbf{k}'}+\tfrac{1-q'}{2}}\nonumber\\
&&\cdot\sum_{ m_s}\frac{\langle
    f_s|Q^{q'}_{n'\mathbf k'}|m_s\rangle\langle
   m_s|Q^{q}_{n\mathbf k}|i_s\rangle}{E_{i_s}-E_{m_s}-q\omega_{\mathbf{k},n}+i\eta}
   ~\mathds{I}({ i_p}\overset{q\cdot n\mathbf k}{\underset{q'\cdot n'\mathbf k'}\longrightarrow}{ f_p}),\nonumber\\
    &&
\end{eqnarray}
where the upper index indicates that within Born's expansion,
$T^{[i,j]}\sim \frac{\langle f|H'_{[j]}|m\rangle\langle m|H'_{[i]}|i\rangle}{E_i-E_m+i\eta}$.

The squared $T$-matrix elements now include
cross-terms between the first and second orders of Born's
expansion (although we keep only terms of order ${\lambda}^4$ at
most). Here we give details of the calculation of one term, the
square of Eq.~\eqref{eq:app23},
$\left|T^{[1,1]}_{\mathtt{i}\rightarrow\mathtt{f}}\right|^2$.

In the numerator, the matrix elements of the $Q$ operators can combine themselves in two
different ways, which we denote in the following as $(a)$: $\langle
    i_s|Q^{q}_{n\mathbf k}|m_s\rangle\langle
    m_s|Q^{q'}_{n'\mathbf k'}|f_s\rangle \langle
    f_s|Q^{-q'}_{n'\mathbf k'}|m'_s\rangle\langle
    m'_s|Q^{-q}_{n\mathbf k}|i_s\rangle$, and $(b)$: $\langle
    i_s|Q^{q}_{n\mathbf k}|m_s\rangle\langle
    m_s|Q^{q'}_{n'\mathbf k'}|f_s\rangle \langle
    f_s|Q^{-q}_{n\mathbf k}|m'_s\rangle\langle
    m'_s|Q^{-q'}_{n'\mathbf k'}|i_s\rangle$.

We use the following time integral representation of each of the denominators (using a regularized definition of the sign function),
\begin{eqnarray}
\label{eq:app22}
    \frac 1 {x \pm i\eta} &=& {\rm PP} \frac 1 x \mp i\pi \delta(x)\\
    &=& \frac 1 {2i} \int_{-\infty}^{+\infty}\text d t_1 e^{it_1 x}\text{sign}(t_1)\pm \frac 1 {2i} \int_{-\infty}^{+\infty}\text d t_1 e^{it_1 x}.\nonumber
\end{eqnarray}
and a introduce a third time integral to enforce the energy conservation $
E_{\mathtt{f}}-E_{\mathtt{i}}=q'\omega_{n'\mathbf{k}'}+q\omega_{n\mathbf{k}}+E_{f_s}-E_{i_s}$.
The product of the denominators (cf.\ Eq.~\eqref{eq:app22}) leads to
four terms, which can be labeled by two signs $s,s'=\pm$, and we
define, for convenience,
\begin{equation}
\Theta_{ss'}(t_1,t_2):=\left[-\text{sign}(t_1)\right]^{\frac{1-s}{2}}\left[\text{sign}(t_2)\right]^{\frac{1-s'}{2}}.
\end{equation}
\begin{widetext}
Then, the transition rate coming from this part of the total squared matrix
element can be written as a sum of eight terms:
\begin{eqnarray}
\label{eq:app24}
  \Gamma^{[1,1];[1,1]}_{ i_p \rightarrow  f_p}
  &=& \sum_{n\mathbf k, n'\mathbf k'}\sum_{q,q'}
      \left(N_{n\mathbf{k}}^i+\tfrac{q+1}2\right) \left(N_{n'\mathbf{k}'}^i+\tfrac{q'+1}2\right)
      \cdot\sum_{s,s'=\pm} \sum_{i=a,b}W^{[1,1];[1,1],(i),ss'}_{n\mathbf{k}q,n'\mathbf{k}'q'}
         ~\mathds{I}({ i_p}\overset{q\cdot n\mathbf k}{\underset{q'\cdot n'\mathbf k'}\longrightarrow}{ f_p}),
\end{eqnarray}
where we defined (notice the order of the
first two operators in the correlator and the sign $t_1\pm t_2$ in the exponential):
\begin{eqnarray}
\label{eq:app25}
    W^{[1,1];[1,1],{(a)},ss'}_{n\mathbf{k}q,n'\mathbf{k}'q'} &=& \int \text d t\text d t_1 \text d t_2 \Theta_{ss'}(t_1,t_2) e^{i(q\omega_{n\mathbf{k}}+q'\omega_{n'\mathbf{k}'})t} e^{i(t_1+t_2) (q\omega_{n\mathbf{k}}-q'\omega_{n'\mathbf{k}'})}\nonumber\\
    &&\cdot \left\langle Q_{n\mathbf k}^{-q}(-t-t_2)Q_{n'\mathbf k'}^{-q'}(-t+t_2)Q_{n'\mathbf k'}^{q'}(-t_1)Q_{n\mathbf k}^{q}(+t_1)\right\rangle_\beta\\
     W^{[1,1];[1,1],{(b)},ss'}_{n\mathbf{k}q,n'\mathbf{k}'q'} &=& \int \text d t\text d t_1 \text d t_2 \Theta_{ss'}(t_1,t_2) e^{i(q\omega_{n\mathbf{k}}+q'\omega_{n'\mathbf{k}'})t} e^{i(t_1-t_2) (q\omega_{n\mathbf{k}}-q'\omega_{n'\mathbf{k}'})}\nonumber\\
     &&\cdot \left\langle Q_{n'\mathbf k'}^{-q'}(-t-t_2)Q_{n\mathbf k}^{-q}(-t+t_2)Q_{n'\mathbf k'}^{q'}(-t_1)Q_{n\mathbf k}^{q}(+t_1)\right\rangle_\beta.
\end{eqnarray}
\end{widetext}
We will investigate the symmetries of these terms in
Sec.~\ref{sec:detailed-balance-2}, and show that only some combinations contribute
to the thermal Hall conductivity. In fact the eight terms from
Eq.~\eqref{eq:app24} can be rewritten as products of
(anti-)commutators. Meanwhile, defining the symmetrized in $(n\mathbf k q\leftrightarrow n'\mathbf k'q')$ collision rate,
\begin{equation}
\label{eq:app26}
\mathcal W^{[1,1];[1,1],ss'}_{n\mathbf kq,n'\mathbf k'q'}=\sum_{i=a,b}  W^{[1,1];[1,1],(i),ss'}_{n\mathbf k q,n'\mathbf k'q'} 
 + (n\mathbf k q\leftrightarrow n'\mathbf k'q'),
\end{equation}
we obtain components of the part of the collision matrix due to $\left|T^{[1,1]}_{\mathtt{i}\rightarrow\mathtt{f}}\right|^2$:
\begin{eqnarray}
\label{eq:app27}
 O^{[1,1];[1,1]}_{n,\mathbf k} &=&
\sum_{n'\mathbf k'} \sum_{q,q'} q~\sum_{s,s'}\mathcal W^{[1,1];[1,1],ss'}_{n\mathbf kq,n'\mathbf k'q'}\\
&&\cdot \left ( N^{\rm eq}_{n,\mathbf k}
 + \tfrac{q+1}{2}\right )\left ( N^{\rm eq}_{n'\mathbf k'} 
 + \tfrac{q'+1}{2} \right ),\nonumber
 \end{eqnarray}
 \begin{equation}
 \label{eq:app272}
 -D^{[1,1];[1,1]}_{n,\mathbf k} =
 \sum_{n'\mathbf k'}\sum_{q,q'}q \left ( 
 N^{\rm eq}_{n',\mathbf k'} 
 + \tfrac{q'+1}{2} \right )\sum_{s,s'}\mathcal W^{[1](2),ss'}_{n\mathbf
 kq,n'\mathbf k'q'},
\end{equation}
 \begin{equation}
 \label{eq:app274}
 M^{[1,1];[1,1]}_{n\mathbf k, n'\mathbf k'} = \sum_{q,q'}
 q~\left ( 
 N^{\rm eq}_{n,\mathbf k}
 + \tfrac{q+1}{2} \right )\sum_{s,s'}\mathcal W^{[1](2),ss'}_{n\mathbf kq,n'\mathbf k'q'}.
\end{equation}

\subsubsection{Commutators and anticommutators}
\label{sec:comm-antic}

In what follows, we write  $\llbracket A,B\rrbracket_\pm = AB\pm BA$.

\begin{widetext}
Upon replacing $Q^{-q}_{n\mathbf{k}}(-t-t_2)Q^{-q'}_{n'\mathbf k'}(-t+t_2)$
by $Q^{-q}_{n\mathbf{k}}(-t-t_2)Q^{-q'}_{n'\mathbf k'}(-t+t_2)-\llbracket Q^{-q}_{n\mathbf{k}}(-t-t_2),Q^{-q'}_{n'\mathbf k'}(-t+t_2)\rrbracket_{s'}$
in the definition of $ W^{[1,1];[1,1],{(a)},ss'}_{n\mathbf{k}q,n'\mathbf{k}'q'}$ hereabove,
 (after changing $t_2 \leftrightarrow -t_2$ if $s'=-$) 
one obtains $- W^{[1,1];[1,1],{(b)},ss'}_{n\mathbf{k}q,n'\mathbf{k}'q'}$.
Therefore,
\begin{eqnarray}
  \label{eq:71}
  W^{[1,1];[1,1],(a),ss'}_{n\mathbf kq,n'\mathbf k'q'}+W^{[1,1];[1,1],(b),ss'}_{n\mathbf kq,n'\mathbf k'q'}
&=& \int \text d t\text d t_1 \text d t_2 \Theta_{ss'}(t_1,t_2) e^{i(q\omega_{n\mathbf{k}}+q'\omega_{n'\mathbf{k}'})t} e^{i(t_1+t_2) (q\omega_{n\mathbf{k}}-q'\omega_{n'\mathbf{k}'})}\nonumber\\
  &&\cdot \left\langle \llbracket Q^{-q}_{n\mathbf{k}}(-t-t_2),Q^{-q'}_{n'\mathbf k'}(-t+t_2)\rrbracket_{s'} Q_{n'\mathbf k'}^{q'}(-t_1)Q_{n\mathbf k}^{q}(+t_1)\right\rangle_\beta.
\end{eqnarray}
Similarly, upon replacing $Q^{q'}_{n'\mathbf k'}(-t_1) Q^{q}_{n\mathbf k}(t_1) $
by $Q^{q'}_{n'\mathbf k'}(-t_1),Q^{q}_{n\mathbf k}(t_1) - \llbracket Q^{q'}_{n'\mathbf k'}(-t_1),Q^{q}_{n\mathbf k}(t_1)\rrbracket_s$
into $W^{[1,1];[1,1],(a),ss'}_{n\mathbf kq,n'\mathbf k'q'}+W^{[1,1];[1,1],(b),ss'}_{n\mathbf kq,n'\mathbf k'q'}$ i.e.\,Eq.\eqref{eq:71},
(after changing $t_1 \leftrightarrow -t_1$ if $s=-$) one obtains
$-W^{[1,1];[1,1],(b),ss'}_{n'\mathbf k'q',n\mathbf kq}-W^{[1,1];[1,1],(a),ss'}_{n'\mathbf k'q',n\mathbf kq}$.
Therefore, the only nonzero contribution to
$W^{[1,1];[1,1],(a),ss'}_{n\mathbf kq,n'\mathbf k'q'}+W^{[1,1];[1,1],(b),ss'}_{n'\mathbf k'q',n\mathbf kq}+(n\mb k q\leftrightarrow n'\mb k'q')$,
i.e.\, Eq.\,\eqref{eq:app26}, takes the form
\begin{eqnarray}
\label{eq:app36}
   \mathcal W^{[1,1];[1,1],ss'}_{n\mathbf k q,n'\mathbf k'q'}&=& \int \text d t\text d t_1 \text d t_2 \Theta_{ss'}(t_1,t_2) e^{i(q\omega_{nk}+q'\omega_{n'k'})t} e^{i(t_1+t_2) (q\omega_{nk}-q'\omega_{n'k'})}\\
    &&\cdot \left\langle \llbracket Q_{n\mathbf k}^{-q}(-t-t_2),Q_{n'\mathbf k'}^{-q'}(-t+t_2)\rrbracket_{s'} \llbracket Q_{n'\mathbf k'}^{q'}(-t_1),Q_{n\mathbf k}^{q}(+t_1)\rrbracket_s \right\rangle_\beta,
\end{eqnarray}
\end{widetext}
where $s,s'=+$ corresponds to an energy conservation constraint,
i.e.\ to on-shell scattering event, while $s,s'=-$ corresponds to a ${\rm
  PP}(E_{\mathtt i}-E_{\mathtt n})^{-1}$ term, i.e.\ off-shell
scattering (with $\mathtt {i},\mathtt{n},\mathtt{f}$ the initial, intermediate, and
final states in the second-order process).

Note that, in a time-reversal symmetric system, these satisfy the symmetry
\begin{equation}
\label{eq:app42}
  \mathcal W^{[1,1];[1,1],ss'}_{n\mathbf k q,n'\mathbf k' q'} =
  ss'\mathcal W^{[1,1];[1,1],s's}_{n-\mathbf k q,n'-\mathbf k' q'},
\end{equation}
reflecting the role of $\pm i\eta$ in the denominators in terms of causality.

\subsubsection{Detailed balance}
\label{sec:detailed-balance-2}

Using the method of the previous subsection Sec.~\ref{sec:detailed-balance-1}, we
show the following (``anti-'')detailed-balance relations
\begin{equation}
\label{eq:app37}
 W^{[1,1];[1,1],(a),ss'}_{n\mathbf{k}q,n'\mathbf{k}'q'} = ss'~W^{[1,1];[1,1],(a),s's}_{n'\mathbf{k}'-q',n\mathbf{k}-q} e^{-\beta(q\omega_\mathbf{k}+q'\omega_{\mathbf{k}'})},
 \end{equation}
 \begin{equation}
 \label{eq:app371}
 W^{[1,1];[1,1],(b),ss'}_{n\mathbf{k}q,n'\mathbf{k}'q'} = ss'~W^{[1,1];[1,1],(b),s's}_{n\mathbf{k}-q,n'\mathbf{k}'-q'} e^{-\beta(q\omega_\mathbf{k}+q'\omega_{\mathbf{k}'})}.
\end{equation}
From this, the same holds for the symmetrized in $n\mathbf
k q\leftrightarrow n'\mathbf k' q'$ scattering rate:
\begin{equation}
\label{eq:app38}
\mathcal W^{[1,1];[1,1],ss'}_{n\mathbf k q,n'\mathbf k'q'} = ss'~e^{-\beta(q\omega_\mathbf{k}+q'\omega_{\mathbf{k}'})}
\mathcal W^{[1,1];[1,1],ss'}_{n\mathbf k-q,n'\mathbf k'-q'}.
\end{equation}

We now identify
\begin{eqnarray}
\label{eq:app39}
  \mathfrak{W}^{\ominus, [1,1];[1,1],qq'}_{n\mathbf{k},n'\mathbf{k}'}
  &=& N_{\rm uc}\sum_{s=\pm}  \mathcal{W}^{[1,1];[1,1],s,-s}_{n\mathbf{k}q,n'\mathbf{k}'q'} ,\\
\mathfrak{W}^{\oplus, [1,1];[1,1],qq'}_{n\mathbf{k},n'\mathbf{k}'}
   &=& N_{\rm uc}\sum_{s=\pm}  \mathcal{W}^{ [1,1];[1,1],ss}_{n\mathbf{k}q,n'\mathbf{k}'q'}, 
\end{eqnarray}
complete expressions of which are given in the main text.

By construction, these enforce 
\begin{equation}
  \label{eq:41}
  \mf W^{\sigma, [1,1];[1,1],qq'}_{n\mathbf{k},n'\mathbf{k}'} =
  \sigma~e^{-\beta(q\omega_\mathbf{k}+q'\omega_{\mathbf{k}'})}
  \mf W^{\sigma, [1,1];[1,1],-q-q'}_{n\mathbf{k},n'\mathbf{k}'},
\end{equation}
where $\sigma=\oplus$ (resp. $\sigma=\ominus$) indicates
that $\mf W$ enforces detailed balance (resp. ``anti-detailed balance'').

\subsubsection{All contributions}
\label{sec:all-contributions}

As mentioned at the beginning of this subsection, other terms of order
${\lambda}^4$ contribute to the thermal conductivity at
second Born's order. In the following, we use the shorthand
$\llbracket A(-t),B(t')\rrbracket={\rm  sign}(t+t')[A(-t),B(t')]$
and $\fint_{t,\{t_j\}}$ (resp.\
$\fint'_{t,\{t_j\}}$), $j=1,..,l$, denotes the set of $1+l$
inverse Fourier transforms evaluated once at
$\Sigma_{n\mb k q}^{n'\mb k'
  q'}=q\omega_{n\mathbf{k}}+q'\omega_{n'\mathbf{k}'}$ and $l$ times at
$\Delta_{n\mb k q}^{n'\mb k'
  q'}=q\omega_{n\mathbf{k}}-q'\omega_{n'\mathbf{k}'}$, i.e.\ $\fint_{t,\{t_j\}}=\int \text dt
\text dt_1.. \text dt_l e^{i\Sigma_{n\mb k q}^{n'\mb k' q'} t}e^{i\Delta_{n\mb k q}^{n'\mb k' q'}(t_1+..+t_l)}$,
(resp.\ once at $q\omega_{n\mathbf{k}}$ and $l$ times at
$q'\omega_{n'\mb k'}$,
i.e.\ $\fint_{t,\{t_j\}}'=\int \text dt
\text dt_1.. \text
dt_l e^{iq\omega_{n\mathbf{k}}t}e^{iq'\omega_{n'\mb
    k'}(t_1+..+t_l)}$). 

\begin{widetext}
One contribution comes from the ``cross-term''
$2\mf{Re}\{(T^{[2]}_{\mathtt{i}\rightarrow\mathtt{f}})^*T^{[1,1]}_{\mathtt{i}\rightarrow\mathtt{f}}\}$, which contributes
to the scattering rates in the form of
\begin{eqnarray}
  \label{eq:105}
  \mathfrak{W}^{\ominus,[1,1];[2],qq'}_{n\mathbf{k}n'\mathbf{k}'}
  &=&\frac{2 N_{\rm uc}^{1/2}}{\hbar^4}~ \mathfrak{Im}\fint_{t,t_1}\left\langle Q^{-q,-q'}_{n\mathbf{k}n'\mathbf{k}'}(-t)
      \{Q_{n'\mathbf{k}'}^{q'}(-t_1),Q_{n\mathbf{k}}^{q}(t_1)\}\right\rangle\\
\label{eq:105b}
  \mathfrak{W}^{\oplus,[1,1];[2],qq'}_{n\mathbf{k}n'\mathbf{k}'}
  &=&-\frac{2 N_{\rm uc}^{1/2}}{\hbar^4}~ \mathfrak{Im}\fint_{t,t_1}\left\langle Q^{-q,-q'}_{n\mathbf{k}n'\mathbf{k}'}(-t)
      \llbracket Q_{n'\mathbf{k}'}^{q'}(-t_1), Q_{n\mathbf{k}}^{q}(t_1)\rrbracket\right\rangle,
\end{eqnarray}
The last contribution comes from considering the second
Born's order matrix element
  \begin{eqnarray}
  \label{eq:42}
  T^{[2,1]}_{\mathtt{i}\rightarrow\mathtt{f}}
  &=& \frac 1{\sqrt{N_{\rm uc}}} 
      \sum_{n\mathbf{k},n'\mathbf{k}'}\sum_{q,q'=\pm}
      \left (N^i_{n'\mathbf{k}'}+\tfrac{1+q'}{2}\right
      )\sqrt{N^i_{n\mathbf{k}}+\tfrac{1+q}{2}}~\mathds{I}({ \mt
   i_p}\overset{q\cdot n\mathbf k}{\longrightarrow}{ \mt f_p})\\
&&\sum_{\mt m_s} \left \{
    \frac{\langle
   \mt f_s|Q^{-q'}_{n'\mathbf k'}|\mt m_s\rangle\langle
    \mt m_s|Q^{qq'}_{n\mathbf k,n'\mb k'}|\mt i_s\rangle}{E_{\mt f_s}-E_{\mt
   m_s}-q'\omega_{n'\mathbf{k}'}+i\eta} +  \frac{\langle
   \mt f_s|Q^{qq'}_{n\mathbf k,n'\mb k'}|\mt m_s\rangle\langle
    \mt m_s|Q^{-q'}_{n'\mathbf k'} |\mt i_s\rangle}{E_{\mt i_s}-E_{\mt
   m_s}+q'\omega_{n'\mathbf{k}'}+i\eta} \right
   \}, \nonumber
\end{eqnarray}
which contains two-phonon operators, of order ${\lambda}^2$.
At order ${\lambda}^4$, it is thus involved in the ``cross-term'' 
$2~\mf{Re}\{(T^{[1]}_{\mathtt{i}\rightarrow\mathtt{f}})^*T^{[2,1]}_{\mathtt{i}\rightarrow\mathtt{f}}\}$,
which contributes to scattering rates in the form of
$\mathfrak{W}^{[2,1];[1],qq'}_{n\mathbf{k}n'\mathbf{k}'}
=\sum_{s=\pm}\mathfrak{W}^{[2,1];[1],qq'}_{n\mathbf{k}n'\mathbf{k}'|s}$, where
\begin{eqnarray}
  \label{eq:34}
  \mathfrak{W}^{[2,1];[1],qq'}_{n\mathbf{k}n'\mathbf{k}'|s}
  &=&\frac{N_{\rm uc}^{1/2}}{\hbar^4}\mathfrak{Im}\fint_{t,t_1}'\left\langle Q^{-q}_{n\mathbf{k}}(-t)
 [ {\rm sign}(t_1)]^{\frac{1-s}{2}}\llbracket   Q^{-q'}_{n'\mathbf{k}'}(-t_1),Q^{qq'}_{n\mathbf{k}n'\mathbf{k}'}(0)\rrbracket_s \right\rangle,
\end{eqnarray}
and we recall the shorthand $\llbracket A,B\rrbracket_\pm =AB\pm BA$.
\end{widetext}

The first two terms enforce the usual, ``two-phonon'', (anti-)detailed balance relations
\begin{equation}
  \label{eq:61}
  \mf W^{\sigma, [2];[1,1],qq'}_{n\mathbf{k}q,n'\mathbf{k}'q'} =
  \sigma~e^{-\beta(q\omega_\mathbf{k}+q'\omega_{\mathbf{k}'})}
  \mf W^{\sigma, [2];[1,1],qq'}_{n\mathbf{k}q,n'\mathbf{k}'q'},
\end{equation}
with $\sigma=\oplus,\ominus$.
Meanwhile, the last term satisfies ``one-phonon'' (anti-)detailed balance,
\begin{equation}
  \label{eq:50}
  \mf W^{[2,1];[1],qq'}_{n\mathbf k, n'\mathbf k'|s}
 = s \,e^{-q\beta\omega_{n\mb k}}  \mf W^{[2,1];[1],-q-q'}_{n\mathbf k, n'\mathbf k'|s},
\end{equation}
where we used a different notation ($s=\pm$ as a lower index) to emphasize
the difference with the other terms derived hereabove.

\subsection{Computation at third Born's order}
\label{sec:computation-at-third}

The only third-order element of $T_{\mt i \mapsto \mt f}$ which can appear in a
term of order $\lambda^4$ in $|T_{\mt i \mapsto \mt f}|^2$ is
\begin{widetext}
\begin{eqnarray}
  \label{eq:362}
   T^{[1,1,1]}_{\mathtt{i}\rightarrow\mathtt{f}}
  &=&   \sum_{n\mathbf{k}q,n'\mathbf{k}'q'}
      \sqrt{N^i_{n\mathbf{k}}+\tfrac{1+q}{2}}\left (N^i_{n'\mathbf{k}'}+\tfrac{1+q'}{2}\right
      )~\mathds{I}({
   i_p}\overset{q\cdot n\mathbf k}{\longrightarrow}{ f_p})
  \times\sum_{m_s,m'_s}\left \{
 \tfrac{\langle
   f_s|Q^{-q'}_{n'\mathbf k'}|m'_s\rangle\langle
    m'_s|Q^{q'}_{n'\mathbf k'}|m_s\rangle\langle
    m_s|Q^{q}_{n\mathbf
     k}|i_s\rangle}{(E_{i_s}-E_{m_s}-q\omega_{n\mathbf{k}}+i\eta)
     (E_{f_s}-E_{m'_s}-q'\omega_{n'\mathbf{k'}}+i\eta)} \right .\nonumber\\
    &&\qquad
     +\tfrac{\langle
   f_s|Q^{-q'}_{n'\mathbf k'}|m'_s\rangle\langle
    m'_s|Q^{q}_{n\mathbf k} |m_s\rangle\langle
    m_s|Q^{q'}_{n'\mathbf
       k'}|i_s\rangle}{(E_{i_s}-E_{m_s}-q'\omega_{n'\mathbf{k'}}+i\eta)
       (E_{f_s}-E_{m'_s}-q'\omega_{n'\mathbf{k'}}+i\eta)}
 \left .
       +\tfrac{\langle
   f_s|Q^{q}_{n\mathbf k} |m'_s\rangle\langle
    m'_s| Q^{-q'}_{n'\mathbf k'} |m_s\rangle\langle
    m_s|Q^{q'}_{n'\mathbf
       k'}|i_s\rangle}{(E_{i_s}-E_{m_s}-q'\omega_{n'\mathbf{k'}}+i\eta)
       (E_{f_s}-E_{m'_s}+q\omega_{n\mathbf{k}}+i\eta)}
 \right \},
\end{eqnarray}
which is involved in the scattering rate
\begin{equation}
  \label{eq:390}
  \mathfrak{W}^{[1,1,1];[1],qq'}_{n\mathbf{k}n'\mathbf{k}'}
 = 2\mathfrak{Re}\left [
      (T^{[1,1,1]}_{\mathtt{i}\rightarrow\mathtt{f}})^*T^{[1]}_{\mathtt{i}\rightarrow\mathtt{f}}\right
      ] = \sum_{s,s'=\pm}\mathfrak{W}^{[1,1,1];[1],qq'}_{n\mathbf{k}n'\mathbf{k}'|ss'},
    \end{equation}
    where we denote
    \begin{eqnarray}
      \label{eq:398}
      \mathfrak{W}^{[1,1,1];[1],qq'}_{n\mathbf{k}n'\mathbf{k}'|ss'}
      &=& \frac{-1}{2\hbar^4}\mathfrak{Re}\fint_{t,t_1,t_2}'
\left \langle Q^{-q}_{n\mathbf k}(-t)\left \lgroup Q^{-q'}_{n'\mathbf k'}(-t_1),\,Q^{q}_{n\mathbf k}(0),\,Q^{q'}_{n'\mathbf   k'}(t_2)
                                                                       \right \rgroup_{s,s'}\right \rangle,\\
   \left  \lgroup A(t),B(t'),C(t'')\right \rgroup_{s,s'}
     &=&
    [\text{sign}(t-t')]^{\tfrac{1-s}{2}}[\text{sign}(t'-t'')]^{\tfrac{1-s'}{2}}
    \left (
      \vphantom{\int}A(t)B(t')C(t'')+sB(t')A(t)C(t'')+s'A(t)C(t'')B(t')\right   ).\nonumber
\end{eqnarray}
\end{widetext}

This term enforces an unusual version of (anti-)detailed balance,
namely
\begin{equation}
  \label{eq:388}
\mathfrak{W}^{[1,1,1];[1],qq'}_{n\mathbf{k}n'\mathbf{k}'|ss'} = ss'
e^{-q\beta\omega_{n\mb k}}\mathfrak{W}^{[1,1,1];[1],-q,q'}_{n\mathbf{k}n'\mathbf{k}'|ss'},
\end{equation}
where the index $q'$ remains inchanged.

\section{Generalizations}
\label{sec:generalizations}

\subsection{Generalized model and higher perturbative orders}
\label{sec:gener-model-high}

To describe
the interaction between phonon and another degree of freedom, we
introduce general coupling terms between phonon annihilation
(creation) 
operators $a^{(\dagger)}_{n\mathbf k}$ and general, for now
unspecified, fields $Q^{\{q_j\}}_{\{n_j,\mathbf k_j\}}$ which
are operators acting in their own Hilbert
space, i.e.\ we write $H'=\sum_{l}H'_{[l]}$, with 
\begin{equation}
  \label{eq:322}
  H'_{[l]}=\frac 1 {\sqrt{N_{\rm uc}^{l}}}\sum_{\{n_i\mathbf{k}_i\}}\sum_{\{q_i=\pm\}}\Big (\prod_{j=1}^m a_{n_j\mathbf{k}_j}^{q_j}\Big )
Q^{\{q_j\}}_{\{n_j\mathbf{k}_j\}}.
\end{equation}
In this expression, $m$ is the number of phonon creation-annihilation operators
coupled to $Q^{\{q_j\}_{i=1..l}}_{\{n_j\mathbf{k}_j\}_{i=1..l}}$. In terms of the perturbative expansion introduced
in the main text and the other appendices, this means
$Q^{\{q_j\}_{i=1..l}}_{\{n_j\mathbf{k}_j\}_{i=1..l}}\sim \lambda^l$;
note that since the perturbative expansion is considered (formally) up to \emph{infinite} order in this appendix,
we make this specification only for the sake of clarity.
To avoid ambiguities, we assume that all the $n_j\mathbf{k}_j$ indices involved in a given term of $H'_{[l]}$
are distinct; this is correct in the thermodynamic limit.
Note also that, for the sake of clarity in the following developments,
the normalization factors of $N_{\rm uc}$ are not
defined following the same convention as in the rest of the paper.

In what follows, we take special notations for the first two indices:
$n_1\mb k_1 \equiv n\mb k$, $n_2\mb k_2 \equiv n'\mb k'$.
Using the model Eq.~\eqref{eq:22} and following the general procedure
described in Sec. \ref{sec:from-interaction-terms} and in the main text, one can then
(at least formally) derive the collision integral which always takes the form
  \begin{align}
    \label{eq:88}
    \mc C_{n\mb k}
    &= \sum_{p=1}^\infty \frac
                     1 {N_{\rm uc}^{p}}\sum_{\{n_i\mb
                       k_i\}_{i=2,..,p}}\sum_{\{q_i\}_{i=1,..,p}} q_1\\
    &\qquad \qquad \times\left [ \prod_{i=1}^{p} \left ( \overline N_{n_i\mathbf k_i}
      +\tfrac {q_i+1}2\right )\right ] W^{(p),\{q_i\}}_{\{n_i\mb k_i\}}\nonumber ,
  \end{align}
  where the index $(p)$ denotes a term of order $\lambda^{2p}$.
The scattering rate $W^{(p),\{q_i\}}_{\{n_i\mb k_i\}}=W^{(p),\{q_i\}_{i=1,..,p}}_{\{n_i\mb k_i\}_{i=1,..,p}}$
is the sum of all the scattering rates of the $[l_1,..,l_{m}];[l'_1,..,l'_{m'}]$ kind (according to the nomenclature introduced
in the main text) such that $\sum_{i=1}^m l_i +\sum_{j=1}^{m'} l'_j=2p$.
In terms of physical process, each of these terms corresponds to the interference between two
scattering channels, $[l_1,..,l_{m}]$ and $[l'_1,..,l'_{m'}]$, such that in all, $2p$ phonon creations or annihilations occur
between the initial $\mt i$ and final $\mt f$ states.
Note that in the present paper, we compute explicitly this expansion up to $p=2$.

We then expand the phonon average populations as
$\overline N_{n_i\mathbf k_i}= N^{\rm eq}_{n_i\mathbf k_i}+\delta \overline N_{n_i\mathbf k_i}$.
Following Eq.\eqref{eq:main2}, the diagonal scattering rate is obtained as
$D_{n\mb k}=-\partial_{\overline N_{n\mb k}} \mc C_{n\mb k} \Big |_{\overline N = N^{\rm eq}}$. It can be decomposed
as $D_{n\mb k}=\sum_{p=1}^\infty D_{n\mb k}^{(p)}$, where 
  \begin{align}
    \label{eq:60}
    D_{n\mb k}^{(p)}
    &=    \frac {-1} {N_{\rm uc}^{p}} \sum_{\{n_i\mb  k_i\}_{i=2,..,p}}\sum_{\{q_i\}_{i=1,..,p}} q_1\\
    &\qquad \qquad\times \left [ \prod_{i =2}^p\left ( N^{\rm  eq}_{n_i\mathbf k_i}
      +\tfrac {q_i+1}2\right )\right ] W^{(p),\{q_i\}}_{\{n_i\mb k_i\}}.\nonumber
  \end{align}

  Similarly, the off-diagonal scattering rate is obtained as
  $M_{n\mb k,n'\mb k'}=\partial_{\overline N_{n'\mb k'}} \mc C_{n\mb k}\Big |_{\overline N = N^{\rm eq}}$. It can be decomposed
  as $M_{n\mb k,n'\mb k'}=\sum_{p=2}^\infty M_{n\mb k,n'\mb k'}^{(p)}$, where
\begin{align}
  \label{eq:62}
  M^{(p)}_{n\mb k n'\mb k'}
  &= \frac
  1 {N_{\rm uc}^{p}}\sum_{\{n_i\mb  k_i\}_{i=3,..,p}} \sum_{\{q_i\}_{i=1,..,p} } q_1\\
  &\times \left ( N^{\rm eq}_{n\mathbf k}
    +\tfrac {q_1+1}2\right ) \left [ \prod_{i=3}^p \left ( N^{\rm eq}_{n_i\mathbf k_i}
    +\tfrac {q_i+1}2\right )\right] W^{(p),\{q_i\}}_{\{n_i\mb k_i\}}.\nonumber
\end{align}
 
Like in the equations for $p=1,2$ derived explicitly in Appendix~\ref{sec:from-interaction-terms},
$q_1$ always factorizes in the collision integral, as the change in number of $n\mb k$ phonons
due to the scattering event.

\subsection{Special properties of first Born's order}
\label{sec:spec-prop-first}

\subsubsection{Definitions and basic results}
\label{sec:defin-conv}

At first order of the Born expansion, all contributions to the collision integral are ``semiclassical'',
in the sense defined in Sec.~\ref{sec:scattering-channels};
i.e.\ an operator $Q^{q_1,..,q_l}_{n_1\mb k_1,..,n_l \mb k_l}$ does only appear
in the collision integral as $|Q^{q_1,..,q_l}_{n_1\mb k_1,..,n_l \mb k_l}|^2$.

To make this statement more precise, we rewrite 
\begin{align}
  \label{eq:12}
H'_{[l]} &= \frac 1 {\sqrt{N_{\rm uc}^{l}}}\sum_{r=0}^l\sum_{\{n_i\mathbf{k}_i\}_{i=1..l}}
  \Big (\prod_{j=1}^r a_{n_j\mathbf{k}_j}^{+}\Big ) \Big (\prod_{j=r+1}^l a_{n_j\mathbf{k}_j}^{-}\Big )\nonumber\\
&\times \frac 1 {\sqrt{r!(l-r)!}} \, Q^{+,..,+|-,..,-}_{n_1\mb k_1..n_r\mb k_r|n_{r+1}\mb k_{r+1}..n_l\mb k_l},
\end{align}
where the upper indices of $Q$ are $r$ times '$+$' and $l-r$ times '$-$',
and $Q$ is by definition symmetric under permutation of its lower indices in the two blocks
$\{n_i\mb k_i\}_{i=1,..,r}$ and $\{n_i\mb k_i\}_{i=r+1,..,l}$ separately.
Hermiticity is guaranteed by $Q^{+,..,+|-,..,-}_{n_1\mb k_1..n_r\mb k_r|n_{r+1}\mb k_{r+1}..n_l\mb k_l}
= (Q^{+,..,+|-,..,-}_{n_{r+1}\mb k_{r+1}..n_l\mb k_l|n_1\mb k_1..n_r\mb k_r })^\dagger$.
Note that at first Born's order, distinct scattering channels $[l]$ and $[l']$ do not interfere for $l\neq l'$;
one can thus study independently the contribution of each $H'_{[l]}$ to the collision integral.

\begin{widetext}
The contribution to the squared T-matrix obtained from $H'_{[l]}$ at first Born's order is
\begin{align}
  \label{eq:25}
   \Big |T^{[l]}_{\mathtt{i}\rightarrow\mathtt{f}}\Big |^2
  &= \frac 1 {N_{\rm uc}^{l}}\sum_{r=0}^l \frac 1 {r!(l-r)!} \sum_{\{n_i\mathbf{k}_i\}_{i=1..l}}
  \left [\prod_{j=1}^r  (N^{\mt i}_{n_j\mathbf{k}_j}+1) \right ] \left [\prod_{j=r+1}^l  (N^{\mt i}_{n_j\mathbf{k}_j}) \right ]\\
  &\times \mathds{I}({i_p}\overset{+\{n_j\mathbf k_j\}_{1}^{r}}{\underset{-\{n_j\mathbf k_j\}_{r+1}^{l}}
    {\xrightarrow{\hspace*{1.5cm}}}}{f_p})
  \, \langle i_s|Q^{+,..,+|-,..,-}_{\{n_j\mb k_j\}_{r+1}^l|\{n_j\mb k_j\}_1^r}|f_s\rangle
  \langle f_s|Q^{+,..,+|-,..,-}_{\{n_j\mb k_j\}_1^r|\{n_j\mb k_j\}_{r+1}^l}|i_s\rangle ,\nonumber
\end{align}

Summing over all scattering channels,
the first Born's order contribution to the collision integral is
\begin{align}
  \label{eq:8}
    \mc C^{\mt {1 B}}_{n\mb k}
  &= \sum_{l=1}^\infty \frac  1 {N_{\rm uc}^{l}}\sum_{r=0}^l \sum_{\{n_i\mb k_i\}_{i=1,..,l}}
   \prod_{j=1}^r  (\overline N_{n_j\mathbf{k}_j}+1) \prod_{j=r+1}^l  (\overline N_{n_j\mathbf{k}_j}) 
  \left ( \frac {r \delta_{n,n_1}\delta_{\mb k,\mb k_1}} {r!(l-r)!} 
    - \frac {(l-r) \delta_{n,n_{r+1}} \delta_{\mb k,\mb k_{r+1}} } {r!(l-r)!}\right ) W^{[l];[l]}_{\{n_j\mb k_j\}_1^r|\{n_j\mb k_j\}_{r+1}^l},
\end{align}
where 
    \begin{align}
      \label{eq:27}
 W^{[l];[l]}_{\{n_j\mb k_j\}_1^r|\{n_j\mb k_j\}_{r+1}^l}
  &= \int \text dt e^{-i \left ( \sum_{j=1}^r \omega_{n_j\mb k_j} - \sum_{j=r+1}^l \omega_{n_j\mb k_j} \right )t}
    \left \langle Q^{+,..,+|-,..,-}_{\{n_j\mb k_j\}_{r+1}^l|\{n_j\mb k_j\}_1^r}(t) \,
    Q^{+,..,+|-,..,-}_{\{n_j\mb k_j\}_1^r|\{n_j\mb k_j\}_{r+1}^l} \right \rangle . 
\end{align}
\end{widetext}

Following the same steps as in Sec. \ref{sec:detailed-balance-1},
it is easy to see that $W^{[l];[l]}_{\{n_j\mb k_j\}_1^r|\{n_j\mb k_j\}_{r+1}^l}$
\emph{always} enforces detailed-balance, namely
\begin{align}
  \label{eq:28}
  W^{[l];[l]}_{\{n_j\mb k_j\}_1^r|\{n_j\mb k_j\}_{r+1}^l}
  &= e^{-\beta \left ( \sum_{j=1}^r \omega_{n_j\mb k_j} - \sum_{j=r+1}^l \omega_{n_j\mb k_j}\right ) } \nonumber\\
 &\qquad \times  W^{[l];[l]}_{\{n_j\mb k_j\}_{r+1}^l|\{n_j\mb k_j\}_1^r}.
\end{align}

We now prove two important properties of the collision integral,
as obtained from first Born's order, which derive therefrom.

\subsubsection{No equilibrium current}
\label{sec:no-energy-shift}

The equilibrium current is due to
$O_{n\mb k}=\mc C_{n\mb k}[\overline N_{n'\mb k'} \mapsto N^{\rm eq}_{n'\mb k'}]$, the constant
term in the collision integral.

In the present case, by performing a change of index $r\mapsto l-r$ in the second term of
$( \dots - \dots)$ in Eq.\ \eqref{eq:8}, and resorting to the detailed balance relation Eq.\ \eqref{eq:28}
and taking $N^{\rm eq}_{n'\mb k'}=\frac 1 {e^{\beta \omega_{n'\mb k'}}-1}$, one can easily
show that
\begin{equation}
  \label{eq:29}
  O^{\mt{1B}}_{n\mb k}=\mc C^{\mt{1B}}_{n\mb k}[\overline N_{n_j\mb k_j} \mapsto N^{\rm eq}_{n_j\mb k_j}] = 0.
\end{equation}
This means that no shift of the phonons' energies
is needed at first Born's order to guarantee cancellation of the equilibrium current.

\subsubsection{No phonon Hall effect }
\label{sec:no-phonon-hall}

The off-diagonal contribution to the collision matrix at first Born's order,
$M^{\mt{1B}}_{n\mb k,n'\mb k'}=\partial_{\overline N_{n'\mb k'}} \mc C^{\mt{1B}}_{n\mb k}$, reads
\begin{widetext}
  \begin{align}
  \label{eq:26}
  M^{\mt{1B}}_{n\mb k,n'\mb k'}
  &= \sum_{l=1}^\infty \frac  1 {N_{\rm uc}^{l}}\sum_{r=0}^l \sum_{\{n_i\mb k_i\}_{i=1,..,l}}
   \prod_{j=1}^r  (N^{\rm eq}_{n_j\mathbf{k}_j}+1) \prod_{j=r+1}^l  (N^{\rm eq}_{n_j\mathbf{k}_j}) 
   \,  W^{[l];[l]}_{\{n_j\mb k_j\}_1^r|\{n_j\mb k_j\}_{r+1}^l}\\
    &\qquad \times \frac 1 {r!(l-r)!}  \left \lgroup r \delta_{n,n_1}\delta_{\mb k,\mb k_1} \nonumber
      \Big( (r-1) \delta_{n',n_2}\delta_{\mb k',\mb k_2} + (l-r) \delta_{n',n_{r+1}}\delta_{\mb k',\mb k_{r+1}} \Big ) \right .\\
   &\left . \qquad \qquad \qquad \qquad \qquad   - (l-r) \delta_{n,n_{r+1}} \delta_{\mb k,\mb k_{r+1}} \nonumber
      \Big ( r \delta_{n',n_1}\delta_{\mb k',\mb k_1} + (l-r-1) \delta_{n',n_{r+2}}\delta_{\mb k',\mb k_{r+2}}\Big ) \right \rgroup
\end{align}
\end{widetext}
After some algebra, following essentially the same steps as outlined hereabove,
it is possible to show that
\begin{equation}
  \label{eq:33}
  M^{\mt{1B}}_{n\mb k,n'\mb k'} e^{\beta\omega_{n'\mathbf{k}'}}(N^{\rm eq}_{n'\mathbf{k}'})^2
  =  M^{\mt{1B}}_{n'\mb k',n\mb k} e^{\beta\omega_{n\mathbf{k}}}(N^{\rm eq}_{n\mathbf{k}})^2.
\end{equation}
This, as was illustrated several times in the main text and the appendices,
entails that $M^{\mt{1B}}$ does not contribute to $\kappa_{\rm H}$ -- see Eq.\ \ref{eq:app6}.
We have thus shown that no contribution to the thermal Hall conductivity can possibly come
from first Born's order, regardless of the number of phonon operators in the Hamiltonian and of the
nature of the operators $Q$ to which they are coupled.

\section{Application---further technical details}
\label{sec:appl-:-furth}

\subsection{Solving the delta functions}
\label{sec:solv-delta-funct}

In order to solve the two simulaneous delta functions, we use the
following rewriting of $\mathfrak{W}^\ominus$,
\begin{widetext}
\begin{eqnarray}
  \label{eq:46}
  \mathfrak{W}^{\ominus,qq'}_{n\mathbf{k},n'\mathbf{k}'}&=&\frac{64\pi^2}{\hbar^4 N_{\rm
     uc}^{2d}}\sum_{\mathbf{p}}\sum_{\{\ell_i,q_i\}} \mathcal{F}^{\ell_3,\ell_1,\ell_2|q_4,q_1,q_2}_{\mathbf{p},q\mathbf{k},q'\mathbf{k}'}\mathfrak{Im}\Bigg\{{\mathcal
    B}^{n\ell_2\ell_3|q_2q_3q}_{\mb k,\mb p +\frac{1}{2}q\mb k + q'\mb
    k'} {\mathcal
    B}^{n'\ell_3\ell_1|-q_3q_1q'}_{\mb k',\mb p+ \frac{1}{2}q'\mb
     k'}\\
  &&{\rm PP}\Bigg[\frac{{\mathcal B}^{n\ell_1\ell_4|-q_1q_4 -q}_{\mb k,\mb p +
    \frac{1}{2}q\mb k} {\mathcal B}^{n'\ell_4\ell_2|-q_4-q_2-q'}_{\mb
    k',\mb p + q\mb k +\frac{1}{2}q'\mb
     k'}}{2v_{\rm m}q_1\left(\frac{q_1q\omega_{n\mathbf{k}}}{v_{\rm m}}+
    \hat{\Omega}_{\ell_1,\mb p}-q_1q_4 \hat{\Omega}_{\ell_4,\mb p +q\mb k}\right)}+(n,q,\mathbf{k},q_4)\leftrightarrow (n',q',\mathbf{k}',-q_4)\Bigg]\Bigg\}
    \nonumber\\
   && \delta\left(v_{\rm m}q_1\left[
    \frac{q_1\Sigma_{n\mathbf{k}n'\mathbf{k}'}^{qq'}}{v_{\rm m}}+\hat{\Omega}_{\ell_1,\mb
     p}+q_1q_2\hat{\Omega}_{\ell_2,\mb p +q\mb k +  q'\mb  k'}\right]\right) \delta
    \left(-2v_{\rm m}q_1\left[\frac{q_1q'\omega_{n'\mathbf{k}'}}{v_{\rm m}}+\hat{\Omega}_{\ell_1,\mb   p}-q_1q_3\hat{\Omega}_{\ell_3,\mb  p+q'\mb  k'}\right]\right),\nonumber
\end{eqnarray}
where $\hat{\Omega}_{\ell,\mathbf{p}}=\Omega_{\ell,\mathbf{p}}/v_{\rm
  m}$ and
\begin{eqnarray}
  \label{eq:344}
  \mathcal{F}^{\ell_3,\ell_1,\ell_2|q_4,q_1,q_2}_{\mathbf{p},q\mathbf{k},q'\mathbf{k}'}&=&q_4
                                                                                           \left(2n_{\rm
                                                                                           B}(\Omega_{\ell_3,\mb
                                                                                           p+q'\mb
                                                                                           k'})+1\right)
\left(2n_{\rm B}(\Omega_{\ell_1,\mb
  p})+q_1+1\right)\left(2n_{\rm B}(\Omega_{\ell_2,\mb p+q\mb k +q'\mb
    k'})+q_2+1\right)
\end{eqnarray}
is a product of thermal factors (note
$\hat{\mathcal{F}}=q_1\mathcal{F}$, cf.\ Eq.~\eqref{eq:44}). Now collapsing the delta functions, we
can write:
\begin{eqnarray}
  \label{eq:347}
 \mathfrak{W}^{\ominus,qq'}_{n\mathbf{k},n'\mathbf{k}'}&=&\frac{8\mf a^2}{\hbar^4 v_{\rm m}^2}\sum_{j}\sum_{\ell,\{q_i\}} \mathfrak{Im}\Bigg\{{\mathcal
    B}^{n\ell\ell|q_2q_3q}_{\mb k,\mb p_j +\frac{1}{2}q\mb k + q'\mb
    k'} {\mathcal
    B}^{n'\ell\ell|-q_3q_1q'}_{\mb k',\mb p_j+ \frac{1}{2}q'\mb
     k'}{\rm PP}\Bigg[\frac{{\mathcal B}^{n\ell\ell|-q_1q_4 -q}_{\mb k,\mb p_j +
    \frac{1}{2}q\mb k} {\mathcal B}^{n'\ell\ell|-q_4-q_2-q'}_{\mb
    k',\mb p_j + q\mb k +\frac{1}{2}q'\mb
     k'}}{2v_{\rm m}q_1\left(\frac{q_1q\omega_{n\mathbf{k}}}{v_{\rm m}}+
    \hat{\Omega}_{\ell,\mb p_j}-q_1q_4 \hat{\Omega}_{\ell,\mb p_j
                                                           +q\mb
                                                           k}\right)}\nonumber\\
  &&\qquad\qquad\qquad\qquad\qquad\qquad+(n,q,\mathbf{k},q_4)\leftrightarrow (n',q',\mathbf{k}',-q_4)\Bigg]\Bigg\}
  ~ J_{\mathfrak{W}}(\mathbf{p}_j)\mathcal{F}^{\ell,\ell,\ell|q_4,q_1,q_2}_{\mathbf{p}_j,q\mathbf{k},q'\mathbf{k}'},
\end{eqnarray}
where, when they exist, the solutions, $j=0,..,3$ take the form
(recall $\mathbf{v}_i=\ul{\mathbf{v}}_i$,
$\mathbf{w}_i=\ul{\mathbf{w}}_i$, $\mathbf{p}_j=\ul{\mathbf{p}}_j$)
\begin{equation}
  \label{eq:348}
  \mathbf{p}_j=t_{\lfloor j/2\rfloor}\mathbf{v}_{\lfloor j/2\rfloor}+u_{\lfloor j/2\rfloor}^{(\widetilde{j\;[2]})}\mathbf{w}_{\lfloor j/2\rfloor},
\end{equation}
where, for $i=0,1$
\begin{eqnarray}
  \label{eq:54}
  \mathbf{v}_i&=&a_2\ul{\mathbf{k}}_1+(-1)^ia_1\ul{\mathbf{k}}_2,
  \hspace{6cm}\mathbf{w}_i = \mathbf{\hat{z}}\times \mathbf{v}_i,\\
  [\smallskipamount]
  t_i&=&\frac{a_2\ul{\mathbf{k}}_1^2+(-1)^ia_1\ul{\mathbf{k}}_2^2-a_1a_2(a_1+(-1)^ia_2)}{2\mathbf{v}_i^2}, \hspace{2.5cm}  u_i^{(\pm)}=\frac{-B_i\pm\sqrt{B_i^2-4A_iC_i}}{2A_i}, \\
  [\smallskipamount]
  A_i&=&4a_{i+1}^2(\mathbf{v}_i^2-(\ul{\mathbf{k}}_1\wedge \ul{\mathbf{k}}_2)^2), \\
  B_i&=&(-1)^i4a_{i+1} (\ul{\mathbf{k}}_1\wedge
         \ul{\mathbf{k}}_2)(a_{i+1}^2-\ul{\mathbf{k}}_{i+1}^2+2(\mathbf{v}_i\cdot\ul{\mathbf{k}}_{i+1})t_i), \nonumber\\
  C_i&=&-\left(a_{i+1}(a_{i+1}-2\delta_\ell)-\ul{\mathbf{k}}_{i+1}^2\right)
         \left(a_{i+1}(a_{i+1}+2\delta_\ell)-\ul{\mathbf{k}}_{i+1}^2\right) -4(a_{i+1}^2-\ul{\mathbf{k}}_{i+1}^2)(\mathbf{v}_i\cdot\ul{\mathbf{k}}_{i+1})t_i +4(a_{i+1}^2\mathbf{v}_i^2-(\mathbf{v}_i\cdot\ul{\mathbf{k}}_{i+1})^2)t_i^2,\nonumber
\end{eqnarray}
and $J_\mathfrak{W}(\mathbf{p}_j)$ is given in the main text, Eq.~\eqref{eq:49}.
\end{widetext}

\subsection{Choice of polarization vectors}
\label{sec:choice-polar-vect}

Below, we enumerate possible explicit choices for a basis of polarization vectors
$(\bs \varepsilon_{0,\mb k}, \bs \varepsilon_{1,\mb k}, \bs \varepsilon_{2,\mb k})$.
In the numerical calculations, we use choice 2.

\subsubsection{Choice 1}
\label{sec:choice-1}

A simple choice is that of momentum-independent polarization vectors,
which can be, for example:
$\boldsymbol{\varepsilon}_0(\mb k)=\mathbf{\hat{x}}$,
$\boldsymbol{\varepsilon}_1(\mb k)=\mathbf{\hat{y}}$,
$\boldsymbol{\varepsilon}_2(\mb k)=\mathbf{\hat{z}}$.

\subsubsection{Choice 2}
\label{sec:choice-2}

Below, we describe the choice of polarization vectors used in the numerical implementation.
Its polarization vectors $ \bs \varepsilon_{n,\mb k}$ form an orthonormal basis in which
$\mb k$ points along the $[1,1,1]$ axis, so that $\mb k \cdot  \bs \varepsilon_{n,\mb k} = \frac{|\mb k|}{\sqrt 3} ~ \forall n$.
This, as explained in the main text, ensures that the structure factor
$\bs {\mc S}^{\alpha,\beta}$ does not vanish for $\alpha=\beta$,
corresponding to the largest coupling constants $\Lambda_{1..5}$ (as opposed to anisotropic $\Lambda_{6,7}$ for which $\alpha\neq\beta$).

The starting point is the orthonormal basis made of three vectors $\mb e_n, n=0,1,2$, defined as
\begin{equation}
   \label{eq:pol1}
\begin{cases}
\mb e_0=\frac{1}{\sqrt{3}}\left(\sqrt{2}\mathbf{\hat{x}}+\mathbf{\hat{z}}\right)\\
\mb e_1=\frac{1}{\sqrt{6}}\left(-\mathbf{\hat{x}}+\sqrt{3}\mathbf{\hat{y}}+\sqrt{2}\mathbf{\hat{z}}\right)\\
\mb e_2=\frac{1}{\sqrt{6}}\left(-\mathbf{\hat{x}}-\sqrt{3}\mathbf{\hat{y}}+\sqrt{2}\mathbf{\hat{z}}\right)
\end{cases};
\end{equation}
in this basis, $\mb{\hat z}=[1,1,1]$. 
To rotate the $\mb{\hat z}$ axis into $\mb{\hat k}$'s direction, we
define the polar angles of $\mb{\hat k}$,
  \begin{eqnarray}
    \label{eq:pol5}
    \theta &=& \arccos(k_z/|\mb k|),\\
    \phi &=& \text{Arg}(k_x + \text i k_y),
  \end{eqnarray}
  so that a good choice for the three polarization vectors is
  \begin{equation}
    \label{eq:pol6}
  \bs \varepsilon_{n,\mb k}   =
           \text i \,
         R_{\mathbf{\hat{z}}}(\phi)\cdot
        R_{\mathbf{\hat{y}}}(\theta)   \cdot
          {\rm diag}\left [\ms s(\mb k), 1, 1\right ]\cdot \mb e_n 
  \end{equation}
for $n=0,1,2$. 
In the above, we defined $R_{\boldsymbol{\hat{\mu}}}(\gamma)$ to be
the direct rotation matrix around the $\mu$ axis by an angle $\gamma$,
and we used the ``sign'' function
    \begin{equation}
    \label{eq:pol2}
    \ms s(\mb k) =
    \begin{cases}
      +1 & \text {if}  \quad \mb k \in \mc D_+\\
      -1 & \text {if}  \quad \mb k \in \mc D_-
    \end{cases}
  \end{equation}
  with respect to two domains $\mc D_\pm$, corresponding (up to unimportant details
    in a set of null measure contained in the $k_z=0$ plane) to the
    ``upper'' ($k_z>0$) and ``lower'' ($k_z<0$) halves of $\mathbb R^3$, and more precisely
  defined by 
  \begin{eqnarray}
    \label{eq:pol4}
    \mc D_+ &=& \Big \{ \mb k \; \big | \; k_z>0 \text{ or }
               \big (k_z=0  \\
    &&\text{ and } (k_y>0 \text{ or } (k_y=0  \text{ and } k_x>0)) \big ) \Big \},\nonumber\\
    \mc D_- &=& \Big \{ \mb k \; \big | \; k_z<0 \text{ or }
                \big (k_z=0 \\
    &&\text{ and } (k_y<0 \text{ or } (k_y=0  \text{ and } k_x<0)) \big ) \Big \}\nonumber
  \end{eqnarray}
  such that $\mathbb R^3= \mc D_+ \cup \mc D_- \cup \{\bs 0 \}$.
  The role of this $\ms s(\mb k)$ function is to help ensure that this choice
  of polarizations enforces $\boldsymbol{\varepsilon}_{n}(-\mb k) = \boldsymbol{\varepsilon}_{n}(\mb k)^*$, as
  well as all the tetragonal symmetry group of the crystal.
  This last statement means that under a symmetry operation $g$ belonging
  to $D_{4h}$ the symmetry group of the crystal, they transform as
  \begin{equation}
    \label{eq:64}
    \bs \varepsilon_n(g\cdot \mb k) = \sum_{n'}c^g_{nn'}(\mb k)\, g\cdot\bs \varepsilon_{n'}(\mb k),
  \end{equation}
  where $g\cdot$ denotes the action of $g$ on a vector,
  and most importantly the $c^g_{nn'}$ coefficient either is $\delta_{nn'}$ or
  exchanges the $n=1$ and $n=2$ polarizations, depending on whether $\mb k$ is in a high-symmetry position. Indeed $n=1,2$ are constructed degenerate
  (as eigenvectors of the dynamical matrix) at the high-symmetry
  planes and axes of the Brillouin zone. See Ref.~\cite{maradudin} for details and further discussion on the
  behavior of polarization vectors under symmetry operations.

\subsubsection{Choice 3}
\label{sec:choice-3}

One may also
use the Hall-plane-dependent basis for the polarization vectors and
label $\bs \varepsilon_{n\mb k}$, assuming $\mu\nu$ is the Hall
plane, $\rho$ is the direction transverse to the plane, and
$\mu\nu\rho$ forms a direct orthonormal basis:
\begin{eqnarray}
  \label{eq:138}
  \bs \varepsilon_{0,\mb k}  
&=& i\mb k/|\mb
  k|\quad\mbox{(longitudinal)},\nonumber\\
\bs \varepsilon_{1,\mb k}  &=&i \frac{\hat{\mb u}_\rho \times \mb k}{|\hat{\mb u}_\rho
  \times \mb k|} \quad\mbox{(transverse, in Hall plane)}, \nonumber\\
  \bs \varepsilon_{2,\mb k} &=& \bs \varepsilon_{0,\mb k}
  \times \bs \varepsilon_{1,\mb k}=\frac{k^\rho\mathbf{k}-\mathbf{k}^2\mathbf{\hat{u}}_\rho}{|\mathbf{k}||\hat{\mb u}_\rho
  \times \mb k|}\quad \mbox{(transverse)}.\nonumber\\
\end{eqnarray}
Then, {\em if}
$\alpha,\beta,\mu,\nu,\rho\in\{x,y,z\}$, we can write:
\begin{eqnarray}
  \label{eq:120}
  \mathcal{S}^{q;\alpha\beta}_{0\mathbf{k}}&=&\frac{1}{\sqrt{\omega_{0\mathbf{k}}}}\frac{2qi}{|\mathbf{k}|}k^\alpha
                                                                                             k^\beta,\\
\mathcal{S}^{q;\alpha\beta}_{1\mathbf{k}}&=&\frac{1}{\sqrt{\omega_{1\mathbf{k}}}}\frac{qi}{|\mathbf{\hat{u}}_\rho\times\mathbf{k}|}\left(k^\mu(k^\alpha\delta_{\beta,\nu}+k^\beta\delta_{\alpha,\nu})\right.\nonumber\\
&&\qquad\qquad\qquad\qquad\left.-k^\nu(k^\alpha\delta_{\beta,\mu}+k^\beta\delta_{\alpha,\mu})\right),\nonumber\\
\mathcal{S}^{q;\alpha\beta}_{2\mathbf{k}}&=&\frac{1}{\sqrt{\omega_{2\mathbf{k}}}}\frac{-1}{|\mathbf{k}||\mathbf{\hat{u}}_\rho\times\mathbf{k}|}\left(\mathbf{k}^2(k^\alpha\delta_{\beta,\rho}+k^\beta\delta_{\alpha,\rho})\right.\nonumber\\
&&\qquad\qquad\qquad\qquad\qquad\left.-2k^\alpha
                                                                                           k^\beta
                                                                                           k^\rho\right),\nonumber
\end{eqnarray}
so that
$\mathcal{L}_0\propto\mathbf{k}\cdot\boldsymbol{\lambda}\cdot\mathbf{k}$,
$\mathcal{L}_1\propto[\mathbf{k}\times(\boldsymbol{\lambda}\cdot\mathbf{k})]^\rho$,
$\mathcal{L}_2\propto[\mathbf{k}^2(\boldsymbol{\lambda}\cdot\mathbf{k})-(\mathbf{k}\cdot\boldsymbol{\lambda}\cdot\mathbf{k})\mathbf{k}]^\rho$.

For this choice of polarization vectors, the phonon-magnon coupling constants can be decomposed in such a way that
their behavior under operations of the $D_{4h}$ point-group defined in the $(\mu,\nu,\rho)$ basis becomes transparent,
in other words in terms of the basis harmonics of the ``Hall geometry'' point-group.
Note that, because the magnetic space group of the system is a priori independent of the symmetries associated
with the choice of ``Hall geometry,'' the coefficients of the harmonics need not be
  independent. ({\em In the the square lattice case discussed here}, some of
  the symmetries of the system coincide with those of the Hall geometry, so that these coefficients are not entirely
  independent. Note that this causes additional constraints for the existence of a nonzero Hall effect.)

\subsection{Numerical implementation}
\label{sec:numer-impl}

We define $\hat{\lambda}_{\xi,\xi'}^{\ell_1,\ell_2;\alpha\beta}=\lambda^{\alpha\beta}_{\tilde{\xi}\ell_1+\bar{\xi}+1,\tilde{\xi}'\ell_2+\bar{\xi}'+1;\xi\xi'}$,
so that, in particular, 
\begin{eqnarray}
  \label{eq:156}
  \hat{\lambda}_{1,1}^{\ell_1,\ell_2;\alpha\beta}&=&\lambda^{\alpha\beta}_{\ell_1+1,\ell_2+1;11},\nonumber\\
  \hat{\lambda}_{0,0}^{\ell_1,\ell_2;\alpha\beta}&=&\lambda^{\alpha\beta}_{\bar{\ell}_1+1,\bar{\ell}_2+1;00},\nonumber\\
  \hat{\lambda}_{1,0}^{\ell_1,\ell_2;\alpha\beta}&=&\lambda^{\alpha\beta}_{\ell_1+1,\bar{\ell}_2+1;10},
\end{eqnarray}
(note the bars) and 
\begin{equation}
  \label{eq:21}
  \mathcal{L}_{n\mathbf{k};\xi,\xi'}^{q,\ell_1,\ell_2}={\rm Tr}\left[({\boldsymbol{\hat{\lambda}}_{\xi\xi'}^{\ell_1\ell_2}})^T\cdot\boldsymbol{\mathcal{S}}_{n\mathbf{k}}^q\right]=\sum_{\alpha,\beta=x,y,z}
  \hat{\lambda}_{\xi,\xi'}^{\ell_1,\ell_2;\alpha\beta}\mathcal{S}_{n\mathbf{k}}^{q;\alpha\beta}.
\end{equation}
Moreover, given {\em (i)} our choice of isotropic elasticity, {\em
  (ii)} a given Hall plane $\mu\nu$ and perpendicular Hall axis,
$\rho$, {\em (iii)} $\ell_1=\ell_2=\ell$, $\hat{\lambda}$ is a function of
$\Lambda_{1,..,7}^{\xi(')}$ contains $72$ values, which can be
parametrized by a single index $i=0,..,71$ through, e.g.\
$i=36\xi+18\xi'+9\ell+3\alpha+\beta$ if we identify $(x,y,z)$ with
$(0,1,2)$ for $\alpha$ and $\beta$, $\mathcal{S}$ is a
complex function of $n,\rho,\mathbf{k},q,\ell,\alpha,\beta$.

\subsection{Details of the derivation of the general forms of the scaling relations}
\label{sec:gener-forms-scal}

Here we give details about the results and calculations in Sec.~\ref{sec:new}.

\subsubsection{Dimensionless functions}
\label{sec:dimens-funct}

The functional forms of the scaling functions introduced in
Eq.~\eqref{eq:77} are
\begin{eqnarray}
  \label{eq:106}
  \tilde{c}_\eta(\tilde{y})&=&\frac{1}{2}\left(|\sin\theta|+\eta\upsilon^{-1}X(\tilde{y})\right),\\
  \tilde{\Omega}^{\pm\eta}_\ell(\tilde{y})&=&\frac{1}{2}\sqrt{\sin^2\theta
                                              X^2(\tilde{y})+\upsilon^{-2}\pm2\eta|\sin\theta|\upsilon^{-1}X(\tilde{y})},\nonumber
\end{eqnarray}
with
  \begin{equation}
    \label{eq:108}
    X(\tilde{y})=\sqrt{1+4\frac{\tilde{y}^2}{\sin^2\theta-\upsilon^{-2}}},
  \end{equation}
and
  \begin{eqnarray}
    \label{eq:107}
  \tilde{{\rm
  f}}_\eta^{s=1}(\tilde{y})&=&\Theta\left(\upsilon^{-2}-\sin^2\theta-4|\tilde{y}|^2\right),\nonumber\\
  \tilde{{\rm f}}_\eta^{s=-1}(\tilde{y})&=&\delta_{\eta,1}\Theta\left(\sin^2\theta-\upsilon^{-2}\right),\nonumber\\
  \tilde{J}_D^s(\tilde{y})&=&\left|\sum_{r=\pm1}\frac{s^{(r-1)/2}r\tilde{c}_r(\tilde{y})}{\sqrt{\tilde{c}_r(\tilde{y})^2+\tilde{y}^2}}\right|^{-1}.
\end{eqnarray}
Inserting the expressions for $\mathcal{L}$ and ${\rm F}$ into that of
$\mathcal{B}$, we find
\begin{eqnarray}
  \label{eq:91}
  &&\mathcal{B}^{n,\ell\ell|+s-}_{\mathbf{k}; -\mb p^{(\eta)}_{\ell,\mathbf{k}}(y)+\frac{\mathbf{k}}{2}}\\
 && =
 \frac{-i}{2\sqrt{2M_{\rm uc}}} 
  \sum_{\xi\xi'} n_0^{-\xi-\xi'} \sum_{\alpha,\beta=x,y,z}
  \hat{\lambda}^{\ell\ell;\alpha\beta}_{\xi\xi'}\;
    i^{\overline{\xi}}(si)^{\overline{\xi}'}(-1)^{(\overline{\xi}+\overline{\xi}')\ell}\nonumber\\
  &&\quad\frac{k^\alpha\varepsilon_{n\mathbf{k}}^\beta+k^\beta\varepsilon_{n\mathbf{k}}^\alpha}
  {\sqrt{\omega_{n\mathbf{k}}}} (\chi\Omega_{\ell\mathbf{p}_{\ell\mathbf{k}}^{(\eta)}}(y))^{\xi-\frac{1}{2}}(\chi\Omega_{\ell\mathbf{p}_{\ell\mathbf{k}}^{(-\eta)}}(y))^{\xi'-\frac{1}{2}}\nonumber
\end{eqnarray}
because $F_{\xi q\ell}(-\mathbf{p})=F_{\xi q\ell}(\mathbf{p})$. Then,
for $\delta_\ell=0$, and defining
$\hat{k}^\alpha=(\mathbf{\hat{k}})_\alpha$, the
$\tilde{\mathcal{B}}_{\xi\xi'}$ introduced in Eq.~\eqref{eq:77} are:
\begin{eqnarray}
  \label{eq:98}
\tilde{\mathcal{B}}_{nn}&=&
 \frac{is \chi^{-1}v_{\rm m}^{-1}}{2\sqrt{2M_{\rm uc}v_{\rm ph}}} 
  \sum_{\alpha,\beta=x,y,z}
  \hat{\lambda}^{\ell\ell;\alpha\beta}_{00}\;(\hat{k}^\alpha\varepsilon_{n\mathbf{k}}^\beta+\hat{k}^\beta\varepsilon_{n\mathbf{k}}^\alpha)
    \nonumber\\
                        &&\qquad\qquad\qquad\qquad(\tilde{\Omega}^{\eta}(\tilde{y})\tilde{\Omega}^{-\eta}(\tilde{y}))^{-\frac{1}{2}},
\end{eqnarray}
\begin{eqnarray}
  \label{eq:100}
\tilde{\mathcal{B}}_{mm}&=&
 \frac{-i n_0^{-2}\chi v_{\rm m}}{2\sqrt{2M_{\rm uc}v_{\rm ph}}} 
  \sum_{\alpha,\beta=x,y,z}
  \hat{\lambda}^{\ell\ell;\alpha\beta}_{11}\;(\hat{k}^\alpha\varepsilon_{n\mathbf{k}}^\beta+\hat{k}^\beta\varepsilon_{n\mathbf{k}}^\alpha)
  \nonumber\\
  &&\qquad\qquad\qquad\qquad
(\tilde{\Omega}^{\eta}(\tilde{y})\tilde{\Omega}^{-\eta}(\tilde{y}))^{\frac{1}{2}},
\end{eqnarray}
and
\begin{eqnarray}
  \label{eq:103}
 \tilde{\mathcal{B}}_{mn}&=&
 \frac{(-1)^{\ell} n_0^{-1}}{2\sqrt{2M_{\rm uc}v_{\rm ph}}} 
  \sum_{\alpha,\beta=x,y,z}
  (\hat{k}^\alpha\varepsilon_{n\mathbf{k}}^\beta+\hat{k}^\beta\varepsilon_{n\mathbf{k}}^\alpha) 
    \nonumber\\
  &&\left[\hat{\lambda}^{\ell\ell;\alpha\beta}_{01}\;
     (\tilde{\Omega}^{\eta}(\tilde{y}))^{-\frac{1}{2}}(\tilde{\Omega}^{-\eta}(\tilde{y}))^{\frac{1}{2}}\right.\nonumber\\
  &&\left.\quad+s\;\hat{\lambda}^{\ell\ell;\alpha\beta}_{10} \;(\tilde{\Omega}^{(\eta)}(\tilde{y}))^{\frac{1}{2}}(\tilde{\Omega}^{(-\eta)}(\tilde{y}))^{-\frac{1}{2}}\right].
\end{eqnarray}

\subsubsection{Details of the behavior of the scaling function $\rm
  F$}
\label{sec:deta-behav-scal}

Here we derive the behavior of the scaling functions ${\rm
  F}_x(\varkappa,\upsilon,\theta)$ defined in Eq.~\eqref{eq:78} at
small and large $\varkappa$.  

Let us first consider the large-$\varkappa$ limit.   The hyperbolic sines in the
denominator of the integrand grow exponentially in this limit (because
the two
$\tilde\Omega^{\pm\eta}_\ell(\tilde{y})$ functions cannot be
simultaneously be made to vanish), so that they can be approximated by
their leading exponential forms.  An application of the saddle point
method then shows that the integral is dominated by region around
$\tilde{y}=0$, and is exponentially suppressed for large $\varkappa$.
For $\upsilon|\sin\theta|>1$, i.e.\ when the azimuthal angle of
$\mathbf{k}$ is smaller than $\upsilon^{-1}$, this suppression exceeds
the exponential growth of the $\sinh \varkappa/2$ prefactor, and the
scaling function decays exponentially:
\begin{equation}
  {\rm
    F}_x(\varkappa,\upsilon>
  |\sin\theta|^{-1})\underset{\varkappa\gg1}{\sim}\exp\left[\frac{\varkappa}{2}(1-\upsilon|\sin\theta|)\right].
  \label{eq:97}
  \end{equation} 
For smaller angles where $\upsilon|\sin\theta|<1$, ${\rm F}$ does not
decay exponentially in the large $\kappa$ limit.  Here
Eq.~\eqref{eq:97} is correct to exponential accuracy, i.e.\ it is
asymptotically correct for $\ln ({\rm F})$ at large $\varkappa$.  To
this accuracy, the asymptotics are independent of $x$.  

Now consider the small $\varkappa$ limit.  The na\"ive result for the
scaling function is obtained by 
expanding both the hyperbolic sine in the numerator and the two in the
denominator of the integrand around zero leads to
\begin{eqnarray}
  \label{eq:82}
    &&{\rm  F}^{(s)}_{x}(\varkappa\ll 1,\upsilon,\theta)  \underset{?}{\sim}
       \frac{1}{\varkappa} \frac{(3-s)
       \mathfrak{a}^2}{2\pi v_{\rm m}\hbar^2
       \upsilon^2 }
  \\
  &&  \int_{-\infty}^{+\infty} \text d\tilde{y}\sum_\eta\tilde{\rm
     f}_\eta^s(\tilde{y})\tilde{J}_D^s(\tilde{y}) \sum_{\ell}  \frac{\tilde{\mathcal C}_{x}(\tilde{y})}
           {\tilde{\Omega}^{+\eta}_\ell(\tilde{y}) \tilde{\Omega}^{-\eta}_\ell(\tilde{y})}.\nonumber
\end{eqnarray}
This expression is correct provided the integral in the second line
converges.  The convergence is problematic only at large $\tilde{y}$
for the case $s=-1$ (in the case $s=1$, the integral is confined by the
$\tilde{{\rm f}}^{s=1}$ factor to a finite domain).
In this limit the Jacobean $\tilde{J}^{s=-1}$ grows linearly in
$\tilde{y}$ as does $\tilde\Omega$, while the factor
$\tilde{\mathcal{C}}_x(\tilde{y})$ behaves as $\tilde{y}^{2x}$.    As
a result, the integral converges for the case $x=-1$ and the
$1/\varkappa$ scaling is correct in this case.  In the cases $x=0,1$, the
integral is logarithmically and quadratically divergent at large
$\tilde{y}$, respectively.

In the latter two cases, we must reconsider the na\"ive result in
Eq.~\eqref{eq:82}.  The divergence in this equation is an artificial
result because the hyperbolic sines in the original expression in Eq.~\eqref{eq:78} grow rapidly once $\tilde{\Omega} > \varkappa^{-1}$ and ensure convergence
of the integral (i.e. the large $\tilde{y} > |\upsilon\varkappa|^{-1}$
contribution is negligible).  Proper behavior is restored for small $\varkappa$ by
simply using the expanded form of Eq.~\eqref{eq:82} but only
integrating up to an upper cutoff $|\tilde{y}|<
|\varkappa\upsilon|^{-1}$.   This regulates the divergences and one
obtains additional $\ln(1/\varkappa)$ and $\varkappa^{-2}$ factors
multiplying the $1/\varkappa$ form for the cases $x=0,1$,
respectively.   Collecting the above results we see that
\begin{equation}
  \label{eq:96}
  {\rm F}^{(-1)}_x(\varkappa)
  \underset{\varkappa\ll1}{\sim}
  \left\{
    \begin{array}{ll} \frac{1}{\varkappa^3}
      & x=1 \\
      \frac{\ln(1/\varkappa)}{\varkappa}
      & x=0 \\ \frac{1}{\varkappa}
      & x=-1 \end{array}.
  \right.
\end{equation}
This function is non-zero for $|\upsilon\sin\theta|>1$, while $F^{(+1)}$
is non-zero when $|\upsilon\sin\theta|<1$.

In the latter case, as mentioned above, the integral over $\tilde y$ always converges because
  the set of integration is an ellipse instead of a half-hyperbola. Therefore the na\"ive scaling is the correct one and the
$ {\rm F}^{(+1)}_x(\varkappa) \underset{\varkappa\ll1}{\sim} 1/\varkappa$ behavior holds for all $x$.

\section{Application---further physical details}
\label{sec:appl-further-physical}

  \subsection{Microscopic derivation of the coupling constants}
\label{sec:micr-deriv-coupl}

We consider the most general coupling between the strain tensor and bilinears of
the $\mb m,\mb n$ fields, exhibiting all the symmetries allowed by
the crystal symmetry group in the paramagnetic phase:
 the $D_{4h}$ tetragonal point-group; translations of one unit cell
---which forbids interactions of the ${\rm m}_a {\rm n}_b$ type;
and time-reversal.
The corresponding hamiltonian density (where for the sake of readability we have replaced $n_0\rightarrow 1$) reads:
\begin{widetext}
\begin{eqnarray}
  \label{eq:13}
 \mc H'_{\rm tetra}
  &=&  \Lambda^{(\rm m)}_{1} \Big ({\rm m}_x {\rm m}_x \mathcal E^{xx}
         +   {\rm m}_y {\rm m}_y \mathcal E^{yy} \Big )
        + \Lambda^{(\rm n)}_{1} \Big ({\rm n}_x {\rm n}_x\mathcal E^{xx}
      + {\rm n}_y {\rm n}_y\mathcal   E^{yy} \Big )
  + \Lambda^{(\rm m)}_{5}  {\rm m}_z  {\rm m}_z \mathcal E^{zz}
     + \Lambda^{(\rm n)}_{5} {\rm n}_z {\rm n}_z \mathcal E^{zz} \\
  &&
     +    \Lambda^{(\rm m)}_{2} \Big ({\rm m}_x {\rm m}_x\mathcal E^{yy}
     +{\rm m}_y {\rm m}_y\mathcal E^{xx}\Big )
     + \Lambda^{(\rm n)}_{2} \Big ({\rm n}_x {\rm n}_x\mathcal E^{yy}
     +{\rm n}_y {\rm n}_y\mathcal E^{xx}\Big ) \\
  &&   + \Big (\Lambda^{(\rm m)}_{3} {\rm m}_z {\rm m}_z
     + \Lambda^{(\rm n)}_{3} {\rm n}_z {\rm n}_z \Big ) \Big (\mc
     E^{xx}+\mc E^{yy}\Big )
     + \Lambda^{(\rm m)}_{4} \Big ({\rm m}_x {\rm m}_x+{\rm m}_y {\rm m}_y
     \Big )\mathcal E^{zz}
     + \Lambda^{(\rm n)}_{4} \Big ({\rm n}_x {\rm n}_x +{\rm n}_y {\rm   n}_y \Big )
     \mathcal E^{zz} \\
  &&+ 4\Lambda^{(\rm m)}_{6} {\rm m}_x {\rm m}_y \mathcal E^{xy} +
                 4\Lambda^{(\rm n)}_{6} {\rm n}_x {\rm n}_y \mathcal E^{xy}
               + 4\Lambda^{(\rm m)}_{7} \Big ({\rm m}_x {\rm m}_z \mathcal E^{xz}
                 +  {\rm m}_y {\rm m}_z\mathcal E^{yz}\Big )
                 + 4\Lambda^{(\rm n)}_{7} ({\rm n}_x {\rm n}_z \mathcal E^{xz}
                 + {\rm n}_y {\rm n}_z \mathcal E^{yz}).
\end{eqnarray}
\end{widetext}
We now propose a microscopic origin to the $\Lambda^{(\xi)}_{I}$ coefficients appearing in it.

We start from a generic spin exchange hamiltonian of the form
\begin{equation}
  \label{eq:6}
 H_{\rm ex}= \sum_{\mb R,\mb R'}\sum_{a,b}S_{\mb R}^aJ_{\mb R-\mb R'}^{ab}S_{\mb R'}^b,
\end{equation}
where $\mb R,\mb R'$ indicate the actual locations of the sites in the
\emph{distorted} lattice, and each sum spans the whole distorted lattice.

We then express $\mb R = \mb r + \bs u_{\mb r}$, where
$\mb r$ belongs to the undistorted lattice and $\bs u_{\mb r}$ is the displacement field
at site $\mb r$. Taylor-expanding the coefficients $J_{\mb R-\mb R'}^{ab}$ with respect to the displacement
field (and identifying $S^a_{\mb R}=S^a_{\mb r}$), we thus obtain
$H_{\rm ex} =  H_{\rm ex}^{0}+ H'_{\rm ex}+O(u^2)$,
where $H_{\rm ex}^{0}=H_{\rm ex} |_{\mb R,\mb R'\mapsto \mb r,\mb r'}$ and 
\begin{equation}
  \label{eq:7}
     H'_{\rm ex}
= \sum_{\mb r,\mb r'}\sum_{ab,\nu} S^a_{\mb r}
     \left ( u^\nu_{\mb r} -  u^\nu_{\mb r'}  \right )
\partial_{\eta^\nu} J_{\bs \eta}^{ab} \Big |_{\bs \eta = \mb r-\mb r'}S^b_{\mb r'}.
\end{equation}
Then, identifying $\mc E_{\alpha\beta}=\frac 1 2 (\partial_\alpha u_\beta +
\partial_\beta u_\alpha)$ the symmetric rank-2 elasticity tensor
(i.e.\ strain tensor), we identify $H'_{\rm ex}=H'_{ss\varepsilon}+\dots$, where
\begin{equation}
  \label{eq:51}
  H'_{ss\varepsilon} = \frac 1 2 \sum_{\mb r,\bs \eta} S_{\mb r+\bs \eta}^a S_{\mb  r}^b 
\left [ \eta_\alpha
    \partial_{\eta^\beta}
  + \eta_\beta
  \partial_{\eta^\alpha} \right ] J^{ab}\Big |_{\bs \eta} \,
\mc E_{\alpha\beta}(\mb r) + \dots,
\end{equation}
where $a,b=x,y,z$ is a spin axis index, $\alpha,\beta=x,y,z$
is a spatial index, and ``$+\dots$'' encompasses terms featuring $\omega_{\alpha\beta}$
the anti-symmetric rank-2 elasticity tensor, as well as higher-order derivatives
of the displacement field.

Note that in this microscopic derivation, we identify $S^a_{\mb R} \mapsto S^a_{{\mb r}}$.
In fact, also expanding the \emph{magnetization fields} (and not only the magnetic exchange)
with respect to displacement yields an interaction term which is \emph{formally} of the same order as that derived here.
However, magnetization in an ordered magnet is a slow variable, while $J$ varies over distances of the order of the lattice parameter $\mf a$,
therefore such terms are \emph{quantitatively} much smaller by a factor $O(k_B T \mf a/v_{\rm m})$,
both within and beyond the Born-Oppenheimer approximation.

Finally, we take the particular case of a square lattice with tetragonal symmetry,
and describe the spins in terms of $\rm m, n$ fields as in the main text,
namely $\mb S_{\mb r}  =  (-1)^{\mb r} \mu_0 \mb n(\mb r)
+ \mathfrak{a}^2 \mb m(\mb r) $.
We identify $H'_{ss\varepsilon} = \sum_{\mb r} \mc H'_{\rm tetra}(\mb r)+\dots$
where ``$+\dots$'' is made of rapidly oscillating (time-reversal breaking) terms,
and $\mc H'_{\rm tetra}$ is as displayed in Eq.~\eqref{eq:13}, with identification
\begin{eqnarray}
      \label{eq:67}
      \Lambda_{ab}^{({\rm m}),\alpha\beta}
      &=& \frac 1 2 \sum_{\bs \eta} \left ( \eta_\alpha\partial_\beta +
          \eta_\beta \partial_\alpha \right ) J^{ab} \Big |_{\bs
          \eta} ,\\
    \Lambda_{ab}^{({\rm n}),\alpha\beta}
      &=& \frac 1 2 \sum_{\bs \eta} e^{i\bs \pi \bs \eta}
          \left ( \eta_\alpha\partial_\beta + \eta_\beta
          \partial_\alpha \right )
          J^{ab} \Big |_{\bs \eta},
\end{eqnarray}
    where the sum over $\bs \eta$ spans the whole direct (two-dimensional square)
    lattice, and
    $\bs \pi = \left ( \frac{\pi}{\mf a} , \frac{\pi}{\mf a}  \right )$ with $\mf a$
      the square lattice parameter.

\subsection{Contributions to intervalley couplings}
\label{sec:contrib-to-intervalley}

In the main text, the $n_a,m_a$ fields live in the valleys identified by:
\begin{align}
    \ell = 0 : &\qquad n_y, m_z\nonumber\\
\ell = 1 : &\qquad n_z, m_y.
\end{align}
Therefore, intervalley couplings are of the form
$\lambda_{ab;\xi\xi'}$ with
$\delta_{\xi\xi'}+\delta_{ab}=1$. More explicitly, using
Eq.~\eqref{eq:15}, they are: 
\begin{eqnarray}
\lambda_{yz;00}^{\alpha\beta}&=&\Lambda_{yz}^{({\rm n}),\alpha\beta},\\
\lambda_{yz;11}^{\alpha\beta}&=&\Lambda_{yz}^{({\rm m}),\alpha\beta},\nonumber\\
 \lambda_{yy;01}^{\alpha\beta} & = &
\frac{-1}{n_0}\left[ m^y_0 \Lambda^{({\rm m}), \alpha\beta}_{yx} +
   m^{z}_0
  \Lambda_{zx}^{({\rm m}), \alpha\beta} + m^y_0
 \Lambda^{({\rm n}), \alpha\beta}_{yx}\right],\nonumber\\
  \lambda_{zz;01}^{\alpha\beta} & = &
\frac{-1}{n_0}\left[ m^z_0 \Lambda^{({\rm m}), \alpha\beta}_{zx} +
  m^{y}_0
  \Lambda_{y x}^{({\rm m}), \alpha\beta} + m^z_0
 \Lambda^{({\rm n}), \alpha\beta}_{zx}\right] .\nonumber
\end{eqnarray}
Also recall from Eq.~\eqref{eq:134} that
\begin{align}
   \Lambda_{yx}^{(\xi),xy}=\Lambda_{yx}^{(\xi),yx}= \Lambda_{xy}^{(\xi),xy}=\Lambda_{xy}^{(\xi),yx}  &=  \Lambda_6^{(\xi)},\nonumber\\
  \Lambda_{zx}^{(\xi),xz}=\Lambda_{zx}^{(\xi),zx}
  =\Lambda_{yz}^{(\xi),yz}=\Lambda_{yz}^{(\xi),zy} &= \Lambda_7^{(\xi)},
\end{align}
and all other values of $\alpha,\beta$ yield 0 for this set of lower
indices. From this, it is clear that the $\Lambda_7^{(\xi)}$ couplings
always mix valleys, regardless of $\bs m_0$, and contribute a
$\lambda_{yz;\xi\xi}$ term. This intervalley coupling is a
small contribution which does not contribute to $\mc T$
breaking. Meanwhile, the $\mc T$-odd $\lambda_{yy;01}$ and
$\lambda_{zz;01}$ intervalley couplings both contain contributions
from both $\Lambda_7^{(\xi)} m_0^z$ and $\Lambda_6^{(\xi)} m_0^y$.

\subsection{Derivation of the gaps from a sigma model}
\label{sec:derivation-gaps-from}

Here we provide a heuristic microscopic argument for expressing the gaps in terms of spin-spin couplings.
We ignore spin-lattice coupling, and just consider corrections to the isotropic Heisenberg model.  We
assume the addition of a term of the XXZ anisotropy form:
\begin{equation}
  \label{eq:gap2}
  H_{\rm XXZ} = g J \sum_{\langle ij\rangle} \left( 2S_i^z S_j^z - S_i^x S_j^x
    -S_i^y S_j^y\right).
\end{equation}
This is to be added to the isotropic Heisenberg model,
along with a Zeeman coupling to the transverse field.

Carrying out the long-wavelength expansion in terms of $\mb{m}$ and $\mb{n}$
fields, we obtain the corrected potential part (i.e.\,without gradient terms) of the
nonlinear sigma-model Eq.\,\eqref{eq:hfreeboson}:
\begin{align}
  \label{eq:gap3}
  \mathcal{H}_{\rm NLS}^{g,h} &= \frac{1}{2\chi} |\mb{m}|^2 + 2 g J
                            \mathfrak{a}^2  \left(2 \text m_z^2 - \text m_x^2 - \text m_y^2\right)  \\
  & - 2 g J
  \frac{\mu_0^2}{\mathfrak{a}^2} \left(2 \text n_z^2 -\text n_x^2 - \text n_y ^2\right)
  -h_y \text m_y - h_z \text m_z .\nonumber
\end{align}
Note that the first term includes an $\text m_x^2$ term, which is
absent in the quadratic expansion describing linear spin waves in the main
text. Indeed this term is higher order in the small fluctuations around an
$x$-ordered state when carrying out a zero field spin
wave expansion, which was the case in the main text where the external field
had already been integrated out to yield the $\text n_a \text n_b$
mass term. We also included an external uniform field which
lies in the $y-z$ plane.

\begin{widetext}
Expanding around the $x$-ordered state, using that $\text m_x = - \ms m_y \ms n_y - \ms m_z\ms n_z$ and
$\text n_x = 1 - \frac{1}{2} (\ms n_y^2+\ms n_z^2 + \frac{1}{n_0^2} (\ms m_y^2+\ms m_z^2))$, yields
\begin{align}
  \label{eq:gap4}
  \mathcal{H}_{\rm NLS}^{g,h}& = \frac{1}{2\chi} \left(\ms m_y^2+\ms m_z^2\right) + 2 g J
  \mathfrak{a}^2  \left(2 \ms m_z^2  - \ms m_y^2\right)  - 2 g J
  \frac{\mu_0^2}{\mathfrak{a}^2} \left(2 \ms n_z^2 - \ms n_y ^2\right)
                                 -h_y \ms m_y - h_z \ms m_z  \nonumber\\
                                 & +\left(\frac{1}{2\chi} - 2 g J
  \mathfrak{a}^2 \right) \left( \ms m_y^2 \ms n_y^2 + \ms m_z^2 \ms n_z^2 + 2 \ms m_y \ms m_z \ms n_y
                                   \ms n_z\right) + 2 g J
  \frac{\mu_0^2}{\mathfrak{a}^2}\left[1 - \frac{1}{2} (\ms n_y^2+\ms n_z^2 +
\frac{1}{n_0^2} (\ms m_y^2+\ms m_z^2))\right]^2.
\end{align}
\end{widetext}
Note that the first term on the second line is of the form
$(\mb{m}_\perp\cdot\mb{n}_\perp)^2$, where the $\perp$ indicates the
components of the vectors normal to the ordering direction. Since we
in the next step shift the magnetization by its value induced by the
field, this is proportional to $(\mb{h}\cdot\mb{n})^2$, as is postulated
in the main text on symmetry grounds.

We now show this explicitly. We shift the definition
$\ms m_a = m_a + \chi_a h_a$ for $a=y,z$,
and expand the result to quadratic order in $m, n$.  Here
$\chi_z = (1/\chi + 4 g J \mathfrak{a}^2)^{-1}$ and
$\chi_y = (1/\chi - 2 g J \mathfrak{a}^2)^{-1}$.
\begin{widetext}
  This gives
\begin{align}
  \label{eq:gap5}
 \mathcal{H}_{\rm NLS}^{g,h} & = \frac{1}{2\chi} \left(m_y^2+m_z^2\right) + 2 g J
  \mathfrak{a}^2  \left(2 m_z^2  - m_y^2\right)  - 2 g J
  \frac{\mu_0^2}{\mathfrak{a}^2} \left(2 n_z^2 - n_y ^2\right)
                                \nonumber \\
  & + \left(\frac{1}{2\chi} - 2 g J
  \mathfrak{a}^2 \right) \left( \chi_y^2 h_y^2 n_y^2 + \chi_z^2 h_z^2
    n_z^2 +2 \chi_y \chi_z h_y h_z n_y n_z\right) - 2 g J
  \frac{\mu_0^2}{\mathfrak{a}^2} \left[ n_y^2+ n_z^2 + \cdots\right],
\end{align}
where the `$\cdots$' in the last brackets account for terms higher order in
field, magnetization fluctuations, etc.

The anisotropy coefficients, denoted by $\Gamma_{ab}$ in the text, can now be extracted.
The terms in $\mathcal{H}_{\rm NLS}^{g,h}$ which are quadratic in the $n_y,n_z$ fields read
\begin{align}
  \label{eq:gap6}
  \mathcal{H}_{nn} = \chi_y^2 h_y^2  \left(\frac{1}{2\chi} - 2 g J
  \mathfrak{a}^2 \right) n_y^2 + \left[  \chi_z^2 h_z^2  \left(\frac{1}{2\chi} - 2 g J
  \mathfrak{a}^2 \right) - 6 g J \frac{\mu_0^2}{\mathfrak{a}^2} \right]
  n_z^2 + 2 \chi_y \chi_z h_y h_z \left(\frac{1}{2\chi} - 2 g J
  \mathfrak{a}^2 \right) n_y n_z.
\end{align}
\end{widetext}
Note that the two terms proportional to $n_y^2$ from the right-most
contributions on each line of Eq.~\eqref{eq:gap5} above canceled.  That
means the the coefficient of $n_y^2$ in Eq.~\eqref{eq:gap6} vanishes if
$h_y=0$.  This occurs because of Goldstone's theorem and the assumed
XXZ form of the anisotropy: if the field is purely along the $z$
direction, XY symmetry of the Hamiltonian under rotations about the
$z$ axis is preserved, and this makes one of the spin wave modes
remain gapless.  Conversely, for a field along the $y$ direction, and
in the presence of anisotropy, both modes are generally gapped.  

We can simplify the above expression if we assume $|g| \ll 1$, which
means $\chi_y^{-1} \approx \chi_z^{-1} \approx \chi^{-1} = 4\mf a^2J$
and therefore $1/\chi \gg g J\mathfrak{a}^2$; hence
\begin{align}
  \label{eq:gap7}
   \mathcal{H}_{nn} \approx \frac{\chi h_y^2}{2}  n_y^2 + \left[\frac{\chi h_z^2}{2} - 6 g J \frac{\mu_0^2}{\mathfrak{a}^2}\right]
  n_z^2 +  \chi h_y h_z  n_y n_z.
\end{align}
The above shows that if $h_z$ is small or zero, stability
requires $g<0$.  This can be understood from the fact that, if the field
is along $y$, then $H_{\rm XXZ}$ is the only term, in the pure spin hamiltonian,
breaking explicitly the $O(2)$ symmetry in the $x-z$ plane.
It should therefore favor antiferromagnetic alignment along the $x$ axis,
which is the initial assumption of this derivation. It also proves the $\frac{\chi}{2}$
prefactor used in the main text.

The coefficients in Eq.~\eqref{eq:gap7} give contributions to $\Gamma_{yy}$, $\Gamma_{zz}$
and $\Gamma_{yz}$, respectively. In this Appendix, as opposed to the more general expressions
given in the main text, we assume they are the only contribution.

Since taking the magnetic field purely along one of the two axes $y,z$ guarantees that $\Gamma_{yz}=0$,
so that (as explained in the main text) the two magnon valleys are independent, 
let us assume that the field is along the $y$ axis.
Then one gap is $\Delta_1= |h_y|$, the Zeeman energy associated
with the field along $y$.  The other gap gets contributions both from
the anisotropy and the Zeeman energy associated
with the field along $z$.

Note that the anisotropy-induced gap involves the square
root of the anisotropy, i.e.\ $\Delta_0\big |_{h_z=0} = 4\sqrt{3|g|} J \mu_0$,
which is not necessarily {\em very} small for reasonably small values of $g$.

\newpage 
\begin{widetext}
  
\section{Application---Supplementary figures}
\label{sec:app-figures-1}

Here we present further calculations of scattering rates and (diagonal) thermal conductivity for the model of Sec.~\ref{sec:application}, as supplemental figures.  

\begin{figure}[htbp]
  \centering
  $|\mb k|={\mt {0.0625}}/\mf a$ 
  \qquad \qquad \qquad \qquad \qquad\qquad$|\mb k|={\mt {0.125}}/\mf a$
  \qquad \qquad \qquad \qquad \qquad\qquad$|\mb k|={\mt {0.25}}/\mf a$ \\
  [\smallskipamount]
  (1) \includegraphics[width=.26\columnwidth]{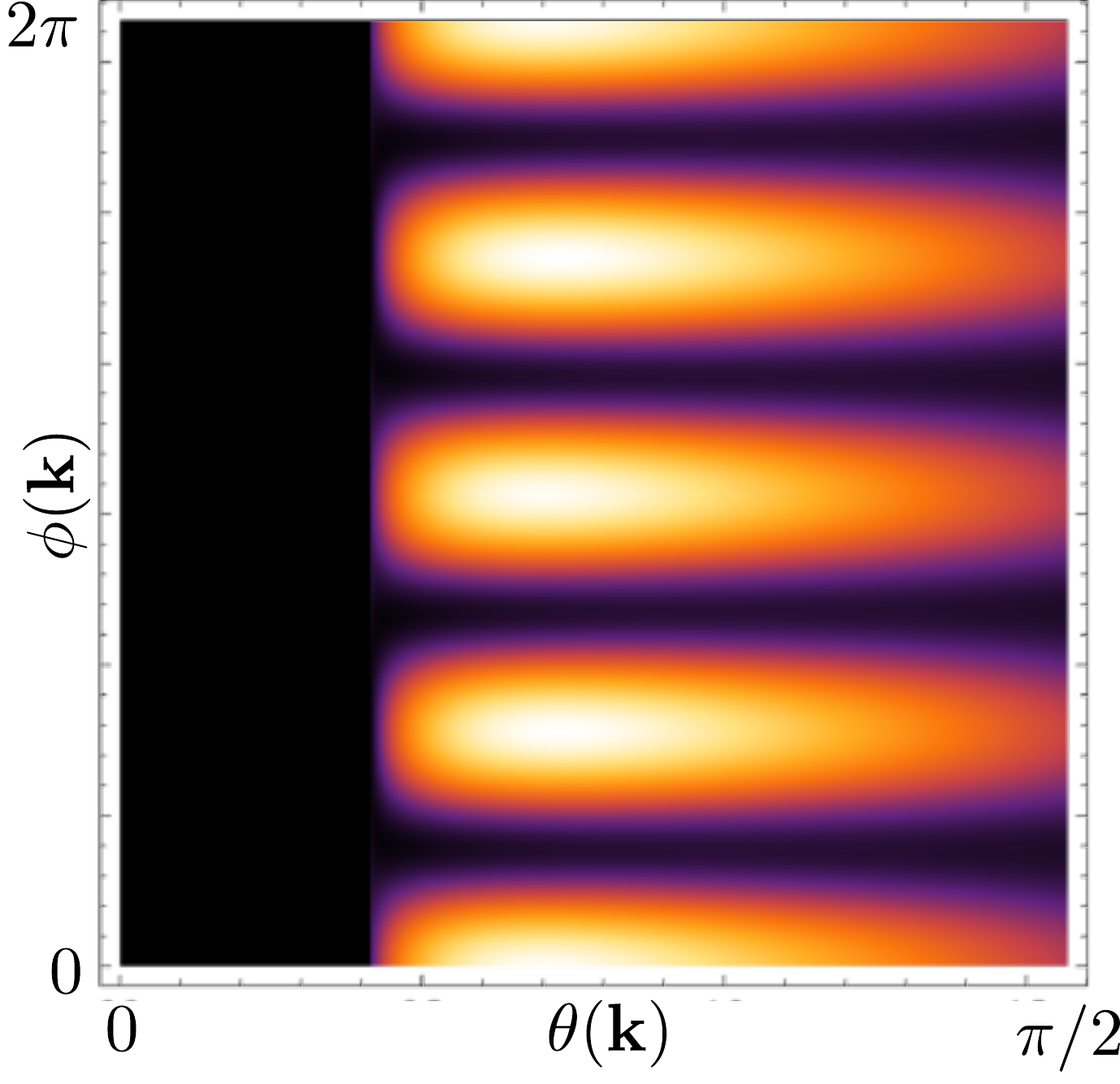}
   \includegraphics[width=.04\columnwidth]{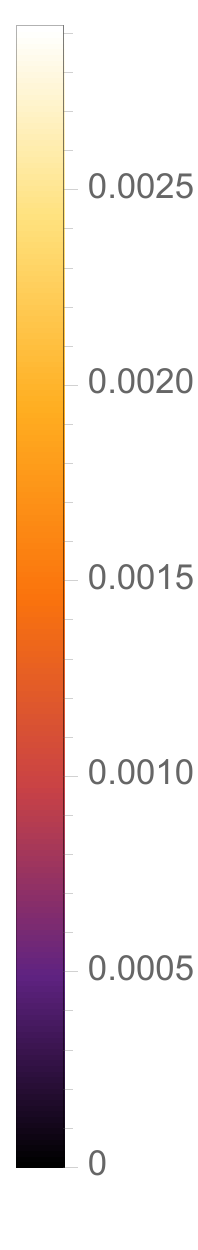}\hfill
   (2) \includegraphics[width=.26\columnwidth]{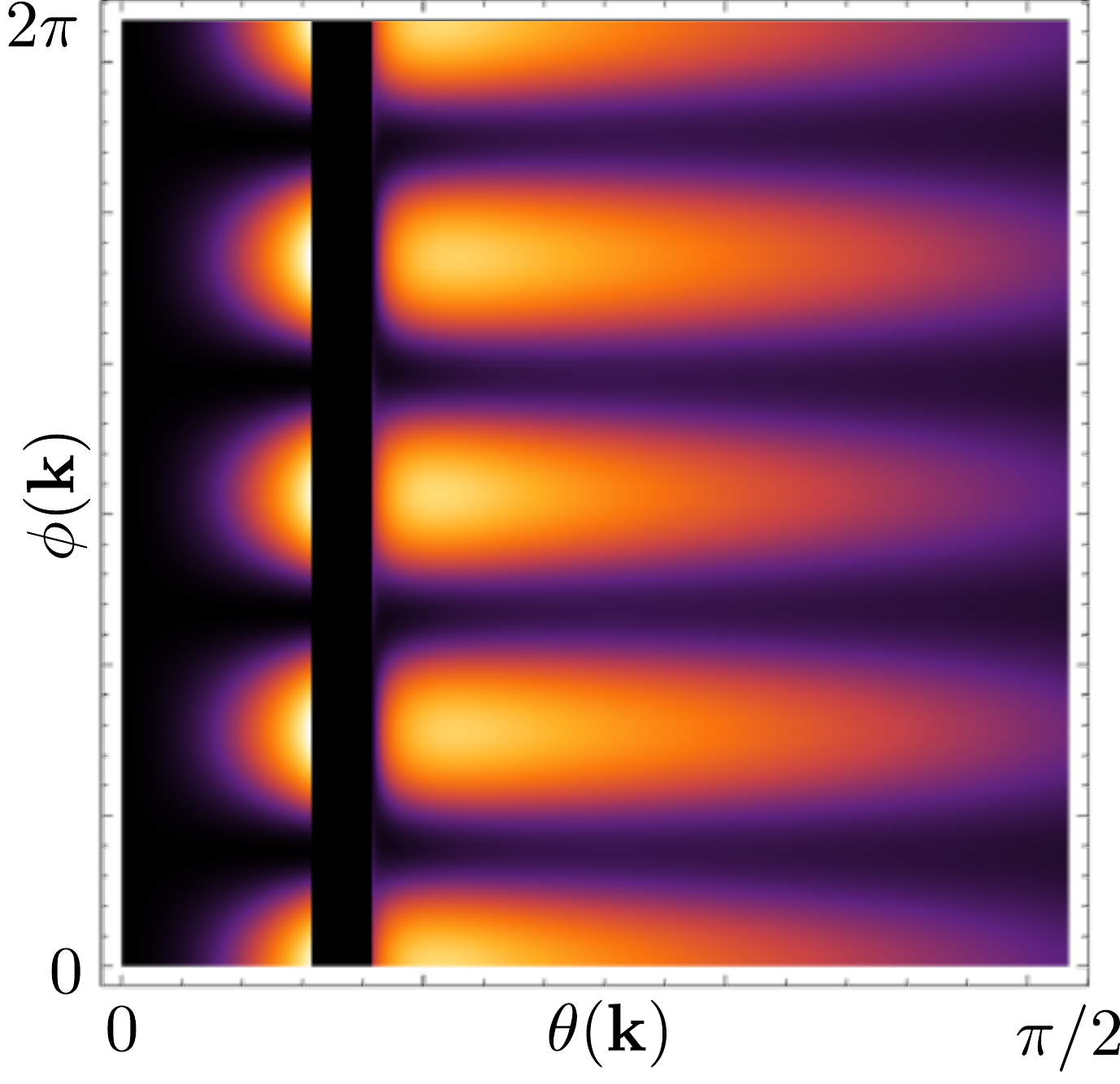}
   \includegraphics[width=.035\columnwidth]{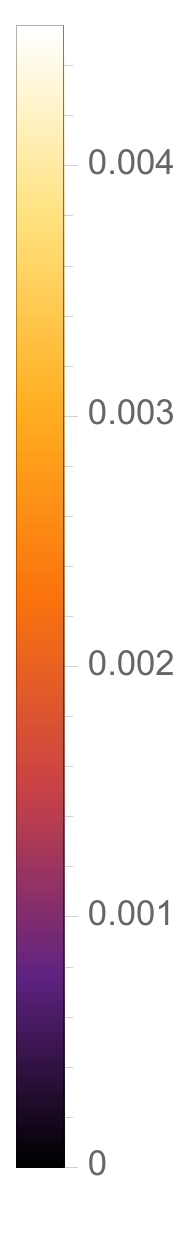}\hfill
   (3) \includegraphics[width=.26\columnwidth]{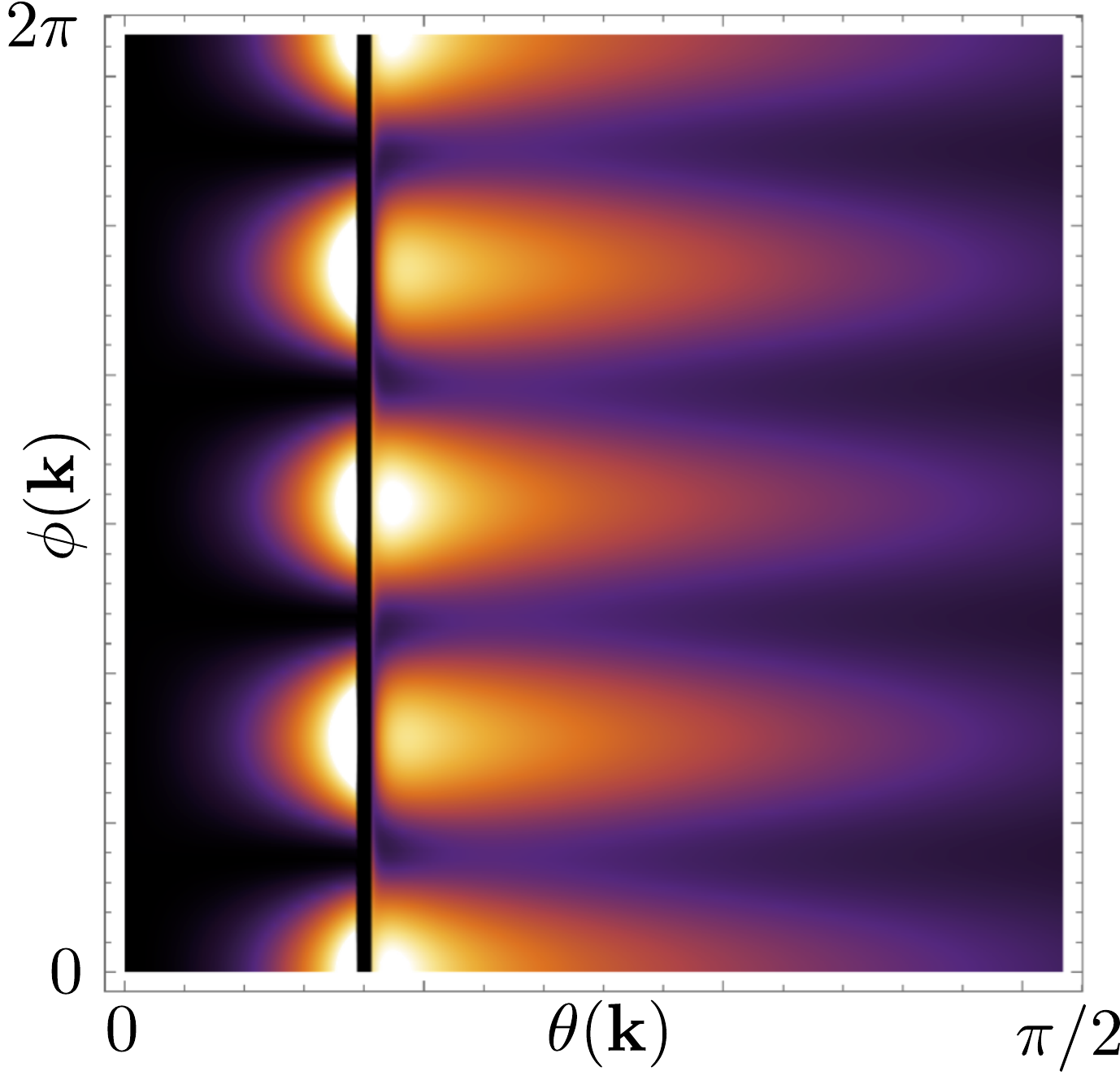}
   \includegraphics[width=.035\columnwidth]{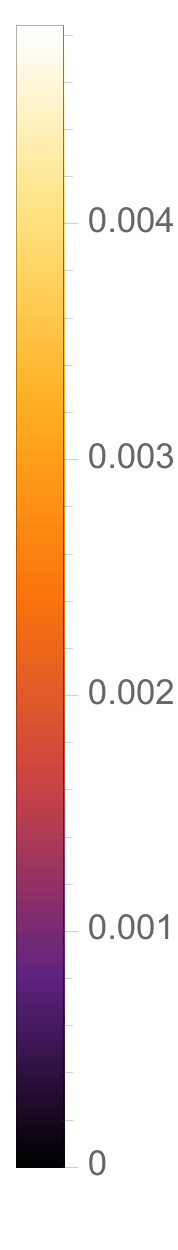}\hfill
   \\[\smallskipamount]
    $|\mb k|={\mt {0.5}}/\mf a$ \qquad \qquad \qquad \qquad
    \qquad\qquad \qquad$|\mb k|={\mt {1.0}}/\mf a$
    \qquad \qquad \qquad \qquad \qquad\qquad \qquad $|\mb k|={\mt {2.0}}/\mf a$\\
    [\smallskipamount]
 (4) \includegraphics[width=.26\columnwidth]{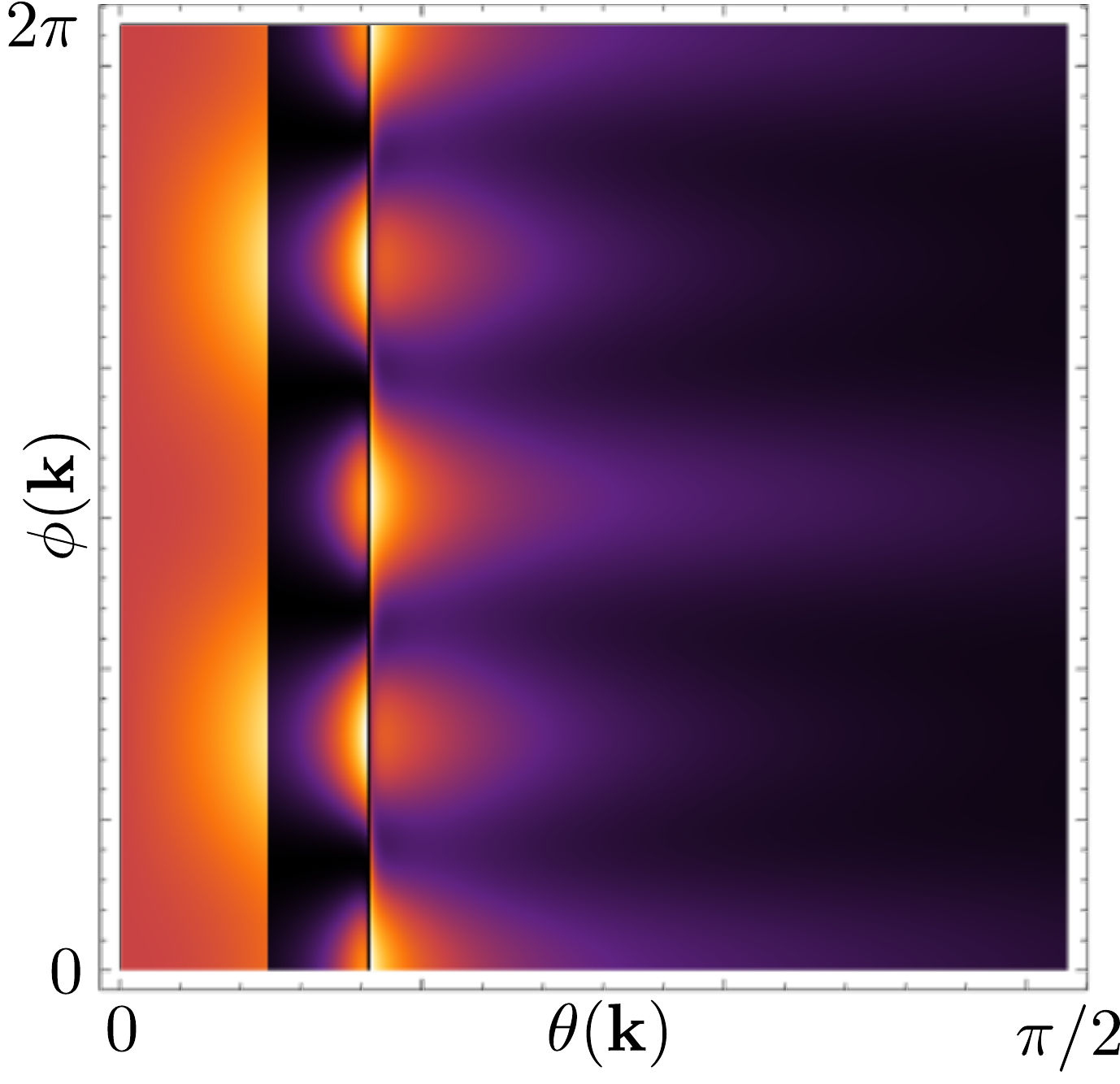}
   \includegraphics[width=.035\columnwidth]{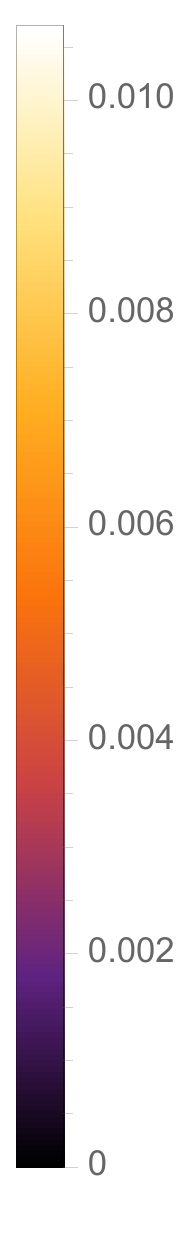}\hfill
   (5) \includegraphics[width=.26\columnwidth]{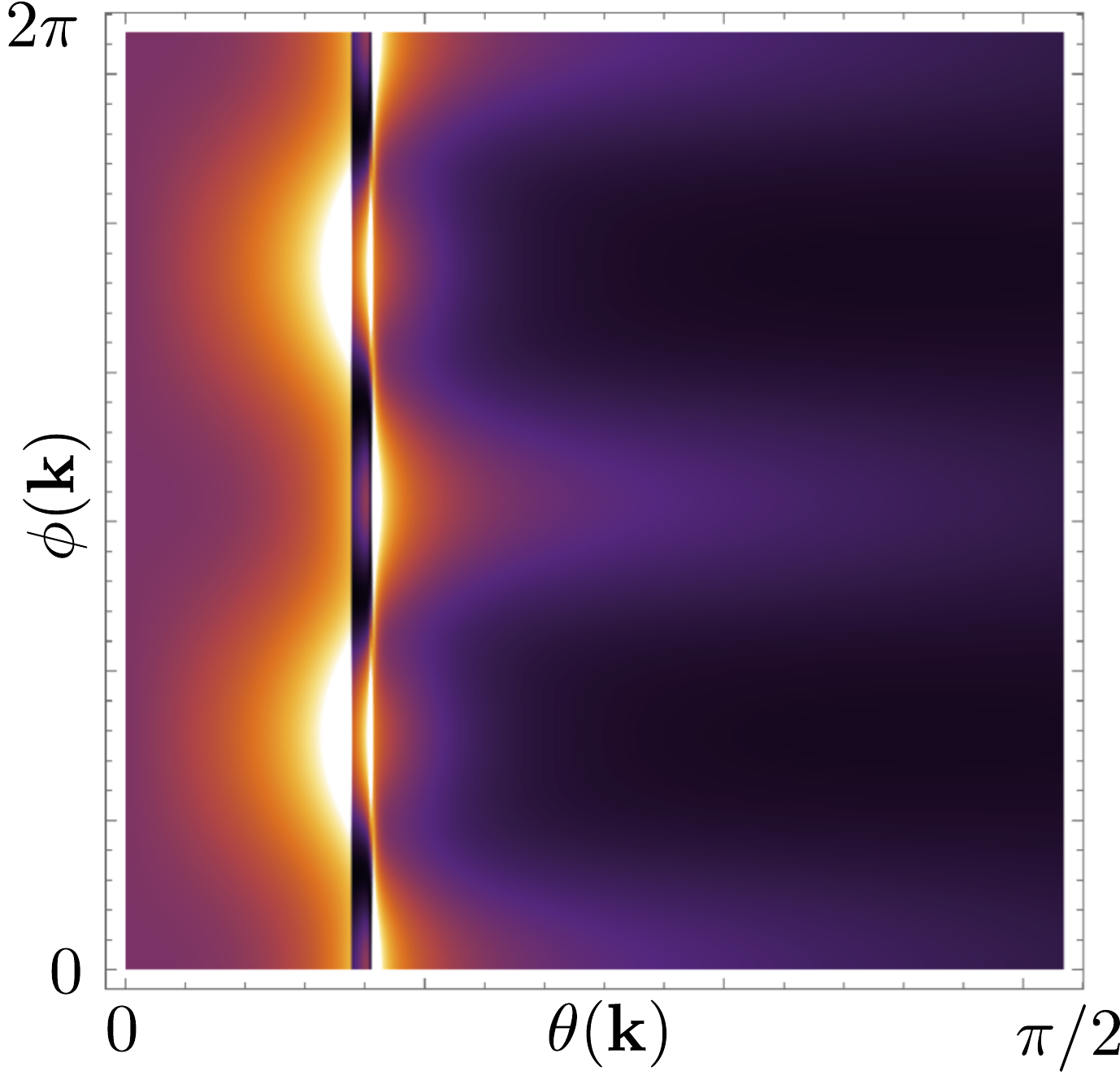}
   \includegraphics[width=.04\columnwidth]{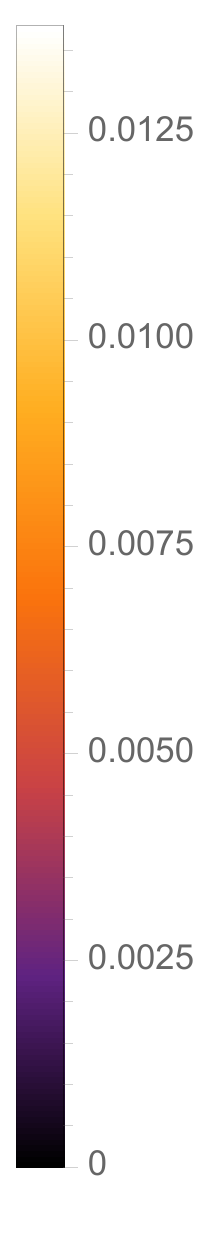}\hfill
   (6) \includegraphics[width=.26\columnwidth]{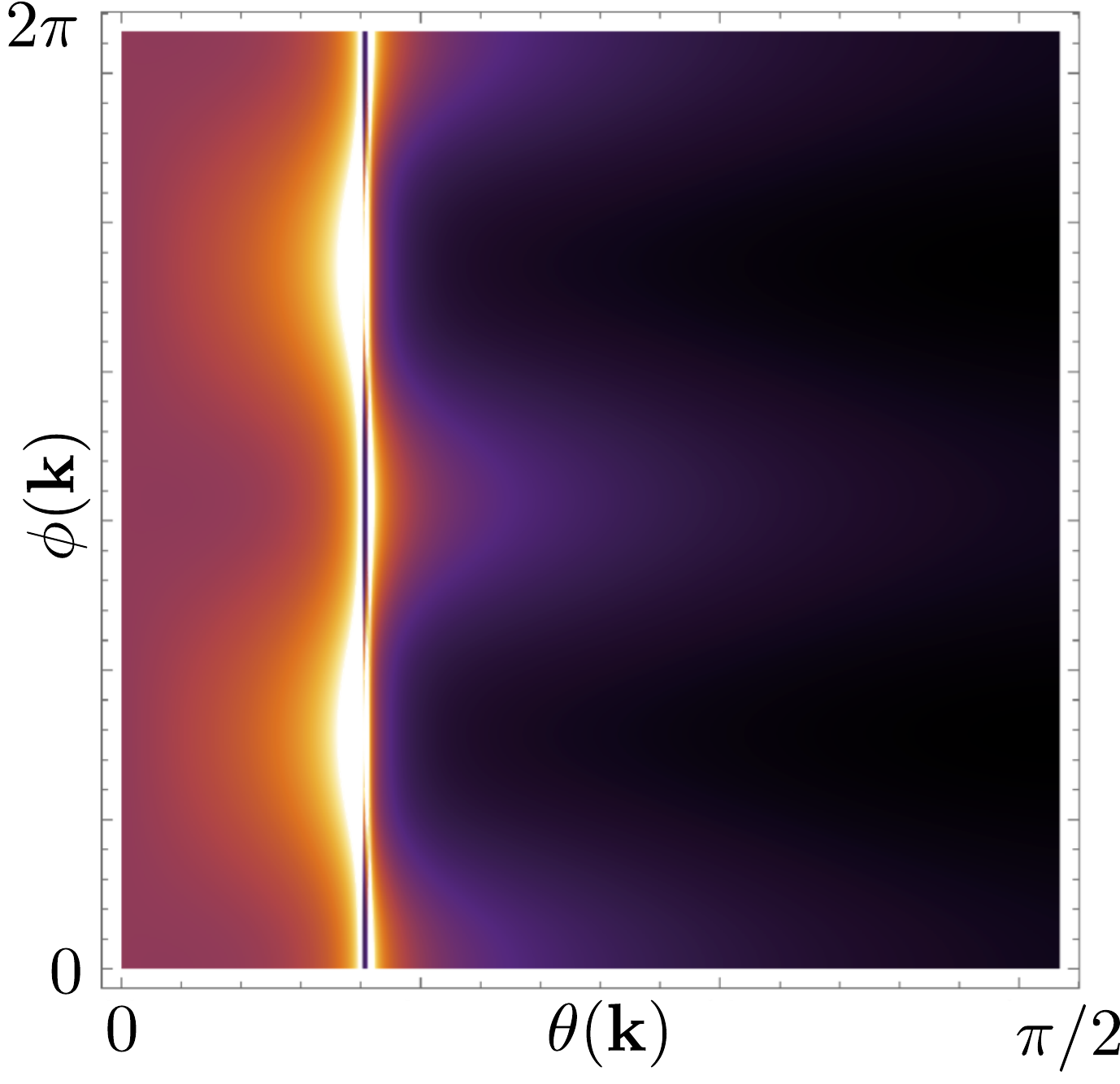}
   \includegraphics[width=.035\columnwidth]{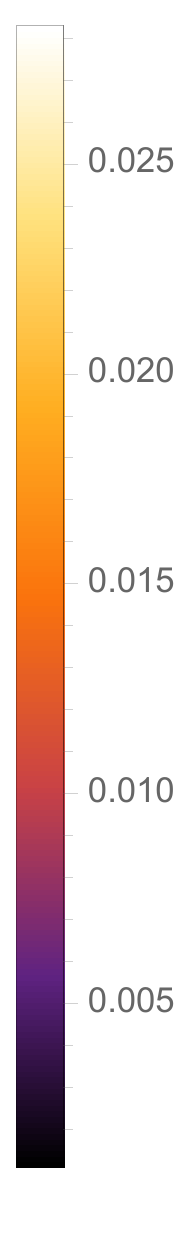}\hfill
   \caption{Diagonal scattering rate $D_{n\mb k}$ with respect to $\theta(\mb k)\in [0,\pi/2]$
     (horizontal axis)
    and $\phi(\mb k) \in [0,2\pi]$ (vertical axis) for fixed temperature $T=0.5\,T_0$, polarization
    $n=0$, and momentum (1) $|\mb k|={\mt {0.0625}}/\mf a$, (2) $|\mb k|={\mt {0.125}}/\mf a$,
    (3) $|\mb k|={\mt {0.25}}/\mf a$, (4) $|\mb k|={\mt {0.5}}/\mf a$, (5) $|\mb k|={\mt {1.0}}/\mf a$,
    (6) $|\mb k|={\mt {2.0}}/\mf a$. Color scales vary from figure to figure. Note that the $C_4$ symmetry is
    approximately preserved for small $|\mb k|$ but broken at large $|\mb k|$, as stated in the main text.
    Also note how scattering processes at $\theta(\mb k)<\theta_-$
    become allowed for $\omega_{n\mb k}\geq 2\Delta,2\Delta'$,
    then dominant at large $|\mb k|$. }
  \label{fig:suppfig1}
\end{figure}

\begin{figure}[htbp]
  \centering
  $n=0$ \qquad \qquad \qquad \qquad \qquad \qquad\qquad \qquad $n=1$ \qquad \qquad \qquad \qquad \qquad\qquad \qquad \qquad  $n=2$\\[\smallskipamount]
  (1) \includegraphics[width=.26\columnwidth]{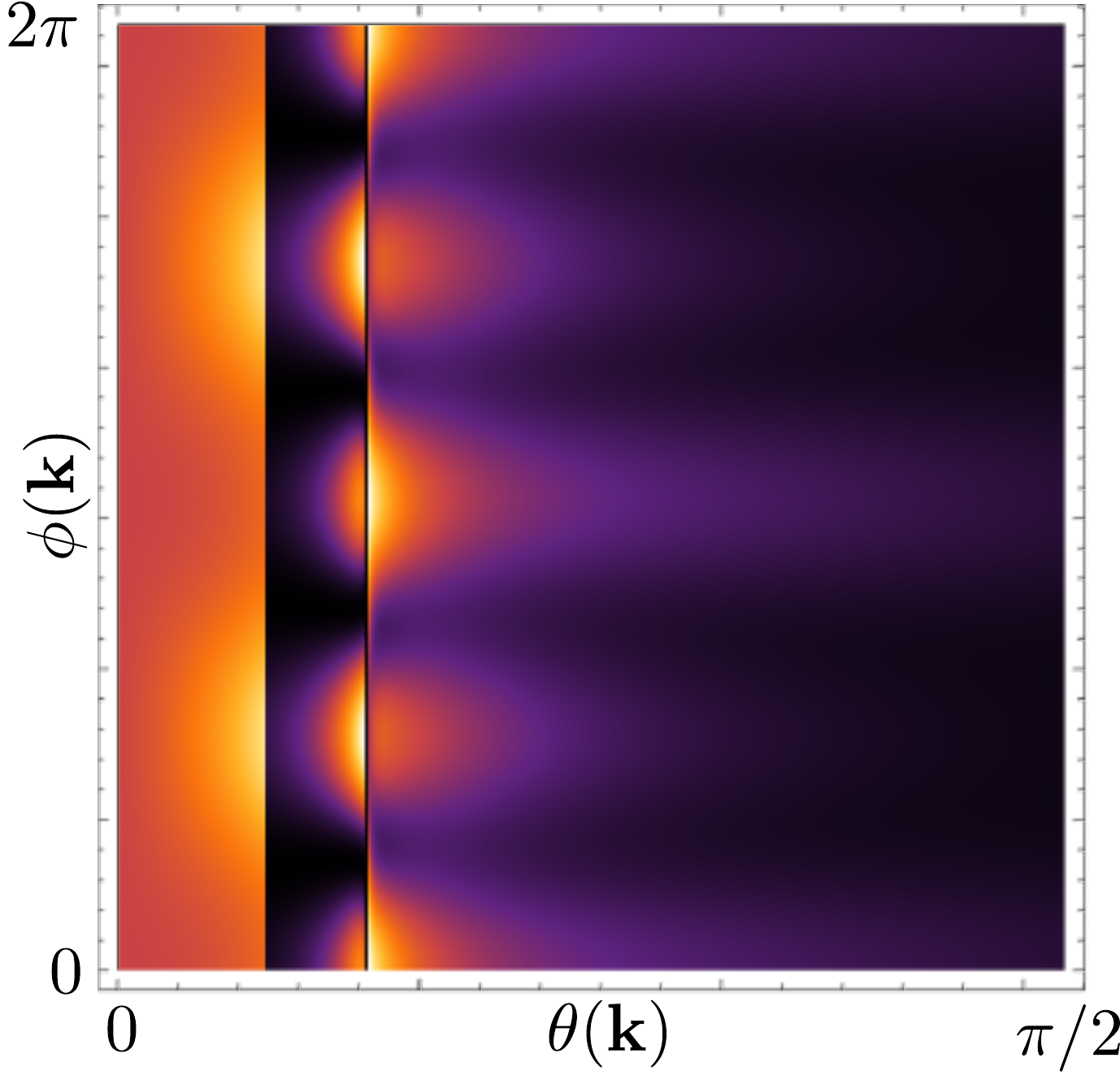}
   \includegraphics[width=.035\columnwidth]{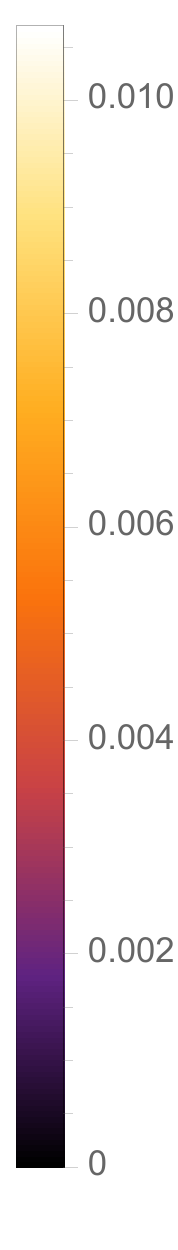}\hfill
   (2) \includegraphics[width=.26\columnwidth]{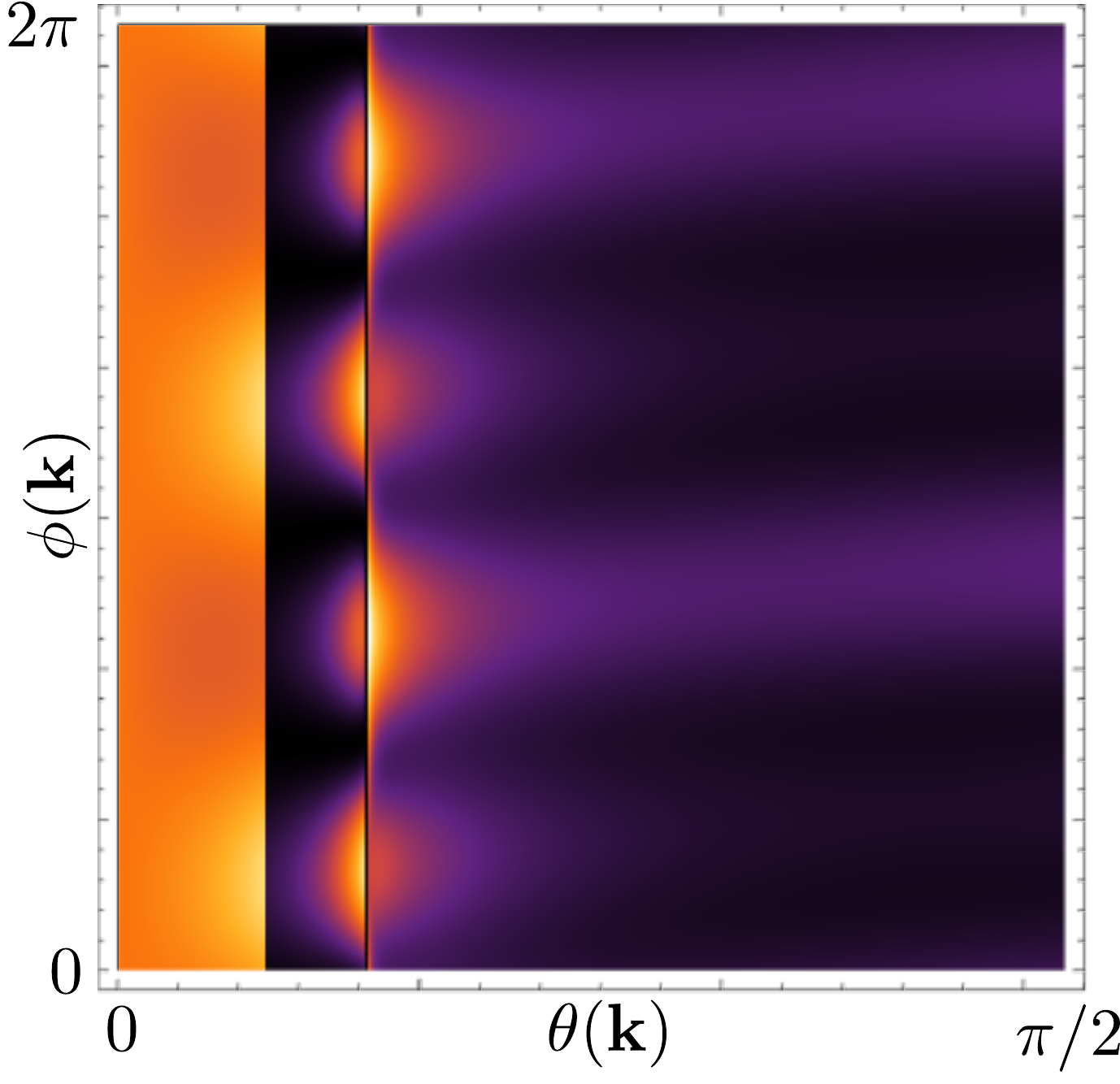}
   \includegraphics[width=.035\columnwidth]{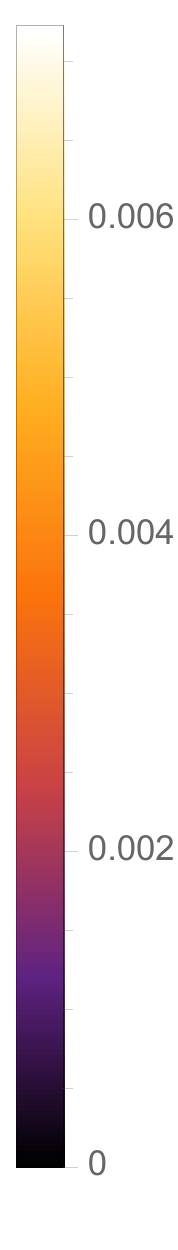}\hfill
   (3) \includegraphics[width=.26\columnwidth]{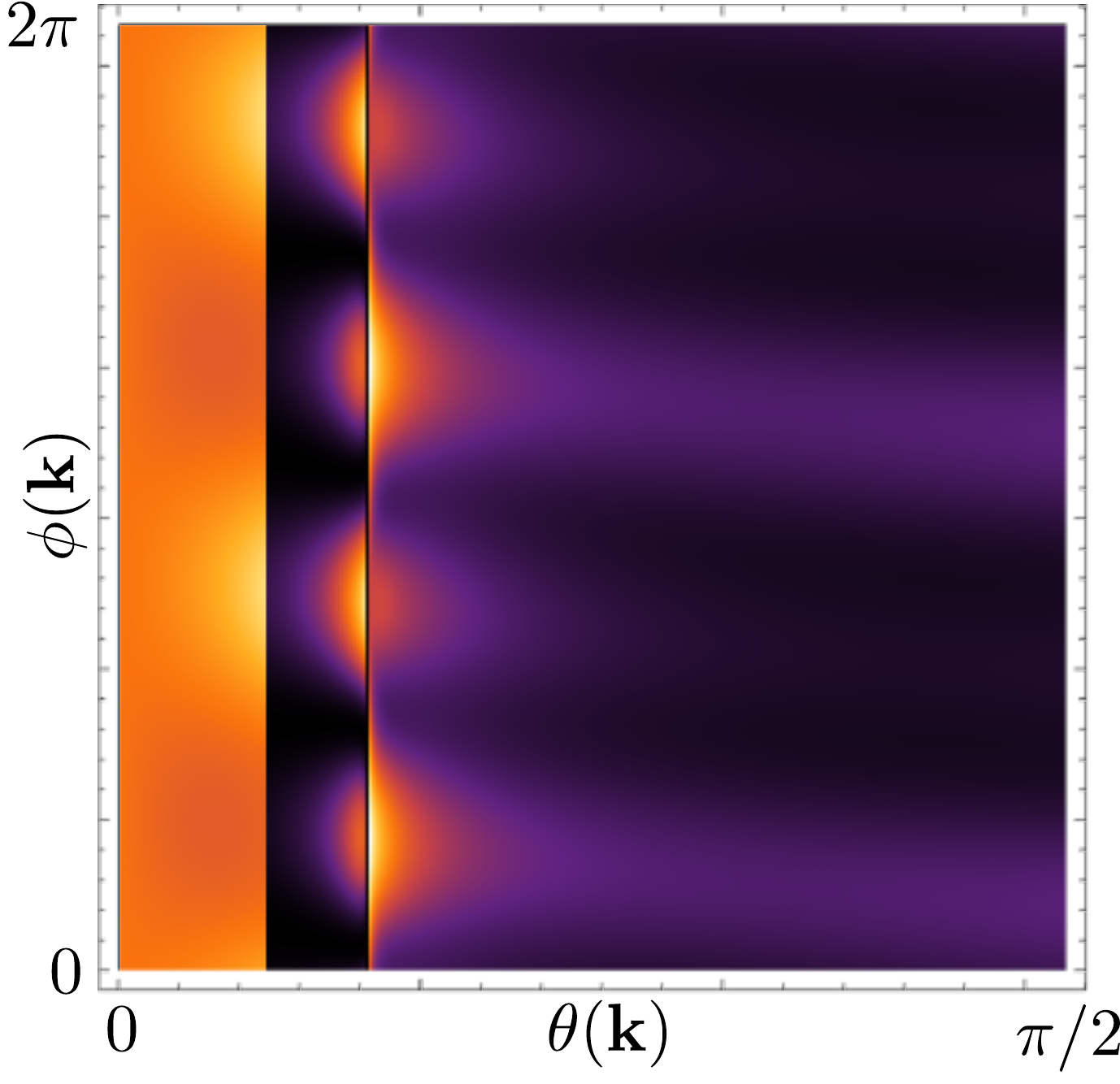}
   \includegraphics[width=.035\columnwidth]{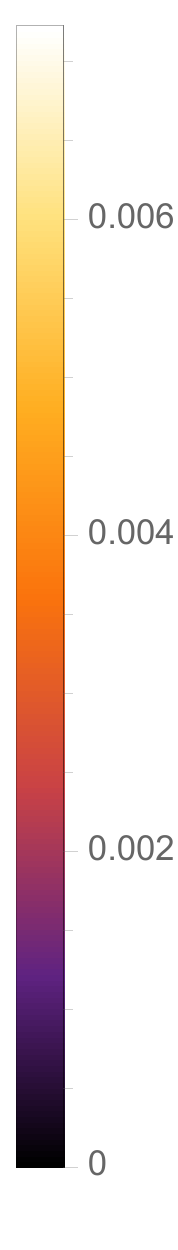}\hfill
  \caption{Diagonal scattering rate $D_{n\mb k}$ with respect to $\theta(\mb k)\in [0,\pi/2]$ (horizontal axis)
    and $\phi(\mb k) \in [0,2\pi]$ (vertical axis) for fixed temperature $T=0.5T_0$, momentum
    $|\mb k|={\mt {0.5}}/\mf a$, and polarizations (1) $n=0$, (2) $n=1$, (3) $n=2$.
    Color scales are different in (1) and (2,3). Subfigure (1) is reproduced from the main text.
    Note that with our choice of polarization vectors $\bs \varepsilon_{n,\mb k}$, results for $n=1$ and $n=2$
    are simply related by the mirror symmetry $\phi \mapsto \pi -
    \phi$. }
  \label{fig:suppfig2}
\end{figure}

\begin{figure}[htbp]
  \centering
  $\phi(\mb k)=0$ ; $n=1$ \qquad \qquad \qquad \qquad \qquad $\phi(\mb k)=\pi/2$ ; $n=0$ \qquad \qquad \qquad \qquad \qquad $\phi(\mb k)=\pi/2$ ; $n=1$\\[\smallskipamount]
  (1) \includegraphics[width=.26\columnwidth]{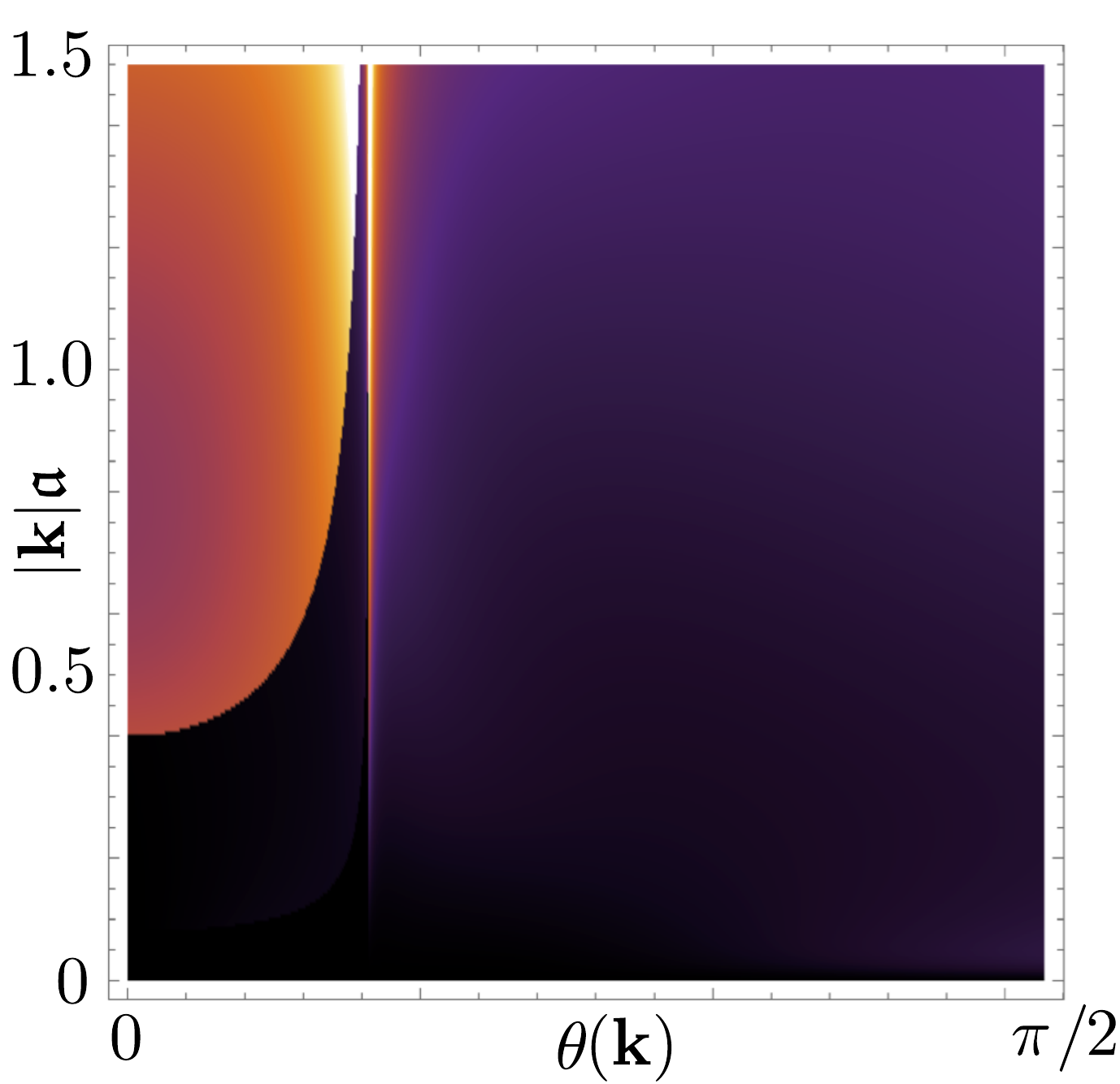}
   \includegraphics[width=.035\columnwidth]{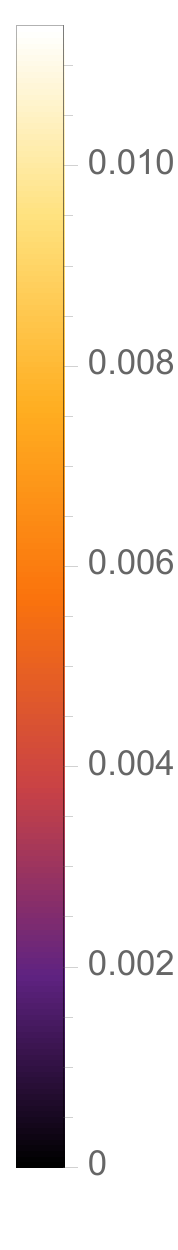}\hfill
   (2) \includegraphics[width=.26\columnwidth]{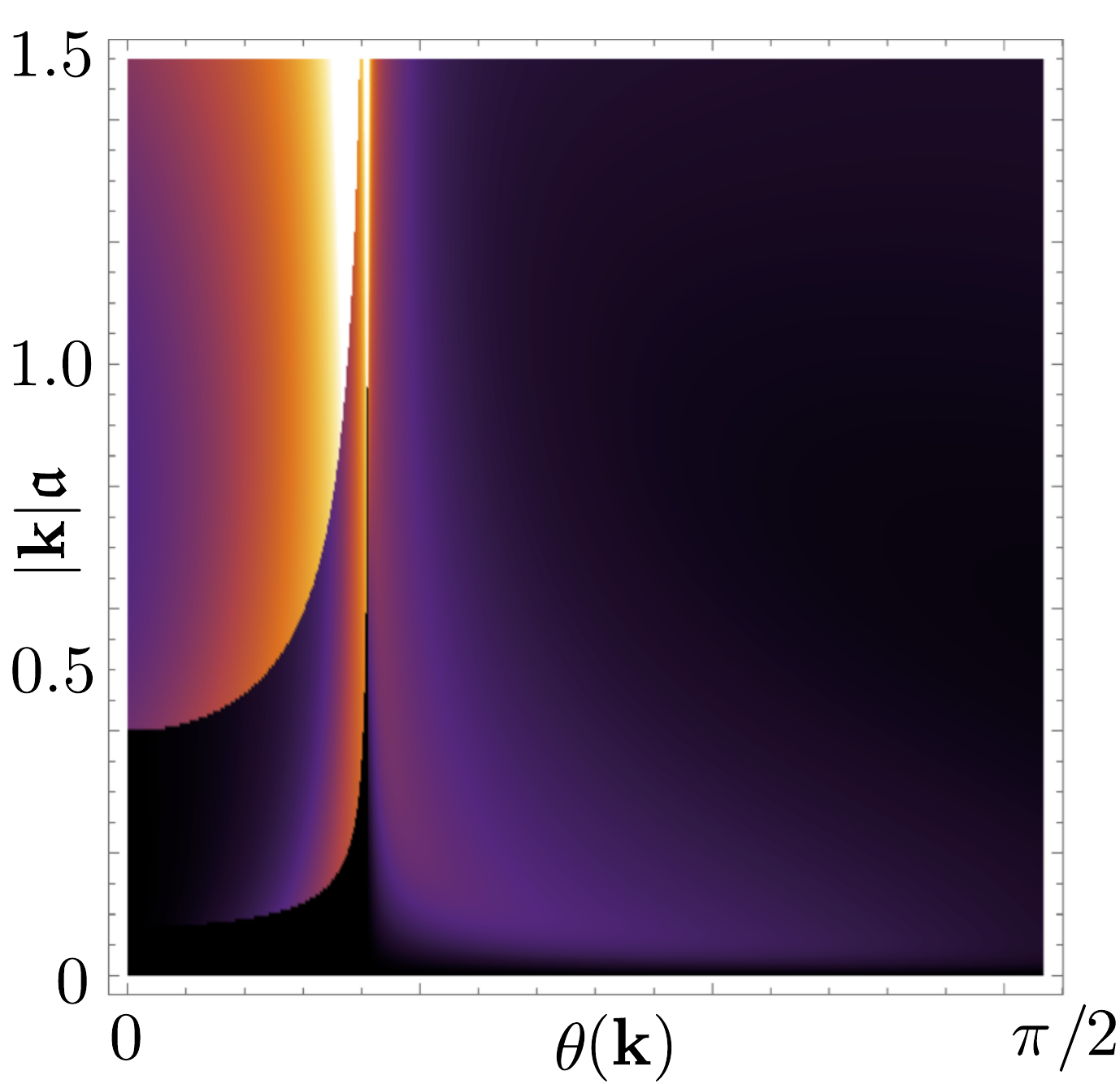}
   \includegraphics[width=.035\columnwidth]{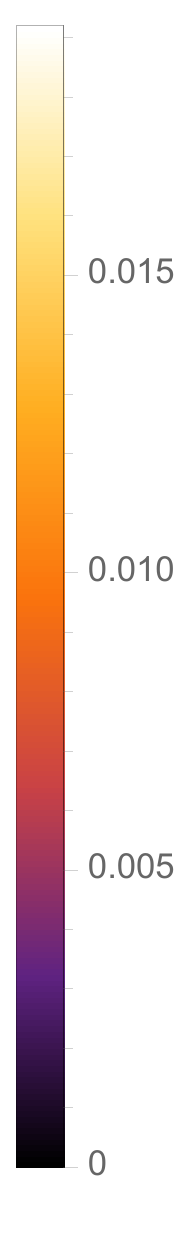}\hfill
   (3) \includegraphics[width=.26\columnwidth]{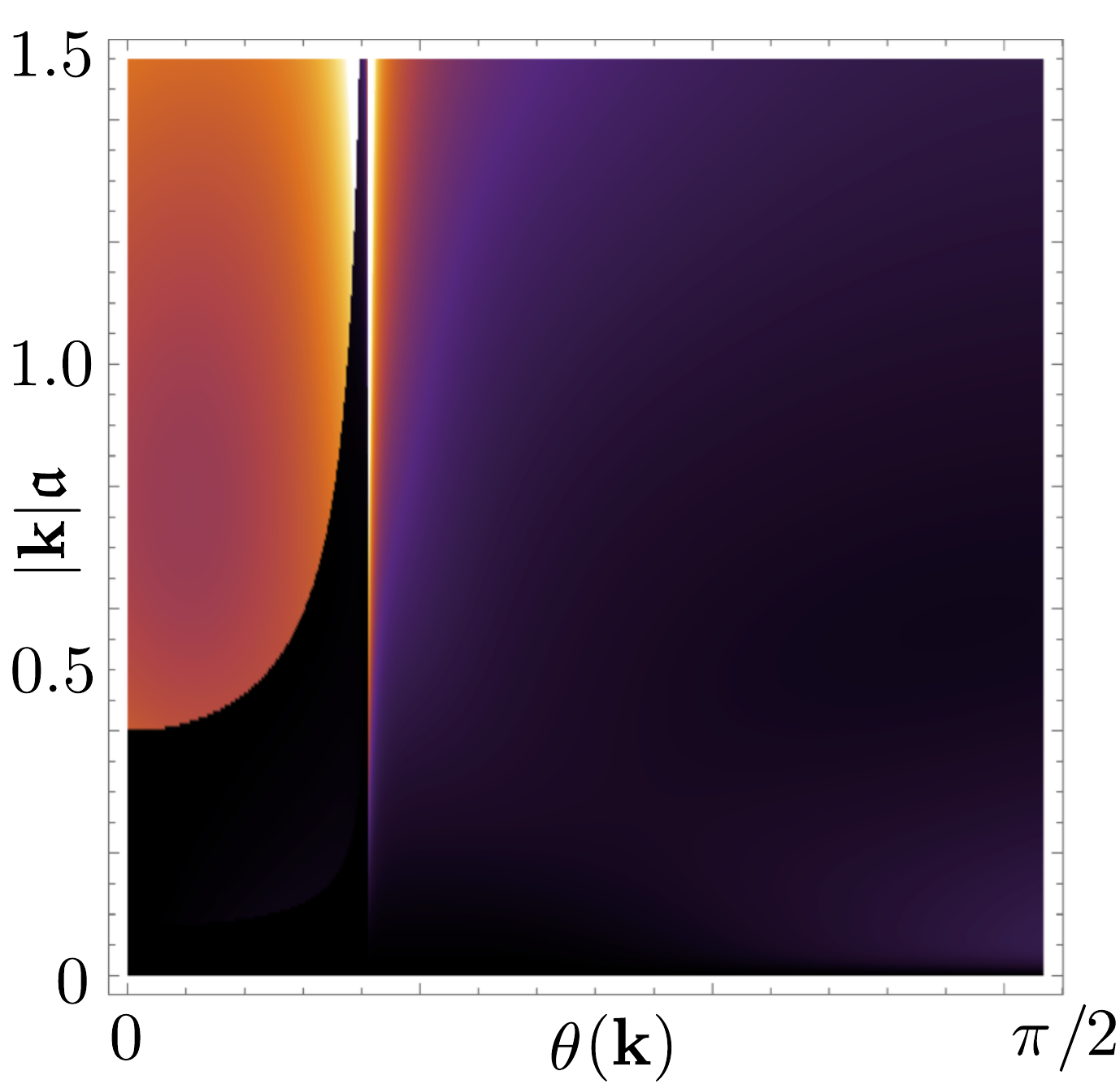}
   \includegraphics[width=.035\columnwidth]{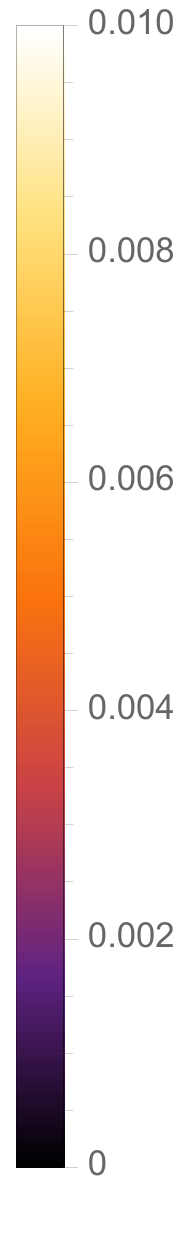}\hfill
  \caption{Diagonal scattering rate $D_{n\mb k}$ with respect to $\theta(\mb k)\in [0,\pi/2]$ (horizontal axis)
    and $|\mb k|\mf a$ (vertical axis) for fixed temperature $T= \mt{0.5} T_0$ and (1) $\phi(\mb k)=0$ and $n=1$,
    (2) $\phi(\mb k)=\pi/2$ and $n=0$, (3) $\phi(\mb k)=\pi/2$ and $n=1$.
    Color scales are different for the three subfigures. The $\phi(\mb k)=0$ and $n=0$ case
    is displayed in the main text. Note that polarizations $n=1$ and $n=2$ yield the same results here.
    Note also that the general features are the same for polarizations $n=1,2$ as for $n=0$: although the scattering rates
    of $n=1,2$ polarizations
    for energies $\omega_{n\mb k}\gtrsim 2\Delta$ are not as clearly visible as they are for $n=0$, they are finite (of order $10^{-4}$
    in our units) and are only {\em parametrically} smaller than those for the $n=0$ polarization, due to purely geometrical factors
  ($\mc S_{n\mb k}^{q;\alpha\beta}$ in the main text). }
  \label{fig:suppfig3}
\end{figure}

\begin{figure}[htbp]
  \centering
  \includegraphics[width=.48\columnwidth]{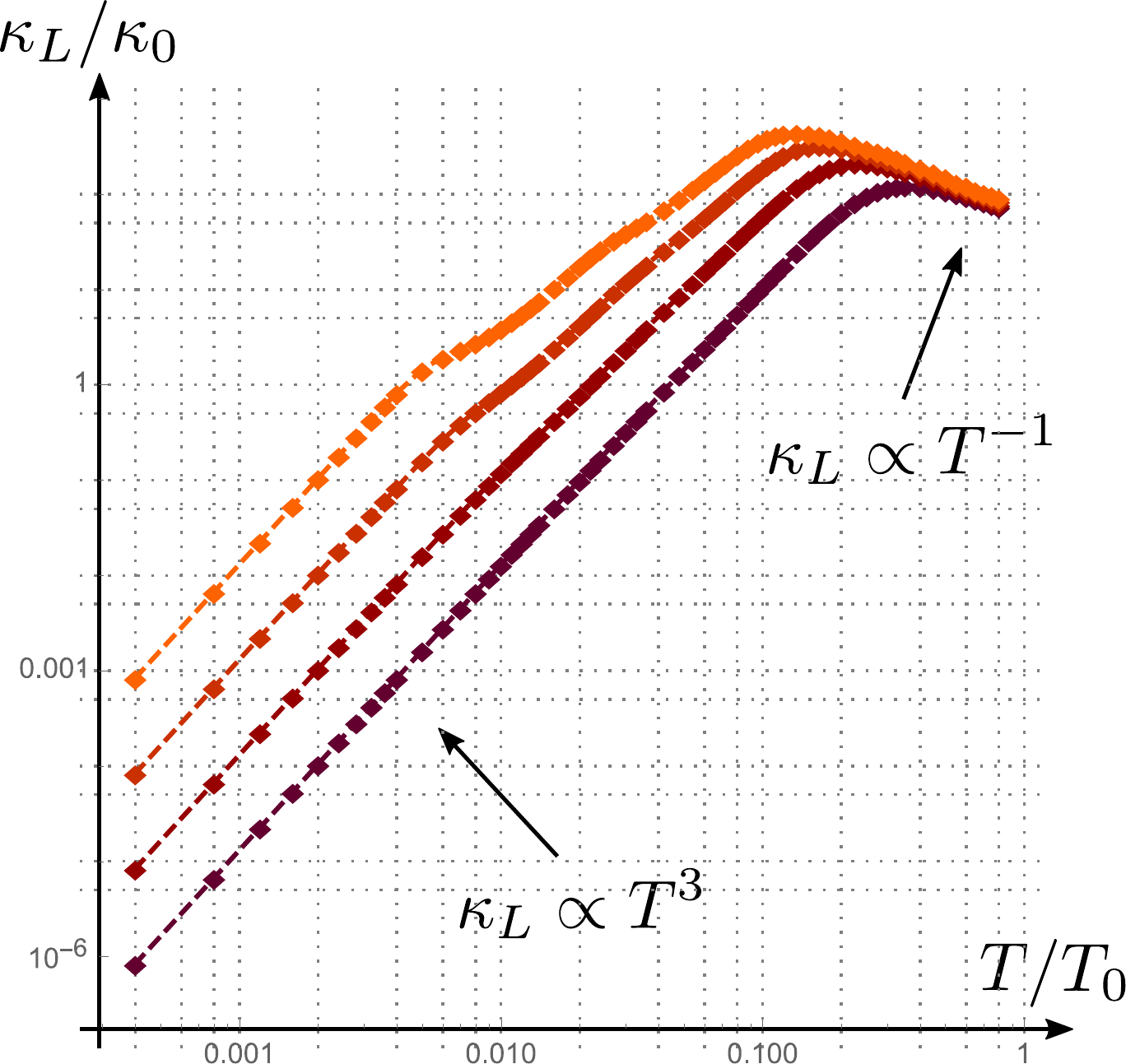}\hfill
   \includegraphics[width=.48\columnwidth]{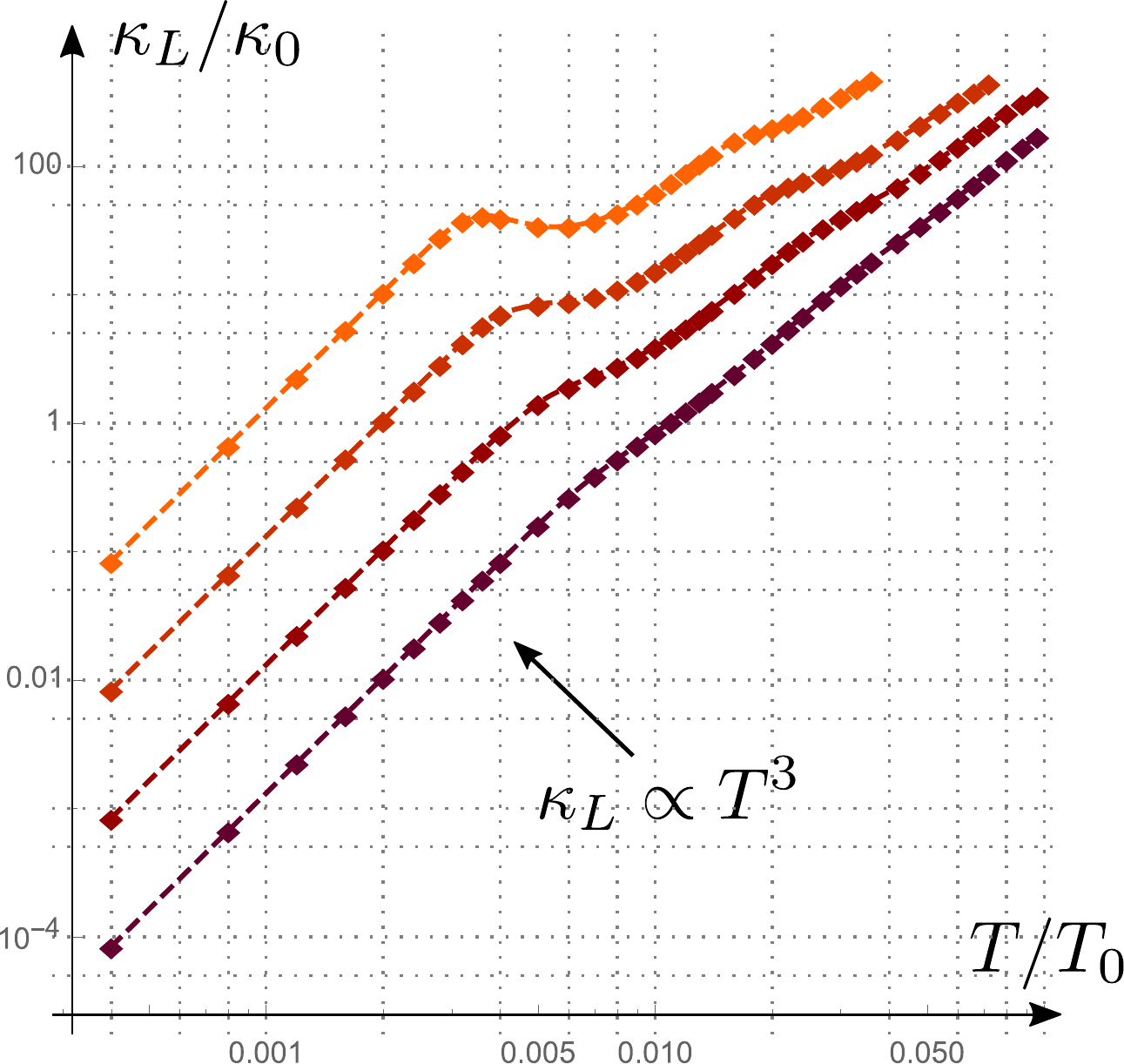}
  \caption{ Longitudinal thermal conductivity $\kappa_L$ with respect to temperature $T$, in log-log scale,
    {\em (left)} for four different values of
    $\gamma_{\rm ext}= 1\cdot 10^{-z}(v_{\rm ph}/\mathfrak{a}), z \in \llbracket 4, 7\rrbracket$,
    from darker $(z=4)$ to lighter $(z=7)$ shade, {\em (right)} for four different values 
    $\gamma_{\rm ext}= 1\cdot 10^{-z}(v_{\rm ph}/\mathfrak{a}), z \in \llbracket 6, 9\rrbracket$,
    from darker $(z=6)$ to lighter $(z=9)$ shade. Note that the two ``bumps'' come from the competition
  between $\gamma_{\rm ext}$ and $D_{nn,\ell}$ for valley index $\ell = 0,1$, as explained in the main text.}
  \label{fig:suppfig4}
\end{figure}

\begin{figure}[htbp]
  \centering
   $(q,q')=(-,-)$ ; $\phi(\mb k) = 0$  \qquad \qquad \qquad
   $(q,q')=(-,+)$ ; $\phi(\mb k) = 0$  \qquad \qquad \qquad
   $(q,q')=(-,+)$ ; $\phi(\mb k) = \pi/2$\\
   [\smallskipamount]
  (1) \includegraphics[width=.25\columnwidth]{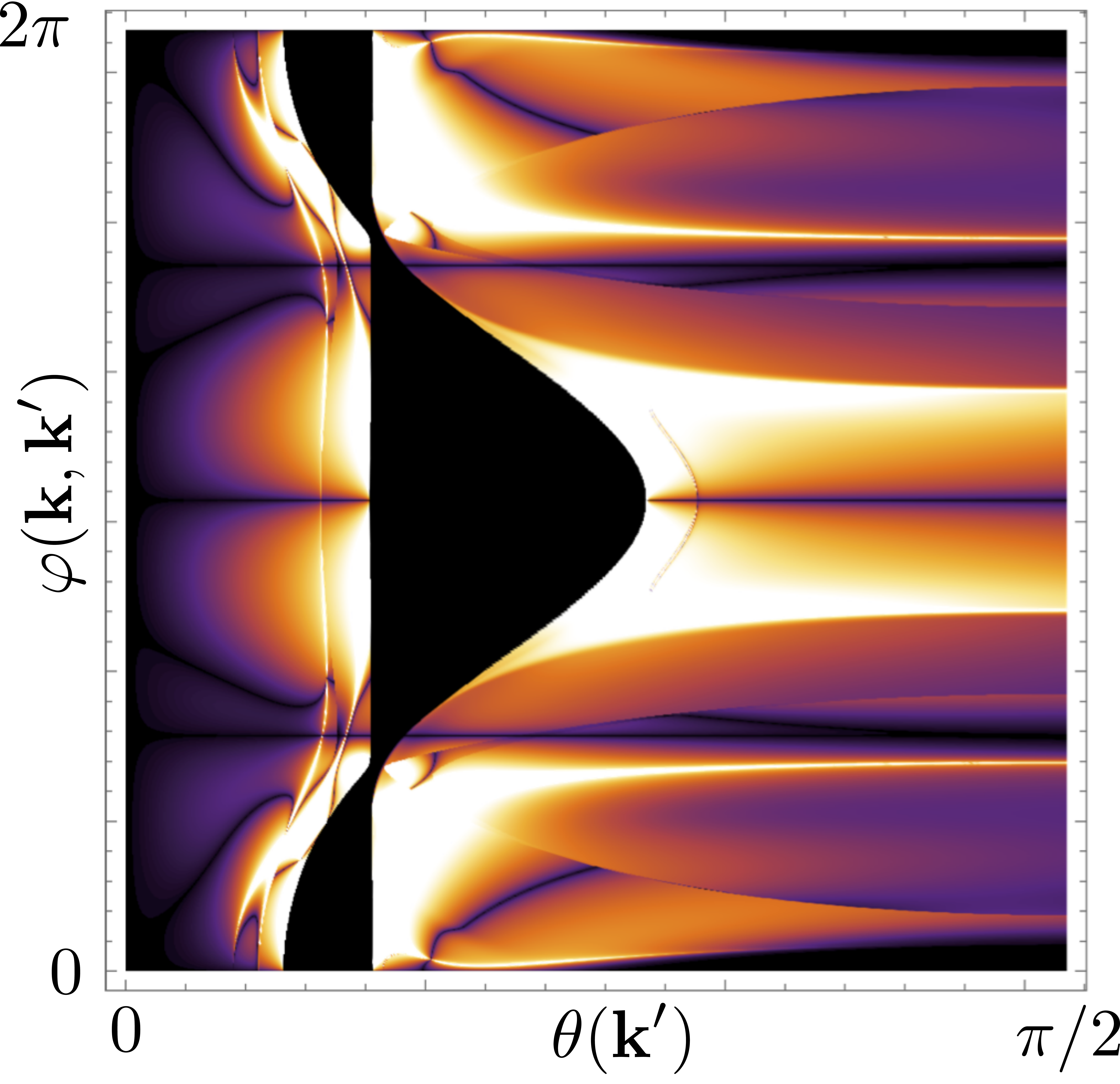}
   \includegraphics[width=.047\columnwidth]{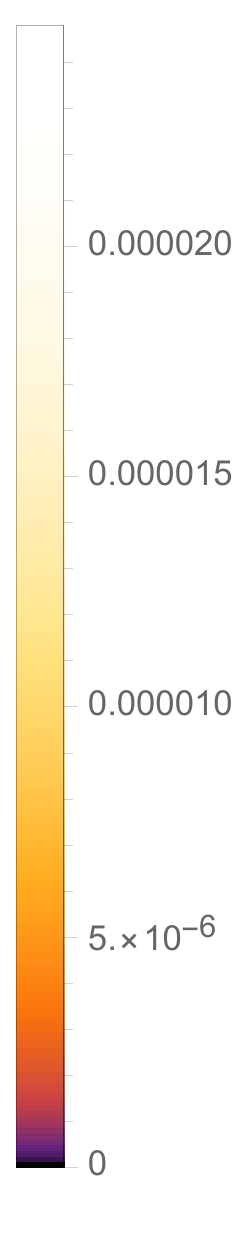}\hfill
   (2) \includegraphics[width=.25\columnwidth]{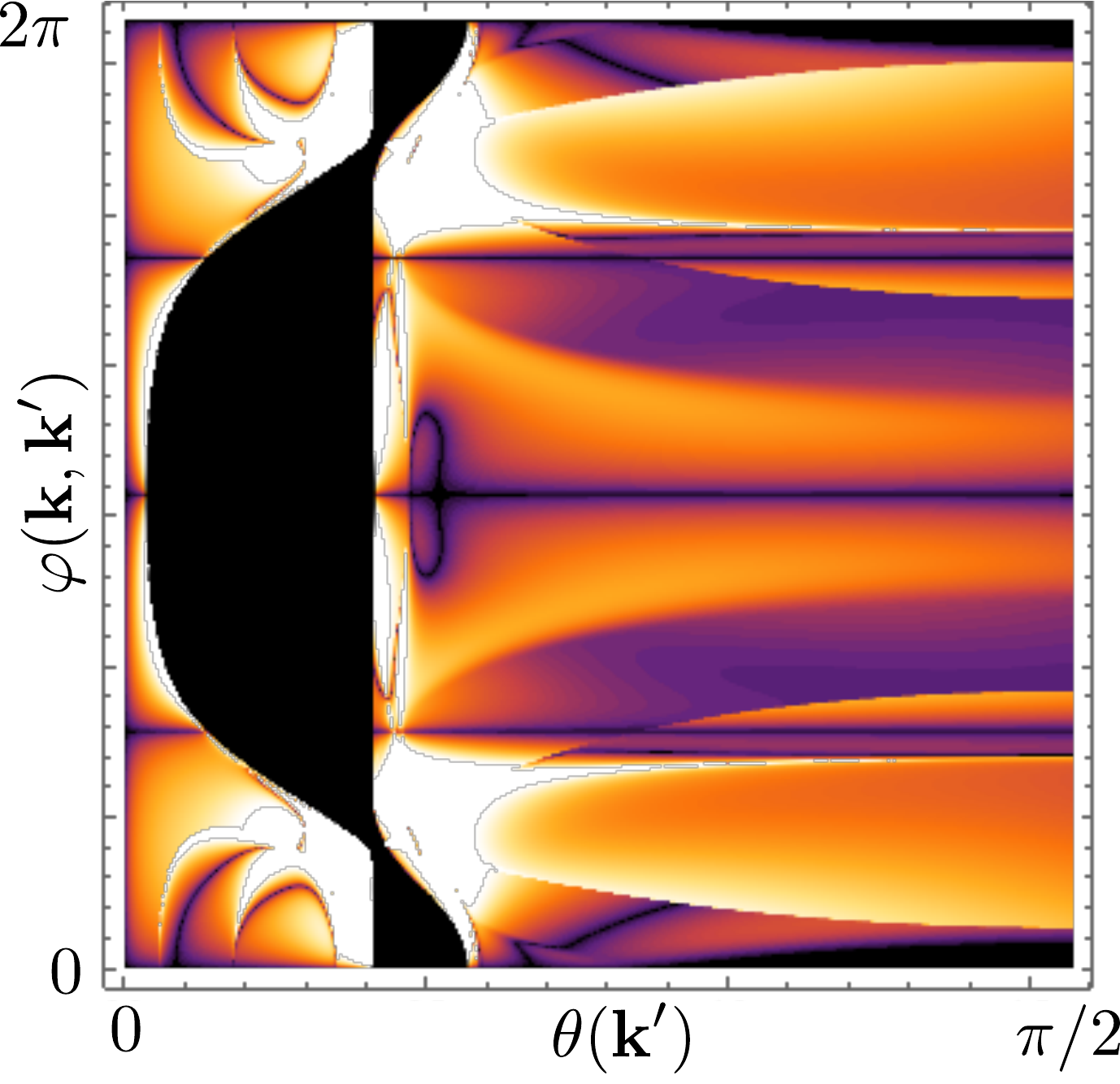}
   \includegraphics[width=.049\columnwidth]{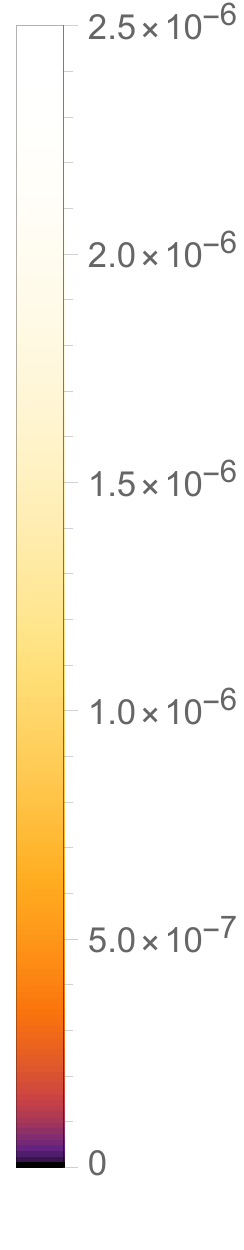}\hfill
   (3) \includegraphics[width=.25\columnwidth]{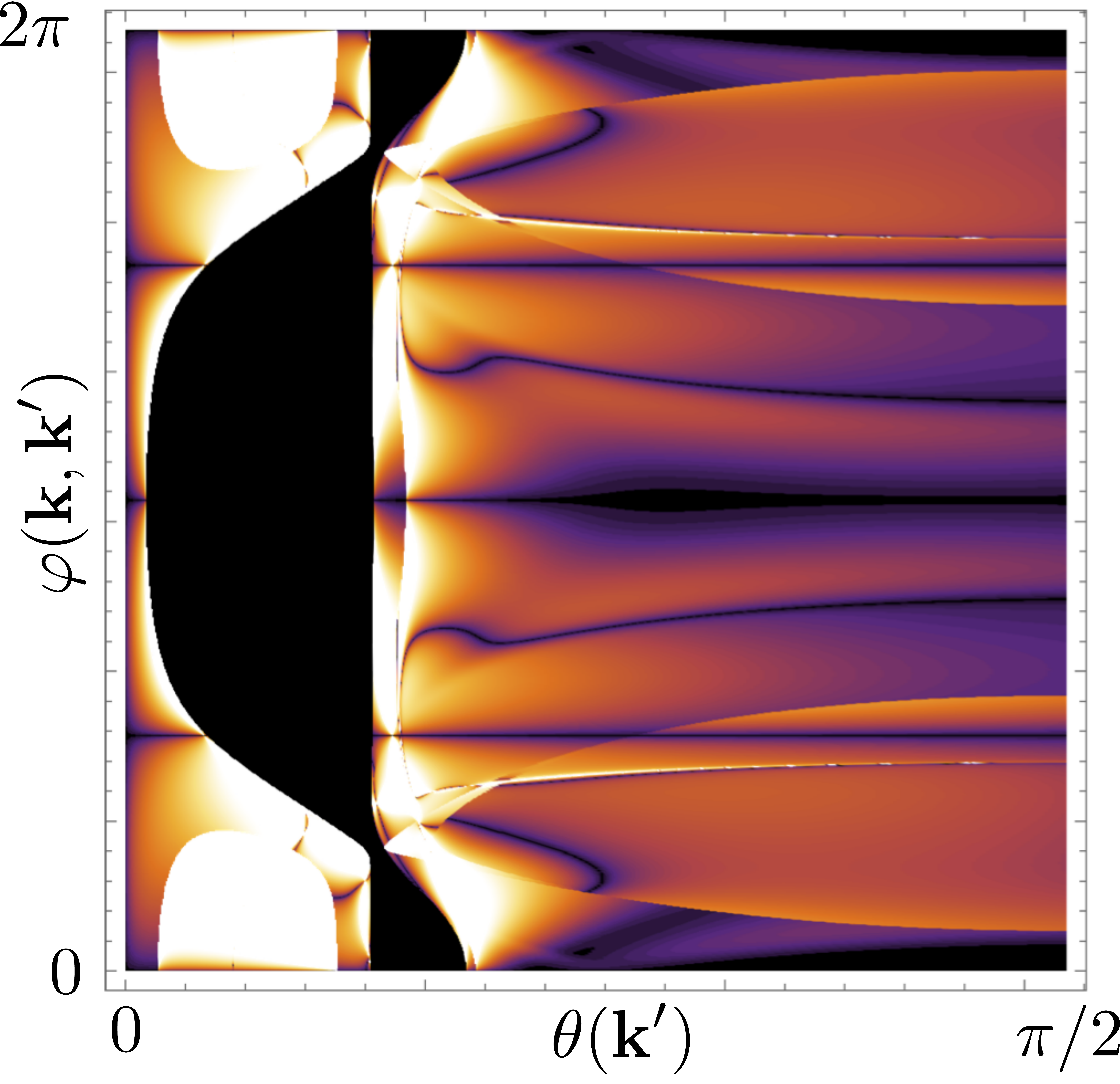}
   \includegraphics[width=.045\columnwidth]{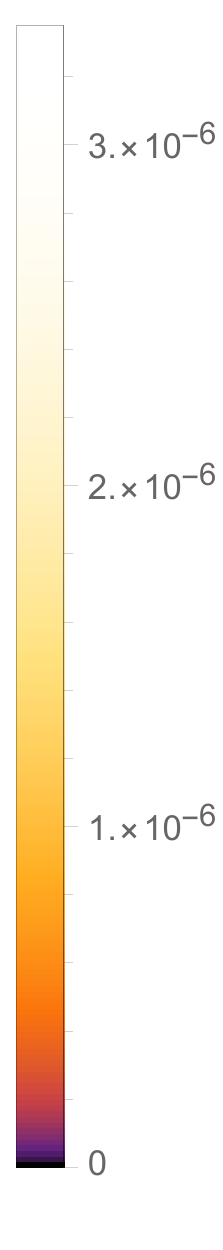}\hfill
  \caption{Skew-scattering rates (1) $\mf W^{\ominus,--}_{n\mb k n'\mb k'}$ and (2,3) $\mf W^{\ominus,-+}_{n\mb k n'\mb k'}$, 
  with respect to $\theta(\mb k')\in [0,\pi/2]$ (horizontal axis) and 
  $\varphi(\mb k,\mb k')=\phi(\mb k')-\phi(\mb k)$ (vertical axis), for fixed magnetization $\bs m_0=\mt{0.05} \hat{\bs z}$, temperature $T=\mt{0.5} T_0$, 
  momentum $|\mb k'|=\mt{0.8}/\mf a$, $k_z = \mt{0.1}/\mf a$, and (1,2) $k_x=\mt{0.2}/\mf a$,
  $k_y = 0$, (3) $k_x = 0$, $k_y=\mt{0.2}/\mf a$. The case $\mf
  W^{\ominus,--}_{n\mb k n'\mb k'}, k_x=0, k_y = \mt{0.2}/\mf a$ is in the main text.
  The colorbars are different for each figure and not linearly scaled. Note that thanks to anti-detailed-balance,
  angular dependences of $\mf W^{\ominus,++}_{n\mb k n'\mb k'},\mf W^{\ominus,+-}_{n\mb k n'\mb k'}$
  are identical to those of $\mf W^{\ominus,--}_{n\mb k n'\mb k'},\mf W^{\ominus,-+}_{n\mb k n'\mb k'}$,
  respectively, for an isotropic phonon dispersion. }
  \label{fig:suppfig5}
\end{figure}

\end{widetext}

\end{document}